\documentclass[11pt]{article}

\usepackage{graphicx}
\usepackage{amssymb}

\usepackage[margin=2.5cm]{geometry}
\usepackage{graphicx}
\usepackage{dcolumn}
\usepackage{bm}
\usepackage{braket}
\usepackage{epstopdf}
\usepackage{mwe}    
\usepackage[T1]{fontenc}
\usepackage{caption}
\usepackage{subcaption}
\usepackage{amssymb}
\usepackage{amsmath}
\usepackage{booktabs}
\usepackage{tabularx}
\usepackage{multirow}
\usepackage{floatrow}
\usepackage{mathtools}
\usepackage{empheq}
\usepackage[most]{tcolorbox}
\usepackage{xcolor}
\usepackage{wrapfig}
\usepackage{enumitem}

\DeclareMathOperator{\sech}{sech}
\newcommand{\tensor}[1]{\bm{\mathsf{#1}}} 
\newcommand{\ms}{\scriptscriptstyle}

\newcommand{\subscript}[1]{_{\ms #1}}
\newcommand{\eqp}{eq\prime}

\newcommand{\Kps}[1]{k^{'}\subscript{#1}}

\newcommand{\widefbox}[1]{\fbox{\hspace{2em}#1\hspace{2em}}}

\definecolor{olivedrab}{rgb}{0.42,0.56,0.14}
\definecolor{oxfordblue}{rgb}{0.0, 0.13, 0.28}

\newcommand{\Sipstar}[1]{\expandafter\hat#1}

\newcommand{\sss}[1]{\scriptscriptstyle{#1}}
\newcommand{\sbs}[1]{_{\sss{#1}}}
\newcommand{\sps}[1]{^{\sss{#1}}}

\providecommand{\keywords}[1]{%
  \textbf{\textit{Keywords---}} #1
}

\title{\vspace{-2.0cm}Lattice Boltzmann Methods for Navier-Stokes Equations in General Orthogonal Coordinates for Efficient Flow Simulations using Nonuniform Clustered Grids}
\author{{Eman Yahia, Kannan Premnath\footnotemark}, William Schupbach\\Department of Mechanical Engineering\\ College of Engineering, Design and Computing\\University of Colorado Denver\\1200 Larimer Street, Denver, CO 80204, U.S.A.}

\begin{document}

\maketitle

\begin{abstract}
\noindent
Resolving fluid flow structures with multiscale features or boundary layers effectively requires the use of spatially nonuniform meshes with localized clustering of grids. The lattice Boltzmann (LB) method, which is based on models from kinetic theory and involves the use of collide-and-stream steps of the particle distribution functions, is a modern computational technique for fluid dynamics. However, its standard formulation restricts it to the use of uniform Cartesian grids that involve high computational resource requirements in simulating such flows. In this paper, we present new and improved formulations of LB methods that accommodate the use of continuously varying spatial grids via coordinate transformations to simulate the Navier-Stokes equations (NSE) in the general orthogonal curvilinear coordinates (GOC). They are constructed using a Chapman-Enskog analysis to specify the equilibrium moments of the distribution functions and the geometric force terms used in the collision step to be dependent on the local metric factors and their spatial derivatives, which parameterize the local variations in grids as a tensor, along with the density, pressure, momentum and their fluxes, and some correction terms related to the normal velocity gradients so as to represent the conservative form of the NSE in the GOC in a consistent manner. The latter enables recovering the normal viscous stress tensor accurately while the shear stress components emerge naturally from the formulations through appropriate choices for the third order moment equilibria. The resulting LB algorithm using the GOC, and referred to as the GOC-LBM, importantly maintains the simplicity of the collide-and-stream approach in the computational domain while allowing grid clustering/stretching in the physical domain, and is Galilean invariant that is free of the cubic velocity artifacts. Our approach for the GOC-LBM is general and modular in that it can be used with any collision model with appropriate modifications to equilibria and forcing terms. We present the implementation details of the GOC-LBM using a variety of collision models while the central moments-based model using multiple relaxation times was found to be the most robust in practical implementations. We validate the accuracy of the GOC-LBM through numerical simulations for various canonical benchmark flow problems, including those with body-fitted curvilinear grids. Moreover, we demonstrate significant computational advantages of our approach over the standard uniform grids-based LBM for a case study on simulating boundary layer flows efficiently that involves coupling the GOC-LBM for the NSE with another newly constructed GOC-LB scheme for solving the magnetic induction equation for magnetohydrodynamics (MHD) that involves thin Hartmann layers near walls.
\end{abstract}


\let\thefootnote\relax\footnote{*Corresponding author (Email: Kannan.Premnath@ucdenver.edu)}

\newpage

\section{Introduction} \label{sec:1}
Computational fluid dynamics (CFD) methods have traditionally been based on the direct discretization of the Navier-Stokes equations (NSE). By contrast, the lattice Boltzmann (LB) methods, which represent fluid dynamics as an emergent behavior, evolved as improvements to the lattice gas cellular automata (LCGA)~\cite{mcnamara1988use,higuera1989boltzmann,qian1992lattice,benzi1992lattice} and with their connection to the Boltzmann equation~\cite{he1997theory}, they have undergone a series of refinements and applications to a diverse range of complex flows, including multiphase and multicomponent flows, magnetohydrodynamics, and turbulent flows~\cite{chen1998lattice,lallemand2021lattice}. The LBM essentially involves computing the spatial and temporal variations of the distributions of the particle populations caused by their collisions and advection on a lattice which is designed based on symmetry and isotropy requirements of the emergent flow physics. The effect of collisions is often modeled as a relaxation of the distributions to its equilibria~\cite{qian1992lattice}, or in terms of relaxations of their raw moments~\cite{d2002multiple}, or their central moments~\cite{geier2006cascaded}, or their cumulants~\cite{geier2015cumulant}, with the construction of equilibria typically informed by the Maxwell distribution function. The relaxations can be performed using a single relaxation time or via multiple relaxation times. Moreover, the collision step of the method can also be made to comply with the H-theorem~\cite{hosseini2023entropic}.  In addition, more recently, from a different perspective, the drift-diffusion nature of the equilibrating process as exemplified by the Fokker-Planck kinetic model and augmented with a renormalization principle has been introduced as a robust approach to represent the effect of collisions in terms of its central moments~\cite{schupbach2024fokker}. With its underpinnings from kinetic theory that provides opportunities for the use of a variety of mesoscopic kinetic models and its simple Lagrangian nature of the collide-and-stream steps arising from its LCGA origins, the resulting LB algorithms naturally scale well on modern parallel computer architectures and have attracted much interest in tackling many different computational flow physics and engineering problems.

Despite these promising developments, some challenges remain with the use of the LBM. In particular, the LBM uses a uniform square lattice (or a cubic lattice in three-dimnensions), which can result in significant inefficiencies in resolving boundary or shear layer flows with their structures varying across multiple spatial scales. Naturally, it is highly desirable to have an ability to use spatially nonuniform grids that can conform with the local changes in the fluid flow. In order to address this issue, efforts from different directions have been made. On the one hand, with the realization that the particle velocity space discretization (i.e., the lattice) can be decoupled from the spatial discretization (i.e., the grid)~\cite{cao1997physical}, modified LB methods involving the use of interpolations~\cite{he1996some,lu2002large} and the use of the conventional discretization procedures such as the finite difference (e.g.,~\cite{reider1995accuracy}), finite volume (e.g.,~\cite{peng1998lattice}), and finite element methods (e.g.,~\cite{li2004least}) with attendant grid strategies were introduced. However, analysis of such methods have revealed the introduction of additional numerical dissipation~\cite{lallemand2000theory} and increased complexity, leading to a compromise in the accuracy along with incurring additional computational overhead of the resulting LB schemes. From a different standpoint, a hierarchical arrangement of Cartesian square/cubic grids of different sizes (i.e., multiblock grids) with attendant procedures for inter-grid transfer were introduced to tackle grid refinement in LBM (see e.g.,~\cite{filippova1998grid,yu2002multi,rohde2006generic}). While this is a useful strategy, it inherently uses the same grid size within each block, which changes only across different blocks and hence lacks generality in that it does not allow continuous and arbitrary spatial variations in grids, which are natural for more efficiently resolving thin boundary layer flows.

On the other hand, as a step towards using more general forms of lattices beyond the square (or cubic) lattice sets, LB schemes based on rectangular lattices were more recently formulated using different collision models, such as those based on the single relaxation times~\cite{hegele2013rectangular}, multiple relaxation times using raw moments~\cite{bouzidi2001lattice,zhou2012mrt,zong2016designing,peng2016hydrodynamically}, as well as those based on the more robust central moments in our previous work~\cite{yahia2021central}, which was later extended to three-dimensions using cuboid lattices~\cite{yahia2021three} as well as to achieve convergence acceleration via preconditioning~\cite{yahia2022preconditioned}. These models were constructed to recover the NSE and are suitable for representing geometrically inhomogeneous and anisotropic flows. While they are usually based on the collide-and-stream steps like in the standard LBM, they lack an ability to vary the grid sizes nonuniformly in space and hence have limited applicability for simulating general class of multiscale/boundary layer flows effectively, which, as discussed below, will be addressed in this work via using transformations based on general orthogonal coordinates. Here, we note that, as a way of accommodating body-fitted grids for flow simulations in curved geometries, the  use of curvilinear coordinates in LBM have been investigated in some prior work (see~\cite{mei1998finite,guo2003explicit,budinski2014mrt,hejranfar2014implementation,mendoza2014lattice,busuioc2019lattice,barraza2020towards}). However, these existing strategies inherit certain significant limitations: they are often formulated based on the finite-difference approximations of the advection operator of the discrete velocity Boltzmann equation that introduce additional numerical dissipation while not being able to perform the standard collision step followed by the lock-step streaming on lattices or involve cumbersome numerical implementations or are limited to the use of certain specific collision models.
Previously, a collide-stream-based LBM in curvilinear coordinates was constructed for the solution of the convection-diffusion equation of a scalar field~\cite{yoshida2014lattice}. More recently, a one-to-one advection-based volumetric LBM for the Navier-Stokes equations (NSE) in curvilinear coordinates was derived in~\cite{chen2021volumetric} and then implemented and tested numerically for some benchmarks in~\cite{chekhlov2023lattice}. Our present work differs from them~\cite{chen2021volumetric,chekhlov2023lattice} in a number of ways, which will be discussed next before outlining the main contributions of this paper.

First, via a moments-based Chapman-Enskog analysis, we develop and validate several different variants of the LBMs with the collide-and-stream algorithmic structure in the general orthogonal coordinates (GOC) for a more general form of the NSE expressed in a conservative form with adjustable weakly compressible effects involving two transport coefficients, viz., shear and bulk viscosities, whereas the derivation given in~\cite{chen2021volumetric} is restricted to the NSE with the shear viscosity only. Second, the previous formulation~\cite{chen2021volumetric} is limited to the single-relaxation-time (SRT) collision model, while our formulation, from the outset, is constructed to be more general and applicable to any collision model, and we have developed GOC-based LBM for a variety of collision models, including the SRT, as well as more robust raw moments-based as well as central moments-based collision models based on multiple relaxation times (MRT), with the latter implemented in a moving frame with the local fluid velocity; moreover, our approach of representing the diagonal and off-diagonal components of the viscous stress tensor for the NSE in the GOC is distinct from that derived in~\cite{chen2021volumetric}. Third, the use of the standard D2Q9 lattice results in cubic non-Galilean invariant errors in the diagonal viscous stress components due to aliasing effects, which we eliminate via corrections for the equilibrium moments in the GOC and also confirm numerically, whereas the GI-related issues on such standard lattices were not analyzed or discussed in~\cite{chen2021volumetric,chekhlov2023lattice}; moreover, unlike in~\cite{chekhlov2023lattice} we provide a consistent parametrization of the speed of sound in terms of the GOC-based metric factors via the Courant-Friedrichs-Lewy (CFL) condition. This along with our distinguishing developments for the above three factors contributes to making more robust LBM in the GOC with good stability properties, especially with the use of more advanced collision models. Then, fourth, we enforce the orthogonality of the curvilinear coordinates system, whereas the previous work~\cite{chen2021volumetric} considers curvilinear meshes without requiring them to be orthogonal; however, such a generality associated with the latter~\cite{chen2021volumetric} comes at the cost of needing to use extended lattice sets, such as the D2Q21 lattice, to properly represent the viscous stress and correctly reproduce the analytical solution of even a simple circular Taylor-Couette flow since the use of the D2Q9 lattice using the scheme given in~\cite{chen2021volumetric} is unable to converge to the associated analytical solution~\cite{chekhlov2023lattice}; on the other hand, importantly, our GOC-based LBM is developed and implemented on the standard D2Q9 lattice without requiring extended lattice sets and is thus relatively easier to implement and more efficient for simulation of flows using clustered/curvilinear orthogonal meshes, and will be shown to accurately reproduce such canonical flows as well as more complex and fully 2D curvilinear flows consistent with other numerical results and experimental data. Fifth, all the flow test cases in~\cite{chekhlov2023lattice} for validating the previous derivation~\cite{chen2021volumetric} were restricted to effectively 1D steady flows with analytical solutions, whereas apart from considering those, we also assess and validate our formulation for an unsteady case as well as more challenging 2D shear flows with corner singularities (such as square cavity flow as well as polar cavity flow) using clustered/curvilinear grids, which reveal the necessity of using advanced collision models based on central moments and MRT in the GOC for cases at significantly higher Reynolds numbers than that is possible with the SRT-based GOC-LB scheme. Finally, sixth, we validate our GOC-LBM for the NSE by going further beyond the previous work~\cite{chen2021volumetric,chekhlov2023lattice} and demonstrate its clear advantages over the standard Cartesian uniform-grid based LBM in terms of savings in the number of grid points for resolving truly thin boundary layer flows in a multiphysics context such as those arising in magnetohydrodynamics by coupling it with another GOC-LB scheme for the magnetic induction equation. 

In light of the above considerations, we now summarize the specific approaches taken and the new contributions made in this paper as follows. We construct new LB schemes for the solution of the 2D NSE based on the GOC. The choice of the GOC (see e.g.,~\cite{morse1946methods,hildebrand2012methods,malvern1969introduction,fitzpatrick2017theoretical,yousefi2020boundary}) enables the use of nonuniform clustered grids with continuous spatial variations achieved via algebraic stretching functions or numerical grid generators (see e.g.,~\cite{vinokur1983one,thompson1998handbook,liseikin1999grid,anderson2020computational}) for resolving the flow physics in the physical domain via mapping it onto a uniform computational domain where the collide-and-stream based steps are executed. In addition, if need be, body-fitted orthogonal grids for curved geometries can be naturally incorporated in this framework. Such a formulation enjoys the advantages of the flexibility generally associated with the conventional CFD methods while maintaining the simplicity of the Lagrangian structure of the collide-and-stream-based LBM and preserving its natural parallelization capabilities. The design of a GOC-LBM is based on a top-down approach via a Chapman-Enskog analysis. By using the standard two-dimensional, nine velocity (D2Q9) lattice, this involves a careful construction of the moment equilibria as well as the geometric forcing terms dependent on the local metric factors of the GOC as well as their derivatives parameterized as a tensor that specifies spatial variations in grids (i.e., their clustering/stretching) and their curvatures, along with the flow variables such as the fluid density, velocity, and inviscid and viscous momentum fluxes. In particular, the off-diagonal components of the viscous stresses are represented exactly and locally by a careful choice of the third order equilibrium moments depending on the metric factors and fluid velocities and densities, while certain corrections to the second order moment equilibria that include the normal velocity derivatives are introduced to recover the NSE in the GOC~\cite{yousefi2020boundary} expressed in a conservative form (see e.g.~\cite{vinokur1974conservation}). The GOC-LBM uses a speed of sound that is parameterized by the metric factors so as to satisfy the CFL condition~\cite{courant1967partial} so that it consistently resolves flow physics in time-explicit computations. With its attendant correction terms, the GOC-LBM is constructed to be Galilean invariant at least up to the third order in fluid velocities for the case of general orthogonal clustered/stretched grids, including curvilinear effects, which generalizes such corrections to the standard LB formulations based on uniform square grids prescribed in previous investigations (see e.g.,~\cite{dellar2014lattice,hajabdollahi2018galilean}). As such, when the metric factors are unity everywhere, the GOC-LBM reduces to the standard LBM for uniform grids as a special case. Moreover, importantly, our GOC-based LB approach is modular and general in that it can be used with any collision model. While the analysis and the construction procedure is based on the use of raw moments with a multiple relaxation times (MRT) model to highlight the main ideas involved, its extensions to the more robust central moments-based collision model as well as the simplest case involving the single relaxation time (SRT) model are also presented. The implementations of the GOC-LBM are numerically assessed and validated for accuracy against a variety of standard benchmark flow problems to demonstrate its ability to resolve small-scale flow structures effectively and its ability to maintain Galilean invariance in numerical simulations; moreover, we also show the capability of the GOC-LBM to simulate flows in a curvilinear grids by considering two benchmark case studies. Furthermore, we demonstrate the significant advantages of using the GOC-LBM when compared to the standard uniform grid-based LBM for computing thin boundary-layer flows. In this regard, by extending a prior work based on uniform grids~\cite{dellar2002lattice}, we construct a new GOC-LBM for solving the magnetic induction equation which is then coupled with the GOC-LBM for the NSE to efficiently simulate wall-bounded magnetohydrodynamics (MHD) involving thin Hartmann layers.

This paper is organized as follows. Section~\ref{sec:NSE} presents the coordinate transformations that map from the Cartesian coordinates to the general orthogonal curvilinear coordinates (GOC) system based on the metric factors as free parameters along with some useful vector and tensor identities to recast the Navier-Stokes equations (NSE) written in a symbolic form (i.e., the Gibbs notation) into the GOC-based NSE in the index notation. The resulting fluid dynamical equations in the GOC are further transformed into a conservative form with compressible effects that is suitable for developing new LB formulations in the GOC in subsequent sections. Then, Secs.~\ref{sec:MRT-LBM} and~\ref{sec:GOC-LBM} form the crux of the mathematical development of the GOC-LBE that is expressed in terms of the raw moments in a multiple relaxation times (MRT) framework for the D2Q9 lattice via a Chapman-Enskog (C-E) analysis. We have chosen the use of raw moments in this regard as it serves as a natural setting to express the main ideas involved in formulating the mesoscopic kinetic equations in the GOC. In particular, Sec.~\ref{sec:MRT-LBM} prescribes the choices for the raw moment equilibria and source terms given in terms of fluid density and velocities and parameterized by the metric factors to recover the conserved moments as well as the inviscid momentum fluxes appearing in the conservative form of the NSE in the GOC and identifies the remaining viscous flux errors using the C-E analysis. This is followed by extending this C-E analysis to eliminate such remaining errors in the viscous fluxes via the introduction of certain moment equilibria corrections in order to consistently recover the NSE in the GOC in Sec.~\ref{sec:GOC-LBM}. A matrix formulation of the resulting GOC-LBM using the raw moments is presented in Sec.~\ref{sec:Reformulation GOC-LB} with its implementation details further elaborated in an appendix (see below). Then, Sec.~\ref{sec:Transformations} discusses some of the most common coordinate transformations and orthogonal grid stretching functions for grid clustering useful in implementations. A detailed numerical study of the GOC-LBM for accuracy validations against various benchmark flow problems, including a numerical test for Galilean invariance, as well as for showing its advantages over the uniform grid-based standard LBM for more effectively resolving thin boundary layers is presented in Sec.~\ref{sec:Results}. Furthermore, demonstration case studies of the use of GOC-LBM for simulating flows using body-fitted grids around curved geometries via two classical benchmark problems are presented in Sec.~\ref{sec: curvedbenchmark}. Finally, the key developments and conclusions are summarized in Sec.~\ref{sec: summary}.

Further attendant details of the GOC-LB algorithms, including their various extensions, are given in various appendices (see Appendices~\ref{sec:Appendix_shear_stress_local} through~\ref{sec:Appendix_E}). Specifically, Appendix~\ref{sec:Appendix_shear_stress_local} presents a brief derivation of the shear stress in the GOC-LBM in terms of a non-equilibrium moment, which is useful for its local computation in numerical implementations. Then, Appendix~\ref{sec:Appendix_B} discusses some of the simplified forms of the coefficients of the corrections terms and the effective body force associated with the GOC-LBM under various limiting special cases. Implementation details of initial and boundary conditions in the GOC-LBM are provided in Appendix~\ref{sec:initial_boundary_conditions}. This is followed by a summary of the algorithmic details of the raw moment-based GOC-LBM using MRT in Appendix~\ref{sec: Appendix_A}. Then, extending this approach to other collision models, Appendices~\ref{sec:Appendix_C} and~\ref{sec:Appendix_D} present the construction procedures and implementation details of the GOC-LBM based on central moments and MRT and the GOC-LBM based on the velocity space and SRT, respectively. Finally, in Appendix~\ref{sec:Appendix_E}, a derivation and implementation details of another new GOC-LBM for solving the magnetic induction equation for magnetohydrodynamics are given.

%
%
%


\section{Navier-Stokes Equations in General Orthogonal Curvilinear Coordinates(GOC): Conservative Form with Compressible Effects}\label{sec:NSE}
In this work, the use of coordinate transformations based on the general orthogonal curvilinear coordinates (GOC) forms the basis for introducing nonuniform and continuously variable grids as well as  body-fitted curvilinear grids in LBM. In this regard, we need to first transform an appropriate version of the governing equations of the fluid motion in such a coordinates system. Originally, the GOC was introduced by Lam{\'e}~\cite{lame1859leccons}, and it has since been widely used to express the partial differential equations representing the physical laws such as those pertaining to continuous mechanics including fluid dynamics (e.g.,~\cite{morse1946methods,hildebrand2012methods,black1968conservation,malvern1969introduction,fitzpatrick2017theoretical,yousefi2020boundary}).
It has been utilized in various numerical methods such as the finite-difference methods for the solution of the Navier-Stokes equations (NSE) (see e.g.,~\cite{hung1977implicit,raithby1986prediction,nikitin2006finite}). As such, the use of GOC offers some distinct advantages over the generalized non-orthogonal curvilinear coordinates (GNOC). In particular, as noted more recently in~\cite{yousefi2020boundary}, the governing equations in GOC preserve much of the simplicity and interpretability of their corresponding versions in the Cartesian coordinates system whereas the GNOC gives rise to significantly more complicated forms of the governing equations and requiring to distinguish between covariant and contravariant vectors and tensor fields. As a result, the GOC provides a natural setting to develop a GOC-LBM that, as shown later in this work, is able to maintain the simplicity of the algorithmic structure associated with the standard LBM for uniform grids, viz., the collide-then-stream steps.

\subsection{Incompressible Navier-Stokes Equations in GOC}
Since the regime of application of the LBM is in the incompressible limit of the fluid motion, we start with the incompressible mass and momentum equations, i.e., the NSE, which in the coordinates-invariant symbolic form (i.e., the Gibbs notation) reads as
\begin{subequations}\label{eq:NSE}
\begin{eqnarray}\label{eq:conservativemass_vector}
& \bm{\nabla}\cdot \bm{u} = 0,
\end{eqnarray}
\begin{eqnarray}\label{eq:momentum_vector}
&\rho \left(\partial_t \bm{u}+ \bm{u}\cdot \bm{\nabla}\bm{u}\right)= -\bm{\nabla}P + \bm{\nabla}\cdot \tensor{\tau}+ \bm{F}_{ext},
\end{eqnarray}
\end{subequations}
Here, $\rho$ is the fluid density, $\bm{u}$ is the fluid velocity, $P$ is the pressure, $\tensor{\tau}$ is the viscous stress tensor, and $\bm{F}_{ext}$ is an external body force. To develop the main ideas, algorithms and to study the numerical features of the resulting GOC-LBM, we will restrict our focus to two-dimensional flows in this work, while its extension to three-dimensions will be discussed in a follow-up work. Moreover, we limit ourselves here to the athermal case with uniform temperature and considerations of energy transport will be reported in a future study.

Let the Cartesian coordinates represented by $\bm{x}_i=(x_1, x_2)$ be mapped to the general orthogonal curvilinear coordinates (GOC) system $\bm{\xi}_i=(\xi_1, \xi_2)$. If $\hat{\bm{x}}_k = (\hat{\bm{x}}_1, \hat{\bm{x}}_2)$ are the \emph{orthonormal} basis vectors of the Cartesian coordinates, then the \emph{orthogonal} basis vectors of the GOC induced by such a mapping can be expressed as (see e.g.,~\cite{morse1946methods,hildebrand2012methods})
\begin{equation}
\bm{\xi}_i= \frac{\partial x_k}{\partial \xi_i} \hat{\bm{x}}_k, \quad i = 1, 2,
\end{equation}
where the usual convention of the summation of the repeated indices is assumed, i.e., here $k = 1, 2$ and the associated terms are summed for every $i$. Then, the \emph{orthonormal} basis vectors of the GOC, designated by $\bm{\hat{\xi}}_i$ is obtained from $\bm{\xi}_i$ as follows (see e.g.,~\cite{morse1946methods,hildebrand2012methods}):
\begin{equation}
\bm{\hat{\xi}}_i= \frac{ \bm{\xi}_i}{h_i}.
\end{equation}
Here, $h_i$ are known variously as the metric factors or the scale factors or the Lam{\'e} coefficients of the GOC and they are obtained from the spatial variations in the coordinates mapping relation $x_i=x_i(\xi_1,\xi_2)$ for $i = 1, 2$ via
\begin{equation}\label{eq:metricfactor}
  h_1 = \sqrt{\left(\frac{\partial x_1}{\partial \xi_1}\right)^2+ \left(\frac{\partial x_2}{\partial \xi_1}\right)^2}, \quad\quad
  h_2 = \sqrt{\left(\frac{\partial x_1}{\partial \xi_2}\right)^2+ \left(\frac{\partial x_2}{\partial \xi_2}\right)^2}.
\end{equation}

To express the incompressible NSE in Eq.~(\ref{eq:NSE}) in the GOC, we require the use of the differential vector operators involving the gradient, divergence, curl, and the Laplacian in terms of the GOC parameterized by the scale factors. In this regard, to achieve a more compact and elegant expressions of the resulting identities (which significantly avoids the use of cumbersome expressions given in previous works), we adopt a new notation recently proposed by~\cite{yousefi2020boundary}, which modifies the Einstein summation convention: \emph{No summation is performed whenever any index is enclosed within parentheses}, and that index can only take the value of the associated dummy or free indices appearing in the concerned expression; all other usual implied summation convention of the repeated indices is preserved; that is, the dummy indices, viz., those indices repeated twice in any term are to be summed over the range of its values, and the free indices, viz., those that are not repeated, are not summed and can take on any value within its range. Thus, for the case with a free index $i$ together with a term involving a parenthesized index $(i)$, we interpret the following expression as $u_i u_j \tensor{\tau}_{(i)j} = u_1 u_1 \tensor{\tau}_{11} + u_1 u_2 \tensor{\tau}_{12}$ when $i=1$ (and similarly a separate expression with $i=2$), and for another case with a dummy index $j$ together with a term involving a parenthesized index $(j)$, we have $u_j u_j \tensor{\tau} _{(j)i} = u_1 u_1 \tensor{\tau}_{1i} + u_2 u_2 \tensor{\tau}_{2i}$. Then, we can express the following differential operator identities in the GOC for any arbitrary scalar $s$ and vector $\bm{V}$ as~\cite{yousefi2020boundary}:
\begin{subequations}\label{eq:Identity}
\begin{eqnarray}
\bm{\nabla} s &=& \frac{1}{h_{(i)} }\frac{\partial s}{\partial {\xi_i}} \hat{\xi_i}, \qquad\qquad\qquad\qquad
\bm{\nabla}\cdot\bm{V} = \frac{1}{h}\frac{\partial}{\partial {\xi_i}} \left(\frac{h}{h_{(i)}}v_i\right) \label{eq:Identity2}\\
\bm{\nabla} \times \bm{V} &=& \varepsilon_{ijk} \frac{h_{(i)}}{h} \frac{\partial}{\partial {\xi_j}} \left(h_{(k)} v_k \right) \hat{\bm{\xi}}_i, \quad\quad \bm{\nabla}^2 s = \frac{1}{h}\frac{\partial}{\partial {\xi_i}} \left(\frac{h}{h_{(i)}h_{(i)}} \frac{\partial s}{\partial {\xi_i}}\right).\label{eq:Identity4}
\end{eqnarray}
\end{subequations}
Here, $h=h_1 h_2$ and $\left(\bm{V}\right)_i= v_i$.

With the above considerations and introducing a curvature coefficient matrix $\tensor{\theta}$ representing the local curvature of the GOC that essentially accounts for nonuniform spatial variations in the grids and/or body-fitted orthogonal curvilinear grids in the context of GOC-LBM developed in the subsequent sections, by defining its elements as
\begin{equation}\label{eq:curvaturematrix}
\theta_{ij} = \frac{1}{h_i h_j} \frac{\partial h_i}{\partial \xi_j},
\end{equation}
or equivalently in the matrix form as
\begin{equation}\label{eq:4}
\tensor{\theta}=
  \begin{bmatrix}
  \theta_{11} &  \theta_{12} \\
   \theta_{21} &  \theta_{22}\\
  \end{bmatrix}
  =
   \begin{bmatrix}
  \cfrac{1}{h_1^2}\cfrac{\partial h_1}{\partial\xi_1} &  \cfrac{1}{h_1 h_2}\cfrac{\partial h_1}{\partial\xi_2} \\
   \cfrac{1}{h_1 h_2}\cfrac{\partial h_2}{\partial\xi_1}  &  \cfrac{1}{h_2^2}\cfrac{\partial h_2}{\partial\xi_2},\\
  \end{bmatrix}.
\end{equation}
we also make a note of the following tensorial identities (which will play a crucial role in transforming the nonlinear advection operator and as well as viscous flux terms in the NSE)~\cite{yousefi2020boundary}:
\begin{eqnarray}
  \left(\bm{A} \cdot \bm{\nabla}\right) \bm{B} &=& \left[ \frac{a_j}{h_{(j)}}\frac{\partial b_i}{\partial \xi_j} + \left( a_i b_j \theta_{(i)j} - a_j b_j \theta_{(j)i}   \right) \right] \hat{\bm{\xi}}_i \label{eq:vectoridentity}\\
  \bm{\nabla}\cdot \tensor{T} &=& \left[ \frac{1}{h}\frac{\partial}{\partial \xi_j}\left(  \frac{h}{h_{(j)}} \tensor{T}_{ij}\right) + \left( T_{ij} \theta_{(i)j} - T_{jj} \theta_{(j)i}   \right) \right] \hat{\bm{\xi}}_i,\label{eq:vectoridentityB}
\end{eqnarray}
where the components of the tensor $\tensor{T}$ and the vectors $\bm{A}$ and $\bm{B}$ are given by $\left(\tensor{T}\right)_{ij}= T_{ij}$, and $\left(\bm{A}\right)_i = a_i$ and $\left(\bm{B}\right)_i = b_i$, respectively.
Then, for a velocity vector field $\bm{u}$ with components $(u_x, u_y)$ in the Cartesian coordinates and with components $(U_1, U_2)$ in the GOC, i.e., $\bm{u}=u_x\hat{\bm{e}}_1+u_y\hat{\bm{e}}_2= U_1\hat{\bm{\xi}}_1+U_2\hat{\bm{\xi}}_2$, by using the above identities, and with the viscous stress tensor $\tensor{\tau}=2\mu\tensor{S}$, where $\mu$ and $\tensor{S}$ respectively being the dynamic viscosity and the strain rate tensor ($\tensor{S}=(\bm{\nabla}\bm{u}+(\bm{\nabla}\bm{u})^\dag)/2$), the incompressible NSE can be rewritten in terms of the GOC as~\cite{yousefi2020boundary}
\begin{subequations}\label{eq:20}
   \begin{eqnarray}
     \frac{1}{h} \frac{\partial}{\partial \xi_i} \left(\frac{h}{h_{(i)}}U_i\right)&=& 0 \\
     \frac{\partial U_i}{\partial t}+ \frac{U_j}{h_{(j)}} \frac{\partial U_i}{\partial \xi_j}+ \left( U_i U_j \theta_{(i)j} - U_j U_j \theta_{(j)i}\right ) &=& - \frac{1}{\rho} \frac{1}{h_{(i)}} \frac{\partial P}{\partial \xi_i} + \frac{F_{ext, i}}{\rho}+ \\\nonumber
     && \frac{1}{\rho} \left[  \frac{1}{h} \frac{\partial}{\partial \xi_j} \left(\frac{h}{h_{(j)}} \tau_{ij}\right) + \left(\tau_{ij} \theta_{(i)j} -  \tau_{jj} \theta_{(j)i}\right) \right],
   \end{eqnarray}
\end{subequations}
where the constitutive relation now reads as $\tau_{ij}=2\mu S_{ij}$ with  $S_{ij}= \frac{1}{2}\Big[ \frac{1}{h_{(j)}} \frac{\partial U_i}{\partial \xi_j}+ \frac{1}{h_{(i)}} \frac{\partial U_j}{\partial \xi_i}- (U_{(i)}\theta_{ij} + U_{(j)}\theta_{ji}) + 2 U_m \theta_{(i)m} \delta_{ij} \Big]$. Moreover, the components of the body force in the GOC are related to the respective components in the Cartesian coordinates via $\bm{F}_{ext}=F_{ext, x}\hat{\bm{x}}_1+F_{ext, y}\hat{\bm{x}}_2= F_{ext, 1}\hat{\bm{\xi}}_1+F_{ext, 2}\hat{\bm{\xi}}_2$ and accordingly used above in Eq.~(\ref{eq:20}). These equations in the GOC system are broadly similar to those in the conventional Cartesian coordinate system except for the additional terms involving the curvature coefficient matrix components that, in fact, account for the curvature of the GOC and arise due to the spatial dependence of the base vectors. Note that when $h_i =1$ everywhere, the standard incompressible NSE in the Cartesian coordinates are recovered.

\subsection{Extension to Conservative Form of NSE with Compressible Effects}
While the above governing equations in Eq.~(\ref{eq:20}) are generally adequate for developing the conventional CFD methods such as those based on finite-differences, additional considerations are necessary for constructing corresponding schemes based on the LBM. This is because (i) owing to the conserved collision invariants of density and momentum in the LBM, their emergent macroscopic fluid dynamical equations naturally appear in their conservative forms via appropriate moments of the distribution function, and this is consistent with the notions proposed in a different context earlier~\cite{vinokur1974conservation}, and (ii) the LBM, owing to its kinetic nature, is inherently weakly-compressible and the bulk viscosity effects naturally arise which is usually independently adjusted to enhance its stability. Incorporating these two additional features, we then introduce our target NSE in the conservative form with compressible effects by replacing Eq.~(\ref{eq:NSE}) as follows:
\begin{subequations}
\begin{eqnarray}\label{eq:conservativemass}
 \frac{\partial \rho}{\partial t}+\bm{\nabla}\cdot (\rho \bm{u}) &=& 0, \\
\frac{\partial}{\partial t}( \rho \bm{u})+\bm{\nabla}\cdot (\rho \bm{u}\bm{u}) &=& -\bm{\nabla}P + \bm{\nabla}\cdot\tensor{\tau} + \bm{F}_{ext},\label{eq:conservativemomentum}
\end{eqnarray}
\end{subequations}
where the constitutive relation between the viscous stress tensor and the strain rate is now modified as
\begin{equation}\label{eq:viscousstress_bulkv}
  \tensor{\tau}=2 \rho \nu \tensor{S} - \frac{2}{D} \rho \nu (\bm{\nabla}\cdot \bm{u}) \tensor{I} + \rho \zeta (\bm{\nabla}\cdot \bm{u})\tensor{I}.
\end{equation}
Here, $\zeta$ is the bulk viscosity that is independently adjustable from the shear viscosity $\nu$, $\tensor{I}$ is the identity matrix of dimension $2\times 2$, and $D$ is the number of spatial dimensions ($D=2$ in this paper). Then, based on the various differential operator identities given in the previous subsection, we can transform the continuity (Eq.~(\ref{eq:conservativemass}) and the momentum equations (Eq.~(\ref{eq:conservativemomentum})) into the corresponding forms in the GOC, which then read as
\begin{equation}\label{eq:conservativemass_OCC}
 \frac{\partial \rho}{\partial t}+\frac{1}{h}\frac{\partial}{\partial \xi_i}\left( \frac{h}{h_{{(i)}}} \rho U_i\right) = 0,
\end{equation}
and
\begin{eqnarray}\label{eq:conservativemomentum_OCC}
\frac{\partial \rho U_i}{\partial t}+\frac{1}{h}\frac{\partial}{\partial \xi_j}\left( \frac{h}{h_{(j)}} \rho U_i U_j\right)+\left( \rho U_i U_j \theta_{(i)j} - \rho U_j U_j \theta_{(j)i} \right) &=& -\frac{1}{h_{(i)}} \frac{\partial P}{\partial \xi_i} + F_{ext,i}+\\
&&\frac{1}{h} \frac{\partial}{\partial \xi_j} \left(\frac{h}{h_{(j)}} \tau_{ij}\right) + (\tau_{ij} \theta_{(i)j} -  \tau_{jj} \theta_{(j)i}),\nonumber
\end{eqnarray}
where again $h=h_1h_2$. Moreover, the constitutive equation given in Eq.~(\ref{eq:viscousstress_bulkv}) then transforms into the following expression in the GOC:
\begin{equation}\label{eq:viscousstress_OCC}
  \tau_{ij}=2 \rho \nu S_{ij} - \frac{2}{D} \rho \nu (\bm{\nabla}\cdot \bm{U}) \delta_{ij} + \rho \zeta (\bm{\nabla}\cdot \bm{U}) \delta_{ij},
\end{equation}
where the strain rate tensor $S_{ij}$ is given as follows:
\begin{equation}\label{eq:strain_tensor}
S_{ij}= \frac{1}{2}\Big[ \frac{1}{h_{(j)}} \frac{\partial U_i}{\partial \xi_j}+ \frac{1}{h_{(i)}} \frac{\partial U_j}{\partial \xi_i}- (U_{(i)}\theta_{ij} + U_{(j)}\theta_{ji}) + 2 U_m \theta_{(i)m} \delta_{ij} \Big],
\end{equation}
and the divergence of the velocity field $\bm{\nabla}\cdot \bm{U}$ follows from an identity given earlier in Eq.~(\ref{eq:Identity}) as
\begin{equation}\label{eq:divergenceU}
\bm{\nabla}\cdot \bm{U} = \frac{1}{h}\frac{\partial}{\partial \xi_i}\left( \frac{h}{h_{(i)}} U_i \right).
\end{equation}

While the hydrodynamic variables, i.e., the fluid density and momentum as well as the inviscid and viscous momentum fluxes all appear inside the spatial and time derivatives in the coordinates-free symbolic form of the governing equations (see Eqs.~(\ref{eq:conservativemass}) and (\ref{eq:conservativemomentum})), such a conservative structure is disrupted under the GOC transformations as seen in Eqs.~(\ref{eq:conservativemass_OCC}) and (\ref{eq:conservativemomentum_OCC}) with the appearance of the metric factors as coefficients outside the derivatives together with some additional terms dependent on the curvature coefficient matrix components. Fortunately, they can be recast into conservative form by some further rearrangement as shown next.

First, noting that the metric factors are time-invariant, the factor $h=h_1h_2$ appearing as the coefficient to the spatial derivative in the second term of the continuity equation in Eq.~(\ref{eq:conservativemass_OCC}) can be moved inside its time derivative so that we obtain it in a desired form.  Moreover, by evaluating the spatial derivatives of the mass flux components $(h/h_{(i)})\rho U_i$, we can rewrite Eq.~(\ref{eq:conservativemass_OCC}) as
\begin{equation}\label{eq:conservativemass_OCC_modified}
\frac{\partial}{\partial t} \left(\rho h_1 h_2\right)+ \frac{\partial}{\partial\xi_1} \left(h_2 \rho U_1\right) + \frac{\partial}{\partial\xi_2} \left(h_1 \rho U_2\right)= 0,
\end{equation}
Next, inspecting this last equation,  we have the mass flux components $h_2\rho U_1$ and $h_1\rho U_2$, along the directions $i=1$ and $i=2$, respectively; these should then effectively appear inside the time derivatives of the respective components of the momentum equation (Eq.~\ref{eq:conservativemomentum_OCC})) in place of $\rho U_1$ and $\rho U_2$, respectively, to partly achieve a conservative structure of the emergent macroscopic equations of the GOC-LBE formulated in the next section via suitable choices of the moments of the distribution function. These can be accomplished by multiplying by $h_2$ the Eq.~(\ref{eq:conservativemomentum_OCC}) for $i=1$ and by $h_1$ the same equation for the case with $i=2$. However, the modified Eq.~(\ref{eq:conservativemomentum_OCC}) resulting from these operations would still contain the metric factors outside the spatial derivatives; such metric factors can then be brought inside the spatial derivative via the product rule of calculus, which would then introduce additional terms involving the spatial derivatives of the metric factors that can be identified in terms of the components of the curvature coefficient matrix given in Eq.~(\ref{eq:curvaturematrix}); such additional terms along with the other extra terms involving the curvature coefficient matrix components appearing in Eq.~(\ref{eq:conservativemomentum_OCC}) can all be lumped together and identified as the components of the geometric body force, which can then be added together with the external body force $F_{ext, i}$ to obtain an effective body force $F_i$. These operations then finally put the momentum equation in GOC in a fully conservative form. For better clarity, we now illustrate these considerations by applying them for Eq.~(\ref{eq:conservativemomentum_OCC}) taking the $i=1$ case as an example. Thus, starting with multiplying Eq.~(\ref{eq:conservativemomentum_OCC}) by $h_2$ and then applying the product rule for rearranging the resulting metric factors-dependent spatial derivative terms associated with the inviscid and viscous momentum flux terms, we get
\begin{align} \label{eq:momentum1}
 &\frac{\partial}{\partial t}(\rho U_1 h_2)+\frac{\partial}{\partial\xi_1}\left( \frac{ h_2}{h_1} \rho U_1^2\right)-\underline{ h_2 \rho U_1^2 \frac{\partial}{\partial\xi_1}\left( \frac{1}{h_1}\right)} + \frac{\partial}{\partial\xi_2}\left( \rho U_1 U_2\right)- \underline{h_1 \rho U_1 U_2 \frac{\partial}{\partial\xi_2}\left( \frac{1}{h_1}\right)} = \nonumber\\
  &- \frac{h_2}{h_1} \frac{\partial P}{\partial \xi_1}  +  \frac{\partial}{\partial \xi_1} \left(\frac{h_2}{h_1} \tau_{11}\right)- \underline{h_2 \tau_{11}\frac{\partial}{\partial \xi_1} \left(\frac{1}{h_1}\right)}
  +  \frac{\partial \tau_{12}}{\partial \xi_2}  - \underline{h_1 \tau_{12} \frac{\partial}{\partial \xi_2} \left(\frac{1}{h_1}\right)} + \nonumber \\
  &h_2 \bigg[(\tau_{1j} \theta_{(1)j} -  \tau_{jj} \theta_{(j)1})- (\rho U_1 U_j\theta_{(1)j} - \rho U_j U_j \theta_{(j)1})\bigg] + h_2 F_{ext,1}.
\end{align}
The spatial derivatives in the underlined terms in this last equation can be identified as components of the curvature coefficient matrix via Eq.~(\ref{eq:curvaturematrix}); moreover, the pressure gradient term in this last equation can also be rewritten in terms of the product rule to put it in a conservative form. That is, we have
\begin{align*}
  & \frac{h_2}{h_1^2}\frac{\partial h_1}{\partial\xi_1}= h_2 \theta_{11}  \quad \quad \quad -\frac{h_2}{h_1}\frac{\partial P}{\partial\xi_1}= -\frac{\partial}{\partial\xi_1}\left( \frac{h_2}{h_1} P\right)+ h_2 \left( \theta_{21}- \theta_{11}\right)P\\
  &\frac{h_1}{h_1^2}\frac{\partial h_1}{\partial\xi_2}= h_2 \theta_{12} \quad \quad \quad \quad   \frac{\partial}{\partial_{\xi_1}}\left(\frac{h_2}{h_1}\right) = h_2 \left( \theta_{21}- \theta_{11}\right).
\end{align*}
Moreover, applying the summation rules of the repeated indices in the additional terms appearing in Eq.~(\ref{eq:momentum1}) and simplifying, we get
\begin{eqnarray*}
  & \left(\tau_{1j} \theta_{(1)j} -  \tau_{jj} \theta_{(j)1}\right)= \tau_{11}\theta_{11} + \tau_{12}\theta_{12} - \tau_{11}\theta_{11}- \tau_{22}\theta_{21} = \tau_{12}\theta_{12} - \tau_{22}\theta_{21}\\
  &\rho \left( U_1 U_j\theta_{(1)j} - U_j U_j \theta_{(j)1}\right) = \rho \left(U_1^2 \theta_{11}+ U_1 U_2 \theta_{12} - U_1^2 \theta_{11}- U_2^2 \theta_{21}\right)= \rho \left(U_1 U_2 \theta_{12}-U_2^2 \theta_{21}\right).
\end{eqnarray*}
Then, lumping together all the extra terms associated with the curvature coefficient matrix components as a component of the `geometric' body force with the external body force $F_{ext, 1}$ and identifying the resulting expression as the effective body force $F_1$ as
\begin{equation}\label{eq:bodyforce1}
   F_1 = h_2 \bigg[F_{ext,1}+ \left( \theta_{21}- \theta_{11}\right)P + \theta_{11}\left(\tau_{11} -  \rho U_1^2\right)+ 2 \theta_{12}\left(\tau_{12} -  \rho U_1 U_2\right)-  \theta_{21}\left(\tau_{22} -  \rho U_2^2\right)\bigg],
\end{equation}
we can finally recast Eq.~(\ref{eq:momentum1}) in a conservative form for the momentum equation along the direction $i=1$, which reads as
\begin{equation}\label{eq:conservmomentum_1}
    \frac{\partial}{\partial t}( h_2\rho U_1)+\frac{\partial}{\partial\xi_1}\left( \frac{ h_2}{h_1} \rho U_1^2\right) + \frac{\partial}{\partial\xi_2}\left( \rho U_1 U_2\right) =- \frac{\partial}{\partial \xi_1}\left(\frac{h_2}{h_1} P\right) +   \frac{\partial}{\partial \xi_1}\left( \frac{h_2}{h_1} \tau_{11}\right)+ \frac{\partial \tau_{12}}{\partial \xi_2} + F_1.
\end{equation}
Along similar lines, we can get the conservative expression for the momentum equation along the direction $i=2$. In this case, the effective body force $F_2$ is given by
\begin{equation}\label{eq:bodyforce2}
   F_2 = h_1 \bigg[F_{ext,2}+ \left( \theta_{12}-  \theta_{22}\right)P + \theta_{22}\left(\tau_{22} -  \rho U_2^2\right)+ 2 \theta_{21}\left(\tau_{21} -  \rho U_1 U_2\right)-  \theta_{12}\left(\tau_{11} -  \rho U_1^2\right)\bigg],
\end{equation}
which appears in the corresponding momentum equation that reads as
\begin{equation}\label{eq:conservmomentum_2}
   \frac{\partial}{\partial t}\left( h_1\rho U_2\right)+ \frac{\partial}{\partial\xi_1}\left( \rho U_1 U_2\right)+\frac{\partial}{\partial\xi_2}\left( \frac{ h_1}{h_2} \rho U_2^2\right) =- \frac{\partial}{\partial \xi_2}\left(\frac{h_1}{h_2} P \right) + \frac{\partial \tau_{21}}{\partial \xi_1} + \frac{\partial}{\partial \xi_2}\left( \frac{h_1}{h_2} \tau_{22}\right) +F_2,
\end{equation}
It now remains to explicitly write down the viscous stress tensor components $\tau_{ij}$ appearing in Eqs.~(\ref{eq:bodyforce1}), (\ref{eq:conservmomentum_1}), (\ref{eq:bodyforce2}), and (\ref{eq:conservmomentum_2}). In this regard, evaluating the constitutive relations appearing in Eqs.~(\ref{eq:viscousstress_OCC}) and (\ref{eq:strain_tensor}) yields the GOC forms of the components of the viscous stresses and strain rates as
\begin{equation*}\label{eq:viscousstress_OCC_modified}
       \tau_{11}= 2\rho \nu S_{11} + \rho (\zeta -\nu)  \bm{\nabla}\cdot \bm{U}, \qquad
       \tau_{22}= 2\rho \nu S_{22}  + \rho (\zeta -\nu)  \bm{\nabla}\cdot \bm{U}, \qquad \tau_{21} = \tau_{12}= 2\rho \nu S_{12}.
\end{equation*}
and
\begin{equation*}
S_{11}= \frac{1}{h_1} \frac{\partial U_1}{\partial \xi_1}+  U_2 \theta_{12}, \qquad S_{22}= \frac{1}{h_2} \frac{\partial U_2}{\partial \xi_2}+  U_1 \theta_{21}
\end{equation*}
\begin{equation*}
S_{12}= \frac{1}{2} \bigg[ \frac{\partial}{\partial \xi_2}\left(\frac{U_1}{h_2}\right)+ \frac{\partial}{\partial \xi_1}\left(\frac{U_2}{h_1}\right)+  U_1 \left(\theta_{22}- \theta_{12}\right)+  U_2 \left(\theta_{11}- \theta_{21}\right) \bigg].
\end{equation*}
Moreover, the divergence of the velocity field now reads via Eq.~(\ref{eq:divergenceU}) as
\begin{equation*}\label{eq:divergenceU_simplified2}
   \bm{\nabla}\cdot \bm{U} = \frac{1}{h_1} \frac{\partial U_1}{\partial \xi_1} + \frac{1}{h_2} \frac{\partial U_2}{\partial \xi_2}+ U_1\theta_{21}+ U_2\theta_{12}.
\end{equation*}
Putting these together, we can express the relations for the viscous stress tensor components in the GOC in more convenient forms in the following:
\begin{subequations} \label{eq:viscousstress_OCC_modified3}
\begin{align}
       &\tau_{11}= \rho \left(\zeta + \nu\right) \frac{1}{h_1} \frac{\partial U_1}{\partial \xi_1} + \rho \left( \zeta- \nu  \right) \frac{1}{h_2} \frac{\partial U_2}{\partial \xi_2} + \tau^c_{11}, \label{eq:viscousstress_OCC_modified3_normal11}\\
       &\tau_{22}= \rho \left(\zeta + \nu\right) \frac{1}{h_2} \frac{\partial U_2}{\partial \xi_2} + \rho \left( \zeta- \nu \right) \frac{1}{h_1} \frac{\partial U_1}{\partial \xi_1} + \tau^c_{22}, \label{eq:viscousstress_OCC_modified3_normal22}\\
       &\tau_{21} = \tau_{12}= \rho \nu  \bigg[ \frac{\partial}{\partial \xi_2}\left(\frac{U_1}{h_2}\right)+ \frac{\partial}{\partial \xi_1}\left(\frac{U_2}{h_1}\right)+  U_1(\theta_{22} - \theta_{12})+U_2(\theta_{11} - \theta_{21}) \bigg], \label{eq:viscousstress_OCC_modified3_shear}
\end{align}
\end{subequations}
where $\tau^c_{11}$ and $\tau^c_{22}$ represent the curvature effects of the GOC on the normal viscous stress components and are given by
\begin{equation}\label{eq:tauc}
\tau^c_{11}= \rho \left(\zeta + \nu\right) U_2\theta_{12}+ \rho \left( \zeta -\nu \right) U_1\theta_{21}, \qquad
\tau^c_{22}= \rho \left(\zeta + \nu\right) U_1\theta_{21}+ \rho \left( \zeta -\nu \right) U_2\theta_{12}.
\end{equation}

In summary, the conservative forms of the continuity equation given in Eq.~(\ref{eq:conservativemass_OCC_modified}) and the momentum equations shown in Eqs.~(\ref{eq:conservmomentum_1}) and  (\ref{eq:conservmomentum_2}) in GOC and supplemented with the effective body force components appearing in Eq.~(\ref{eq:bodyforce1}) and (\ref{eq:bodyforce2}) and together with the  viscous stress tensor components appearing in Eqs.~(\ref{eq:viscousstress_OCC_modified3}) and (\ref{eq:tauc}) form the target macroscopic equations for developing the new LBE in the GOC in the subsequent sections.

\subsection{Mappings of Scalars, Vectors, and Tensors between Cartesian Coordinates and the General Orthogonal Curvilinear Coordinates}
Before we discuss the construction procedure of the GOC-LBE, we conclude this section by noting that any scalar (e.g., density), vector (e.g., velocity field), and tensor (e.g., viscous stress) known in the computational domain in the GOC can be mapped to those in the physical domain in the Cartesian coordinates via appropriate transformation relations. Given a mapping of coordinates $x_i = x_i(\xi_1, \xi_2)$ for $i = 1, 2$, any scalar $\phi$ remains invariant under such a transformation, i.e., $\phi(x_1, x_2) = \phi(\xi_1, \xi_2)$. However, the components of vectors and tensors between the two coordinate systems are related with one another that is consistent with the mappings of the coordinates.

Thus, for a vector $\bm{v} = v_i \hat{x}_i = V_i \hat{\xi}_i$, its components in the Cartesian coordinates $v_i$ are related to those in the GOC $V_i$ via (see e.g.,~\cite{morse1946methods})
\begin{equation}\label{eq:vectormapping}
v_i = \frac{1}{h_j}\frac{\partial x_i}{\partial \xi_j} V_j, \quad \mbox{for}\; i = 1, 2.
\end{equation}
Similarly, for a tensor $\tensor{T} = \tau_{ij} \hat{x}_i\hat{x}_j = T_{pq}\hat{\xi}_p\hat{\xi}_q$, its components in the Cartesian coordinates $\tau_{ij}$ are related to those in the GOC $T_{pq}$ via (see e.g.,~\cite{morse1946methods})
\begin{equation}\label{eq:tensormapping}
\tau_{ij} = \frac{1}{h_p}\frac{1}{h_q}\frac{\partial x_i}{\partial \xi_p}\frac{\partial x_j}{\partial \xi_q} T_{pq}.
\end{equation}
Since the calculations using the GOC-LBE are actually performed in the computational domain in the GOC with the use of orthogonal clustered and curvilinear grids, the computed vectors and tensors such as $V_i$ and $T_{pq}$, respectively, can be readily mapped back using these relations to the corresponding $v_i$ and $\tau_{ij}$ in the physical domain in the Cartesian coordinates, which can then be used for the post-processing or visualization of simulation results. If the clustered grids are orthogonal but not curvilinear, such transformations are further simplified.

\section{Chapman-Enskog Analysis of Raw Moment-based LBE in GOC: Construction of Moment Equilibria and Source Terms, and Identification of Remaining Errors in Viscous Fluxes}\label{sec:MRT-LBM}
\subsection{GOC-based LBE using Raw Moments}
The Cartesian coordinates $(x_1, x_2)$ representing the \emph{physical domain} and discretized using the general orthogonal curvilinear grids (i.e., using clustered/body-fitted grids) are mapped into a uniform rectangular \emph{computational domain} $(\xi_1, \xi_2)$ via transformations based on the GOC given by $x_i=x_i(\xi_1,\xi_2)$ for $i=1,2$. A lattice Boltzmann equation (LBE) that is implementable using the standard collide-stream steps, will now be constructed to evolve in this computational domain discretized with grid spacings $\Delta \xi_1 = \Delta \xi_2 = 1$ (i.e., set to unity, as usual, in lattice units) using the D2Q9 lattice such that its emergent dynamics at time scales longer the collision time scales recover the NSE in the GOC derived in the previous section. The components of the particle velocities $\mathbf{e}_{\xi_1}$ and $\mathbf{e}_{\xi_2}$ for the D2Q9 lattice are given by
\begin{subequations}
\begin{eqnarray}
\ket{\mathbf{e}_{\xi_1}} &=& (0,1,0,-1,0,1,-1,-1,1)^\dag, \label{eq:1a}\\
\ket{\mathbf{e}_{\xi_2}} &=& (0, 0, 1, 0, -1, 1, 1, -1, -1)^\dag,\label{eq:1b}
\end{eqnarray}
\end{subequations}
Here, the 'ket' operator $\ket\cdot$ based on the Dirac notation is used to indicate a column vector of any variable defined for the lattice velocity set, while the 'dagger' operator $\dagger$ indicates the transpose operation. Defining the moment basis on which to express the LBE below will also require the following 9-dimensional vector consisting entirely of the unit elements:
\begin{eqnarray}
\ket{1}  = (1,1,1,1,1,1,1,1,1)^\dag. \label{eq:4}
\end{eqnarray}
The distribution function along a direction $\alpha$ for the LBE is denoted by $f_\alpha$, which is taken to relax to certain equilibrium distribution function $f_\alpha^{eq}$ under collision, and augmented with a source term $\S_\alpha$ to account for the effective body force. In practical implementations as well as in the Chapman-Enskog analysis of the resulting scheme in the GOC, it is more effective to express the effects of collision as well as the body forces in terms of their various moments. Thus, we define the \emph{raw moments} $k_{mn}^\prime$ of $f_\alpha$ of order ($m+n$), its corresponding equilibria $k_{mn}^{eq\prime}$, and the source term $\sigma_{mn}^\prime$, respectively, based on the monomials of the particle velocity components $e_{\alpha \xi_1}^m  e_{\alpha \xi_2}^n$ as
\begin{equation}\label{eq:9A}
k_{mn}^\prime= \sum_{\alpha=0}^{8} f_{\alpha} \;e_{\alpha \xi_1}^m  e_{\alpha \xi_2}^n, \quad k_{mn}^{eq\prime}= \sum_{\alpha=0}^{8} f_{\alpha}^{eq} \;e_{\alpha \xi_1}^m  e_{\alpha \xi_2}^n, \quad \sigma_{mn}^\prime= \sum_{\alpha=0}^{8} S_{\alpha} \;e_{\alpha \xi_1}^m  e_{\alpha \xi_2}^n.
\end{equation}
As in our previous work~\cite{yahia2021central,yahia2021three}, we consider a non-orthogonal moment basis from which to construct the GOC-LBE with the following nine independent basis vectors:
\begin{align}\label{eq:8}
\ket{T_0}&=\ket{1}, & \ket{T_1}&=\ket{e_{\xi_1}},& \ket{T_2}&=\ket{e_{\xi_2}}, & \ket{T_3}&=\ket{e_{\xi_1}^2 +e_{\xi_2}^2}, & \ket{T_4}&=\ket{e_{\xi_1}^2-e_{\xi_2}^2}, \nonumber\\
\ket{T_5}&=\ket{{e_{\xi_1}}{e_{\xi_2}}},& \ket{T_6}& =\ket{e_{\xi_1}^2 {e_{\xi_2}}},& \ket{T_7}&=\ket{{e_{\xi_1}} e_{\xi_2}^2}, & \ket{T_8}&=\ket{e_{\xi_1}^2 e_{\xi_2}^2}.
\end{align}
Here, we note that as usual the basis vectors for the second order diagonal moments $\ket{T_3}$ and $\ket{T_4}$ involve the use of $\ket{e_{\xi_1}^2}$ and $\ket{e_{\xi_2}^2}$ in terms of the combinations so as to independently evolve the bulk and shear viscous effects in the emergent dynamics of the LBE. Then the basis vectors in Eq.~(\ref{eq:8}) can be conveniently grouped together as the following transformation matrix
\begin{equation}\label{eq:transform_matrix_combined_moments}
\tensor{T}= \Big[\ket{T_{0}},\ket{T_{1}},\ket{T_{2}},\ldots,\ket{T_{8}} \Big]^{\dag},
 \end{equation}
which will be used to express the mappings between the raw moments and the distribution functions in the velocity space as follows. That is, we first collect the distribution functions, their equilibria, as well as the source terms for the nine different particle velocity directions in the form of the following vectors:
\begin{eqnarray}\label{eq:8A}
&\mathbf{f}=\left(f_{0},f_{1},f_{2},\ldots,f_{8}\right)^{\dag}, \quad
\mathbf{{f}}^{eq}=\left({f}_{0}^{eq},{f}_{1}^{eq},{f}_{2}^{eq},\ldots,{f}_{8}^{eq}\right)^{\dag},\quad &\mathbf{S}=\left({S}_{0},{S}_{1},{S}_{2},\ldots,{S}_{8}\right)^{\dag},
\end{eqnarray}
and similarly for their respective moments:
\begin{subequations}
\begin{align}\label{eq:8B}
&\mathbf{n}=\big(k_{00}^{\prime},k_{10}^{\prime},k_{01}^{\prime},k_{20}^{\prime}+k_{02}^{\prime},  k_{20}^{\prime}-k_{02}^{\prime}, k_{11}^{\prime}, k_{21}^{\prime}, k_{12}^{\prime}, k_{22}^{\prime} \big)^{\dag}, \\
&\mathbf{n}^{eq}=\big(k_{00}^{eq\prime},k_{10}^{eq\prime},k_{01}^{eq\prime},k_{20}^{eq\prime}+k_{02}^{eq\prime},  k_{20}^{eq\prime}-k_{02}^{eq\prime}, k_{11}^{eq\prime}, k_{21}^{eq\prime}, k_{12}^{eq\prime}, k_{22}^{eq\prime} \big)^{\dag},\\
&\mathbf{\Psi}=\big(\sigma_{00}^{\prime},\sigma_{10}^{\prime},\sigma_{01}^{\prime},\sigma_{20}^{\prime}+\sigma_{02}^{\prime},  \sigma_{20}^{\prime}-\sigma_{02}^{\prime}, \sigma_{11}^{\prime}, \sigma_{21}^{\prime}, \sigma_{12}^{\prime}, \sigma_{22}^{\prime} \big)^{\dag}.
\end{align}
\end{subequations}
Then, the various moments are related to the corresponding distribution functions in the velocity space via the following transformations:
\begin{eqnarray} \label{eq:9}
&\mathbf{n}= \tensor{T} \mathbf{f}, \quad  \mathbf{n}^{eq}=  \tensor{T} \mathbf{f^{eq}}, \quad \mathbf{\Psi}= \tensor{T}\mathbf{S}.
\end{eqnarray}

Based on these considerations, we then formally represent the GOC-LBE using the raw moments and multiple relaxation times (MRT) in the computational domain as
\begin{equation}\label{eq:MRT-LBE}
 \mathbf{f} (\bm{\xi}+\mathbf{e} \Delta t, t+\Delta t)- \mathbf{f} (\bm{\xi},t) =  \tensor{T^{-1}}
 \Big[  \tensor{\Lambda} \;  \left(\; \mathbf{n}^{eq}-\mathbf{n} \;\right) + \left(\tensor{I} - \frac{\tensor{\Lambda}}{2}\right)  \mathbf{\Psi} \Delta t \Big],
\end{equation}
where $\Delta t$ is the time step, and $\tensor{\Lambda}$ is the diagonal relaxation rates matrix which reads as
\begin{equation}\label{eq:relaxationratematrix}
  \tensor{\Lambda}= \mbox{diag}(\omega_0, \omega_1, \omega_2, \omega_3,\omega_4, \omega_5, \omega_6, \omega_7, \omega_8),
\end{equation}
in which some of the parameters will be related to the bulk and shear viscosities using the Chapman-Enskog analysis discussed subsequently, and the rest are free to choose to maintain numerical stability. We note here that the implementation of the source term in Eq.~(\ref{eq:MRT-LBE}) is based on the second-order trapezoidal rule, which is effectively made explicit via a variable transformation such as $\bar{f}_\alpha=f_\alpha-S_\alpha\Delta t/2$~\cite{he1998novel}, or more specifically via its moments counterpart $\bar{k}_{mn}^\prime=k_{mn}^\prime-\sigma_{mn}^\prime\Delta t/2$; however, in order to be lighter on the notations for better readability, we have dropped the 'bar' over the symbols for the distribution functions (and their moments), but the contribution due to $\sigma_{mn}^\prime\Delta t/2$ will be appropriately taken into account when evaluating their moments. The key elements in the GOC-LBE in Eq.~(\ref{eq:MRT-LBE}) for consistently recovering the NSE in GOC are the raw moments of the equilibria $k_{mn}^{eq\prime}$ and the source terms $\sigma_{mn}^{\prime}$ appearing in $\mathbf{n}^{eq}$ and $\mathbf{\Psi}$, respectively, which will be constructed in what follows.

\subsection{Determination of Raw Moments of Equilibria and Sources}\label{sec:3}
In general, the raw moments of the equilibria depend on the fluid density $\rho$ and the velocity components $U_1$ and $U_2$ and should be parameterized by the metric factors $h_1$ and $h_2$ carefully by inspection as well as a mathematical analysis based on the Chapman-Enskog (C-E) expansion that determines the long-time emergent behavior of the GOC-LBE so that the consistent fluid dynamical equations derived earlier are recovered. In this regard, as noted before, an inspection of the target continuity equation in GOC in Eq.~(\ref{eq:conservativemass_OCC_modified}) yields the zeroth and first order equilibrium moments, which are the collision invariants, by matching the terms inside the time derivative (i.e., effective density) as well as the spatial derivatives (i.e., effective mass fluxes or momentum components):
\begin{equation*}
k_{00}^{eq\prime}= h_1 h_2 \rho, \qquad k_{10}^{eq\prime}=h_2 \rho U_1, \qquad k_{01}^{eq\prime}=h_1 \rho U_2.
\end{equation*}
Then, since the second order equilibrium moments are related to the inviscid momentum fluxes, they can be determined by an inspection of the momentum equations in the GOC in Eqs.~(\ref{eq:conservmomentum_1}) and  (\ref{eq:conservmomentum_2}), and the results read as
\begin{equation*}
k_{20}^{eq\prime}= \frac{h_2}{h_1}(\rho U_1^2 +  P),\qquad k_{02}^{eq\prime}= \frac{h_1}{h_2}(\rho U_2^2 + P), \qquad k_{11}^{eq\prime}=\rho U_1 U_2.
\end{equation*}
Unlike the case of representing the usual NSE in the Cartesian coordinates, the D2Q9 lattice does not have enough degrees of freedom to fully recover the diagonal or normal stress components $\tau_{11}$ and $\tau_{22}$ for the NSE in the GOC via the choices in its associated higher order moment equilibria. These will need to be compensated for via appropriate moment equilibria corrections based on a C-E analysis later. Nevertheless, one key feature of the GOC-LBE is that due to the orthogonality in the coordinates associated with the GOC just like in the Cartesian coordinates, it preserves the exact relationship between the off-diagonal second order non-equilibrium moments that are related to the spatial gradients of the third-order equilibrium moments $k_{21}^{eq\prime}$ and $k_{12}^{eq\prime}$ to the off-diagonal, shear stresses. This aspect will be confirmed via a C-E analysis later. For now, based on an inspection of the expression for the shear stresses $\tau_{12}=\tau_{21}$ in Eqs.~(\ref{eq:viscousstress_OCC_modified3_shear}), which contain terms such as $U_1/h_2$ and $U_2/h_1$,  and supplemented by a C-E analysis subsequently, we prescribe the third-order moments $k_{21}^{eq\prime}$ and $k_{12}^{eq\prime}$ as follows:
\begin{equation*}
k_{21}^{eq\prime}=\rho \left(c_{s}^2 + U_1^2\right)\frac{U_2}{h_1}, \qquad k_{12}^{eq\prime}=\rho \left(c_{s}^2 + U_2^2 \right)\frac{U_1}{h_2},
\end{equation*}
where they also contain cubic velocity terms $\rho U_1U_2^2/h_2$ and $\rho U_1^2U_2/h_1$, which are necessary to achieve Galilean invariance similar to the standard LBE based on uniform grids, and $c_s$ is the speed of sound which will determine the shear viscosity $\nu$ and its parametrization will be discussed at the end of this section. This leaves us with the one remaining fourth order equilibrium moment $k_{22}^{eq\prime}$ which does not appear in the C-E analysis in determining the emergent macroscopic equations and can be freely chosen. In this work, for simplicity and consistency with the uniform grid-based LBE, we prescribe it to be the same as that for the continuous Maxwell distribution and, based on numerical stability considerations, do not involve any metric factors, which then yields
\begin{equation*}
k_{22}^{eq\prime}=\rho c_{s}^4 + \rho  c_{s}^2\left( U_1^2+ U_2^2 \right)+ \rho U_1^2 U_2^2.
\end{equation*}
Summarizing these choices, we can then write the \emph{base} moment equilibria for the GOC-LBE for the D2Q9 lattice as
\begin{eqnarray}\label{eq:GOCmomentequilibria}
&& n_0^{eq} = k_{00}^{eq\prime}= h_1 h_2 \rho, \qquad\qquad n_1^{eq} = k_{10}^{eq\prime}=h_2 \rho U_1, \qquad\qquad n_2^{eq} = k_{01}^{eq\prime}=h_1 \rho U_2,\nonumber\\
&& n_3^{eq} = k_{20}^{eq\prime}+k_{02}^{eq\prime} =\frac{h_2}{h_1}\rho U_1^2 + \frac{h_1}{h_2}\rho U_2^2+ \left(\frac{h_2}{h_1}+ \frac{h_1}{h_2}\right)P,\nonumber\\
&& n_4^{eq} = k_{20}^{eq\prime}-k_{02}^{eq\prime}=\frac{h_2}{h_1}\rho U_1^2 - \frac{h_1}{h_2}\rho U_2^2+ \left(\frac{h_2}{h_1}- \frac{h_1}{h_2}\right)P,\nonumber\\
&& n_5^{eq} = k_{11}^{eq\prime}=\rho U_1 U_2, \qquad\quad
n_6^{eq} = k_{21}^{eq\prime}=\rho \left(c_{s}^2 + U_1^2\right)\frac{U_2}{h_1}, \qquad\quad
n_7^{eq} = k_{12}^{eq\prime}=\rho \left(c_{s}^2 + U_2^2 \right)\frac{U_1}{h_2}, \nonumber\\
&& n_8^{eq} = k_{22}^{eq\prime}=\rho c_{s}^4 + \rho  c_{s}^2\left( U_1^2+ U_2^2 \right)+ \rho U_1^2 U_2^2.
\end{eqnarray}
Here, we note that based on these choices of moment equilibria, the C-E analysis given later will require the pressure $P$ to satisfy the equation of state $P = \rho c_s^2$ just as in the case of the standard LBE for uniform grids.

Additionally, in order to introduce the effective body force given in Eq.~(\ref{eq:bodyforce1}) and (\ref{eq:bodyforce2}) in the GOC-LBE due to any external force as well as the geometric body force, the raw moments of the source term $\sigma_{mn}^{\prime}$ for various orders ($m+n$) need to be prescribed. Now, an inspection of the continuity and momentum equations (see Eq.~(\ref{eq:conservativemass_OCC_modified}) and Eqs.~(\ref{eq:conservmomentum_1}) and  (\ref{eq:conservmomentum_2})) reveal the choices for their zeroth and first order moments as
\begin{equation*}
\sigma_{00}^{\prime}=0, \qquad \sigma_{10}^{\prime}=F_1, \qquad \sigma_{01}^{\prime}=F_2.
\end{equation*}
Moreover, in order to recover the NSE in the GOC consistently, the second order raw moments of the source terms $\sigma_{20}^{\prime}$, $\sigma_{02}^{\prime}$ and $\sigma_{11}^{\prime}$ need to be carefully chosen with appropriate dependence on the metric factors. These will be determined via a C-E analysis in the next subsection. Here, we simply collect together the results of that analysis for convenience, which read as
\begin{equation*}
\sigma_{20}^{\prime}=2F_1 \dfrac{U_1}{h_1}, \qquad \sigma_{02}^{\prime}=2F_2\dfrac{U_2}{h_2}, \qquad \sigma_{11}^{\prime}=F_1\dfrac{U_2}{h_2}+ F_2\dfrac{U_1 }{h_1}.
\end{equation*}
These, then leave us with selecting the third and higher order moments of the sources, which do not appear in the C-E analysis in determining the fluid dynamical behavior of the GOC-LBE. Here, for simplicity, they are set to zero, that is, $\sigma_{mn}^{\prime} = 0$ for $m+n\geq 3$. Then, summarizing these choices, raw moments of the sources for the GOC-LBE for the D2Q9 lattice can be written as
\begin{eqnarray}\label{eq:GOCmomentsources}
&&\Psi_0 = \sigma_{00}^{\prime}=0, \qquad\qquad \Psi_1 = \sigma_{10}^{\prime}=F_1, \qquad\qquad \Psi_2 = \sigma_{01}^{\prime}=F_2, \nonumber\\
&&\Psi_3 = \sigma_{20}^{\prime} + \sigma_{02}^{\prime} = 2\left(F_1 \dfrac{U_1}{h_1} + F_2\dfrac{U_2}{h_2}\right), \qquad
\Psi_4 = \sigma_{20}^{\prime} - \sigma_{02}^{\prime} = 2\left(F_1 \dfrac{U_1}{h_1} - F_2\dfrac{U_2}{h_2}\right), \nonumber\\
&&\Psi_5 = \sigma_{11}^{\prime}=F_1\dfrac{U_2}{h_2}+ F_2\dfrac{U_1 }{h_1}, \quad \Psi_6 = \sigma_{21}^{\prime}=0, \quad \Psi_7 = \sigma_{12}^{\prime}=0, \quad \quad \Psi_8 = \sigma_{22}^{\prime}=0.
\end{eqnarray}

Then, finally, a solution of the GOC-LBE given in Eq.~(\ref{eq:MRT-LBE}) yields the distribution functions $f_\alpha$ everywhere in the $\xi_1 - \xi_2$ computational domain, from which the hydrodynamic fields, viz., the fluid density and the velocity fields can be obtained respectively as their zeroth and first moments. That is,
\begin{equation}\label{eq:GOChydrodynamicfields}
h_1 h_2 \rho =\sum_{\alpha=0}^{8} f_{\alpha}, \qquad h_2 \rho U_1 =\sum_{\alpha=0}^{8} f_{\alpha} e_{\xi_{1,\alpha}} + \frac{F_1}{2}\Delta t, \qquad h_1 \rho U_2 =\sum_{\alpha=0}^{8} f_{\alpha} e_{\xi_{2,\alpha}} + \frac{F_2}{2}\Delta t.
\end{equation}
In the expressions in the last equation (Eq.~(\ref{eq:GOChydrodynamicfields})) as noted in the paragraph below Eq.~(\ref{eq:relaxationratematrix}), we augment each of the respective moments with a contribution $\sigma_{mn}^{\prime}\Delta t/2$ for $(mn) = (00), (10), (01)$ due to the use of the trapezoidal rule to incorporate the source term in the GOC-LBE.

\subsubsection{Parametrization of the speed of sound $c_s$}
An important consideration in the use of the GOC-LBE is the selection of the speed of sound. In the case of the standard square grid-based LBE, the speed of sound is generally set equal to $1/\sqrt{3}$, which we write as $c_{s*} =1/\sqrt{3}$. With the use of nonuniform clustered/curvilinear grids in the physical (or the Cartesian) domain of the GOC-LBE, the parametrization of the speed of sound should reflect the choice of the metric factors of the GOC when the LBE is applied in the computational domain. On the one hand, it should be held a fixed constant throughout the domain for physical consistency for the case of isothermal flow situation considered in this work. On the other hand, for stable and convergent solution of the explicit schemes such as those based on the LBE, the numerical region of dependence must encompass the analytical region of dependence, or that the scheme should respect the Courant-Friedrichs-Lewy (CFL) condition~\cite{courant1967partial} that limits the fraction of the grid size covered by any quantity (flow or wave) in a single time step of any time-explicit numerical method. In other words, we limit the speed of sound of the GOC-LBE according to $c_s\Delta t/\Delta x_{min} \leq c_{s*}$. Now, $\Delta x_{min} = \mbox{min}(\Delta x_1, \Delta x_2) = \mbox{min} (h_1\Delta\xi_1, h_2\Delta\xi_2)$, and as the LBE is usually applied using the lattice units, i.e., $\Delta \xi_1 = \Delta \xi_2 \Delta t = 1$, it follows that $\Delta x_{min} = \mbox{min}(h_1, h_2)$. In addition, since we require the speed of sound to be held constant spatially as indicated above, we further take the minimum of each of these metric factors across the entire computational domain, i.e. $h_{1min} = \mbox{min}(h_1)$ and $h_{2min} = \mbox{min}(h_2)$, and then select $c_s$ according to
\begin{equation}\label{eq:GOC-LBE_CFL}
c_s = qc_{s*},  \qquad\mbox{where}\qquad q = \mbox{min}(h_{1min}, h_{2min}), \quad c_{s*} = 1/\sqrt{3}.
\end{equation}
Thus, it follows that we set $c_s^2 = q^2/3$ where $q = \mbox{min}(h_{1min}, h_{2min})$, which appears in the expressions of various moment equilibria as seen above and in the relations for the fluid viscosity as well as the various correction terms that need to be applied to the moment equilibria which will be derived via C-E analyses in subsequent sections. In practice, this choice for the speed of sound for the GOC-LBE is found to work quite well in the numerical simulations of various flow case studies whose results are reported in the later sections of this work.

\subsection{Chapman-Enskog Analysis}
Based on the developments in the last subsection we now perform a Chapman-Enskog (C-E) analysis~\cite{chapman1990mathematical} in the moment space (see e.g.,~\cite{hajabdollahi2018galilean,yahia2021central,yahia2022preconditioned}). In this regard, we will apply the C-E multiscale expansions to the GOC-LBE given in Eq.~(\ref{eq:MRT-LBE}), by first expanding the raw moments and the time derivative as
\begin{equation}\label{eq:53}
\mathbf{n}= \sum_{j=0}^{\infty} \epsilon^{j} \mathbf{n}^{(j)}, \quad \partial_t= \sum_{j=0}^{\infty} {\epsilon}^{j} {\partial_{t_j}},
\end{equation}
where $\epsilon= \Delta t$ represents a small bookkeeping perturbation parameter serving in what follows to delineating the terms of different orders in $\epsilon$. Then, we substitute the above equation (Eq.~(\ref{eq:53})) into Eq.~(\ref{eq:MRT-LBE}) and use a Taylor series expansion in the streaming operator, i.e., $\mathbf{f} (\bm{\xi}+\mathbf{e} \Delta t, t+\Delta t)=\sum_{j=0}^{\infty} (\epsilon^{j}/j!)\left(\partial_t+ \mathbf{e}\cdot \mathbf{\nabla}\right)^j\mathbf{f}(\bm{\xi}, t)$ in its left hand side; subsequently, we convert the resulting expression to the moment space via $\mathbf{f}=  \tensor{T}^{-1} \mathbf{n}$ and then account for including the source contribution $\sigma_{mn}^{\prime}\Delta t/2$ or, equivalently, $\mathbf{\Psi}\Delta t/2$ when evaluating the resulting moments $\mathbf{n}$ (see the note in the paragraph below Eq.~(\ref{eq:relaxationratematrix})) to  obtain the evolution equations of the moments of different orders of $\epsilon$, i.e., $O(\epsilon^k)$, where $k=0,1$, and $2$, which read as
\begin{subequations}\label{eq:54}
\begin{eqnarray}
\centering
&&O (\epsilon^0 ):\quad \mathbf{n}^{(0)} =  \mathbf{n}^{eq},  \label{eq:54a}\\
&&O (\epsilon^1 ):\quad \left(\partial_{t_0} + \tensor{E}_i \partial_i\right)  \mathbf{n}^{(0)} =  - \tensor{\Lambda} \; \mathbf{n}^{(1)}+\mathbf{\Psi},  \label{eq:54b} \\
&&O (\epsilon^2 ):\quad \partial_{t_1} \; \mathbf{n}^{(0)} +  \left(\partial_{t_0} + \tensor{E}_i \partial_i\right) \;\left[ \tensor{I} - \frac{\tensor{\Lambda}} {2} \right] \mathbf{n}^{(1)} =  - \tensor{\Lambda} \; \mathbf{n}^{(2)}, \label{eq:54c}
\end{eqnarray}
\end{subequations}
where, as shorthand notations, we use $\tensor{E}_i= \tensor{T} \;( \mathbf{e}_i \; \tensor{I})\tensor{T}^{-1}$ and $ \mathbf{e}_i=\ket{\mathbf{e}_{\xi i}}$ for $i \in (1,2)$ and $\partial_i = \partial_{\xi_i}$.

First, evaluating the moment system $O(\epsilon)$ given in Eq.~(\ref{eq:54}) up to the second order moments (i.e., the first six components of $\mathbf{n}$), which are relevant in recovering the hydrodynamics, we obtain
\begin{subequations}
\begin{align}\label{eq:orderonemomentsystem}
&\partial_{t_0}n_0^{eq} + \partial_{\xi_1} n_1^{eq} + \partial_{\xi_2}n_2^{eq} = \Psi_0,\\
&\partial_{t_0}n_1^{eq} + \partial_{\xi_1} \left[\frac{1}{2} (n_3^{eq}+ n_4^{eq})\right] + \partial_{\xi_2} n_5^{eq} = \Psi_1,\\
&\partial_{t_0}n_2^{eq}+ \partial_{\xi_1} n_5^{eq}+ \partial_{\xi_2} \left[\frac{1}{2} (n_3^{eq}- n_4^{eq})\right] = \Psi_2,\\
& \partial_{t_0}n_3^{eq}+  \partial_{\xi_1} \left[n_1^{eq}+ n_7^{eq} \right] + \partial_{\xi_2} \left[  n_2^{eq}+ n_6^{eq} \right] = - \omega_3\;  n_3^{(1)}+  \Psi_3,\\
& \partial_{t_0}n_4^{eq}+  \partial_{\xi_1} \left[n_1^{eq}- n_7^{eq} \right] + \partial_{\xi_2} \left[ - n_2^{eq}+ n_6^{eq} \right] = - \omega_4\;  n_4^{(1)}+  \Psi_4,\\
&\partial_{t_0}n_5^{eq} + \partial_{\xi_1} n_6^{eq} + \partial_{\xi_2} n_7^{eq}= - \omega_5\;  n_5^{(1)}+ \Psi_5.
\end{align}
\end{subequations}

Then, substituting the raw moments of the equilibria shown in Eqs.~(\ref{eq:GOCmomentequilibria}) into Eq.~(\ref{eq:orderonemomentsystem}) along with writing the known source terms for $\Psi_0$, $\Psi_1$ and $\Psi_2$ in terms of the effective body force ($F_1, F_2$),  i.e., for the conserved moments, while treating the source moments contributions $\sigma_{mn}^{\prime}$ for $(mn) = (20), (02)$ and $(11)$, i.e., for the non-conserved moments that appear in $\Psi_3$, $\Psi_4$, and $\Psi_5$ as unknowns to be determined as part of the outcomes of this analysis, we get
\begin{subequations}
\begin{eqnarray}
&\partial_{t_0} \left(h_1 h_2 \rho\right) + \partial_{\xi_1} \left(h_2 \rho U_1\right) + \partial_{\xi_2} \left(h_1 \rho U_2\right)= 0,\label{eq:55a}\\
&\partial_{t_0}\left(h_2 \rho U_1 \right)+ \partial_{\xi_1}\left[  \dfrac{h_2}{h_1}\left(\rho U_1^2+P\right)\right] + \partial_{\xi_2}\left(\rho U_1 U_2\right)= F_1,\label{eq:55b}\\
&\partial_{t_0} \left(h_1 \rho U_2\right) + \partial_{\xi_1} \left(\rho U_1 U_2\right)+ \partial_{\xi_2}\left[\dfrac{h_1}{h_2}\left(\rho U_2^2+P\right)\right]= F_2,\label{eq:55c}\\
&\partial_{t_0}\left[\dfrac{h_2}{h_1} \rho U_1^2 + \dfrac{h_1}{h_2} \rho U_2^2 + \left(\dfrac{h_2}{h_1}+ \dfrac{h_1}{h_2}\right) P\right]+  \partial_{\xi_1} \left[\left(h_2 + \dfrac{  c_{s}^2}{h_2}\right)\rho U_1 + \rho U_2^2 \dfrac{U_1}{h_2} \right] \nonumber \\
&+\partial_{\xi_2}\left[\left( h_1 + \dfrac{ c_{s}^2}{h_1}\right)\rho U_2 + \rho U_1^2 \dfrac{U_2}{h_1} \right]= - \omega_3\;  n_3^{(1)}+  \left( \sigma_{20}^{\prime} + \sigma_{02}^{\prime} \right),\label{eq:55d}\\
&\partial_{t_0}\left[\dfrac{h_2}{h_1}\rho U_1^2 - \dfrac{h_1}{h_2}\rho U_2^2+\left(\dfrac{h_2}{h_1}- \dfrac{h_1}{h_2}\right) P )\right]+  \partial_{\xi_1} \left[\left(h_2 - \dfrac{  c_{s}^2}{h_2}\right)\rho U_1 - \rho U_2^2 \dfrac{U_1}{h_2}\right]\nonumber \\
&+\partial_{\xi_2}\left[\left(- h_1 + \dfrac{ c_{s}^2}{h_1}\right)\rho U_2 +\rho U_1^2 \dfrac{U_2}{h_1}\right] =
- \omega_4\;  n_4^{(1)}+ \left( \sigma_{20}^{\prime} - \sigma_{02}^{\prime} \right),\label{eq:55e}\\
 &\partial_{t_0}(\rho U_1 U_2)+  \partial_{\xi_1} \left[ ( c_{s}^2  +  U_1^2) \rho \dfrac{U_2}{h_1} \right] + \partial_{\xi_2}\left[( c_{s}^2  + U_2^2) \rho \dfrac{U_1}{h_2} \right] = -\omega_5\;  n_5^{(1)}+ \sigma_{11}^{\prime}.\label{eq:55f}
\end{eqnarray}
\end{subequations}
Similarly, evaluating the $O(\epsilon^2)$ moment system in Eq.~(\ref{eq:54}) relevant to hydrodynamics, i.e., for the first three conserved moments, we obtain
\begin{subequations}\label{eq:56}
\begin{eqnarray}
&\partial_{t_1}(h_1 h_2 \rho)=0,
\label{eq:56a}\\
&\partial_{t_1}\left(h_2 \rho U_1\right)+\partial_{\xi_1} \left[\dfrac{1}{2}\left(1-\dfrac{\omega_3}{2}\right) n_3^{(1)}+\dfrac{1}{2}\left(1-\dfrac{\omega_4}{2}\right)n_4^{(1)}\right]
+\partial_{\xi_2} \left[\left(1-\dfrac{\omega_5}{2}\right)n_5^{(1)}\right]=0, \label{eq:56b}\\
&\partial_{t_1}\left(h_1 \rho U_2 \right) + \partial_{\xi_1} \left[\left(1 - \dfrac{\omega_5}{2}\right) n_5^{(1)}\right]+\partial_{\xi_2} \left[\dfrac{1}{2}\left(1-\dfrac{\omega_3}{2}\right)n_3^{(1)}-\dfrac{1}{2}\left(1-\dfrac{\omega_4}{2}\right)n_4^{(1)}\right]=0. \label{eq:56c}
\end{eqnarray}
\end{subequations}
From Eqs.~(\ref{eq:56a})-(\ref{eq:56c}), it can be seen that evolution of the momentum components $h_2\rho U_1$ and $h_1\rho U_2$ at the slower time scale $t_1$ depend on the non-equilibrium moments $n_3^{(1)}$, $n_4^{(1)}$, and $n_5^{(1)}$, which, as seen in the following, would result in the determination of the viscous fluxes of momentum (or simply the viscous stresses). As such, we will evaluate these three non-equilibrium moments from the $O(\epsilon)$ momentum system for the non-conserved components, viz., Eqs.~(\ref{eq:55d}), (\ref{eq:55e}), and (\ref{eq:55f}). The rest of this section will provide the essential details of the determination of these three non-equilibrium moments.

By rearranging Eqs.~(\ref{eq:55d}), (\ref{eq:55e}), and (\ref{eq:55f}) and as in the standard LBM for uniform grid by setting the pressure in terms of the density as $P=\rho c_s^2$, we can then write the full forms of the expressions that serve as the starting points for the evaluations of $n_3^{(1)}$, $n_4^{(1)}$, and $n_5^{(1)}$, which read as
\begin{align}\label{eq:57}
&n_3^{(1)}=  \frac{1}{\omega_3} \Big\{ -\frac{h_2}{h_1}\boxed{\partial_{t_0}( \rho U_1^2)} - \frac{h_1}{h_2} \boxed{\partial_{t_0} (\rho U_2^2)} - \left(\frac{h_2}{h_1}+ \frac{h_1}{h_2}\right)  c_{s}^2 \boxed{\partial_{t_0} \rho} -  \partial_{\xi_1} \left[\left(h_2 + \frac{  c_{s}^2}{h_2}\right)\rho U_1 + \rho U_2^2 \frac{U_1}{h_2} \right] \nonumber \\
&\qquad\qquad-\partial_{\xi_2}\left[\left( h_1 + \frac{ c_{s}^2}{h_1}\right)\rho U_2 + \rho U_1^2 \frac{U_2}{h_1} \right] + \left( \sigma_{20}^{\prime} + \sigma_{02}^{\prime} \right) \Big\},
\end{align}
\begin{align}\label{eq:58}
&n_4^{(1)}= \frac{1}{\omega_4} \Big\{- \frac{h_2}{h_1} \boxed{ \partial_{t_0}(\rho U_1^2)}+ \frac{h_1}{h_2} \boxed{\partial_{t_0}(\rho U_2^2)}- \left(\frac{h_2}{h_1}- \frac{h_1}{h_2}\right)  c_{s}^2\boxed{ \partial_{t_0} \rho}-
\partial_{\xi_1} \left[\left(h_2 - \frac{  c_{s}^2}{h_2}\right)\rho U_1 - \rho U_2^2 \frac{U_1}{h_2}\right]\nonumber \\
&\qquad\qquad-\partial_{\xi_2}\left[\left(- h_1 + \frac{ c_{s}^2}{h_1}\right)\rho U_2 +\rho U_1^2 \frac{U_2}{h_1}\right] + \left( \sigma_{20}^{\prime} - \sigma_{02}^{\prime} \right)\Big\},
\end{align}
\begin{align}\label{eq:59}
&n_5^{(1)}= \cfrac{1}{\omega_5} \Big\{- \boxed{\partial_{t_0}(\rho U_1 U_2)}-  \partial_{\xi_1} \left[ ( c_{s}^2  +  U_1^2) \rho \frac{U_2}{h_1} \right] - \partial_{\xi_2} \left[ ( c_{s}^2  + U_2^2) \rho \frac{U_1}{h_2} \right]+ \sigma_{11}^{\prime}\Big\}.
\end{align}
In order to simplify these expressions, first we separately need the time derivatives of the terms associated with the inviscid momentum fluxes $\rho U_i U_j$ and $P$, i.e., specifically $\partial_{t_0}(\rho U_1^2)$, $\partial_{t_0}(\rho U_2^2)$, $\partial_{t_0}(\rho U_1U_2)$ and $\partial_{t_0}(\rho c_s^2)$. which are highlighted in the boxes in the above. These can be obtained via appropriate manipulations of the continuity and momentum equations under the faster time scale $t_0$ given in Eqs.~(\ref{eq:55a}), (\ref{eq:55b}), and (\ref{eq:55c}), which would then replace the temporal derivatives in terms of the spatial derivatives of the velocity components and the density. Thus, we first start from Eqs.~(\ref{eq:55a}), (\ref{eq:55b}), and (\ref{eq:55c}) and rearrange them to write as
\begin{subequations}
\begin{eqnarray}
\partial_{t_0}\rho &=& -\frac{1}{h_1 h_2} \big[\partial_{\xi_1} \left(h_2 \rho U_1\right) + \partial_{\xi_2} \left(h_1 \rho U_2\right)\big],\label{eq:td_density}\\
\partial_{t_0} \left(\rho U_1\right) &=& -\frac{1}{h_2}\partial_{\xi_1} \left[\frac{h_2}{h_1}\left(\rho U_1^2 + \rho c_s^2 \right)\right] -\frac{1}{h_2} \partial_{\xi_2}\left(\rho U_1 U_2\right) +\frac{F_1}{h_2},\label{eq:td_momentum1}\\
\partial_{t_0} \left(\rho U_2\right) &=& -\frac{1}{h_1} \partial_{\xi_1}\left(\rho U_1 U_2\right) -\frac{1}{h_1}\partial_{\xi_2} \left[ \frac{h_1}{h_2}\left(\rho U_2^2 + \rho c_s^2\right)\right] + \frac{F_2}{h_1}. \label{eq:td_momentum2}
\end{eqnarray}
\end{subequations}
Then, from the usual product rules, we can write the temporal derivatives of $\rho U_i U_j$ with the appropriate scalings involving the metric factors as seen in Eqs.~(\ref{eq:57}), ~(\ref{eq:58}), and ~(\ref{eq:59}) as
\begin{subequations}
\begin{eqnarray}
-\frac{h_2}{h_1}\partial_{t_0}(\rho U_1^2) &=& -2\frac{h_2}{h_1}U_1\partial_{t_0}(\rho U_1) + \frac{h_2}{h_1} U_1^2 \partial_{t_0}\rho, \label{eq:td_momflux11}\\
-\frac{h_1}{h_2}\partial_{t_0}(\rho U_2^2) &=& -2\frac{h_1}{h_2}U_2\partial_{t_0}(\rho U_2) + \frac{h_1}{h_2} U_2^2 \partial_{t_0}\rho, \label{eq:td_momflux22}\\
-\partial_{t_0}(\rho U_1 U_2) &=& -U_1 \partial_{t_0}(\rho U_2) - U_2 \partial_{t_0}(\rho U_1) + U_1 U_2\partial_{t_0}\rho. \label{eq:td_momflux12}
\end{eqnarray}
\end{subequations}
From these equations, i.e., Eqs.~(\ref{eq:td_momflux11}), (\ref{eq:td_momflux22}), and (\ref{eq:td_momflux12}) by substituting for the time derivatives of the density and the momentum components given in Eqs.~(\ref{eq:td_density}), (\ref{eq:td_momentum1}), and (\ref{eq:td_momentum2}), we can then obtain the required temporal derivatives that appear in Eqs.~(\ref{eq:57}), (\ref{eq:58}), and (\ref{eq:59}) for $n_3^{(1)}$, $n_4^{(1)}$ and $n_5^{(1)}$, respectively. Let's then apply these steps separately for each of these non-equilibrium moments and simplify them as much as possible by combining them with the rest of the terms that appear together in these equations.

\subsubsection{Determination of the non-equilibrium moment $n_3^{(1)} = k_{20}^{(1)\prime} + k_{02}^{(1)\prime}$}
The Eqs.~(\ref{eq:td_momflux11}), (\ref{eq:td_momflux22}) after substituting Eqs.~(\ref{eq:td_momentum1}) and (\ref{eq:td_momentum2}), and using the definition of the curvature coefficient matrix elements $\theta_{ij}$ given in Eq.~(\ref{eq:curvaturematrix}) in the resulting expressions, and upon simplification and rearrangement, become
\begin{eqnarray}\label{eq:60f}
 &-\dfrac{h_2}{h_1}\partial_{t_0} \left( \rho U_1^2\right)= 4 \rho U_1^2 \dfrac{h_2}{h_1^2}\partial_{\xi_1}  U_1 + \dfrac{2 \rho U_1^2}{h_1}\partial_{\xi_2}  U_2 -\dfrac{2 F_1 U_1}{h_1}+ 2 c_{s}^2 U_1 \dfrac{h_2}{h_1^2}\partial_{\xi_1} \rho + \dfrac{2 \rho U_1 U_2}{h_1}\partial_{\xi_2}  U_1  \nonumber\\
   &+ 2 \rho U_1 \left( U_1^2 + c_{s}^2 \right)\dfrac{h_2}{h_1}\left( \theta_{21} - \theta_{11}\right) + U_1^2 \dfrac{h_2}{h_1} \partial_{t_0}\rho,
\end{eqnarray}
and
\begin{eqnarray}\label{eq:60g}
 &-\dfrac{h_1}{h_2}\partial_{t_0} \left( \rho U_2^2\right)= 4 \rho U_2^2 \dfrac{h_1}{h_2^2}\partial_{\xi_2}  U_2 + \dfrac{2 \rho U_2^2}{h_2}\partial_{\xi_1}  U_1 -\dfrac{2 F_2 U_2}{h_2}+ 2 c_{s}^2 U_2 \dfrac{h_1}{h_2^2}\partial_{\xi_2} \rho + \dfrac{2 \rho U_1 U_2}{h_2}\partial_{\xi_1}  U_2 \nonumber \\
   &+ 2 \rho U_2 \left( U_2^2 + c_{s}^2 \right)\dfrac{h_1}{h_2}\left( \theta_{12} - \theta_{22}\right) + U_2^2 \dfrac{h_1}{h_2} \partial_{t_0}\rho.
\end{eqnarray}
Notice that these last two equations when combined as in Eq.~(\ref{eq:57}) contain the effective body force contributions in the form
$-2F_1U_1/h_1-2F_2U_2/h_2$, which can be eliminated if we choose the combinations of the source moments $\sigma_{20}^{\prime}+\sigma_{02}^{\prime}$ that appear in that equation as
\begin{align*}
&\left(\sigma_{20}^{\prime}+\sigma_{02}^{\prime}\right)= \frac{2F_1 U_1}{h_1}+\frac{2F_2 U_2}{h_2},
\end{align*}
which then provides the reasoning behind the choices made earlier in this regard in Eq.~(\ref{eq:GOCmomentsources}). The non-equilibrium moment equation (Eq.~(\ref{eq:57})) also contains various spatial derivatives, which can be further expanded via using the definition of the curvature coefficient matrix and the product rule to the following useful forms:
\begin{eqnarray*}\label{eq:60h}
   &-\partial_{\xi_1} \left(\rho U_2^2 \dfrac{U_1}{h_2}\right)= -\dfrac{\rho U_2^2}{h_2}\partial_{\xi_1} U_1 - \dfrac{2 \rho U_1 U_2}{h_2}\partial_{\xi_1} U_2 + \dfrac{h_1}{h_2} \rho U_1 U_2^2 \theta_{21},\\
 &-\partial_{\xi_2} \left(\rho U_1^2 \dfrac{U_2}{h_1}\right)= -\dfrac{\rho U_1^2}{h_1}\partial_{\xi_2} U_2 - \dfrac{2 \rho U_1 U_2}{h_1}\partial_{\xi_2} U_1 + \dfrac{h_2}{h_1} \rho U_1^2 U_2 \theta_{12},\\
    &-\partial_{\xi_1} \Bigg[ \left(h_2 + \dfrac{c_{s}^2}{h_2}\right) \rho U_1\Bigg]= -\left( h_2 + \dfrac{c_{s}^2}{h_2}\right)\rho \partial_{\xi_1} U_1-\left(h_2 + \dfrac{c_{s}^2}{h_2}\right)U_1 \partial_{\xi_1} \rho -  \rho U_1 \left( h_1 h_2 - \dfrac{h_1}{h_2} c_{s}^2 \right)\theta_{21},\\
  &-\partial_{\xi_2} \Bigg[ \left( h_1 + \dfrac{c_{s}^2}{h_1}\right) \rho U_2\Bigg]= -\left( h_1 + \dfrac{c_{s}^2}{h_1}\right)\rho \partial_{\xi_2} U_2-\left( h_1 +\dfrac{c_{s}^2}{h_1}\right)U_2 \partial_{\xi_2} \rho -\rho U_2 \left(  h_1 h_2 - \dfrac{h_2}{h_1} c_{s}^2 \right)\theta_{12}.
\end{eqnarray*}
Then, using these expressions for the spatial derivatives and also substituting for the time derivatives given in Eqs.~(\ref{eq:60f}) and (\ref{eq:60g}) along with the expression for the temporal derivative of density provided in Eq.~(\ref{eq:td_density}) into Eq.~(\ref{eq:57}), after considerable simplifications and rearrangements that involve cancellations of various cubic velocity terms associated with the truncation errors, we get the final expression for the non-equilibrium moment $n_3^{(1)}$. Here,  we omit the details of such lengthy algebraic manipulations to maintain brevity of the presentation and focus on the end result of such simplifications. We write the final expression in a form that can be compared with that for the standard LBE based on the uniform grid, i.e., express it in the following form
\begin{align}\label{eq:63a}
&n_3^{(1)} = -\frac{2  c_{s}^2}{\omega_3}  \rho \left(\partial_{\xi_1} U_1+ \partial_{\xi_2} U_2\right)+ \mathcal{E}_{3},
\end{align}
where $\mathcal{E}_{3}$ represents the deviation arising from the use of clustered/curvilinear grids via the GOC when compared to the uniform grid and is given by
\begin{eqnarray}\label{eq:63b}
&\mathcal{E}_{3} = \dfrac{1}{\omega_3}\Bigg\{ \left[2 c_{s}^2 - h_2 + \left(3 U_1^2 + c_{s}^2\right)\dfrac{h_2}{h_1^2} \right]\rho \partial_{\xi_1} U_1 + \left[2 c_{s}^2 -  h_1 + \left(3 U_2^2 + c_{s}^2\right)\dfrac{h_1}{h_2^2} \right]\rho \partial_{\xi_2} U_2+\nonumber\\
&\quad\quad\quad\;\; \left[\left(3 c_{s}^2 - U_1^2 \right)\dfrac{h_2}{h_1^2}- h_2 - \dfrac{U_2^2}{h_2}\right] U_1 \partial_{\xi_1} \rho +
\left[\left(3 c_{s}^2 - U_2^2 \right)\dfrac{h_1}{h_2^2}-  h_1 - \dfrac{U_1^2}{h_1}\right] U_2 \partial_{\xi_2} \rho\nonumber\\
&-2 \rho U_1 \left( U_1^2+  c_{s}^2 \right) \dfrac{h_2}{h_1} \theta_{11}-2 \rho U_2 \left( U_2^2+  c_{s}^2 \right) \dfrac{h_1}{h_2} \theta_{22}\nonumber\\
&+\rho U_1 \left[\dfrac{h_2}{h_1}\left(3 c_{s}^2 + U_1^2 \right)+ 2 c_{s}^2 \dfrac{h_1}{h_2}- h_1 h_2 \right]\theta_{21}\nonumber\\
&\;\;\;+\rho U_2 \left[\dfrac{h_1}{h_2}\left(3 c_{s}^2 + U_2^2 \right)+ 2 c_{s}^2 \dfrac{h_2}{h_1}- h_1 h_2 \right]\theta_{12}  \Bigg\}.
\end{eqnarray}

\subsubsection{Determination of the non-equilibrium moment $n_4^{(1)} = k_{20}^{(1)\prime} - k_{02}^{(1)\prime}$}
Comparing Eq.~(\ref{eq:58}) with Eq.~(\ref{eq:57}), it is evident that the terms that appear in $n_4^{(1)}$ have the same form as those in
$n_3^{(1)}$, but with different signs in various places. Then, following the same steps as indicated above for the $n_3^{(1)}$ case, it can be shown that the corresponding combinations of the source moments satisfy
\begin{align*}
&\left(\sigma_{20}^{\prime}-\sigma_{02}^{\prime}\right)= \frac{2F_1 U_1}{h_1}-\frac{2F_2 U_2}{h_2},
\end{align*}
which confirms the expression shown in Eq.~(\ref{eq:GOCmomentsources}) for this case, and the final expression for the non-equilibrium moment $n_4^{(1)}$ after considerable algebraic manipulations and simplifications reads as
\begin{eqnarray}\label{eq:66a}
&n_4^{(1)} = -\dfrac{2  c_{s}^2}{\omega_4}  \rho \left(\partial_{\xi_1} U_1- \partial_{\xi_2} U_2\right)+ \mathcal{E}_{4}.
\end{eqnarray}
Here, the deviation term relative to that of the uniform grid case is $\mathcal{E}_{4}$ and is given by
\begin{eqnarray}\label{eq:66b}
&\mathcal{E}_{4} = \dfrac{1}{\omega_4}\Bigg\{ \left[2 c_{s}^2 - h_2 + \left(3 U_1^2 + c_{s}^2\right)\dfrac{h_2}{h_1^2} \right]\rho \partial_{\xi_1} U_1 + \left[-2 c_{s}^2 +  h_1 - \left(3 U_2^2 + c_{s}^2\right)\dfrac{h_1}{h_2^2} \right]\rho \partial_{ \xi_2} U_2 + \nonumber \\
&\quad\quad\quad\;\;\left[\left(3 c_{s}^2 - U_1^2 \right)\dfrac{h_2}{h_1^2}- h_2 + \dfrac{U_2^2}{h_2}\right] U_1 \partial_{\xi_1} \rho +
\left[-\left(3 c_{s}^2 - U_2^2 \right)\dfrac{h_1}{h_2^2}+  h_1 - \dfrac{U_1^2}{h_1}\right] U_2 \partial_{\xi_2} \rho \nonumber\\
&-2 \rho U_1 \left( U_1^2+  c_{s}^2 \right) \dfrac{h_2}{h_1} \theta_{11}+2 \rho U_2 \left( U_2^2+  c_{s}^2 \right) \dfrac{h_1}{h_2} \theta_{22}\nonumber\\
&+ \rho U_1 \left[ \left(3 c_{s}^2 + U_1^2 \right) \dfrac{h_2}{h_1}- 2 c_{s}^2 \dfrac{h_1}{h_2}- h_1 h_2 \right]\theta_{21}\nonumber\\
&\;\;\;\;\;\;+ \rho U_2 \left[ -\left(3 c_{s}^2 + U_2^2 \right) \dfrac{h_1}{h_2}+ 2 c_{s}^2 \dfrac{h_2}{h_1}+  h_1 h_2 \right]\theta_{12}  \Bigg\}.
\end{eqnarray}

We note here that the non-equilibrium moments derived in Eqs.~(\ref{eq:63a}) and (\ref{eq:66a}) for $n_3^{(1)}$ and $n_4^{(1)}$, respectively, unlike those for the standard LBE for the uniform grid, do not exactly correspond to the expressions for the normal stress components $\tau_{11}$ and $\tau_{22}$ in the GOC given in Eq.~(\ref{eq:viscousstress_OCC_modified3}). Hence, the deviation terms $\mathcal{E}_3$ and $\mathcal{E}_4$ include the errors in achieving this correspondence. Moreover, due to the discrete nature of the D2Q9 lattice used, it is associated with aliasing effects, such as the degeneracy of the third order longitudinal moments to the first order moments, i.e., $\sum_\alpha f_\alpha e_{\xi_i}^3 = \sum_\alpha f_\alpha e_{\xi_i}$, which manifest as cubic velocity non-Galilean invariant (GI) errors and they are also contained in the deviation terms $\mathcal{E}_3$ and $\mathcal{E}_4$ shown in Eqs.~(\ref{eq:63b}) and (\ref{eq:66b}), respectively; in fact, these terms generalize the non-GI cubic velocity errors previously identified for the uniform grid cases using the Cartesian coordinates~\cite{dellar2014lattice,hajabdollahi2018galilean} to the case of clustered/curvilinear grids based on the GOC. In the next section (Sec.~\ref{sec:GOC-LBM}), via a further analysis using the C-E expansions based on introducing extended moment equilibria, we construct the necessary correction terms that eliminate both the errors associated with the correct representation of the normal stress components as well non-GI cubic velocity artifacts and recover the NSE in the GOC consistently.

\subsubsection{Determination of the non-equilibrium moment $n_5^{(1)} = k_{11}^{(1)\prime}$}
Substituting Eqs.~(\ref{eq:td_density}), (\ref{eq:td_momentum1}) and (\ref{eq:td_momentum2}) into Eq.~(\ref{eq:td_momflux12}) as well as invoking the definition of the curvature coefficient matrix elements $\theta_{ij}$ presented in Eq.~(\ref{eq:curvaturematrix}), we can rewrite the time derivative $\partial_{t_0} (\rho U_1 U_2)$ as follows:
\begin{eqnarray*}\label{eq:td_rhoU1U2}
-\partial_{t_0} \left(\rho U_1 U_2\right)& = &\rho\left[ \partial_{\xi_1}\left(\frac{U_1^2U_2}{h_1}\right) + \partial_{\xi_2}\left(\frac{U_1U_2^2}{h_2}\right)\right]  + c_s^2\left[ \dfrac{U_2}{h_1}\partial_{\xi_1} \rho + \dfrac{U_1}{h_2}\partial_{\xi_2}\rho \right] -\left(F_1\dfrac{U_2}{h_2} + F_2\dfrac{U_1}{h_1}\right)\nonumber\\
&&\quad-\rho c_s^2 \left[U_1(\theta_{22}-\theta_{12}) + U_2(\theta_{11}-\theta_{21}) \right].
\end{eqnarray*}
It can be seen that when this last equation is substituted into Eq.~(\ref{eq:59}), the spurious terms associated with the effective body force contributions that appear as $-F_1U_2/h_2 - F_2 U_1/h_1$ can be cancelled if we select the source moment $\sigma_{11}^{\prime}$ as
\begin{align*}
  &\sigma_{11}^{\prime}= F_1 \frac{ U_2}{h_2} + F_2 \frac{U_1}{h_1}.
\end{align*}
which provides the basis for the respective expression given in Eq.~(\ref{eq:GOCmomentsources}) for this case. Then, after considerable simplifications and rearrangements, we finally obtain a compact form of the non-equilibrium moment $n_5^{(1)}$, which reads as
\begin{equation}\label{eq:58a}
n_5^{(1)} = -\dfrac{ \rho c_{s}^2}{\omega_5} \bigg\{ \partial_{\xi_1} \left( \frac{U_2}{h_1} \right) + \partial_{\xi_2} \left( \dfrac{U_1}{h_2} \right) + \left[ U_1 \left(\theta_{22} - \theta_{12}\right) + U_2 \left(\theta_{11}-\theta_{21}\right)\right] \bigg\}.
\end{equation}
This equation for $n_5^{(1)}$ \emph{exactly} recovers the expression for the shear stress components $\tau_{12} = \tau_{21}$ in the GOC shown in  Eq.~(\ref{eq:viscousstress_OCC_modified3_shear}) upon relating its prefactor to the shear viscosity as will be done in the next section (Sec.~\ref{sec:GOC-LBM}). In other words, as one of the advantages maintaining the orthogonality of grids in the GOC (like in the Cartesian coordinates as a special case), the shear stress components are exactly represented by the appropriate choices of the moment equilibria made in Eq.~(\ref{eq:GOCmomentequilibria}) and in particular for the moments $k_{21}^{eq\prime}$ and $k_{12}^{eq\prime}$, and the formulation is free of errors due to cubic velocity non-GI artifacts just like in the case of the uniform grid-based standard LBE, i.e., there is no deviation term arise for this case and we can set $\mathcal{E}_{5} = 0$. Moreover, generalizing the well-known result of using the uniform grid-based LBE, Eq.~(\ref{eq:58a}) provides a way to compute the shear stress components \emph{locally} based on the non-equilibrium moment $n_5^{(1)} = k_{11}^{(1)\prime}$ for clustered/curvilinear grids (see Appendix~\ref{sec:Appendix_shear_stress_local} for details), which will be exploited later in the algorithmic implementations of the resulting GOC-LBM.

\section{Chapman-Enskog Analysis of Raw Moment-based LBE in GOC: Elimination of Remaining Errors in Viscous Fluxes via Extended Moment Equilibria Corrections and Recovery of the NSE in GOC}\label{sec:GOC-LBM}
In order to eliminate the errors in representing the normal components of the stress, i.e., $\tau_{11}$ and $\tau_{22}$, as well as in avoiding the non-GI cubic velocity artifacts as identified in the paragraph below Eq.~(\ref{eq:66b}), we now introduce corrections to the moment equilibria, designated as $\bm{n}^{eq(1)}$, to the base equilbrium moments $\mathbf{n}^\mathit{eq}$ given in Eq.~(\ref{eq:GOCmomentequilibria}), and write the resulting effective moment equilibria as
\begin{eqnarray}\label{eq:67a}
&\mathbf{n}^\mathit{eq,eff}= \mathbf{n}^\mathit{eq}+ \mathbf{n}^\mathit{eq(1)},
\end{eqnarray}
Inspecting the deviation terms $\mathcal{E}_3$ and $\mathcal{E}_4$ given in Eqs.~(\ref{eq:63b}) and (\ref{eq:66b}), respectively, we note that they depend on the normal spatial derivatives along the $\xi_1$ and $\xi_2$ directions of the velocity and density fields, viz., $\partial_{\xi_1}U_1$, $\partial_{\xi_2}U_2$, $\partial_{\xi_1}\rho$, and $\partial_{\xi_2}\rho$ and together with some terms related to the curvature effects of the GOC via $\theta_{ij}$. Hence, the correction terms should depend on these quantities and parameterized by some unknown coefficients. Moreover, importantly, the deviation terms occur only in the diagonal components of the second order non-equilibrium moments, viz., $n_3^{(1)}$ and $n_4^{(1)}$. Hence, the extended equilibria corrections should only need to be associated with the respective second order moments, i.e., $n^{eq(1)}_j$ for $j = 3, 4$ and the rest of them can be set to zero. That is,
\begin{eqnarray}\label{eq:67b}
n^{eq(1)}_j =
\begin{cases}
D_{3,1} \partial_{\xi_1} U_1  +  D_{3,2} \partial_{\xi_2} U_2 + D_{3,3} \partial_{\xi_1} \rho + D_{3,4} \partial_{\xi_2} \rho +C_3 & \quad j =3\\
D_{4,1} \partial_{\xi_1} U_1  +  D_{4,2} \partial_{\xi_2} U_2 + D_{4,3} \partial_{\xi_1} \rho + D_{4,4} \partial_{\xi_2} \rho +C_4 & \quad j =4\\
0 & \quad \mbox{otherwise},
\end{cases}
\end{eqnarray}
where $D_{3,k}$ and $D_{4,k}$ for $k=1,2,3,4$, and $C_{3}$ and $C_{4}$ are the unknown coefficients to be determined in such a way that the above identified errors are eliminated and the NSE in the GOC derived earlier are recovered.

We then apply a modified C-E analysis by replacing the base equilibria with the effective moment equilibria in the expansion given earlier in Eq.~(\ref{eq:53}) as
\begin{eqnarray} \label{eq:68ab}
\mathbf{n} &=& \mathbf{n}^\mathit{eq,eff}+\epsilon \mathbf{n}^{(1)} + \epsilon^2 \mathbf{n}^{(2)}+ \ldots = \mathbf{n}^{(0)}+\underline{\epsilon \mathbf{n}^{eq(1)}} +\epsilon \mathbf{n}^{(1)}+ \epsilon^2 \mathbf{n}^{(2)}+ \ldots \nonumber\\
\partial_t&=&\partial_{t_0} +\epsilon \partial_{t_1} + \epsilon^2 \partial_{t_2}+ \ldots,
\end{eqnarray}
which thus now includes the correction term $\epsilon \mathbf{n}^{eq(1)}$ that is shown as underlined.
As a result of this change, the moment system at $O(\epsilon^k)$ for $k=0, 1$ and $2$ given earlier in Eq.~(\ref{eq:54}) are now replaced with the following:
\begin{subequations}\label{eq:69}
\begin{eqnarray}
\centering
&&\hspace{-1.8em}O (\epsilon^0 ):  \mathbf{n}^{(0)} =  \mathbf{n}^{eq},  \label{eq:69a}\\
&&\hspace{-1.8em}O (\epsilon^1 ): \left(\partial_{t_0} + \tensor{E}_i \partial_i\right)  \mathbf{n}^{(0)} =  - \tensor{\Lambda} \left( \mathbf{n}^{(1)}- \underline{\mathbf{n}^{eq(1)}}\right)+\mathbf{\Psi},  \label{eq:69b} \\
&&\hspace{-1.8em}O (\epsilon^2 ): \partial_{t_1} \; \mathbf{n}^{(0)} +  \left(\partial_{t_0} + \tensor{E}_i \partial_i\right) \;\left[ \tensor{I} - \dfrac{\tensor{\Lambda}} {2} \right] \mathbf{n}^{(1)}+ \partial_{t_0}\left( \dfrac{\tensor{\Lambda}} {2} \underline{\mathbf{n}^{eq(1)}}\right)+ \partial_{i}\left(\tensor{E}_i \dfrac{\tensor{\Lambda}} {2} \underline{\mathbf{n}^{eq(1)}}\right) =  - \tensor{\Lambda} \; \mathbf{n}^{(2)}. \label{eq:69c}
\end{eqnarray}
\end{subequations}
For convenience, the changes relative to Eq.~(\ref{eq:54}) in the previous section are highlighted by the underlined terms in Eq.~(\ref{eq:69}). The derivation performed out in the last section to determine the second order non-equilibrium moments thus carries over here but with the inclusion of the equilibrium moment correction term $\mathbf{n}^{eq(1)}$. In particular, the non-equilibrium moments for the components $j= 3$ and $4$, based on Eqs.~(\ref{eq:63a}) and (\ref{eq:66a}), i.e., $n_3^{(1)}$ and $n_4^{(1)}$ now modify to include the as yet unknown corrections $n_3^{eq(1)}$ and $n_4^{eq(1)}$, respectively, while the $n_5^{(1)}$ given in Eq.~(\ref{eq:58a}) remains the same. Thus, we can summarize the non-equilibrium second order moments due to the inclusion of the corrections as
\begin{subequations}
\begin{eqnarray}
n_3^{(1)}&=& -\dfrac{ 2 c_{s}^2}{\omega_3}  \rho \left(\partial_{\xi_1} U_1 + \partial_{\xi_2} U_2 \right)+ \mathcal{E}_{3}+ \underline{n_3^{eq(1)}},\label{eq:70a}\\
n_4^{(1)}&=& -\dfrac{ 2 c_{s}^2}{\omega_4}  \rho \left(\partial_{\xi_1} U_1 - \partial_{\xi_2} U_2 \right)+ \mathcal{E}_{4}+ \underline{n_4^{eq(1)}},\label{eq:70b}\\
n_5^{(1)}&=& -\dfrac{ \rho c_{s}^2}{\omega_5} \bigg\{ \partial_{\xi_1} \left( \frac{U_2}{h_1} \right) + \partial_{\xi_2} \left( \dfrac{U_1}{h_2} \right) + \left[ U_1 \left(\theta_{22} - \theta_{12}\right) + U_2 \left(\theta_{11}-\theta_{21}\right)\right] \bigg\},\label{eq:70c}
\end{eqnarray}
\end{subequations}
where the deviation terms $\mathcal{E}_{3}$ and $\mathcal{E}_{4}$ are given in Eqs.~\eqref{eq:63b} and \eqref{eq:66b}, respectively.

Next, let's analyze the macroscopic hydrodynamical equations for the conserved moments for the mass density and the momentum in the GOC that arise from these modifications. As such, for the fast scale or the $O (\epsilon^1)$ moment system given in Eq.~(\ref{eq:69b}) when evaluated for the collision invariants or the conserved moments, i.e., for components with $j=0,1$ and $2$, since they do not involve any corrections, they remain the same as those given in Eqs.~(\ref{eq:55a}), (\ref{eq:55b}) and (\ref{eq:55c}), respectively, and are not repeated again here.

On the other hand, for the slow scale or the $O (\epsilon^2)$ moment system given in Eq.~(\ref{eq:69c}), when evaluated for these same conserved moments with components $j=0,1$ and $2$ now modify from Eq.~(\ref{eq:56}) to the following due to the appearance of the matrix $\tensor{E}_i$ (defined earlier in the paragraph below Eq.~(\ref{eq:54})) multiplying the correction $\mathbf{n}^{eq(1)}$:
\begin{subequations}\label{eq:A}
\begin{eqnarray}
&&\partial_{t_1}\left(h_1 h_2 \rho\right)=0, \label{eq:A_a}\\
&&\partial_{t_1}\left(h_2 \rho U_1\right)+\partial_{\xi_1} \left[\dfrac{1}{2}\left(1-\dfrac{\omega_3}{2}\right) n_3^{(1)}+\dfrac{1}{2}\left(1-\dfrac{\omega_4}{2}\right)n_4^{(1)}\right] +\partial_{\xi_2} \left[\left(1-\dfrac{\omega_5}{2}\right)n_5^{(1)}\right] \nonumber\\
&&\qquad\qquad\quad + \partial_{\xi_1} \left[\dfrac{\omega_3}{4} \underline{n_3^{eq(1)}}+\dfrac{\omega_4}{4} \underline{n_4^{eq(1)}}\right] =0, \label{eq:A_b}\\
&&\partial_{t_1}\left(h_1 \rho U_2 \right) + \partial_{\xi_1} \left[\left(1 - \dfrac{\omega_5}{2}\right) n_5^{(1)}\right]+\partial_{\xi_2} \left[\dfrac{1}{2}\left(1-\dfrac{\omega_3}{2}\right)n_3^{(1)}-\dfrac{1}{2}\left(1-\dfrac{\omega_4}{2}\right)n_4^{(1)}\right]\nonumber\\
&&\qquad\qquad\quad + \partial_{\xi_2} \left[\dfrac{\omega_3}{4} \underline{n_3^{eq(1)}}-\dfrac{\omega_4}{4} \underline{n_4^{eq(1)}}\right]=0. \label{eq:A_c}
\end{eqnarray}
\end{subequations}
Thus, the slow time scale evolution of the conserved moments given in Eq.~(\ref{eq:A}) replace the previous Eq.~(\ref{eq:56}) with the appearance of the corrections $n_3^{eq(1)}$ and $n_4^{eq(1)}$ that are scaled by the respective relaxation rates as highlighted by the underline terms. Then, following the standard multiscale analysis procedure associated with the C-E expansions, to obtain the macroscopic hydrodynamical equations, we combine the equations at the fast timescale $t_0$ with $\epsilon$ times the respective equations at the slow timescale $t_1$. That is, combining Eq.~\eqref{eq:55a} + $\epsilon$ $\times$ Eq.~\eqref{eq:A_a}, Eq.~\eqref{eq:55b} + $\epsilon$ $\times$ Eq.~\eqref{eq:A_b} and Eq.~\eqref{eq:55c} + $\epsilon$ $\times$ Eq.~\eqref{eq:A_c}, and then requiring $\partial_ t=\partial_{t_0}+ \epsilon \partial_ {t_1}$ and setting $\epsilon = \Delta t$, we get the evolution equations of the conserved moments $h_1 h_2 \rho$, $h_2 \rho U_1$ and $h_1 \rho U_2 $. The results read as
\begin{subequations}\label{eq:71}
\begin{eqnarray}
&&\partial_t \left(h_1 h_2 \rho\right) + \partial_{\xi_1} \left(h_2 \rho U_1\right) + \partial_{\xi_2}\left(h_1 \rho U_2\right) = 0,\label{eq:71a}\\
&&\partial_t \left(h_2 \rho U_1\right) + \partial_{\xi_1}  \bigg[\dfrac{h_2}{h_1}(P  +\rho U_1^2)\bigg] + \partial_{\xi_2}\left(\rho U_1 U_2\right) = -\partial_{\xi_1} \left[\dfrac{1}{2}\left(1-\dfrac{\omega_3}{2}\right) n_3^{(1)} \Delta t+\dfrac{1}{2}\left(1-\dfrac{\omega_4}{2}\right)n_4^{(1)} \Delta t\right]\nonumber\\
&&\hspace{8.2cm} -\partial_{\xi_2} \bigg[\left(1-\dfrac{\omega_5}{2}\right)n_5^{(1)}\Delta t\bigg]\nonumber\\
&&\hspace{8.2cm} -\partial_{\xi_1} \bigg[\dfrac{\omega_3}{4} n_3^{eq(1)} \Delta t +\dfrac{\omega_4}{4} n_4^{eq(1)} \Delta t\bigg] + F_1,\label{eq:71b}\\
&&\partial_t \left(h_1 \rho U_2\right) + \partial_{\xi_1} \left(\rho U_1 U_2\right)+ \partial_{\xi_2}\Bigg[\dfrac{h_1}{h_2}(P + \rho U_2^2)\Bigg] = -\partial_{\xi_2} \left[\dfrac{1}{2}\left(1-\dfrac{\omega_3}{2}\right)n_3^{(1)}\Delta t-  \dfrac{1}{2}\left(1-\dfrac{\omega_4}{2}\right)n_4^{(1)}\Delta t\right]\nonumber\\
&&\hspace{8.2cm}-\partial_{\xi_1} \bigg[\left(1-\dfrac{\omega_5}{2}\right)n_5^{(1)}\Delta t\bigg]\nonumber\\
&&\hspace{8.2cm}-\partial_{\xi_2} \left[\dfrac{\omega_3}{4} n_3^{eq(1)}\Delta t-\dfrac{\omega_4}{4} n_4^{eq(1)}\Delta t\right] + F_2.\label{eq:71c}
\end{eqnarray}
\end{subequations}


We now compare the above emergent evolution equations in Eqs.~\eqref{eq:71} with the target mass density and momentum conservation equations in the GOC system given in Eqs.~(\ref{eq:conservativemass_OCC_modified}),~(\ref{eq:conservmomentum_1}) and  (\ref{eq:conservmomentum_2}) in the GOC. From Eqs.~\eqref{eq:71a} and (\ref{eq:conservativemass_OCC_modified}), we see that the continuity equation is identically recovered as such by our formulation. Next, in order to recover the components of the momentum equation along the directions $\xi_1$ and $\xi_2$ by comparing Eq.~(\ref{eq:conservmomentum_1}) with Eq.~\eqref{eq:71b}, and similarly, Eq.~(\ref{eq:conservmomentum_2}) with Eq.~\eqref{eq:71c}, we obtain the following key constraint relations on the second order non-equilibrium moments together with appropriate corrections related to the components of the stress tensor in the GOC:
\begin{subequations}\label{eq:72}
\begin{align}
&-\left(1-\frac{\omega_5}{2}\right) n_5^{(1)}\Delta t = \tau_{21}= \tau_{12},\label{eq:72a}\\
&-\frac{1}{2}\left(1-\frac{\omega_3}{2}\right) n_3^{(1)}\Delta t- \frac{1}{2}\left(1-\frac{\omega_4}{2}\right) n_4^{(1)}\Delta t
-\frac{\omega_3}{4} n_3^{eq(1)} \Delta t -\frac{\omega_4}{4} n_4^{eq(1)} \Delta t=\frac{h_2}{h_1}\tau_{11},\label{eq:72b} \\
&-\frac{1}{2} \left(1-\frac{\omega_3}{2}\right) n_3^{(1)}\Delta t + \frac{1}{2} \left(1-\frac{\omega_4}{2}\right) n_4^{(1)}\Delta t
-\frac{\omega_3}{4} n_3^{eq(1)} \Delta t + \frac{\omega_4}{4} n_4^{eq(1)} \Delta t=\frac{h_1}{h_2}\tau_{22}.\label{eq:72c}
\end{align}
\end{subequations}

Then, substituting the non-equilibrium moment $n_5^{(1)}$ given in Eq.~\eqref{eq:70c} into the first of the above equations, i.e., Eq.~\eqref{eq:72a}, we obtain
\begin{equation}\label{eq:shearstress_noeq5}
\rho c_s^2 \left(\dfrac{1}{\omega_5} -\dfrac{1}{2}\right)\Delta t \Bigg\{ \partial_{\xi_1}\left( \dfrac{U_2}{h_1} \right) + \partial_{\xi_2}\left( \dfrac{U_1}{h_2} \right) + U_1 \left( \theta_{22} - \theta_{12} \right)  +U_2 \left( \theta_{11} - \theta_{21} \right) \Bigg\} = \tau_{12} =\tau_{21},
\end{equation}
which when compared with the shear stress components $\tau_{21} = \tau_{12}$ in the GOC given in Eq.~(\ref{eq:viscousstress_OCC_modified3_shear}) become exactly equal to one another if we relate the relaxation rate $\omega_5$ to the kinematic shear viscosity of the fluid $\nu$ by the following expression:
\begin{equation}\label{eq:73}
\nu= c_s^2 \left(\frac{1}{\omega_5} -\dfrac{1}{2}\right)\Delta t.
\end{equation}
Interestingly, this expression is identical to that used in the standard LBE for the uniform grid and is also applicable for the GOC-LBE using clustered/curvilinear grids.

Then, substituting for the non-equilibrium moments $n_3^{(1)}$ and $n_4^{(1)}$ from Eqs.~\eqref{eq:70a} and~\eqref{eq:70b}, respectively, into the remaining constraint expressions shown in Eqs.~\eqref{eq:72b} and \eqref{eq:72c}, we get the required relationships between the necessary corrections and the deviation terms, and their dependence on the target GOC-based normal stress components $\tau_{11}$ and $\tau_{22}$, respectively. The results read as follows:
\begin{eqnarray}\label{eq:74}
&&\hspace{-1.5cm}\rho c_s^2 \left(\dfrac{1}{\omega_3} -\dfrac{1}{2}\right)\Delta t \left( \partial_{\xi_1} U_1 + \partial_{\xi_2} U_2 \right) - \dfrac{1}{2}\left(1-\dfrac{\omega_3}{2}\right) \left[\mathcal{E}_3+  n_3^{eq(1)}\right]\Delta t -\dfrac{\omega_3}{4} n_3^{eq(1)} \Delta t + \nonumber\\
&&\hspace{-1.5cm}\rho c_s^2 \left(\dfrac{1}{\omega_4} -\dfrac{1}{2}\right)\Delta t \left( \partial_{\xi_1} U_1 - \partial_{\xi_2} U_2 \right)- \dfrac{1}{2}\left(1-\dfrac{\omega_4}{2}\right) \left[\mathcal{E}_4+  n_4^{eq(1)}\right]\Delta t  - \dfrac{\omega_4}{4} n_4^{eq(1)} \Delta t  \hspace{0.5cm}=\dfrac{h_2}{h_1}\tau_{11},
\end{eqnarray}
and
\begin{eqnarray}\label{eq:75}
&&\hspace{-1.5cm}\rho c_s^2 \left(\dfrac{1}{\omega_3} -\dfrac{1}{2}\right)\Delta t \left( \partial_{\xi_1} U_1 + \partial_{\xi_2} U_2 \right) - \dfrac{1}{2}\left(1-\dfrac{\omega_3}{2}\right) \left[\mathcal{E}_3+  n_3^{eq(1)}\right]\Delta t  -\dfrac{\omega_3}{4} n_3^{eq(1)} \Delta t - \nonumber\\
&&\hspace{-1.5cm}\rho c_s^2 \left(\dfrac{1}{\omega_4} -\dfrac{1}{2}\right)\Delta t \left( \partial_{\xi_1} U_1 - \partial_{\xi_2} U_2 \right) + \dfrac{1}{2}\left(1-\dfrac{\omega_4}{2}\right) \left[\mathcal{E}_4+  n_4^{eq(1)}\right]\Delta t + \dfrac{\omega_4}{4} n_4^{eq(1)} \Delta t  \hspace{0.5cm}=\dfrac{h_1}{h_2}\tau_{22}.
\end{eqnarray}
For simplicity, in order to make the parametric relationships between the relaxation rates $\omega_3$ and $\omega_4$ and the transport coefficients for the kinematic bulk viscosity $\zeta$ and the kinematic shear viscosity $\nu$ to be of the same forms as those used for the standard LBE based on the uniform grid for the normal stress components given in the above two equations, we set them as
\begin{align}\label{eq:76}
& \zeta= c_s^2 \left(\frac{1}{\omega_3} -\dfrac{1}{2}\right)\Delta t, \quad \quad\quad \nu= c_s^2 \left(\frac{1}{\omega_4} -\dfrac{1}{2}\right)\Delta t.
\end{align}

Then, the two constraint relationships given in Eqs.~\eqref{eq:74} and \eqref{eq:75} needed to recover the NSE in the GOC upon the substitution of Eq.~(\ref{eq:76}) simplify as follows:
\begin{subequations}
\begin{eqnarray}
n_3^{eq(1)} \Delta t + n_4^{eq(1)} \Delta t &=& 2\rho\left( \zeta + \nu \right)\partial_{\xi_1} U_1+ 2\rho\left( \zeta - \nu \right) \partial_{\xi_2} U_2 - 2 \dfrac{h_2}{h_1}\tau_{11}\nonumber\\
&& -\left(1-\dfrac{\omega_3}{2}\right)\mathcal{E}_3 \Delta t- \left(1-\dfrac{\omega_4}{2}\right)\mathcal{E}_4 \Delta t, \label{eq:77a}\\
n_3^{eq(1)} \Delta t - n_4^{eq(1)} \Delta t &=& 2\rho\left( \zeta - \nu \right)\partial_{\xi_1} U_1+ 2\rho\left( \zeta + \nu \right) \partial_{\xi_2} U_2 - 2 \dfrac{h_1}{h_2}\tau_{22}\nonumber\\
&& -\left(1-\dfrac{\omega_3}{2}\right)\mathcal{E}_3 \Delta t+ \left(1-\dfrac{\omega_4}{2}\right)\mathcal{E}_4 \Delta t.\label{eq:77b}
\end{eqnarray}
\end{subequations}
These Eqs.~\eqref{eq:77a} and \eqref{eq:77b} can be readily solved for the necessary corrections $n_3^{eq(1)}$ and $n_4^{eq(1)}$ directly in terms of the deviations $\mathcal{E}_3$ and $\mathcal{E}_4$ and the desired target normal stress components $\tau_{11}$ and $\tau_{22}$ with appropriate scalings along with their dependence of normal velocity derivatives and the transport coefficients so that the momentum equations of the NSE in the GOC given in Eqs.~(\ref{eq:conservmomentum_1}) and~(\ref{eq:conservmomentum_2}) are also recovered. The final expressions read as
\begin{subequations}
\begin{eqnarray}\label{eq:78}
& n_3^{eq(1)} \Delta t =2 \rho \zeta \partial_{\xi_1} U_1+ 2 \rho \zeta \partial_{\xi_2} U_2 -\dfrac{h_2}{h_1}\tau_{11}- \dfrac{h_1}{h_2}\tau_{22} - \left(1-\dfrac{\omega_3}{2}\right)\mathcal{E}_3 \Delta t, \label{eq:78a}\\
& n_4^{eq(1)} \Delta t =2 \rho \nu \partial_{\xi_1} U_1- 2 \rho \nu \partial_{\xi_2} U_2 -\dfrac{h_2}{h_1}\tau_{11}+ \dfrac{h_1}{h_2}\tau_{22} - \left(1-\dfrac{\omega_4}{2}\right)\mathcal{E}_4 \Delta t,\label{eq:78b}
\end{eqnarray}
\end{subequations}
It now remains to substitute for the normal stress components as well as the deviation terms to bring them to forms that are suitable for algorithmic implementations. In this regard, fulfilling the first of these, using Eqs.~(\ref{eq:viscousstress_OCC_modified3_normal11}) and~~(\ref{eq:viscousstress_OCC_modified3_normal22}) for $\tau_{11}$ and $\tau_{22}$, respectively, we get
\begin{subequations}
\begin{eqnarray}\label{eq:79}
n_3^{eq(1)} \Delta t &=&2 \rho \zeta \partial_{\xi_1} U_1+ 2 \rho \zeta \partial_{\xi_2} U_2 -
  \bigg[\rho \left( \nu + \zeta \right) \dfrac{h_2}{h_1^2} \dfrac{\partial U_1}{\partial \xi_1} + \rho \left( \zeta- \nu  \right) \dfrac{1}{h_1} \dfrac{\partial U_2}{\partial \xi_2} + \dfrac{h_2}{h_1}\tau^c_{11}\bigg] \nonumber\\
&& -\bigg[\rho \left( \nu + \zeta \right) \dfrac{h_1}{h_2^2} \dfrac{\partial U_2}{\partial \xi_2} + \rho \left( \zeta- \nu \right) \dfrac{1}{h_2} \dfrac{\partial U_1}{\partial \xi_1} + \dfrac{h_1}{h_2}\tau^c_{22}\bigg]- \left(1-\dfrac{\omega_3}{2}\right)\mathcal{E}_3 \Delta t, \label{eq:79a}\\
n_4^{eq(1)} \Delta t &=& 2 \rho \nu \partial_{\xi_1} U_1- 2 \rho \nu \partial_{\xi_2} U_2 -
 \bigg[\rho \left( \nu + \zeta \right) \dfrac{h_2}{h_1^2} \dfrac{\partial U_1}{\partial \xi_1} + \rho \left( \zeta- \nu \right) \dfrac{1}{h_1} \dfrac{\partial U_2}{\partial \xi_2} + \dfrac{h_2}{h_1}\tau^c_{11}\bigg]  \nonumber\\
&&+ \bigg[\rho \left( \nu + \zeta \right) \dfrac{h_1}{h_2^2} \dfrac{\partial U_2}{\partial \xi_2} + \rho \left( \zeta- \nu \right) \dfrac{1}{h_2} \dfrac{\partial U_1}{\partial \xi_1} + \dfrac{h_1}{h_2}\tau^c_{22}\bigg]- \left(1-\dfrac{\omega_4}{2}\right)\mathcal{E}_4 \Delta t,\label{eq:79b}
\end{eqnarray}
\end{subequations}
Here, $\tau_{11}^c$ and $\tau_{22}^c$ correspond to the curvature related contributions to the normal stress components, which are presented in Eq.~(\ref{eq:tauc}).

Finally, upon substituting for the deviation terms $\mathcal{E}_3$ and $\mathcal{E}_4$ which are derived earlier in Eqs.~\eqref{eq:63b} and \eqref{eq:66b}, respectively, into Eqs.~(\ref{eq:79a}) and (\ref{eq:79b}), respectively, we finally arrive at one of the main results of this paper, viz., the moment equilibria corrections $n_3^{eq(1)}$ and $n_4^{eq(1)}$ necessary to recover the NSE in the GOC consistently including the constitutive relations for the normal stress components and without any non-GI cubic velocity artifacts. The resulting expressions are simplified and rearranged in more convenient forms involving the normal spatial derivatives of the velocity field $\partial_{\xi_1} U_1$ and $\partial_{\xi_2} U_2$ as well as the derivatives of the density field $\partial_{\xi_1} \rho$ and $\partial_{\xi_2} \rho$, which need to computed numerically such as such second order finite difference schemes, along the respective coefficients that depend on the metric factors $h_1$ and $h_2$, curvature coefficient elements $\theta_{11}$, $\theta_{22}$, $\theta_{21}$, and $\theta_{12}$, the transport coefficients $\nu$ and $\zeta$, and the local fluid density $\rho$ and the velocity components $U_1$ and $U_2$. The resulting moment equilibrium correction $n_3^{eq(1)}$ reads as
\begin{equation}\label{eq:82}
n_3^{eq(1)} \Delta t =D_{3,1} \partial_{\xi_1} U_1 + D_{3,2} \partial_{\xi_2} U_2+ D_{3,3} \partial_{\xi_1} \rho +D_{3,4} \partial_{\xi_2} \rho + C_3,
\end{equation}
where the associated coefficients are given by
\begin{subequations}\label{eq:83}
  \begin{eqnarray}
\hspace{-1cm}  D_{3,1}&=& \rho \left[ 2 \zeta - \left(\zeta+ \nu  \right)\frac{h_2}{h_1^2} - \left(\zeta- \nu  \right)\frac{1}{h_2}- \zeta \left\{2 -  \frac{h_2}{c_{s}^2}+ \left(3\frac{U_1^2}{c_{s}^2} +1 \right) \frac{h_2}{h_1^2}  \right\}  \right], \\
\hspace{-1cm}  D_{3,2}&=& \rho \left[ 2 \zeta - \left(\zeta+ \nu  \right)\frac{h_1}{h_2^2} - \left(\zeta- \nu  \right)\frac{1}{h_1}- \zeta \left\{2 -  \frac{h_1}{c_{s}^2}+ \left(3\frac{U_2^2}{c_{s}^2} +1 \right) \frac{h_1}{h_2^2}  \right\}  \right], \\
\hspace{-1cm}  D_{3,3}&=& - \zeta \left[\left(3-\frac{ U_1^2}{c_{s}^2}\right) \frac{h_2}{h_1^2} - \frac{h_2}{c_{s}^2}- \frac{U_2^2}{c_{s}^2} \frac{1}{h_2}\right] U_1, \\
\hspace{-1cm}  D_{3,4}&=& - \zeta \left[\left(3-\frac{ U_2^2}{c_{s}^2}\right) \frac{h_1}{h_2^2} -  \frac{h_1}{c_{s}^2}- \frac{U_1^2}{c_{s}^2} \frac{1}{h_1}\right] U_2, \\
\hspace{-1cm}  C_3 &=& - \left(\frac{h_2}{h_1} \tau_{11}^c + \frac{h_1}{h_2} \tau_{22}^c\right)- \rho \zeta \Bigg\{-2 U_1 \left(\frac{ U_1^2}{c_{s}^2} +1 \right)\frac{h_2}{h_1} \theta_{11}  -2 U_2 \left(\frac{ U_2^2}{c_{s}^2} +1 \right)\frac{h_1}{h_2} \theta_{22} \nonumber\\
\hspace{-1cm}   &&+ U_1 \left[  \left(3+\frac{ U_1^2}{c_{s}^2}\right)\frac{h_2}{ h_1} +2 \frac{h_1}{h_2} - \frac{h_1 h_2}{c_{s}^2} \right] \theta_{21} + U_2 \bigg[ \left(3+\frac{ U_2^2}{c_{s}^2}\right)\frac{h_1}{ h_2} +2 \frac{h_2}{h_1}-  \frac{h_1 h_2}{c_{s}^2} \bigg] \theta_{12}\Bigg\}.
  \end{eqnarray}
\end{subequations}
Similarly, the moment equilibrium correction $n_4^{eq(1)}$ can be written as
\begin{equation}\label{eq:84}
  n_4^{eq(1)} \Delta t =D_{4,1} \partial_{\xi_1} U_1 + D_{4,2} \partial_{\xi_2} U_2+ D_{4,3} \partial_{\xi_1} \rho +D_{4,4} \partial_{\xi_2} \rho + C_4,
\end{equation}
and their coefficients are expressed as
\begin{subequations}\label{eq:85}
  \begin{eqnarray}
\hspace{-1cm}  D_{4,1}&=& \rho \left[ 2 \nu - \left(\zeta+ \nu  \right)\frac{h_2}{h_1^2} + \left(\zeta- \nu  \right)\frac{1}{h_2}- \nu \left\{2 -  \frac{h_2}{c_{s}^2}+ \left(3\frac{U_1^2}{c_{s}^2} +1 \right) \frac{h_2}{h_1^2}  \right\}  \right], \\
\hspace{-1cm}  D_{4,2}&=& \rho \left[ -2 \nu + \left(\zeta+ \nu  \right)\frac{h_1}{h_2^2} - \left(\zeta- \nu  \right)\frac{1}{h_1}- \nu \left\{-2 +  \frac{h_1}{c_{s}^2}- \left(3\frac{U_2^2}{c_{s}^2} +1 \right) \frac{h_1}{h_2^2}  \right\}  \right], \\
\hspace{-1cm}  D_{4,3}&=& - \nu \left[\left(3-\frac{ U_1^2}{c_{s}^2}\right) \frac{h_2}{h_1^2} - \frac{h_2}{c_{s}^2}+ \frac{U_2^2}{c_{s}^2} \frac{1}{h_2}\right] U_1, \\
\hspace{-1cm}  D_{4,4}&=& - \nu \left[-\left(3-\frac{ U_2^2}{c_{s}^2}\right) \frac{h_1}{h_2^2} +  \frac{h_1}{c_{s}^2}- \frac{U_1^2}{c_{s}^2} \frac{1}{h_1}\right] U_2, \\
\hspace{-1cm}  C_4&=& - \left(\frac{h_2}{h_1} \tau_{11}^c - \frac{h_1}{h_2} \tau_{22}^c\right)- \rho \nu \Bigg\{-2 U_1 \left(\frac{ U_1^2}{c_{s}^2} +1 \right)\frac{h_2}{h_1} \theta_{11} +2 U_2 \left(\frac{ U_2^2}{c_{s}^2} +1 \right)\frac{h_1}{h_2} \theta_{22} \nonumber\\
\hspace{-1cm}   &&+ U_1 \bigg[  \left(3+\frac{ U_1^2}{c_{s}^2}\right)\frac{h_1}{ h_2} -2 \frac{h_1}{h_2}- \frac{h_1 h_2}{c_{s}^2} \bigg] \theta_{21} + U_2 \bigg[ -\left(3+\frac{ U_2^2}{c_{s}^2}\right)\frac{h_2}{ h_1} +2 \frac{h_2}{h_1}+  \frac{h_1 h_2}{c_{s}^2} \bigg] \theta_{12}\Bigg\}.
  \end{eqnarray}
\end{subequations}
Here, we remind again that the curvature related contributions to the normal stress components $\tau_{11}^c$ and $\tau_{22}^c$ appearing in the above are given in Eq.~(\ref{eq:tauc}). Note that the corrections terms contain the contributions that eliminate the non-GI cubic velocity artifacts due to the aliasing effects for the D2Q9 lattice in the general case of the orthogonally clustered/curvilinear grids thereby generalizing the previous results in this regard for the uniform grid case~\cite{dellar2014lattice,hajabdollahi2018galilean}.

Moreover, we emphasize that in Eqs.\eqref{eq:83} and~\eqref{eq:85}, when the metric factors are everywhere unity, i.e., $h_1=h_2=1$, when the grid is uniform with the same spacing everywhere and with the speed of sound reducing to the standard value $c_{s}^2=c_{s*}^2=1/3$ (see Eq.~(\ref{eq:GOC-LBE_CFL})), and by droppig any vanishing small terms that are higher than $O(U^3)$, i.e., higher than the cubic velocities, it is evident that the coefficients of the correction terms all vanish self consistently, i.e., $D_{3,k}=D_{4,k}=C_3=C_4=0 $ for $k=1,2,3,4$, or equivalently, the moment equilibria corrections become zero, i.e., $n_3^{eq(1)}=n_4^{eq(1)}=0$, which then reduces the GOC-LBE to the standard LBE for the uniform grids case as desired. This shows that the GOC-LBE is modular in construction and can be constructed from an existing LBE for the uniform grid by including appropriate moment equilibria corrections to the base equilibria and the source moments that are carefully designed from suitable scalings based on the metric factors as shown in detail in these last two sections. Moreover, for convenience, Appendix~\ref{sec:Appendix_B} provides a summary of some of the simplified forms of the corrections to the equilibria as well as the effective body force under various limiting special cases that could be exploited in implementations when appropriate.

\section{Raw Moment-based GOC-LBM: Matrix Formulation and Implementation Details, and Further Extensions}\label{sec:Reformulation GOC-LB}
We will now present an effective implementation of the GOC-LBM based on raw moments and the multiple relaxation times (MRT) that take into account all the developments in the previous sections by formulating it in a compact matrix form following our previous work~\cite{yahia2021central,yahia2021three,yahia2022preconditioned} on rectangular/cuboid lattice-based LB schemes. The previous sections on the application of the C-E analysis involved the use of two basis vectors with combined diagonal moments $\ket{e_{\xi_1}^2+e_{\xi_2}^2}$ and $\ket{e_{\xi_1}^2-e_{\xi_2}^2}$ to independently evolve the bulk and shear viscosity effects, and all the resulting basis vectors for the D2Q9 lattice were grouped together in the form of a transformation matrix $\tensor{T}$ (see Eqs.~(\ref{eq:transform_matrix_combined_moments}) along with (\ref{eq:8})). However, the use of such a combined moments-based transformation matrix is strictly needed only for the step involving collisions based on the relaxations of moments to their equilibria. Since the streaming step is carried out in terms of the distribution functions, the raw moment-based LBM necessarily involves the mappings between the distribution functions and the moments pre- and post-collisions and there is some flexibility in defining such mappings in terms of a transformation matrix simpler than $\tensor{T}$. Based on this consideration, we can introduce the following transformation matrix $\tensor{P}$:
\begin{eqnarray}\label{eq:Q-momentbasis}
\tensor{P}=\Big[ \; \ket{1}, \ket{e_{\xi_1}}, \ket{e_{\xi_2}},\ket{e_{\xi_1}^2},\ket{e_{\xi_2}^2},\ket{e_{\xi_1} e_{\xi_2}}, \ket{e_{\xi_1}^2 e_{\xi_2}}, \ket{e_{\xi_1} e_{\xi_2}^2}, \ket{e_{\xi_1}^2 e_{\xi_2}^2}\; \Big] ^{\dag},
\end{eqnarray}
where the 9-dimensional particle velocity components vectors $\ket{e_{\xi_1}}$ and $\ket{e_{\xi_2}}$ are defined in Eqs.~\eqref{eq:1a} and~\eqref{eq:1b}. As such, it is easier to use $\tensor{P}$  than $\tensor{T}$ especially when performing inverse mapping from moments to the distribution functions. Accordingly, we can define a simpler set of 9 independent bare forms of the raw moments $\mathbf{m}$ of the distribution functions $\mathbf{f}$, as well as the corresponding equilibria $\mathbf{m}^{eq}$, and the source moments $\mathbf{\Phi}$, which we write as
\begin{subequations} \label{baremoments-meq-phi}
\begin{eqnarray}
&\mathbf{m}=\left(k_{00}^\prime,k_{10}^\prime,k_{01}^\prime,k_{20}^\prime,k_{02}^\prime, k_{11}^\prime, k_{21}^\prime, k_{12}^\prime, k_{22}^\prime\right)^{\dag},\\
&\mathbf{m}^{eq}=\left(k_{00}^{eq\prime},k_{10}^{eq\prime},k_{01}^{eq\prime},k_{20}^{eq\prime},k_{02}^{eq\prime}, k_{11}^{eq\prime}, k_{21}^{eq\prime}, k_{12}^{eq\prime}, k_{22}^{eq\prime}\right)^{\dag},\\
&\mathbf{\Phi}=\left(\sigma_{00}^\prime,\sigma_{10}^\prime,\sigma_{01}^\prime,\sigma_{20}^\prime,\sigma_{02}^\prime, \sigma_{11}^\prime, \sigma_{21}^\prime, \sigma_{12}^\prime, \sigma_{22}^\prime\right)^{\dag}.
\end{eqnarray}
\end{subequations}
Then, the mappings between the quantities in the moment space and corresponding velocity space under $\tensor{P}$ can be expressed as
\begin{equation}
\mathbf{m}= \tensor{P} \mathbf{f}, \qquad \mathbf{f}={\tensor{P}}^{-1} \mathbf{m},
\end{equation}
and similarly, $\mathbf{m}^{eq}= \tensor{P} \mathbf{f}^{eq}$ and $\mathbf{\Phi}= \tensor{P} \mathbf{S}$, and their inverse mappings. Clearly, it is easier to apply ${\tensor{P}}^{-1}$ than ${\tensor{T}}^{-1}$. However, the collision step still needs to be formally based on $\tensor{T}$, which can be related to $\tensor{P}$ via introducing a block matrix $\tensor{B}$, i.e.,
\begin{equation}\label{eq:TandQrelation}
\tensor{T}= \tensor{B} \tensor{P},
\end{equation}
where $\tensor{B}$ represents the operation of combining the second order diagonal moments as $\ket{e_{\xi_1}^2+e_{\xi_2}^2}$ and $\ket{e_{\xi_1}^2-e_{\xi_2}^2}$ from $\ket{e_{\xi_1}^2}$ and $\ket{e_{\xi_2}^2}$. While Eq.~(\ref{eq:TandQrelation}) is defined to highlight this in a structural form, in practice, there is no need to implement this by performing matrix products but can rather be performed using some simple algebraic steps as shown later.

Based on the properties of these matrices and under some rearrangement, the raw moment-based GOC-LBE in Eq.~\eqref{eq:MRT-LBE} can be reformulated in an equivalent form given as (see Refs.~\cite{yahia2021central,yahia2021three,yahia2022preconditioned} for details).
\begin{equation}\label{eq:GOC-LBM-equivalent}
 \mathbf{f} (\bm{\xi}+\mathbf{e}\Delta t, t+\Delta t) = \tensor{P}^{-1} \Big[\mathbf{m} + \tensor{B}^{-1}\tensor{\Lambda}\;\left(\; \tensor{B}\mathbf{m}^{eq}-\tensor{B}\mathbf{m} \;\right) + \tensor{B}^{-1}\left(\tensor{I} - \frac{\tensor{\Lambda}}{2}\right)  \tensor{B}\mathbf{\Phi}\Delta t \Big].
\end{equation}
A more convenient way to represent the above equation is to split it into a sequence of sub-steps. Each of these steps can be implemented independently in a sequence according to the following:
\begin{eqnarray}
\mathbf{m}&=&\tensor{P}\mathbf{f},\nonumber\\
\tilde{\mathbf{m}}&=&\mathbf{m} + \tensor{B}^{-1}\left\{\tensor{\Lambda}\;\left(\; \tensor{B}\mathbf{m}^{eq}-\tensor{B}\mathbf{m} \;\right) + \left(\tensor{I} - \frac{\tensor{\Lambda}}{2}\right)  \tensor{B}\mathbf{\Phi}\Delta t\right\},\nonumber\\
\tilde{\mathbf{f}} (\bm{\xi},t) &=&  \tensor{P}^{-1}\tilde{\mathbf{m}},\nonumber\\
 \mathbf{f} (\bm{\xi}+\mathbf{e}\Delta t, t+\Delta t)&=&\tilde{\mathbf{f}} (\bm{\xi},t).\label{eq:LBErawmomentmatrixform}
\end{eqnarray}
Here, we emphasize that the moment equilibria $\tensor{B}\mathbf{m}^{eq}$ equates to $\mathbf{n}^{eq,eff}=\mathbf{n}^{eq}+\mathbf{n}^{eq(1)}$, which thus includes the base equilibria (Eq.~(\ref{eq:GOCmomentequilibria})) along with some associated corrections derived in the previous section to recover the NSE in the GOC, whereas the source moments appearing in $\Phi$ are given in Eq.~(\ref{eq:GOCmomentsources}). Before we discuss the explicit implementation details of each of these steps, since the GOC-LBM involves the use of metric factors to implement orthogonally clustered/curvilinear grids, we need to understand how the initial and boundary conditions in this approach can be specified consistently, since they are done somewhat differently than those with the standard LBM using uniform grids, which are discussed in Appendix~\ref{sec:initial_boundary_conditions}. This is followed by a presentation of the step-by-step algorithmic implementation details of the GOC-LBM based on raw moments, i.e., Eq.~(\ref{eq:LBErawmomentmatrixform}), in Appendix~\ref{sec: Appendix_A}. Moreover, as indicated earlier, our approach is general, in that it can be used with any collision model. Thus, extending the various elements of this section, Appendices~\ref{sec:Appendix_C} and~\ref{sec:Appendix_D} discuss the derivations and implementation details of the more general and robust GOC-LBM based on the central moments and MRT and the simplest GOC-LBM based on SRT with relaxations specified directly in the velocity space or the distribution functions, respectively.

\section{Some Common Stretching Functions for Orthogonal Grid Clustering}\label{sec:Transformations}
An important consideration in the implementation of the various new GOC-LBMs constructed in the previous sections and in the appendices is the generation of general orthogonally clustered/curvilinear grids. In general, there are several ways by which orthogonal grids can be generated~\cite{thompson1998handbook}. These include the use of algebraic stretching functions for regular or straight geometries (see e.g.,~\cite{roberts2005computational,vinokur1983one,liseikin1999grid,anderson2020computational}), conformal mappings for regular curvilinear geometries, and differential numerical grid generation methods based on the solution of certain elliptic partial differential equations (see e.g.,~\cite{thompson1974automatic,pope1978calculation,mobley1980numerical,ryskin1983orthogonal,duraiswami1992orthogonal,ecca19962d,akcelik2001nearly}).
While our approach is general and can be used with any appropriately generated GOC grid, the first two approaches provide useful analytic functions for readily setting up grids for simulating different canonical flow benchmark problems for the validations and the assessments of the efficacy of the GOC-LB schemes and will be our focus in this work. Such a strategy addresses our twofold objectives of the numerical study of this paper: As such, (i) even for flows with regular straight geometries, we aim to show that the GOC-LBM with its flexibility in using grid clustering can result in significant improvements in efficiency in resolving relatively thin boundary layer flows or small-scales flow features when compared to the uniform grid-based standard LBM; and (ii) for curved geometries, the GOC-LBM presents a natural framework for resolving them effectively using orthogonal curvilinear body-fitted grids, while preserving the simplicity and numerical features of the collide-and-stream steps.

In this section, for convenience, we collect together and summarize some common coordinate transformations and associated stretching functions for orthogonal grid clustering/coarsening for regular geometries to resolve boundary layers/shear layer regions effectively (see e.g.,~\cite{roberts2005computational,vinokur1983one,liseikin1999grid,anderson2020computational}). We will also discuss the development of an orthogonal curvilinear grid for flow between concentric rotating cylinders (by considering a sector of it) as part of validation for a body-fitted geometry in the later part of the next section.

\subsection{Grid Clustering Around Both Confining Walls of a Flow Domain: Hyperbolic Stretching Function}\label{sec:Transformation_i}
\begin{figure}[h!]
\centering
    \subfloat[physical plane\; $(L_x, L_y)$] {
        \includegraphics[width=0.48\textwidth]{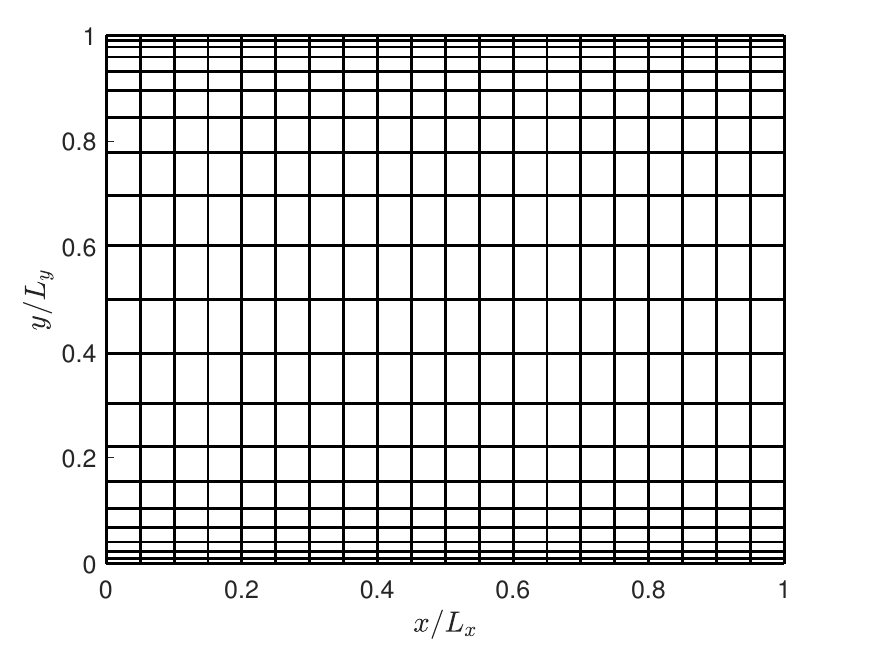}
        \label{fig:domain_a} } 
    \subfloat[computational plane\; $(L_{\xi_1}, L_{\xi_2})$] {
        \includegraphics[width=0.48\textwidth]{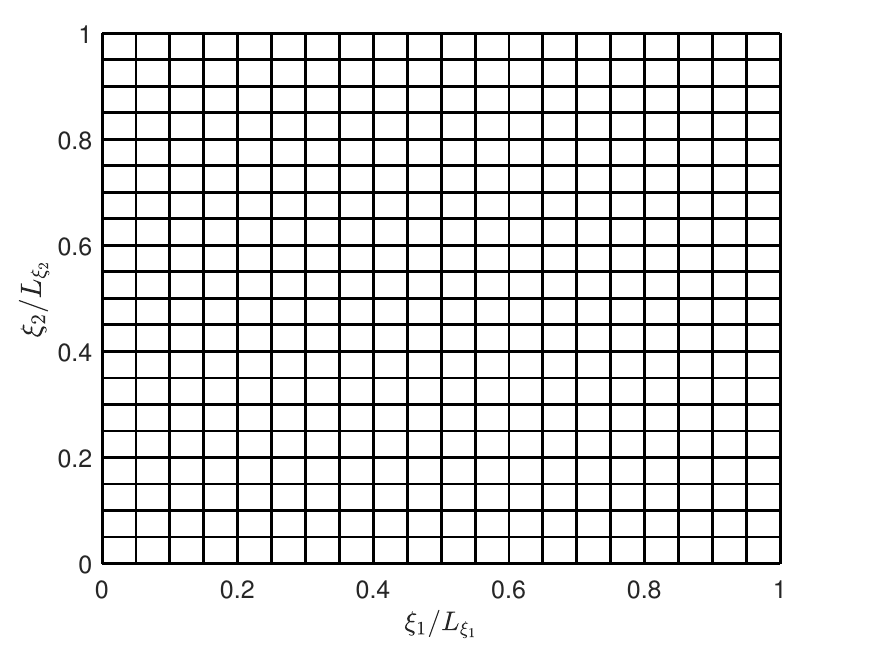}
        \label{fig:domain_b} } \\
    \caption{Grid clustering around both confining walls of a flow domain using a hyperbolic stretching function: (a) Physical plane (Cartesian) and (b) Computational plane (GOC).}
    \label{fig:domain}
\end{figure}

Consider a flow domain resolved with grid clustering around both the confining walls in the physical plane using the Cartesian coordinates ($x,y$) with the side lengths $L_x$ and $L_y$, whereas the computational plane has a uniform grid structure in the GOC ($\xi_1, \xi_2$) with the side lengths $L_{\xi_1}$ and $L_{\xi_2}$ as shown schematically in Fig.~\ref{fig:domain}.
Let $N_x$ and $N_y$ be the number of grid points used in the two directions of the Cartesian coordinates, and likewise $N_{\xi_1}$ and $N_{\xi_2}$, respectively, in the GOC, which can be related to the corresponding lengths as $L_x = N_x - 1$ and $L_y = N_y - 1$ and similarly $L_{\xi_1} = N_{\xi_1} -1$ and $L_{\xi_2} = N_{\xi_2} - 1$. Since the coordinate transformations generally preserve the domain scales and shape, $N_{\xi_1} = N_x$ and $N_{\xi_2} = N_y$, and hence $L_x = L_{\xi_1}$ and $L_y = L_{\xi_2}$. However, to resolve boundary layer flows more effectively, coordinate transformations $y = \varphi(\xi_2)$ can be designed to enable clustering of the grids around $y=0$ and $y=L_y$ while coarsening them around $y=L_y/2$ in the Cartesian coordinates, i.e., the mesh size $\Delta y$ is nonuniform or varies continuously throughout the domain, whereas in the GOC, with the use of the usual lattice units $\Delta \xi_2 = 1$ everywhere, the GOC-LBM is implemented in this computational domain for flow simulations. For simplicity, considering the flow domain to be open along the horizontal direction, we do not apply any stretching/clustering in the coordinate transformation for $x$, i.e., $x=\xi_1$. For this situation, one simple approach that enables the grid clustering around both the confining walls is via using a hyperbolic tangent function with a parameter that controls the degree of clustering. That is,
\begin{equation}\label{eq:hypertanh_transformation1}
x  = \xi_1,  \qquad y = \frac{L_y }{2} \left\{1- \frac{\tanh \left[\beta \left(1- 2(\xi_2/L_{\xi_2}) \right)\right]}{\tanh \beta} \right\},
\end{equation}
where $\beta$ is the stretching or clustering parameter that determines the degree of nonuniformity in the mesh in the physical domain. Here, $\beta = 0$ is the uniform mesh case and as $\beta$ is progressively increased, the thinner the grid size $\Delta y$ becomes closer both the walls and coarser in the center of the domain. In the GOC-LBM, we need the metric factor $h_2$ that indicates the local mesh size as well as the associated curvature coefficient matrix element $\theta_{22}$, which measures the sensitivity of the local variation of the mesh size. They can be obtained analytically via invoking their definitions, i.e, $h_2 = \partial y/\partial \xi_2$ and followed by $\theta_{22} = (1/h_2^2)\partial h_2/\partial \xi_2$ (see Eq.~(\ref{eq:curvaturematrix})). Note that along the $x$ direction, since the grid size is uniform with $h_1 = 1$, $\theta_{11}$ is zero everywhere and so are the mixed elements $\theta_{12}$ and $\theta_{21}$ since the clustered grids are straight and not curvilinear. Hence, for the mesh setup in Fig.~\ref{fig:domain}, the grid parameterizations read as
\begin{align}\label{eq:metriccoefficient}
h_1&=1,   &h_2&= \frac{\beta}{\tanh \beta} \; \sech^2 \bigg[\beta\left( 1- 2 \frac{\xi_2}{L_{\xi_2}}\right) \bigg].
\end{align}
and
\begin{align}\label{curvaturematrixcomponents}
  \theta_{22} =  4 \dfrac{\tanh \beta}{L_y} \dfrac{\tanh \left[ \beta \left( 1 - 2 \dfrac{\xi_2}{L_{\xi_2}}  \right) \right]}{\sech^2 \left[ \beta \left( 1 - 2 \dfrac{\xi_2}{L_{\xi_2}}  \right) \right]}, \qquad \theta_{11} = \theta_{12} = \theta_{21} = 0.
\end{align}
Note that if we instead consider a closed domain with boundary layers on the confining walls in both the directions (such as the lid-driven cavity flow considered in the next section), the above transformation for $y$ can be adapted for the $x$ direction as well. In this situation, we have
\begin{equation*}
x = \frac{L_x}{2} \left\{1- \frac{\tanh \left[\beta \left(1- 2(\xi_1/L_{\xi_1}) \right)\right]}{\tanh \beta} \right\},  \qquad y = \frac{L_y }{2} \left\{1- \frac{\tanh \left[\beta \left(1- 2(\xi_2/L_{\xi_2}) \right)\right]}{\tanh \beta} \right\},
\end{equation*}
\begin{equation*}
h_1= \frac{\beta}{\tanh \beta} \; \sech^2 \bigg[\beta\left( 1- 2 \frac{\xi_1}{L_{\xi_1}}\right) \bigg], \qquad h_2= \frac{\beta}{\tanh \beta} \; \sech^2 \bigg[\beta\left( 1- 2 \frac{\xi_2}{L_{\xi_2}}\right) \bigg],
\end{equation*}
and
\begin{equation*}
\theta_{11} =  4 \dfrac{\tanh \beta}{L_x} \dfrac{\tanh \left[ \beta \left( 1 - 2 \dfrac{\xi_1}{L_{\xi_1}}  \right) \right]}{\sech^2 \left[ \beta \left( 1 - 2 \dfrac{\xi_1}{L_{\xi_1}}  \right) \right]},
\qquad\theta_{22} =  4 \dfrac{\tanh \beta}{L_y} \dfrac{\tanh \left[ \beta \left( 1 - 2 \dfrac{\xi_2}{L_{\xi_2}}  \right) \right]}{\sech^2 \left[ \beta \left( 1 - 2 \dfrac{\xi_2}{L_{\xi_2}}  \right) \right]}, \qquad \theta_{12} = \theta_{21} = 0.
\end{equation*}
Here, if necessary, the stretching parameter $\beta$ can be specified independently in each of the coordinate directions. The above analytic formulas for the metric factors $h_1$ and $h_2$ and the curvature coefficient matrix elements $\theta_{11}$, $\theta_{22}$, $\theta_{12}$, and $\theta_{21}$ are very useful in the numerical implementations of the GOC-LBM by serving as a rapid way of setting up the mesh related parameterizations with orthogonally clustered grids.

\subsection{Grid Clustering Around Both Confining Walls of a Flow Domain: Roberts Boundary Layer Transformation} \label{sec:Transformation_ii}
The hyperbolic tangent function given in the previous subsection (Sec.~\ref{sec:Transformation_i}) is a simple coordinate transformation. A more sophisticated transformation that was specifically developed for boundary layer flows involving relatively thin regions was proposed by Roberts~\cite{roberts2005computational} (see also~\cite{anderson2020computational}). We will summarize this approach here as it will be utilized along with above transformation in evaluating the performance of the GOC-LBM numerically. For the situation illustrated in Fig.~\ref{fig:domain}, the degree of grid clustering in the Roberts transformation approach is controlled by a free parameter $\gamma$, which scales with the boundary layer thickness $\delta_b$ relative to the domain size $L_y$ as $\gamma \sim 1/\sqrt{1-\delta_b/L_y}$. Then, for convenience we introduce
\begin{equation*}
\lambda = \dfrac{(\gamma+1)}{(\gamma-1)}, \qquad
\alpha=
 \begin{cases}
   0, & \mbox{if mesh clustered only near}\; y=L_y \\
   \dfrac{1}{2}, & \mbox{if mesh clustered equally near both}\; y=0 \; \mbox{and} \;y=L_y.
 \end{cases}
\end{equation*}
as well as a scaled local variable in the GOC $\xi_2$ given by
\begin{equation}\label{eq:scaled_local _xi2}
s = \left( \frac{\xi_2}{L_{\xi_2}}-\alpha \right)/\left( 1- \alpha  \right).
\end{equation}
Based on these, the Roberts' boundary layer coordinate transformation for grid clustering/stretching in the physical domain is given by:
\begin{align}\label{eq:y_Roberts}
x&=\xi_1, &y &= L_y \left[ \frac{(\gamma +2 \alpha) \lambda^{s}- \gamma + 2 \alpha}{(2 \alpha + 1)(1 + \lambda^{s})}   \right],
\end{align}
where $s$ is given in Eq.~(\ref{eq:scaled_local _xi2}) and depends on the local coordinate $\xi_2$ in the GOC. Here, $\gamma$ varies in the range $1 < \gamma < \infty$. For large $\gamma$ (or as $\gamma \rightarrow \infty$), the grid becomes uniform, whereas as $\gamma \rightarrow 1$ the clustering of grids near the walls become progressively thinner. In the related numerical results shown later in Sec.~\ref{sec:Results}, we use $\gamma = 1.05$ with the GOC-LBM in the simulation of some canonical flow benchmarks. Moreover, the metric coefficient $h_2$ and the curvature coefficient matrix element $\theta_{22}$ can be obtained analytically via using their definitions and taking the appropriate derivatives based on the above coordinate transformation, which can be expressed as
\begin{align}\label{eq:h2_Roberts}
h_2 = \frac{2 \gamma}{(2 \alpha +1)} \frac{\ln\left(\lambda\right)}{(1- \alpha)} \frac{\lambda^{s}}{ (1+\lambda^s)^2}, \qquad \theta_{22} = \frac{1}{L_y} \left( \frac{1 +2 \alpha}{2 \gamma}\right) \left( \frac{1}{ \lambda^{s}}- \lambda^{s}\right).
\end{align}
where $s$ is given in Eq.~(\ref{eq:scaled_local _xi2}) and $h_1 = 1$, $\theta_{11} = \theta_{12} = \theta_{21} = 0$.


\subsection{Grid Clustering Around Only One of the Confining Walls of a Flow Domain: Hyperbolic Stretching Function} \label{sec:Transformation_iii}
\begin{figure}[h!]
\centering
    \subfloat[physical plane\; $(L_x, L_y)$] {
        \includegraphics[width=0.48\textwidth] {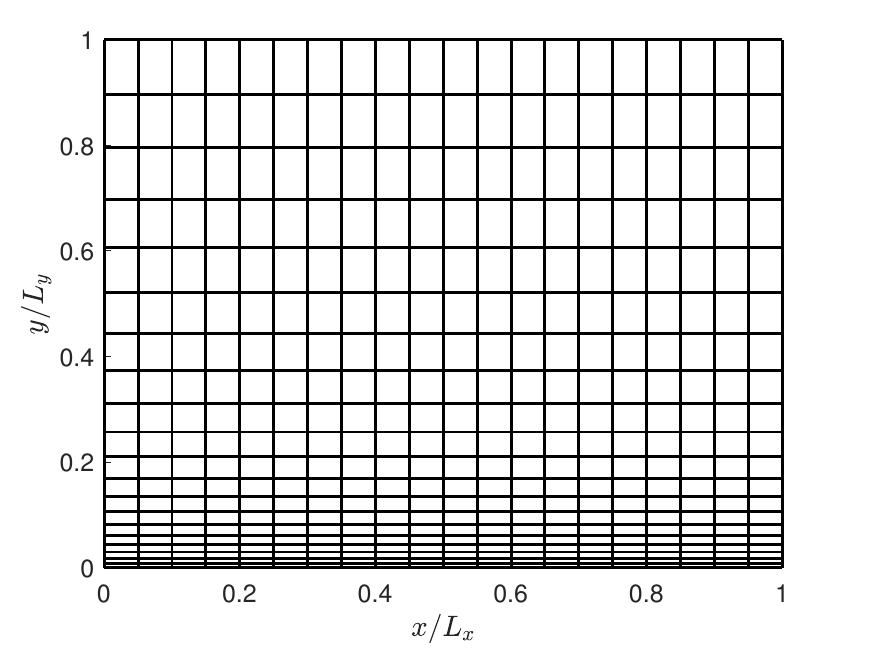}
        \label{fig:domain_2_a} } 
    \subfloat[computational plan \; $(L_{\xi_1}, L_{\xi_2})$] {
        \includegraphics[width=0.48\textwidth] {c_domain}
        \label{fig:domain_2_b} } \\
    \caption{Grid clustering around only one of the confining walls of a flow domain using a hyperbolic stretching function: (a) Physical plane (Cartesian) and (b) Computational plane (GOC).}
    \label{fig:domain_2}
\end{figure}

In some cases, it may be desirable to use clustering around only one of the confining walls such as near the bottom wall at $y=0$ as shown in Fig.~\ref{fig:domain_2} while coarsening around the rest of the regions.
In such situations, the hyperbolic tangent function-based transformation given in Sec.~\ref{sec:Transformation_i} can be slightly modified, and the results read as follows:
\begin{equation}\label{eq:hypertanh_transformation}
x = \xi_1, \qquad y = L_y \left\{1- \frac{\tanh \left[\beta \left(1- (\xi_2/L_{\xi_2}) \right)\right]}{\tanh \beta} \right\}.
\end{equation}
and
\begin{equation}\label{eq:metriccoefficientB}
h_2 = \frac{\beta}{\tanh \beta} \; \sech^2 \left[ \beta \left( 1-  \frac{\xi_2}{L_{\xi_2}} \right) \right], \qquad
\theta_{22} =  2 \frac{\tanh \beta}{L_y} \frac{\tanh \left[ \beta \left( 1 - \dfrac{\xi_2}{L_{\xi_2}}  \right) \right]}{\sech^2 \left[ \beta \left( 1 -  \dfrac{\xi_2}{L_{\xi_2}}  \right) \right]},
\end{equation}
along with $h_1 = 1$, $\theta_{11} = \theta_{12} = \theta_{21} = 0$.

\subsection{Grid Clustering Around an Interior Location of a Flow Domain: Hyperbolic Stretching Function}\label{sec:Transformation_iv}
\begin{figure}[h!]
\centering
\subfloat[physical plane\; $(L_x, L_y)$] {
\includegraphics[width=0.48\textwidth] {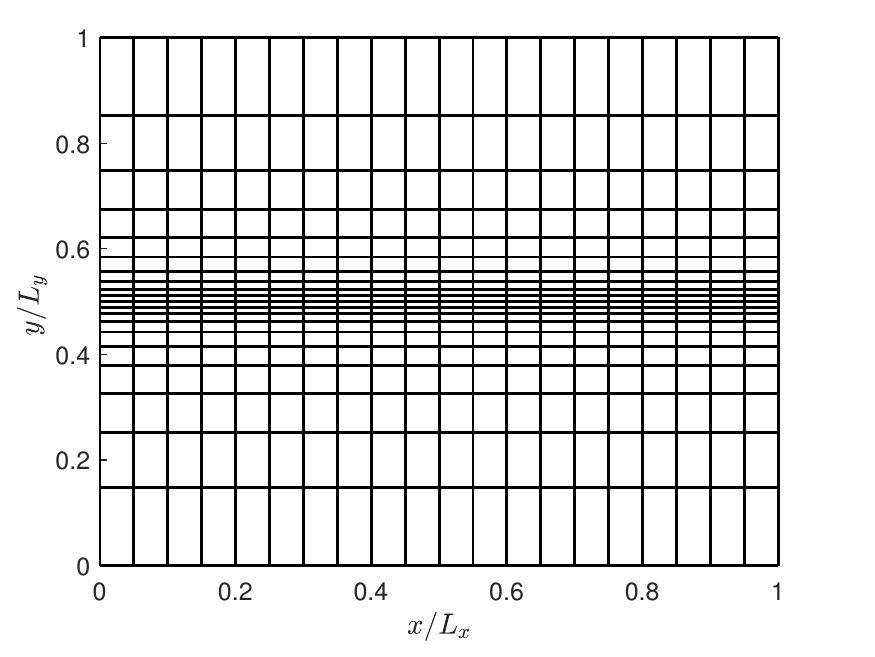}
\label{fig:domain_iv_a} } 
\subfloat[computational plan\; $(L_{\xi_1}, L_{\xi_2})$] {
\includegraphics[width=0.48\textwidth] {c_domain}
\label{fig:domain_iv_b} } \\
\caption{Grid clustering around an interior zone of a flow domain using a hyperbolic stretching function: (a) Physical plane (Cartesian) and (b) Computational plane (GOC).}
\label{fig:domain_iv}
\end{figure}

For completeness, we will conclude this section by summarizing a strategy for clustering the grids around an interior location of a flow domain such as at the location $y=y_c$ in Fig.~\ref{fig:domain_iv}, which may be utilized for resolving thin shear layer zones effectively.
In this case, the coordinate transformation can be expressed as
\begin{align}\label{eq:Holst_transformation}
x&=\xi_1,  &y&= y_c \left\{1+ \frac{\sinh \left] \mu \left( \xi_2/L_{\xi_2}- B\right)\right]}{\sinh (\mu B)} \right\}.
\end{align}
where
\begin{equation}\label{eq:B_Holst}
B=\frac{1}{2 \mu} \ln \left[ \frac{1+ \left(e^{\mu} -1\right)\left( y_c/L_y \right)}{1+ \left(e^{-\mu} -1\right)\left( y_c/L_y\right)} \right].
\end{equation}
Here, $ 0 < \mu < \infty$, and the grid clustering around $y_c$ becomes finer for as $\mu$ becomes progressively larger, i.e., $\mu \gg 0$. By using procedures similar to those used for deriving the metric factor $h_2$ and the curvature coefficient matrix element $\theta_{22}$ for the previous cases, we obtain them for this interior clustering case, which read as
\begin{equation}\label{eq:h_2_Holst}
h_2 = \frac{y_c}{L_{\xi_2}} \frac{\mu}{\sinh (\mu B)}\cosh \bigg[ \mu \left( \frac{\xi_2}{L_{\xi_2}}-B \right) \bigg],
\theta_{22}= \dfrac{\sinh (\mu B)}{y_c}\tanh \left[ \mu \left( \dfrac{\xi_2}{L_{\xi_2}}-B \right) \right] \sech \left[ \mu \left( \dfrac{\xi_2}{L_{\xi_2}}-B \right) \right],
\end{equation}
along with $h_1 = 1$, and $\theta_{11}= \theta_{21}= \theta_{12}=0$.

\section{Results and Discussion} \label{sec:Results}
In this section, we will first validate our GOC-LBM with the attendant orthogonally clustered grids through simulations of various standard flow problems, such as the steady and transient flow between two parallel plates driven by a body force and shear, respectively, and the lid-driven cavity flow at various Reynolds numbers so as to verify its numerical accuracy; we will also numerically test our approach with attendant cubic-velocity corrections for the moment equilibria derived earlier using a Chapman-Enskog analysis for Galilean invariance via simulations of decaying oblique shear wave with an imposed background velocity. Then, we will demonstrate its advantages of efficiently simulating thin boundary layer flows using a MHD benchmark flow case study in this regard, which is followed by showcasing its ability to naturally use body-fitted grids for flow simulations involving two additional benchmark problems with curved geometries, viz., Taylor-Couette flow and shear-driven polar cavity flow. All the different versions of the GOC-LBM using the MRT-based on raw moments (see Appendix~\ref{sec: Appendix_A}), MRT-based on central moments (see Appendix~\ref{sec:Appendix_C}) and the SRT-based on relaxations in the velocity space (see Appendix~\ref{sec:Appendix_D}) were implemented and tested yielding similar accuracy for the flow simulation benchmarks considered here, with the central moments-based GOC-LBM being the most stable and robust in attaining high Reynolds numbers as found through numerical experiments involving the lid-driven cavity flow and hence recommended for its use as a general-purpose GOC-LBM. The coordinate transformations discussed in Sec.~\ref{sec:Transformations} were utilized in the clustering of grids near walls for the different flow benchmarks considered in this study.

\subsection{Steady Flow Driven by a body force Between Two Parallel Plates}
\begin{figure}[ht!]
\centering
\captionsetup{justification=centering}
\includegraphics[scale=0.75]{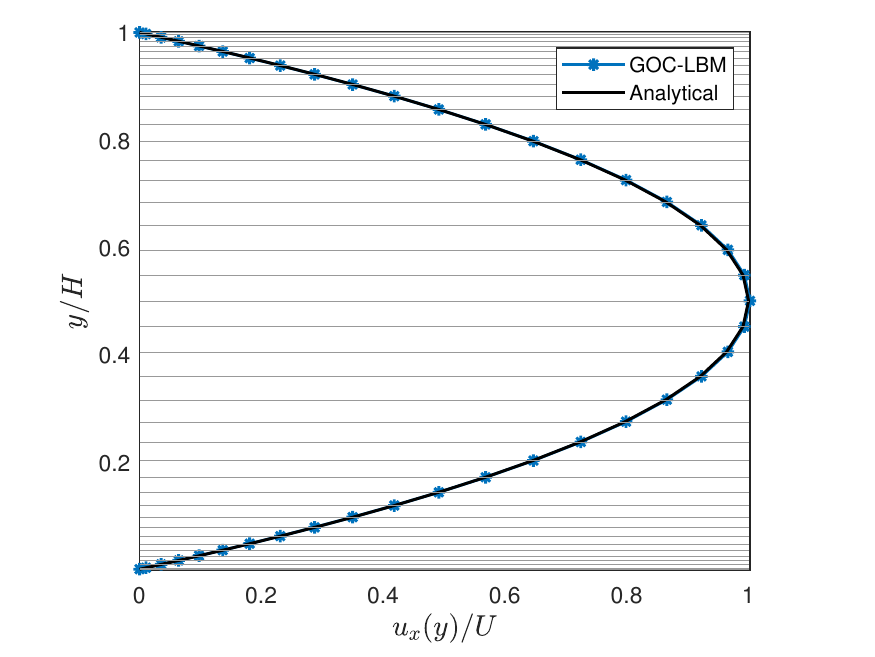}
\caption{Velocity profile computed using the GOC-LBM and compared with the analytical solution for the steady flow between two parallel plates at a Reynolds number $\mbox{Re}=20$ via using the Roberts coordinate transformation to achieve continuously varying grids along the $y$ direction with clustering around both the walls based on setting the grid clustering/stretching parameter $\gamma = 1.06$ (see Sec.~\ref{sec:Transformations}), which is indicated by the horizontal grid lines in the background of this figure.}
\label{fig:1}
\end{figure}

Let's consider a steady flow between two parallel plates separated by a distance $H$ and driven by an externally imposed body force $F_{ext,x}$. The coordinate normal to the plates $y$ is considered to originate from the bottom wall. This well-known Poiseuille flow has an analytical solution given by $u_a(y)=4 U (y/H) (1- (y/H))$, where $U$ is the fully-developed and maximum velocity of the flow that occurs at the midway location between the two plates, which can be related to the parameters of this flow problem as $U = F_{ext, x} (H/2)^2/2\nu$. Here, $\nu$ is the kinematic shear viscosity, which is related to the relaxation rates $\omega_4$ and $\omega_5$ via Eqs.~(\ref{eq:73}) and Eq.~\eqref{eq:76}. Taking the relaxation time $\tau$ to be 0.8, we set these relaxation rates as $\omega_4 = \omega_5=1/\tau$. A reference density of 1.0 is used when setting up the initial condition here, and throughout the rest of this paper, the natural lattice units are used in setting up the grid parameters in the computational domain, i.e., $\Delta \xi_1 =\Delta \xi_2 = \Delta t = 1.0$.
The standard half-way bounce back scheme is used to impose the no-slip boundary conditions at the solid walls and periodic boundary conditions are considered along the $x$ direction. For concreteness, in setting up the parameters for the assessment of the GOC-LBM, we take the Reynolds number $\mbox{Re}= U(H/2)/\nu$ as $20$ and use $N_x = 3$ grid points along the $x$ direction and $N_y = 40$ grid points along the $y$ direction. The Roberts coordinate transformation discussed in Sec.~\ref{sec:Transformations} is utilized for grid clustering around both the walls in the physical domain, which is in the Cartesian coordinates, by setting $\alpha = 1/2$ and $\gamma = 1.06$ for the grid-clustering parameter. Figure~\ref{fig:1} shows the grid lines based on such a clustering in its background along with the velocity profile computed using the GOC-LBM overlaid with the analytical solution. The computed results are seen to agree very well with the analytical solution.

\subsection{Transient Flow Driven by a Shear between Two parallel plates}
As a second test case, we consider the time-dependent flow between two parallel plates separated by a spacing of $H$ and driven by the shearing motion of the top plate at a velocity $U$, where the bottom plate is held fixed. Initially, the fluid is taken to be at rest. For this classical transient Couette flow problem, the analytical solution of the velocity field $u_a$ at any location $y$ from the bottom plate and at time $t$ is given by
\begin{equation*}
 u_a(y,t)= U \frac{y}{H} - \frac{2U}{\pi} \sum_{n=1}^{\infty} \frac{1}{n} \exp{\left[-\frac{n^2 \pi^2 \nu t}{H^2}\right]}\sin
 \left[ n\pi \left( 1- y/H \right)\right].
\end{equation*}
For the purpose of making comparisons of the computed results with this analytical solution at different times, let's first define the characteristic time scale of this problem based on a viscous diffusive process as $T^* = H^2/\nu$. Then for any time $t$, we can obtain an equivalent dimensionless time as $T = t/T^*$. We use a grid resolution of $6\times 50$ and a relaxation time $\tau$ as $1.0$ to set the relaxation rates for the kinematic shear viscosity via $\omega_4 = \omega_5=1/\tau$. As in the previous case, we consider periodic boundary conditions along the flow or $x$ direction, but, by contrast, imposed the momentum-augmented half-way bounce back scheme at the top wall in order to set the plate velocity at $U$, which is specified to be equal to $0.02$ in our numerical simulations. A discussion on the implementation of this moving wall boundary condition that is parameterized by the wall metric factors for consistent implementation with the GOC-LBM is provided in the Appendix~\ref{sec:initial_boundary_conditions}. Again, we use the Roberts coordinate transformation to achieve grid clustering around both the plates by setting $\gamma=1.06$.

\begin{figure}[ht!]
\centering
\captionsetup{justification=centering}
\includegraphics[scale=0.7]{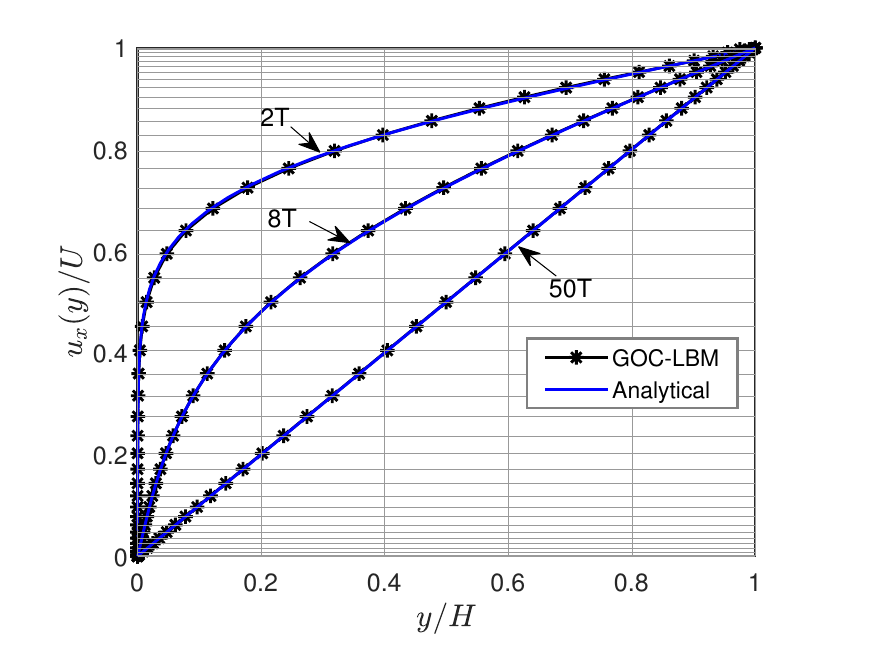}
\caption{Time-dependent velocity profile at the instants $2T$, $8T$, and $50T$ computed using the GOC-LBM and compared with the analytical solution for the transient flow between two parallel plates driven by the shearing motion of the top plate at a velocity $U$. The Roberts coordinate transformation is used to achieve continuously varying grids along the $y$ direction with clustering around both the walls based on setting the grid clustering/stretching parameter $\gamma = 1.06$ (see Sec.~\ref{sec:Transformations}), which is indicated by the horizontal grid lines in the background of this figure.}
\label{fig:couttee_GOC}
\end{figure}

Figure~\ref{fig:couttee_GOC} displays the numerical results of the transient velocity profiles computed using the GOC-LBM at the instants $2T$, $8T$ and $50T$ and compared with the corresponding analytical solution based on the above. Excellent agreement between the numerical results and the analytical solution at all times can be seen thereby confirming the ability of the GOC-LBM in accurately reproducing time-dependent flow behavior.

\subsection{Lid-driven Cavity Flow at Various Reynolds Numbers}
Next, as part of a further validation of the GOC-LBM, we consider the simulation of shear-driven flow in a square cavity of side length $H$ under the motion of the top lid at a velocity $U$. This lid-driven cavity flow is a standard benchmark problem and is characterized by the presence of a main vortex around the center of the cavity and accompanied by various secondary vortices of various scales and shapes around the corners, and the complex vortical flow patterns become richer as the Reynolds number $\mbox{Re} = U H/\nu$ is increased. In particular, with higher $\mbox{Re}$, the near-wall structures become smaller that can be effectively resolved by the use of GOC-LBM and hence serves as a good case study of its application. We note here that when we assessed the stability limits of the different collision models under relatively coarser resolutions, the SRT-formulation (see Appendix~\ref{sec:Appendix_D}) was found to be stable only for the lower $\mbox{Re}$ range, the raw moments-based version of the GOC-LBM (see Appendix~\ref{sec: Appendix_A}) maintains stability for lower and intermediate $\mbox{Re}$, while the central moments-based GOC-LBM (see Appendix~\ref{sec:Appendix_C}) was found to be stable at all $\mbox{Re}$, including $7500$ tested in the following, which is consistent with the prior observations with the standard LBM based on uniform grids~\cite{ning2016numerical}. Hence, due to robustness, the central moments-based GOC-LBM is recommended for its applications to simulating flows under more general and extreme ranges of the parametric conditions.

The no-slip boundary conditions for the top moving lid is implemented using a momentum-augmented half-way bounce-back scheme discussed in Appendix~\ref{sec:initial_boundary_conditions}. Since this problem involves confined flow within four solid boundaries, unlike in the two case studies above, we used the coordinate transformations based on the hyperbolic tangent function to specify continuously varying grids along \emph{both} the $x$ and $y$ directions with clustering around all the walls of the cavity via setting the grid clustering/stretching parameter $\beta = 1.2$ (see Sec.~\ref{sec:Transformations}). Since the boundary layers become progressively thinner as $\mbox{Re}$ increases, we considered different number of grid points for different $\mbox{Re}$, which were determined via performing numerical experiments. In the results reported in the following the choice of the grid resolutions at different $\mbox{Re}$ are as follows: $120\times 120$ at $\mbox{Re} = 1000$, $300\times 300$ at $\mbox{Re} = 3200$, $350\times 350$ at $\mbox{Re} = 5000$, and $400\times 400$ at $\mbox{Re} = 7500$. Simulations were run until the flow field $\bm{u}=u \hat{x}+ v\hat{y}$ attained their steady states for each case via ensuring that the residual errors under the 2-norm between the successive time steps reached less than $10^{-15}$.

\begin{figure}[H]
\centering
\advance\leftskip-1.7cm
        \subfloat[$u$ component, Re=1000] {
        \includegraphics[width=.45\textwidth] {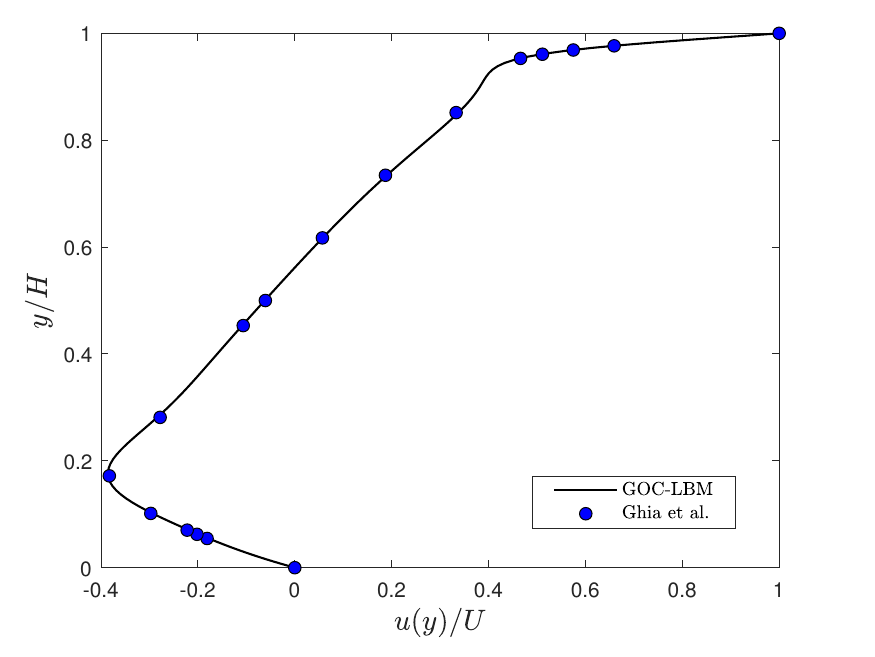}
        \label{cavity_profiles_a} } 
    \subfloat[$v$ component, Re=1000] {
        \includegraphics[width=.45\textwidth] {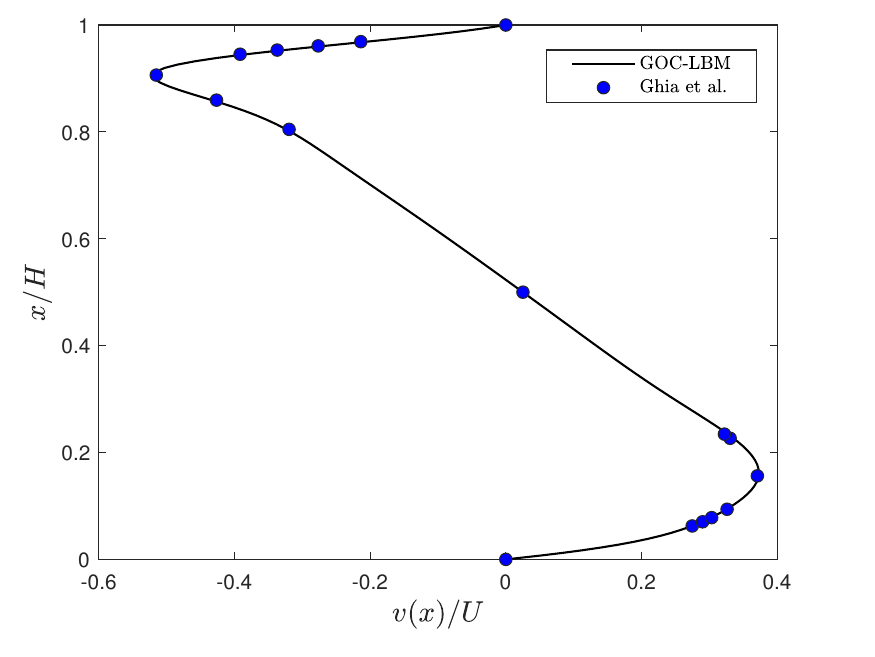}
        \label{cavity_profiles_b} } \\
        \subfloat[$u$ component, Re=3200] {
        \includegraphics[width=.45\textwidth] {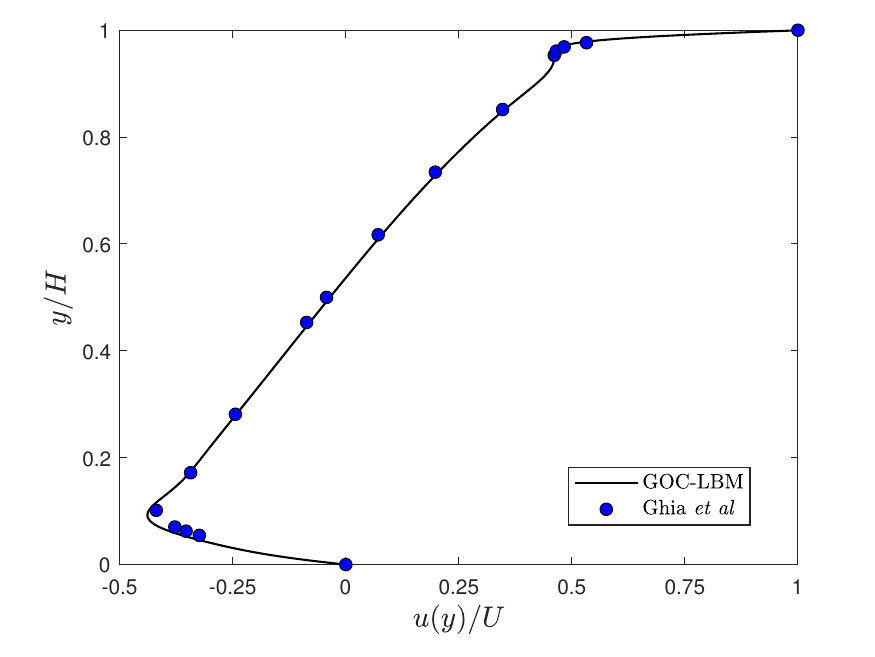}
        \label{cavity_profiles_c} } 
    \subfloat[$v$ component, Re=3200] {
        \includegraphics[width=.45\textwidth] {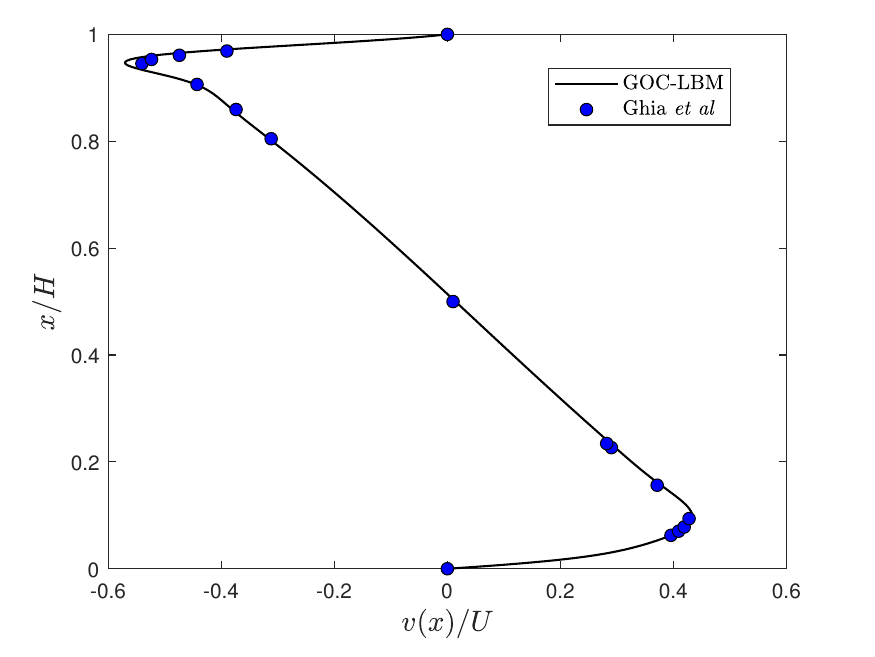}
        \label{cavity_profiles_d} } 
    \caption{Velocity profiles $u(y)$ along the vertical centerlines at $x=H/2$ and $v(x)$ along the horizontal centerlines at $y=H/2$ in a 2D lid driven cavity flow for two different Reynolds numbers of $\mbox{Re}=1000$ and $\mbox{Re}=3200$ computed using the GOC-LBM and compared with the benchmark numerical solutions of Ghia \emph{et al.}~\cite{ghia1982high} (symbols). Coordinate transformations based on the hyperbolic tangent function are used to achieve continuously varying grids along both the $x$ and $y$ directions with clustering around all the walls of the cavity based on setting the grid clustering/stretching parameter $\beta = 1.2$ (see Sec.~\ref{sec:Transformations}) for use with the GOC-LBM.}
    \label{fig:cavity_profiles1}
\end{figure}

\begin{figure}[H]
\centering
\advance\leftskip-1.7cm
    \subfloat[$u$ component, Re=5000] {
        \includegraphics[width=.45\textwidth] {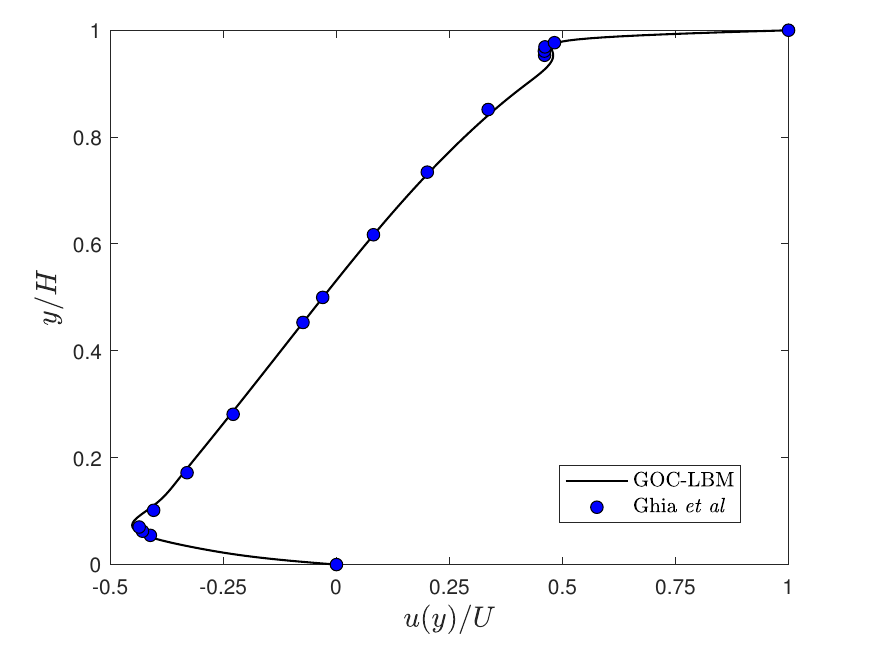}
        \label{cavity_profiles_a2} } 
    \subfloat[$v$ component, Re=5000] {
        \includegraphics[width=.45\textwidth] {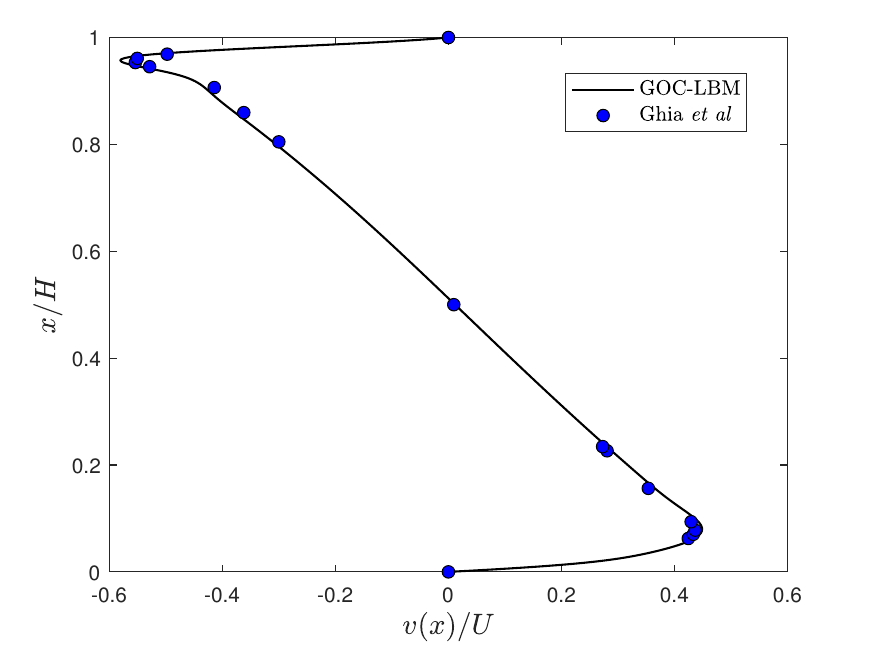}
        \label{cavity_profiles_b2} } \\
    \subfloat[$u$ component, Re=7500] {
        \includegraphics[width=.45\textwidth] {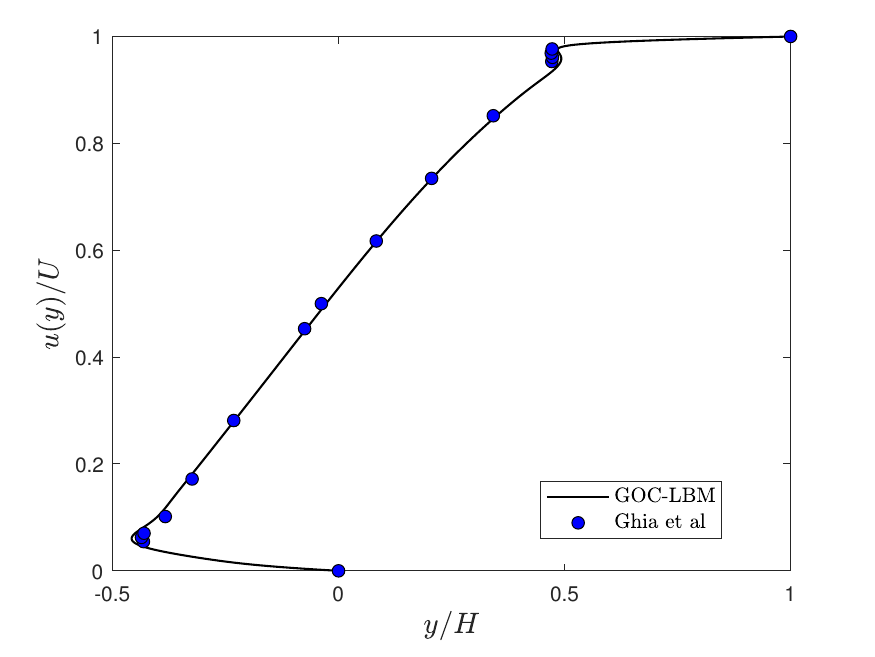}
        \label{cavity_profiles_c2} } 
    \subfloat[$v$ component, Re=7500] {
        \includegraphics[width=.45\textwidth] {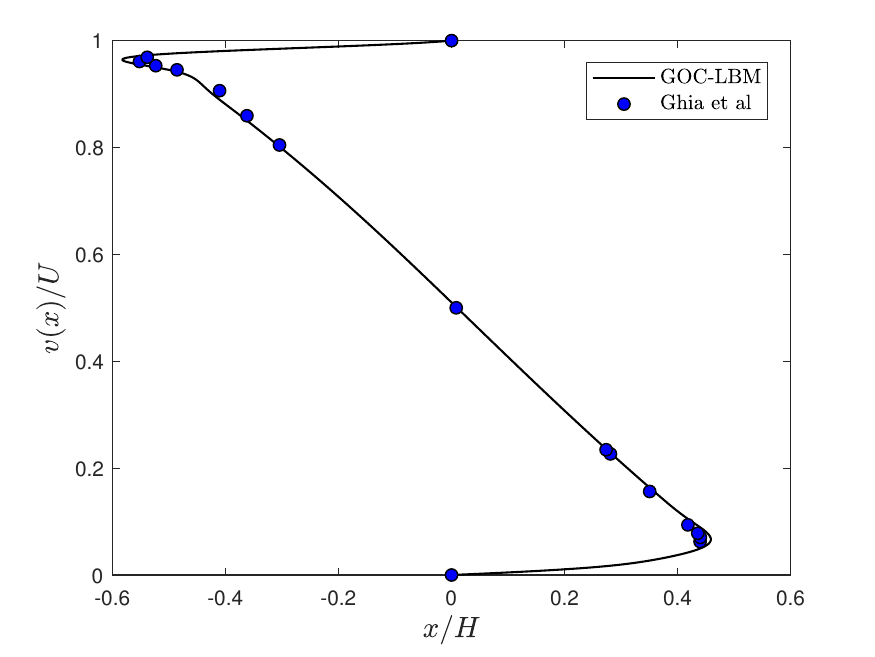}
        \label{cavity_profiles_d2} } 
\caption{Velocity profiles $u(y)$ along the vertical centerlines at $x=H/2$ and $v(x)$ along the horizontal centerlines at $y=H/2$ in a 2D lid driven cavity flow for two different Reynolds numbers of $\mbox{Re}=5000$ and $\mbox{Re}=7500$ computed using the GOC-LBM and compared with the benchmark numerical solutions of Ghia \emph{et al.}~\cite{ghia1982high} (symbols). Coordinate transformations based on the hyperbolic tangent function are used to achieve continuously varying grids along both the $x$ and $y$ directions with clustering around all the walls of the cavity based on setting the grid clustering/stretching parameter $\beta = 1.2$ (see Sec.~\ref{sec:Transformations}) for use with the GOC-LBM.}
    \label{fig:cavity_profiles2}
\end{figure}

Figure~\ref{fig:cavity_profiles1} shows comparisons of the velocity profiles along the vertical and horizontal centerlines of the cavity $\mbox{Re}=1000$ and $3200$ computed using the GOC-LBM against the benchmark numerical solution of Ghia \emph{et al.}~\cite{ghia1982high}, while Fig.~\ref{fig:cavity_profiles2} presents similar comparisons at a higher $\mbox{Re} = 5000$ and $7500$. It can be seen that the GOC-LBM results are in very good agreement with the reference data of Ghia \emph{et al.}~\cite{ghia1982high} for all $\mbox{Re}$.

\begin{figure}[H]
\centering
\includegraphics[width=0.8\textwidth] {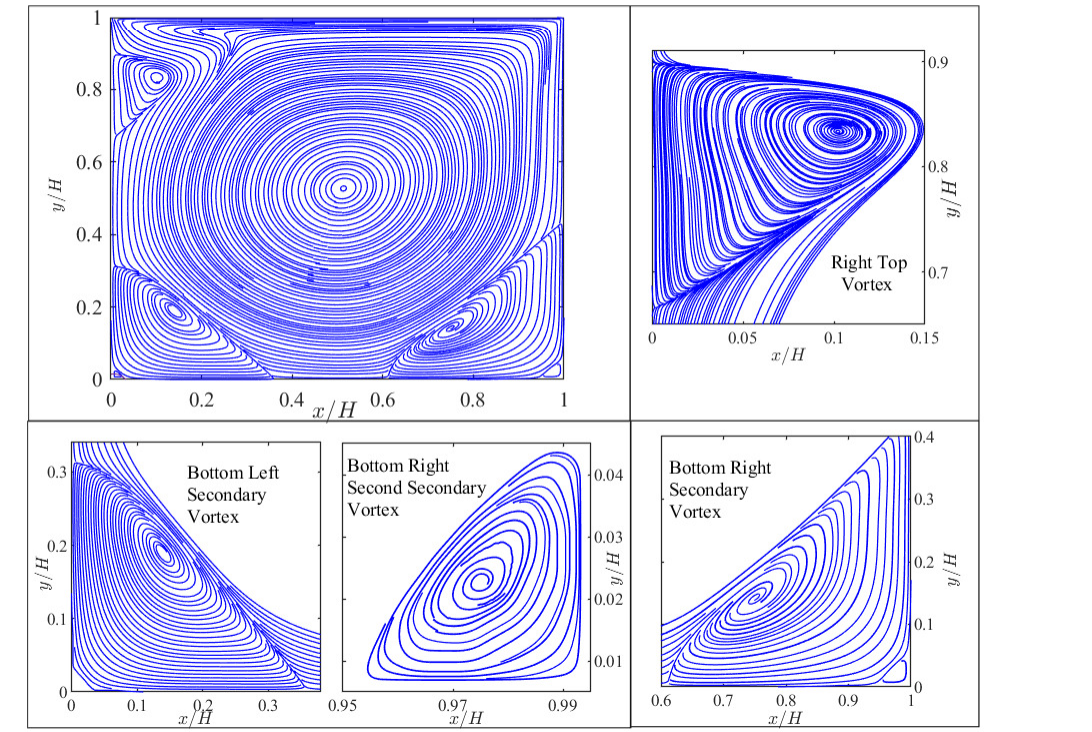}
\label{SV_3200}  \\
\caption{Streamlines in a 2D lid driven cavity flow, including the presence of various corner secondary vortices for the Reynolds numbers of $\mbox{Re}=3200$. Results are computed with GOC-LBM, where coordinate transformations based on the hyperbolic tangent function are used to achieve continuously varying grids along both the $x$ and $y$ directions with clustering around all the walls of the cavity based on setting the grid clustering/stretching parameter $\beta = 1.2$ (see Sec.~\ref{sec:Transformations}).}
\label{fig:cavity_streams_3200}
\end{figure}

\begin{figure}[H]
\centering
\includegraphics[width=0.8\textwidth] {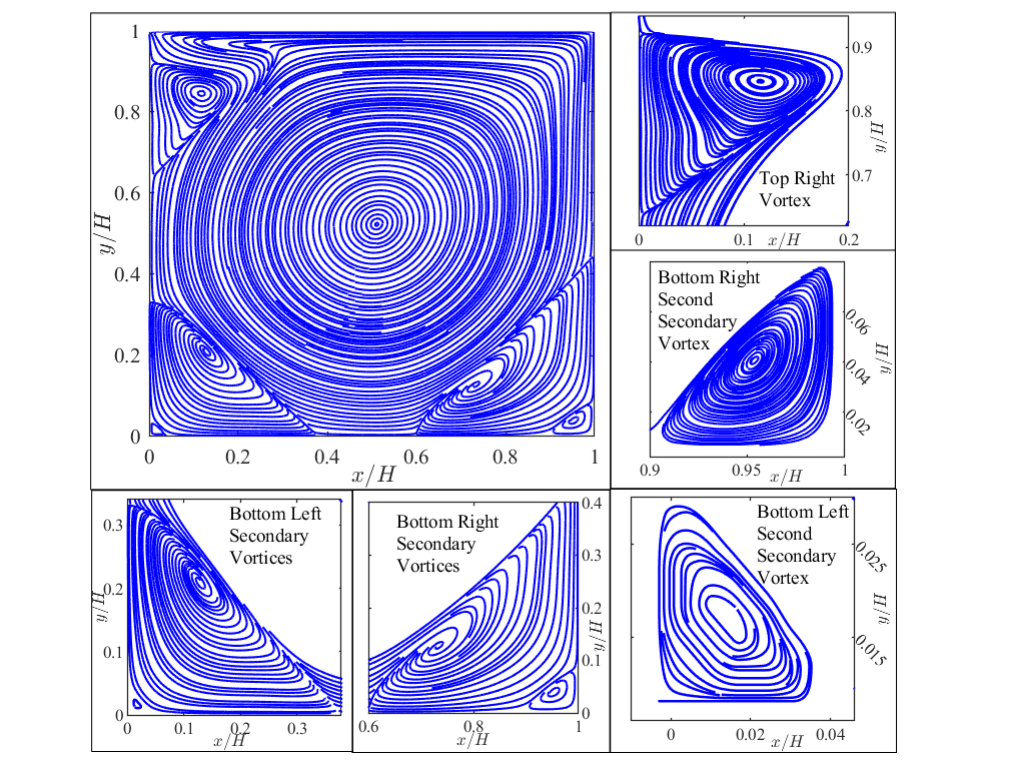}
\label{SV_5000}  \\
\caption{Streamlines in a 2D lid driven cavity flow, including the presence of various corner secondary vortices for the Reynolds numbers of $\mbox{Re}=5000$. Results are computed with GOC-LBM, where coordinate transformations based on the hyperbolic tangent function are used to achieve continuously varying grids along both the $x$ and $y$ directions with clustering around all the walls of the cavity based on setting the grid clustering/stretching parameter $\beta = 1.2$ (see Sec.~\ref{sec:Transformations}).}
\label{fig:cavity_streams_5000}
\end{figure}

\begin{figure}[H]
\centering
\includegraphics[width=0.8\textwidth] {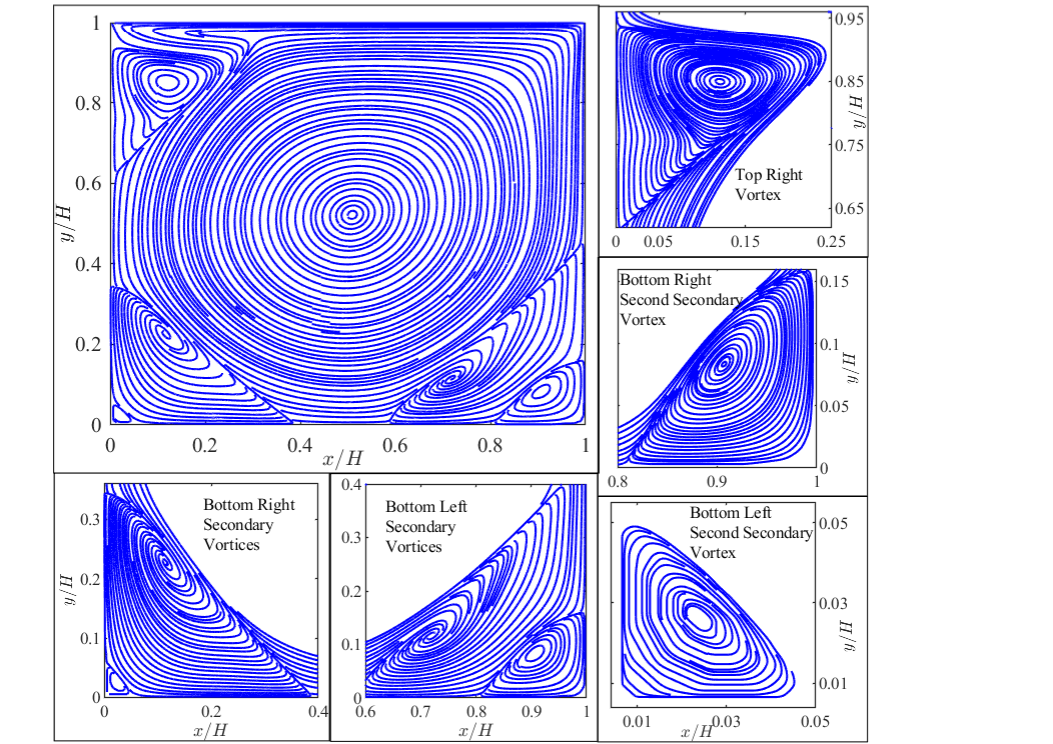}
\label{SV_7500}  \\
\caption{Streamlines in a 2D lid driven cavity flow, including the presence of various corner secondary vortices for the Reynolds numbers of $\mbox{Re}=7500$. Results are computed with GOC-LBM, where coordinate transformations based on the hyperbolic tangent function are used to achieve continuously varying grids along both the $x$ and $y$ directions with clustering around all the walls of the cavity based on setting the grid clustering/stretching parameter $\beta = 1.2$ (see Sec.~\ref{sec:Transformations}).}
\label{fig:cavity_streams_7500}
\end{figure}

In order to further visualize the flow patterns occurring at different $\mbox{Re}$, Figs.~\ref{fig:cavity_streams_3200},~\ref{fig:cavity_streams_5000} and~\ref{fig:cavity_streams_7500} present the streamlines, including the presence of various corner secondary (and/or second-secondary vortices) vortices at $\mbox{Re}= 3200$, $5000$, and $7500$, respectively. The regions where the smaller vortical structures are developed are zoomed and reproduced as insets to these figures for better clarity. As the $\mbox{Re}$ is increased, it is evident that complex and finer vortical structures emerge, which are well captured by the GOC-LBM.
Moreover, these qualitative features are further supplemented with the numerical values of the coordinate locations of the primary and secondary vortices at different $\mbox{Re}$ in Tables~\ref{tab:1} and~\ref{tab:2}, respectively, where the results computed using the GOC-LBM are compared with the benchmark data of Ghia \emph{et al.}~\cite{ghia1982high} and Erturk \emph{et al.}~\cite{erturk2005numerical}. The results of the GOC-LBM are seen to be in very good quantitative agreement with the reference data.
\begin{table}[htbp]
\small
\centering
\captionsetup{justification=centering}
\scalebox{0.95}{ 
\begin{tabular}{c|c|c|c|c}
\hline
  \multirow{2}{*}{}& \multicolumn{4}{c}{Primary Vortex}\\ \cline{2-5}
{Method} &$\mbox{Re}=1000$ &$\mbox{Re}=3200$ & $\mbox{Re}=5000$ & $\mbox{Re}=7500$ \\
 \hline
 \hline
GOC-LBM &  (0.5314, 0.5646)& (0.5178, 0.5404)& (0.5151, 0.5356) & (0.5134, 0.5313) \\
Ref~\cite{ghia1982high} &(0.5313, 0.5625) &(0.5165, 0.5469) &(0.5115, 0.5352) & (0.5117, 0.5322)\\
Ref~\cite{erturk2005numerical}&(0.5300, 0.5650) & NA & (0.5150, 0.5350)& (0.5133, 0.5317)\\
\hline
\hline
\end{tabular}
} 
\caption{Location of the primary vortices in a 2D lid-driven cavity flow at different Reynolds numbers obtained using the GOC-LBM and compared with the results of Ref.~\cite{ghia1982high}(Ghia et al. (1982)) and Ref.~\cite{erturk2005numerical} (Erturk et al. (2005)), which are benchmark solutions obtained from using classical numerical methods for the direct discretization of the NSE.}
\label{tab:1}
\end{table}
\begin{table}[htbp]
\small
\centering
\captionsetup{justification=centering}
\scalebox{0.95}{ 
\begin{tabular}{l c c c c c }
\hline
 Method & \multicolumn{3}{l}{First Secondary Vortex}&  \multicolumn{2}{l}{Second Secondary Vortex} \\
   & Top vortex & Bottom Left & Bottom Right & Bottom Left & Bottom Right\\
 \hline
 \hline
\multicolumn{6}{c}{Re=1000} \\
 \hline
GOC-LBM   & NA & (0.0855, 0.0779) & (0.8581, 0.1104) &NA & (0.9905, 0.0069)   \\
Ref~\cite{ghia1982high}  & NA & (0.0859, 0.0781) & (0.8594, 0.1094)&NA & (0.9922, 0.0078)    \\
Ref~\cite{erturk2005numerical} & NA &(0.0833, 0.0783)  & (0.8633, 0.1117)  &  (0.0050, 0.0050)&  (0.9917, 0.0067)     \\
\hline
\multicolumn{6}{c}{Re=3200} \\
 \hline
GOC-LBM  & (0.05318, 0.89824) & (0.0813, 0.1194)  &  (0.8249, 0.0845)& (0.00189, 0.00189) & (0.9981, 0.0018)  \\
Ref~\cite{ghia1982high} &(0.0547, 0.8984)  & (0.0859, 0.1094)  &  (0.8125, 0.0859)& (0.0078, 0.0078) & (0.9844, 0.0078)     \\
Ref~\cite{erturk2005numerical} &NA  & NA  &  NA& NA & NA     \\
\hline
\multicolumn{6}{c}{Re=5000} \\
 \hline
GOC-LBM  & (0.0646, 0.9060)  &  (0.0726, 0.1383)& (0.8040, 0.0747) & (0.0079, 0.0079)&  (0.9779, 0.0194) \\
Ref~\cite{ghia1982high} & (0.0625, 0.9102)& (0.0703, 0.1367)  &  (0.8086, 0.0742)& (0.0117, 0.0078) & (0.9805, 0.0195)     \\
Ref~\cite{erturk2005numerical}  & (0.0633, 0.9100)& (0.0733, 0.1367)  &  (0.8050, 0.0733)& (0.0083, 0.0083) & (0.9783, 0.0183)     \\
\hline
\multicolumn{6}{c}{Re=7500} \\
 \hline
GOC-LBM  & (0.0625, 0.9193)  & (0.0704, 0.15806)  & (0.7836, 0.0626)& (0.0116, 0.0116)& (0.9516, 0.0397)\\
Ref~\cite{ghia1982high}  & (0.0664, 0.9141)  &  (0.0645, 0.1504) &(0.7813, 0.0625) & (0.0117, 0.0117)& (0.9492, 0.0430) \\
Ref~\cite{erturk2005numerical}   & (0.0667, 0.9133)& (0.0650, 0.1517)  &  (0.7900 , 0.0650)& (0.0117, 0.0117) & (0.9517, 0.0417)     \\
\hline
\hline
\end{tabular}
} 
\caption{Location of the secondary vortices in a 2D lid-driven cavity flow at different Reynolds numbers obtained using the GOC-LBM and compared with the results of Ref.~\cite{ghia1982high}(Ghia et al. (1982)) and Ref.~\cite{erturk2005numerical} (Erturk et al. (2005)), which are benchmark solutions obtained from using classical numerical methods for the direct discretization of the NSE.}
\label{tab:2}
\end{table}

\begin{figure}[H]
\centering
\advance\leftskip-1.5cm
    \subfloat[GOC-LBM] {
        \includegraphics[width=0.52\textwidth] {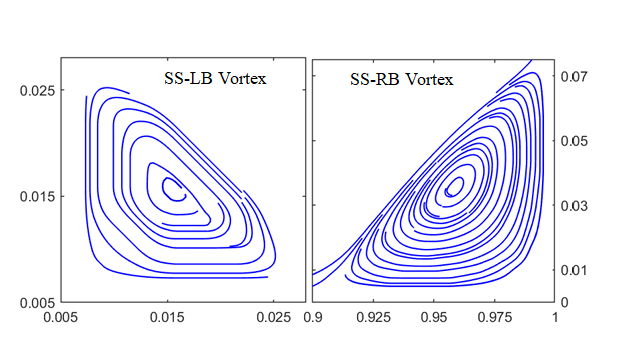}
        \label{fig:5a} } 
    \subfloat[Standard LBM] {
        \includegraphics[width=0.52\textwidth] {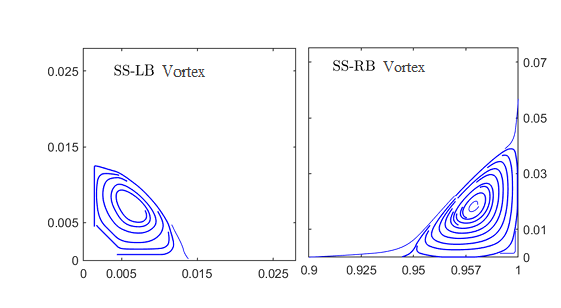}
        \label{fig:5b} } \\
                \advance\leftskip0cm
    \caption{Streamlines of the second-secondary vortices around the left bottom (SS-LB) and right bottom (SS-RB) corners in a 2D lid-driven cavity flow at a Reynolds number $\mbox{Re}=5000$ obtained using the (a) GOC-LBM, where the coordinate transformations based on the hyperbolic tangent function are used to achieve continuously varying grids along both the $x$ and $y$ directions with clustering around all the walls of the cavity based on setting the grid clustering/stretching parameter $\beta = 1.1$ (see Sec.~\ref{sec:Transformations}), are compared with the results of the (b) uniform grid-based standard LBM using the same number of grid points of $350\times350$.}
    \label{fig:Re5000_compare}
\end{figure}

In addition, if the same number of grid points used in the GOC-LBM are also used in the uniform grid-based standard LBM, when simulating flows at relatively high $\mbox{Re}$ with the number of grid points being used is not large enough, the latter may not fully resolve the smaller corner vortices unlike the GOC-LBM, which is able to do so via using adequate grid clustering near the corner wall regions. For example, as seen in Fig.~\ref{fig:Re5000_compare}, at $\mbox{Re}=5000$, the use of $350\times 350$ grid points using the uniform grid-based standard LBM is not able to properly resolve the second secondary left bottom (SS-LB) and second secondary right bottom (SS-RB) vortices, while the GOC-LBM is capable of correctly resolving them with their center locations in very good agreement with the reference data as shown in Table~\ref{tab:2}. Only when the number of grid points are increased further adequately, the uniform grid-based standard LBM can reproduce these features with good quantitative accuracy, which points to  effectiveness of the GOC-LBM. The advantages of using the GOC-LBM will become even more significant in the simulation of thin boundary layers arising in walls-confined magnetohydrodynamic flows discussed later.

\subsection{Simulation of Decaying Oblique Shear Wave as a Test for Galilean Invariance}\label{sec:GI-Test}
In the derivation of the GOC-LBE via a Chapman-Enskog analysis given earlier, we derived corrections to the equilibrium moments which included contributions that were shown to theoretically eliminate the non-Galilean invariant (GI) cubic velocity errors and recover the NSE in the GOC. Such non-GI artifacts arise from the aliasing effects of the discrete nature of the D2Q9 lattice; in particular, the third order longitudinal moments of the LBM on this standard lattice degenerate to the first order moments which in turn leads to cubic velocity defects. Without such corrections, the defect would appear in the diagonal components of the viscous stress so that the effective transport coefficient of the resulting scheme would acquire a dependence on the velocity of the observer or an imposed uniform background velocity. We will now verify these observations numerically by determining whether our GOC-LB formulation with its cubic-velocity corrections to the equilibrium moments measurably removes this defect on a nonuniform grid using a benchmark case involving oblique shear wave decay in the presence of a uniform background velocity, where the answer is known analytically~\cite{dellar2014lattice}.

We consider a doubly periodic square domain $L\times L$, in which the flow generated by an inclined divergence-free shear wave is imposed over a uniform background velocity $\bm{U}_0 = (U_{0,x}, U_{0,y})$ with wavenumbers $k_x = 2\pi m/L$ and $k_y = 2\pi n/L$. By independently adjusting the integers $m$ and $n$ with both $m\neq 0$ and $n\neq 0$, the wave vector $\bm{k} = (k_x, k_y)$ can be made inclined with the principal directions of the coordinates vector $\bm{x} = (x,y)$ so that the full viscous stress tensor (i.e., both the off-diagonal and diagonal components) becomes active (unlike the axis aligned case with either $m$ or $n$ is zero while the other is not, in which case only the off-diagonal components remain active and hence does not test for the non-GI aliasing errors mentioned earlier)~\cite{dellar2014lattice}. This can be achieved by considering the velocity field $\bm{u}=(u_x, u_y)$ as
\begin{equation}\label{eq:shearwavevelocityfield}
u_x(\bm{x}, t) = U_{0,x}-k_y\epsilon A(t)\cos(\bm{k}\cdot\bm{x}),\qquad u_y(\bm{x}, t) = U_{0,y}+k_x\epsilon A(t)\cos(\bm{k}\cdot\bm{x}),
\end{equation}
which follows from defining it from the streamfunction $\psi$ as $\psi = A(t)\sin(\bm{k}\cdot\bm{x})$ via $\bm{u}=\bm{U}_0+\epsilon\hat{z}\times\bm{\nabla}\psi$, where $\epsilon$ is a small parameter and $A(t)$ represents the amplitude of perturbation of the shear wave. Then, by the specifying the initial amplitude as $A(t=0) = A(0)$, where $A(0)$ is given, we can express the time decay of the amplitude function analytically obtained as an exact solution of the NSE as
\begin{equation}\label{eq:anal_decay_rate}
A_{anal}(t) = A(0)e^{-\Omega_{anal}t}, \qquad \Omega_{anal} = (k_x^2 + k_y^2)\nu
\end{equation}
where $\Omega_{anal}$ is the analytical decay rate that is dependent only on the molecular viscosity $\nu$ and the wavenumbers $k_x$ and $k_y$, and is, importantly, independent of the background velocity $\bm{U}_0$, i.e., the NSE solution is Galilean invariant. On the other hand, any measured dependence of the decay rate $\Omega$ on the background velocity field $\bm{U}_0$ in simulations using a given formulation of the LBM is therefore a property of its numerical scheme arising from its discrete effects and not of the flow itself.

In order to test our approach for GI consistently, for the purpose of illustration, we consider nonuniform grids in the physical domain by using the hyperbolic tangent coordinate transformation in both the coordinate directions (see Sec.~\ref{sec:Transformations}) with the clustering parameter $\beta = 0.3$ for resolving a square domain with a side length $L = 512$. Thus, this case study involving local variations in the GOC parameters $h_i$ and $\theta_{ij}$ for the GI test goes beyond the previous work~\cite{dellar2014lattice} which was restricted to using uniform grids. An oblique shear wave is initiated by using $(m,n) = (1,2)$ and setting $\epsilon = 0.01$, $A(0) = 1$ with the shear viscosity $\nu = 0.1$. A background velocity field is directed along the $y$ direction throughout, i.e., $\bm{U}_0 = (0, U_B)$, where we consider the following choices for its magnitude: $U_B\in \{0, \pm 0.05, \pm 0.1, \pm 0.15, \pm 0.20, \pm 0.25\}$. We consider two test cases: (i) the GOC-LBM with all its moment equilibria corrections turned on, and (ii) the GOC-LBM by turning off all the cubic velocity terms (including the density gradients as well) appearing in the corrections; specifically, various associated terms in the coefficients given in Eqs.~(\ref{eq:83}) and (\ref{eq:85}) that manifest as $O(U^3)$ for the viscous stress are set to be zero.

\begin{figure}[htbp]\centering
\includegraphics[width=0.54\textwidth]{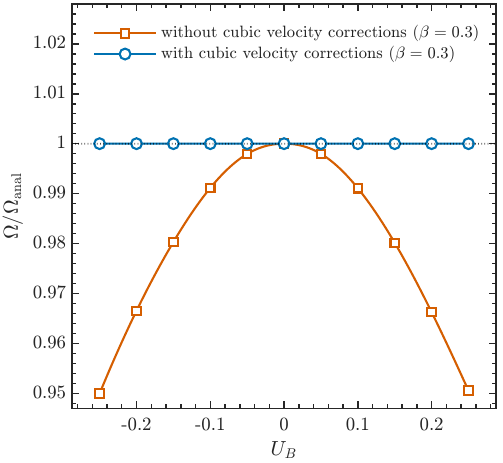}
\caption{Ratio of the measured decay rate to the analytical value of Eq.~(\ref{eq:anal_decay_rate}), $\Omega/\Omega_{\rm anal}$, against the signed background velocity $U_{B}$, for two cases: with cubic velocity corrections and without cubic velocity corrections for the equilibrium moments in the collision step of the GOC-LBM. Doubly periodic square domain of side length $L=512$ with hyperbolic tangent grid clustering in both coordinate directions is used with $\beta = 0.3$ in which oblique shear wave mode is initiated with $(m,n) = (1,2)$ according to Eq.~(\ref{eq:shearwavevelocityfield}) under a uniformly imposed background velocity $\bm{U}_0 = (0, U_B)$.}
\label{fig:GItest}
\end{figure}

Figure~\ref{fig:GItest} presents the ratio of the measured decay rate in simulations to the analytical value of Eq.~(\ref{eq:anal_decay_rate}), $\Omega/\Omega_{\rm anal}$, against the signed background velocity $U_{B}$, for these two cases. Clearly, for the GOC-LBM with the cubic velocity corrections, the computed decay rate is a fixed constant which is consistent with the analytic value and is independent for the entire range of the background velocity. By contrast, for the uncorrected case, the computed decay rate depends on the magnitude of the background velocity $U_B$. In summary, this benchmark study verifies that the GOC-LBM with its attendant cubic velocity corrections in the equilibrium moments preserves the Galilean invariance of its emergent flow behavior, viz., the NSE, as the decay rate of the inclined shear wave is found to be independent of the imposed background velocity. As such, keeping such cubic-velocity related correction terms not only make the scheme to be more accurate physically, but can also help in maintaining its numerical stability.

\subsection{Magnetohydrodynamic Flow Between Two Parallel Plates at High Hartmann Numbers: Resolving Thin Boundary Layers Effectively using GOC-LBM}
A prototypical problem that involves the presence of thin boundary layers is the magnetohydrodynamic (MHD) flow of an electrically conducting fluid with a conductivity $\sigma$ between two parallel plates separated by a distance $2H$ and driven by a body force $F_x$ while being subjected to an imposed magnetic field $B_0$ acting normal to the direction of the fluid motion, i.e., along the $y$ direction. Since the boundary layer thickness can be readily controlled externally by varying the characteristic Hartmann number $\mbox{Ha}=B_0 H \sqrt{\sigma/(\rho\nu)}$, which is the ratio of the Lorentz force to the viscous force, it serves as an excellent test case to study the efficacy of the GOC-LBM in resolving the resulting thin Hartmann layers when compared to the standard LBM based on uniform grids. As such, the Hartmann boundary layer thickness $\delta_{\tiny \mbox{Ha}}$ scales as $1/\mbox{Ha}$ thereby resulting in significant increase in the gridding requirements as $\mbox{Ha}$ increases for its adequate resolution numerically. The analytical solution of this benchmark problem for the velocity field $u_x(y)$ and the induced magnetic field $B_x(y)$ generated in the horizontal or the $x$ direction can be written as follows~\cite{chang1961duct,muller2013magnetofluiddynamics}:
\begin{eqnarray*}
  u_x(y) &=& U_{r}\bigg\{\frac{1}{\mbox{Ha}}\frac{1}{\tanh(\mbox{Ha})}\left[1-\frac{\cosh(\mbox{Ha}y^*)}{\cosh(\mbox{Ha})}\right]\bigg\},\\
  B_x(y) &=& B_{r}\bigg\{-\frac{y^*}{\mbox{Ha}}+\frac{1}{\mbox{Ha}}\frac{1}{\tanh(\mbox{Ha})}\frac{\sinh(\mbox{Ha}y^*)}{\cosh(\mbox{Ha})}\bigg\},
\end{eqnarray*}
where $y^* = y/H$ and $-1\le y^* \le 1$, i.e., the origin of the vertical coordinate $y$ is set at the center of the channel. Here, the characteristic reference velocity $U_{r}$, the reference scale for the magnetic field $B_{r}$, as well as the Hartmann number $\mbox{Ha}$ can be expressed as
\begin{equation*}
  U_{r} = \frac{H^2}{\rho\nu}F_x,\quad\quad B_{r} = H^2F_x\mu_m\left(\frac{\sigma}{\rho\nu}\right)^{1/2},\quad\quad \mbox{Ha} = B_0 H \left(\frac{\sigma}{\rho\nu}\right)^{1/2},
\end{equation*}
where $\mu_m$ is the magnetic permeability, which is taken to be unity in the numerical experiments reported in what follows.

This MHD flow case study requires the solution of the magnetic induction equation (MIE) to compute the induced magnetic field, which is then coupled to the NSE for the fluid motion via the Lorentz force. A derivation and algorithmic implementation of the vector distribution function based GOC-LBM extended from the uniform grid-based formulation~\cite{dellar2002lattice} along with an approach to compute the Lorentz force locally via the current density for such a coupling given in Appendix~\ref{sec:Appendix_E} is used in this regard. This appendix also provides a half-way bounce-back scheme to impose the applied magnetic field on the insulated boundaries. To achieve continuously variable grids with clustering near both the walls, we use the Roberts coordinate transformation with the grid clustering parameter $\gamma = 1.05$. Simulations were then performed with the Hartmann numbers $\mbox{Ha} = 50, 100, 300$, and $500$, which are significantly higher than the $\mbox{Ha}$ around 10 used in a prior work on the uniform grid-based LBM for MHD~\cite{dellar2002lattice}. With the consideration of periodic boundary conditions along the horizontal flow direction, we keep the number of grid points in that direction to be fixed at $N_x = 3$ in all cases. On the other hand, in order to resolve the Hartmann boundary layers in the $y$ direction, we vary the number of grid points in that direction $N_y$ according to $\mbox{Ha}$ as follows: $N_y = 270, 360, 1200$, and $2500$ for $\mbox{Ha} = 50, 100, 300$, and $500$, respectively. For the purpose of making comparisons we keep the same number of grid points $N_x\times N_y$ for both the GOC-LBM and the uniform grid-based standard LBM for each $\mbox{Ha}$.

\begin{figure}[htbp]
\centering
\advance\leftskip-1.2cm
    \subfloat[GOC-LBM] {
        \includegraphics[trim = 0 0 0 0, clip, width =90mm] {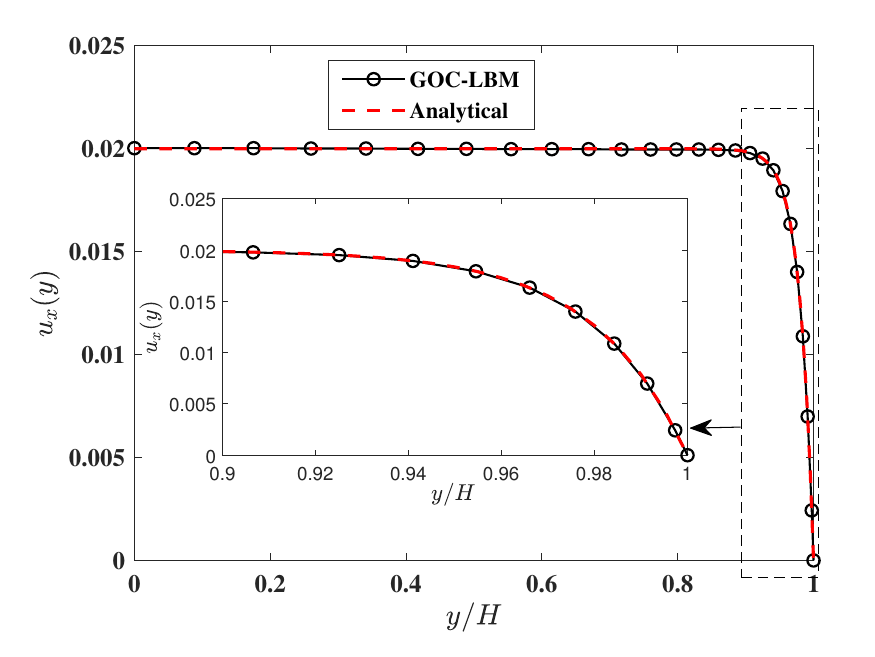}
        \label{fig:u_Ha50_GOC} } 
    \subfloat[Standard LBM] {
        \includegraphics[trim = 0 0 0 0, clip, width =90mm] {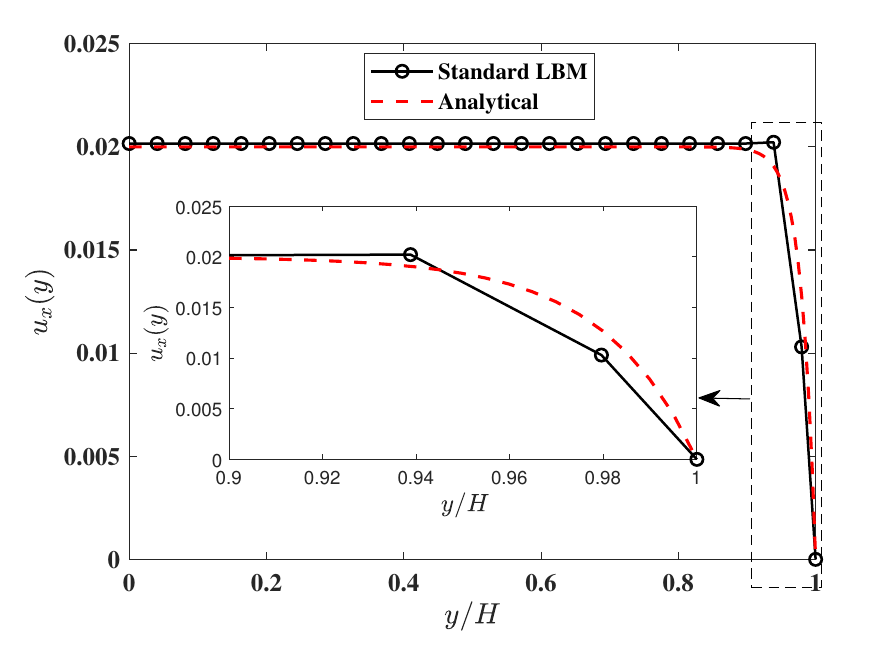}
        \label{fig:u_Ha50_standard} } \\
                \advance\leftskip0cm
    \caption{Velocity profiles across a half of the channel computed using (a) the GOC-LBM and (b) the standard LBM based on uniform grids, and compared with the analytical solution for an electrically conducting fluid flow (i.e., the Hartmann flow) between two insulated parallel plates under an imposed magnetic field at a Hartmann number $\mbox{Ha}=50$. With the GOC-LBM, the Roberts coordinate transformation is to achieve continuously varying grids along the $y$ direction with clustering around both the walls based on setting the grid clustering/stretching parameter $\gamma = 1.05$ (see Sec.~\ref{sec:Transformations}). The insets show magnified views of the variations in the profiles very close to the wall.}
    \label{fig:u_Ha50}
\end{figure}

\begin{figure}[htbp]
\centering
\advance\leftskip-1.2cm
    \subfloat[GOC-LBM] {
        \includegraphics[trim = 0 0 0 0, clip, width =90mm] {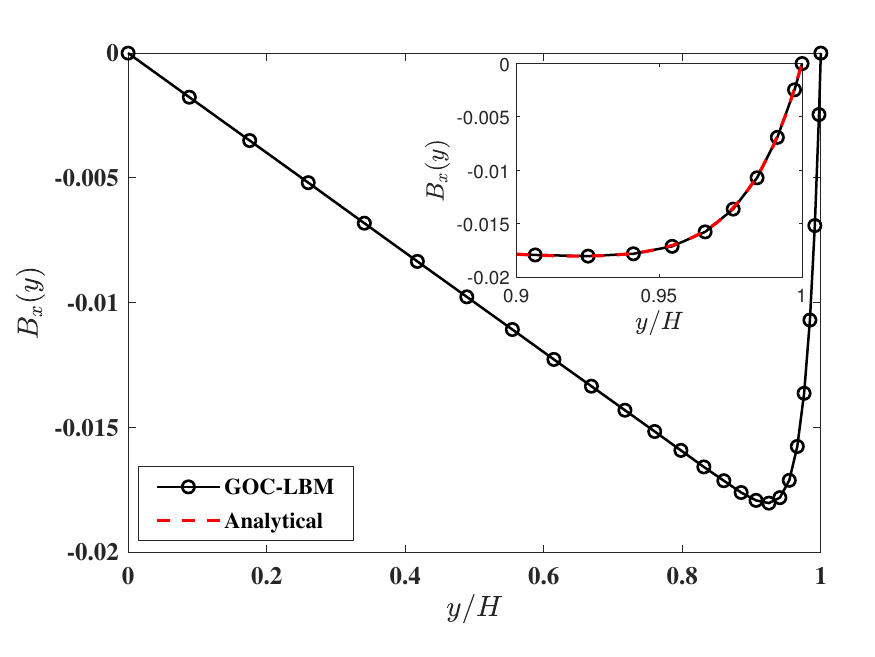}
        \label{fig:Bx_Ha50_GOC} } 
    \subfloat[Standard LBM] {
        \includegraphics[trim = 0 0 0 0, clip, width =90mm] {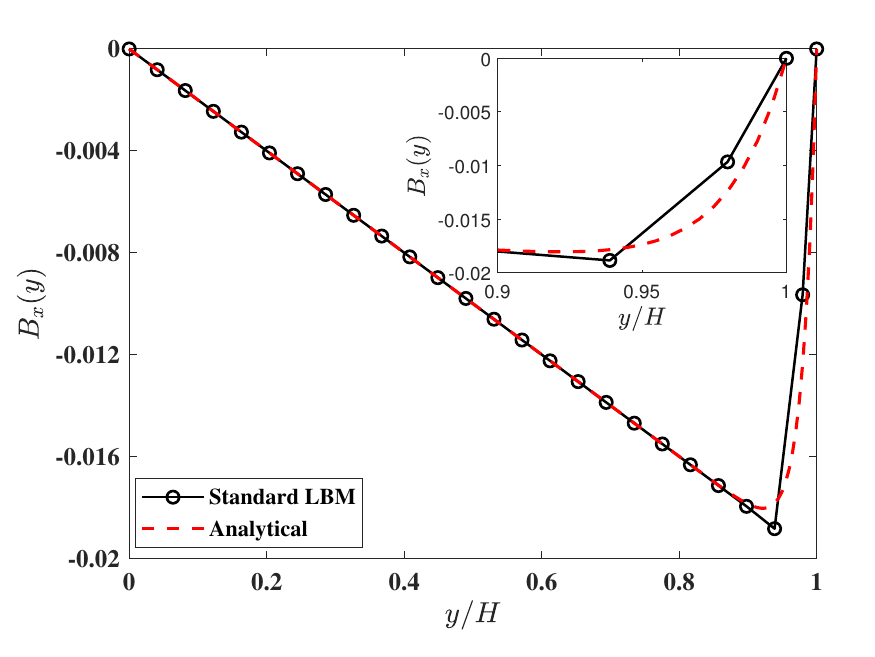}
        \label{fig:Bx_Ha50_standard} } \\
                \advance\leftskip0cm
    \caption{Induced magnetic field profiles across a half of the channel computed using (a) the GOC-LBM and (b) the standard LBM based on uniform grids, and compared with the analytical solution for an electrically conducting fluid flow (i.e., the Hartmann flow) between two insulated parallel plates under an imposed magnetic field at a Hartmann number $\mbox{Ha}=50$. With the GOC-LBM, the Roberts coordinate transformation is to achieve continuously varying grids along the $y$ direction with clustering around both the walls based on setting the grid clustering/stretching parameter $\gamma = 1.05$ (see Sec.~\ref{sec:Transformations}). The insets show magnified views of the variations in the profiles very close to the wall.}
    \label{fig:Bx_Ha50}
\end{figure}

Exploiting the symmetry (anti-symmetry) of the velocity field (magnetic field) profiles about the center of the channel, we will display them based on the results obtained from our simulations for only one-half of the channel, which also helps to focus our attention on their variations in the near-wall regions better. Thus, Figs.~\ref{fig:u_Ha50} and~\ref{fig:Bx_Ha50} show the velocity profiles and magnetic field profiles, respectively, at $\mbox{Ha} = 50$ computed using the GOC-LBM (left) and the uniform grid-based standard LBM (right), where the respective analytical solutions given above are also plotted for assessing the performance of these two LB schemes. In general, the velocity profile is generally flat for the bulk of the channel owing to the relatively strong opposing Lorentz force, which then sharply decays to zero to maintain the no-slip condition in a thin region around the wall. On the other hand, the magnitude of the induced magnetic field $|B_x(y)|$ increases monotonically from zero at the center of the channel to a maximum value at a location very close to the wall and then rapidly decaying to zero within the boundary layer region. With the number of grid points $N_y = 270$, the computed results using the GOC-LBM with near grid clustering as indicated above are in excellent agreement with the analytical solutions for both the velocity and the induced magnetic fields everywhere including, especially, within the boundary layer as shown in the insets of these figures. By contrast, when the same number of grid points are used with the standard LBM based on uniform grids, it is not able to accurately resolve the sharp variations in the profiles of these two fields in the near-wall region, which can be clearly seen in the magnified views given in the insets of these figures. This is due to the fact, unlike the GOC-LBM which uses a nonuniform gridding structure that conforms with the nature of the variations of these fields by clustering adequate number of grid points within the boundary layer, the standard LBM uses too few grid points within the thin layer and too many outside of it as a consequence of its uniform gridding structure.

\begin{figure}[htbp]
\centering
\advance\leftskip-1.2cm
    \subfloat[GOC-LBM] {
        \includegraphics[trim = 0 0 0 0, clip, width =90mm] {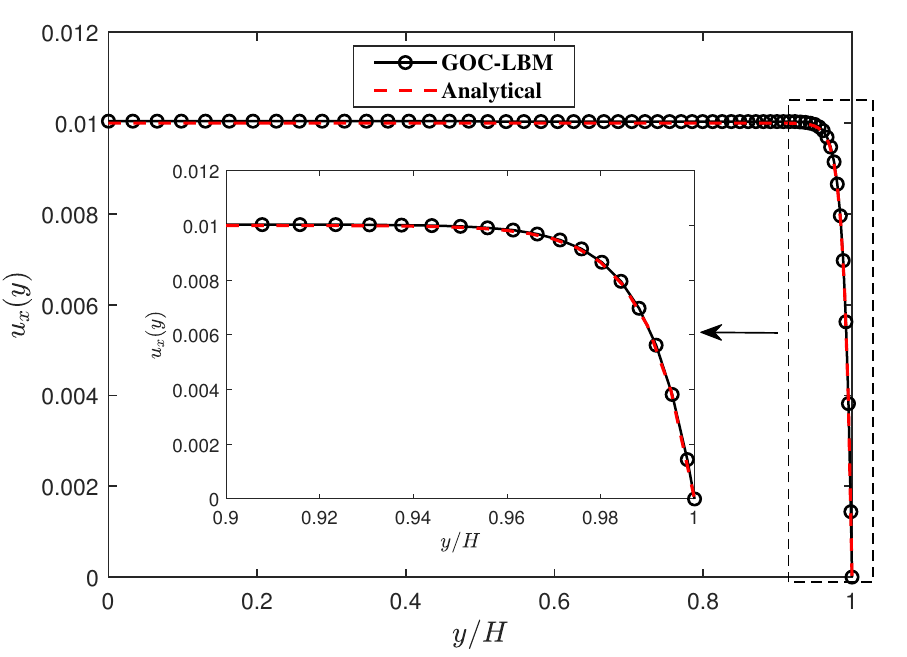}
        \label{fig:u_Ha100_GOC} } 
    \subfloat[Standard LBM] {
        \includegraphics[trim = 0 0 0 0, clip, width =90mm] {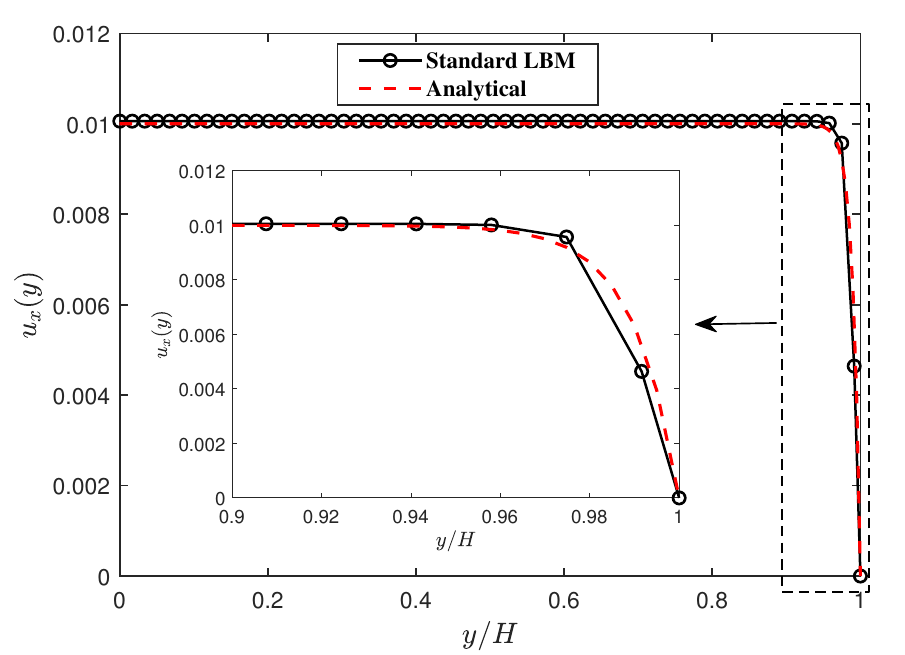}
        \label{fig:u_Ha100_standard} } \\
                \advance\leftskip0cm
    \caption{Velocity profiles across a half of the channel computed using (a) the GOC-LBM and (b) the standard LBM based on uniform grids, and compared with the analytical solution for an electrically conducting fluid flow (i.e., the Hartmann flow) between two insulated parallel plates under an imposed magnetic field at a Hartmann number $\mbox{Ha}=100$. With the GOC-LBM, the Roberts coordinate transformation is to achieve continuously varying grids along the $y$ direction with clustering around both the walls based on setting the grid clustering/stretching parameter $\gamma = 1.05$ (see Sec.~\ref{sec:Transformations}). The insets show magnified views of the variations in the profiles very close to the wall.}
    \label{fig:u_Ha100}
\end{figure}

\begin{figure}[htbp]
\centering
\advance\leftskip-1.2cm
    \subfloat[GOC-LBM] {
        \includegraphics[trim = 0 0 0 0, clip, width =90mm] {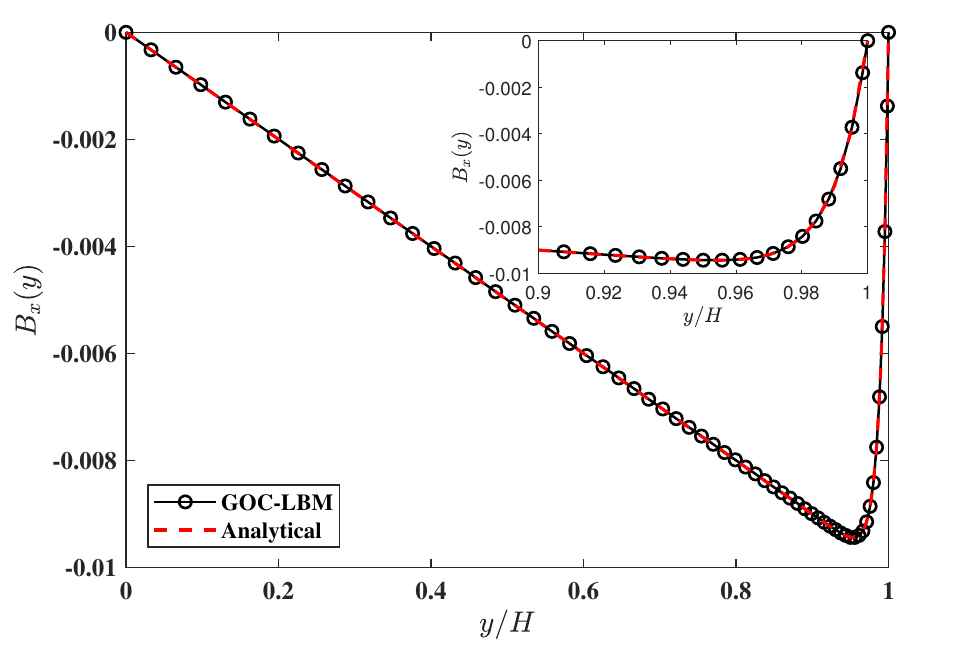}
        \label{fig:Bx_Ha100_GOC} } 
    \subfloat[Standard LBM] {
        \includegraphics[trim = 0 0 0 0, clip, width =90mm] {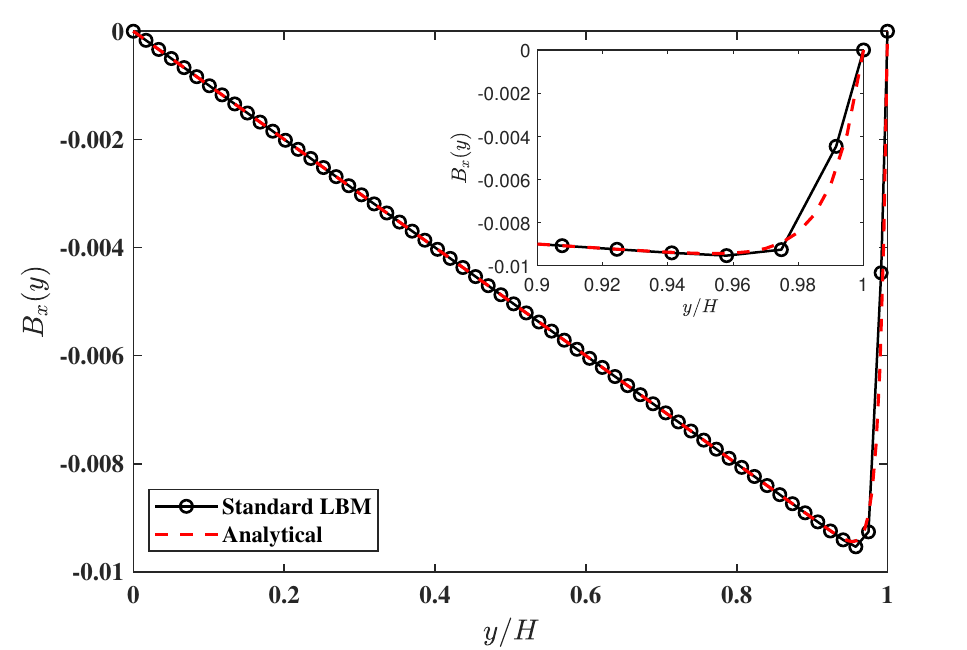}
        \label{fig:Bx_Ha100_standard} } \\
                \advance\leftskip0cm
    \caption{Induced magnetic field profiles across a half of the channel computed using (a) the GOC-LBM and (b) the standard LBM based on uniform grids, and compared with the analytical solution for an electrically conducting fluid flow (i.e., the Hartmann flow) between two insulated parallel plates under an imposed magnetic field at a Hartmann number $\mbox{Ha}=100$. With the GOC-LBM, the Roberts coordinate transformation is to achieve continuously varying grids along the $y$ direction with clustering around both the walls based on setting the grid clustering/stretching parameter $\gamma = 1.05$ (see Sec.~\ref{sec:Transformations}). The insets show magnified views of the variations in the profiles very close to the wall.}
    \label{fig:Bx_Ha100}
\end{figure}

Moreover, as another example, with using $N_y = 360$ at a higher $\mbox{Ha} = 100$, the simulation results with both the GOC-LBM and the standard LBM for the velocity and induced the magnetic field profiles are presented in Figs.~\ref{fig:u_Ha100} and~\ref{fig:Bx_Ha100}, respectively. Clearly, the boundary layer becomes thinner compared to the previous example carried out at $\mbox{Ha} = 50$, which required the use of more number of grid points in the former case. Again, the GOC-LBM results are in excellent agreement with the respective analytical solutions throughout the domain, whereas the standard LBM using the same number of grid points under resolves the variations within the all-important boundary layer region.

\begin{figure}[htbp]
  \centering
    \captionsetup{justification=centering}
    \includegraphics[scale=0.65]{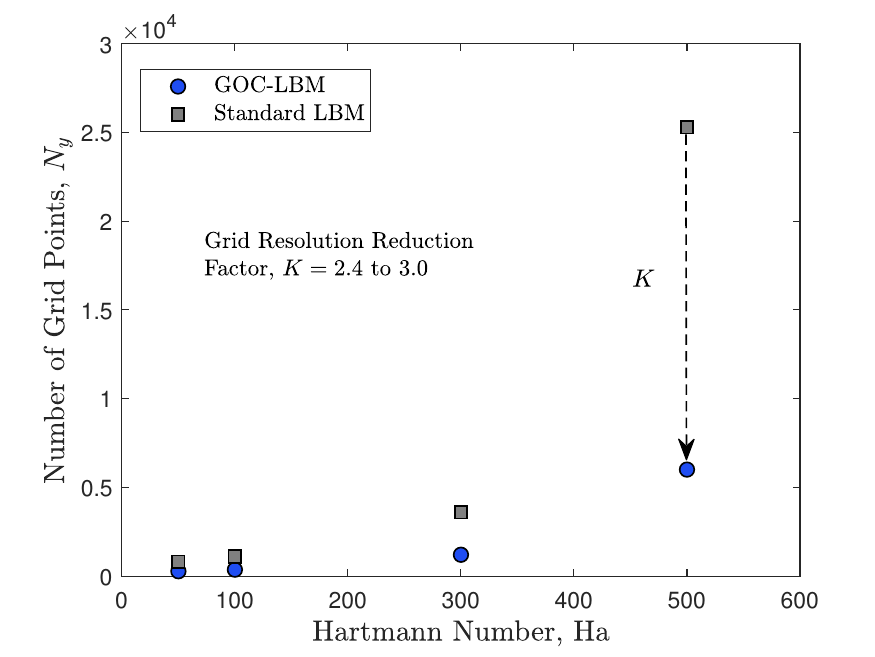}
    \caption{Requirements on the number of grid points across the channel $N_y$ for the GOC-LBM using grid clustering based on the Roberts coordinate transformation with $\gamma = 1.05$ and the uniform grid-based standard LBM for resolving the Hartmann boundary layers to achieve the same quantitative accuracy with these two methods at different Hartmann numbers, viz., $\mbox{Ha} = 50, 100, 300$, and $500$. The grid resolution reduction factor $K$ refers to the savings in the number of grid points with the GOC-LBM when compared to the standard LBM.}
    \label{fig:GOC_preference}
\end{figure}

On the other hand, it would be interesting to know the requirements on the number of grid points $N_x\times N_y$ for resolving the Hartmann boundary layer structures for both the velocity and magnetic fields at different $\mbox{Ha}$ to achieve the \emph{same overall quantitative accuracy} with both the GOC-LBM and the standard LBM. Figure~\ref{fig:GOC_preference} presents a visual representation along with the corresponding quantitative information in the Table~\ref{table:3} on the number of grid points required to achieve a relative error of $1\times 10^{-5}$ between either the GOC-LBM or the standard LBM and the respective analytical solution under the 2-norm for the Hartmann numbers $\mbox{Ha} = 50, 100, 300$, and $500$. From these, the grid resolution reduction factor $K$ providing the savings in the number of grid points with the GOC-LBM when compared to the standard LBM can be inferred. In general, the GOC-LBM is seen to result in savings in the gridding requirements by a factor that ranges between 2.4 and 3.05 for the above considered cases, which is already quite significant despite the fact that the clustering is done only along one of the directions (which is along $y$ here). In practical applications that involve grid clustering in all the three directions, such a savings factor becomes multiplicative and one can expect about an order of magnitude improvements in terms of memory and computational cost in effectively resolving confined boundary layer flows with the GOC-LBM when compared to the standard LBM.
\begin{table}[htbp]
\centering
\begin{tabular}{c | c| c |c}
\hline
\hline
 Hartmann Number &  GOC-LBM  &  Standard LBM & Grid Savings Factor K \\
 \hline\hline
 $50$  &  $3 \times 270$  & $3\times 800 $ & $2.9$ \\
 $100$ & $3 \times 360$  & $3\times 1100$ & $3.05$ \\
 $300 $& $3 \times 1200$ & $3\times 3600$ & $3$  \\
 $500 $& $3 \times 2500$ & $3\times 6000$ & $2.4$\\
\hline
\end{tabular}
\caption{Number of grid points required for resolving the Hartmann boundary layer flows $N_x\times N_y$ to achieve a relative error of $1\times 10^{-5}$ between either the GOC-LBM or the standard LBM and the respective analytical solution under the 2-norm for the Hartmann numbers $\mbox{Ha} = 50, 100, 300$, and $500$. The grid resolution reduction factor $K$ refers to the savings in the number of grid points with the GOC-LBM when compared to the standard LBM.}
\label{table:3}
\end{table}

\section{Demonstration of GOC-LBM for Flow Simulations within Curved Geometries} \label{sec: curvedbenchmark}
Finally, we will demonstrate the capability of the GOC-LBM in accurately simulating flows within curved geometries by considering two benchmark flow problems: (i) the canonical Taylor-Couette flow between two concentric rotating cylinders, and (ii) shear-driven polar cavity flow. Since these require setting up orthogonal curvilinear grids within sections of concentric cylinders, we will first discuss a simple body-fitted grid generation approach based on a conformal mapping in the following and then show some simulation results that compare with velocity profiles computed using the GOC-LBM against the analytical solution, benchmark numerical results or experimental data, as appropriate, for these problems.

\subsection{Body-Fitted Orthogonal Curvilinear (Circular) Grids using Conformal Mapping} \label{sec: curvedbenchmark-conformal-mapping}
\begin{figure}[htbp]
\centering
\begin{subfigure}{0.40\textwidth}
\includegraphics[trim = 0 0 0 0, clip, width = 55mm]{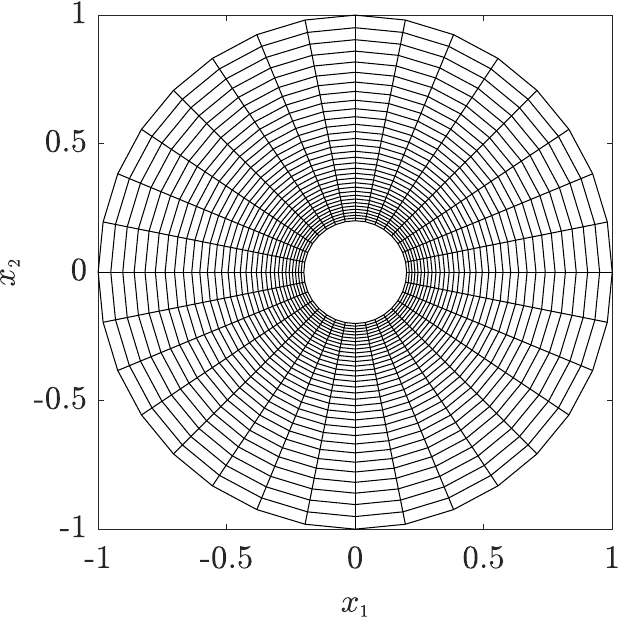}
\caption{Physical Domain}
\end{subfigure}
\begin{subfigure}{0.40\textwidth}
\includegraphics[trim = 0 0 0 0, clip, width = 55mm]{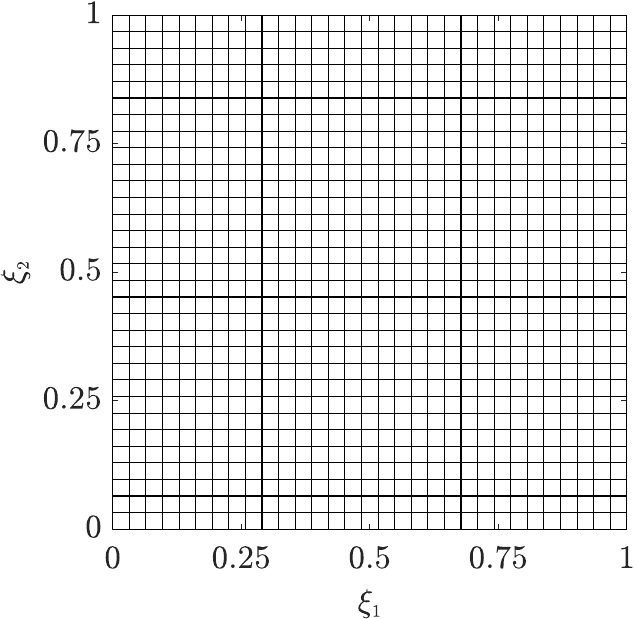}
\caption{Computational Domain}
\end{subfigure}
\caption{Orthogonal curvilinear grids around a circular cylinder via conformal mapping between the Cartesian coordinates ($x_1, x_2$) for the physical domain and the GOC ($\xi_1, \xi_2$) for the computational domain.}
\label{fig:1_cylinder_conformal_mapping}
\end{figure}

\begin{figure}[htbp]
\centering
\begin{subfigure}{0.40\textwidth}
\includegraphics[trim = 0 0 0 0, clip, width = 55mm]{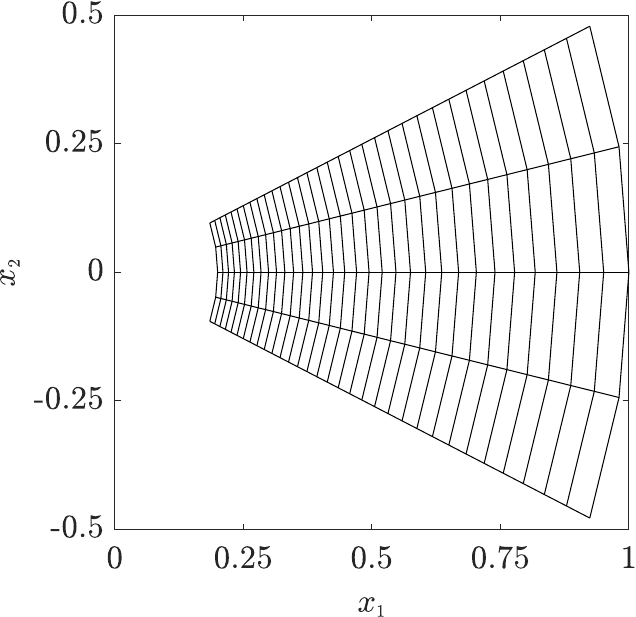}
\caption{Physical Domain}
\end{subfigure}
\begin{subfigure}{0.40\textwidth}
\includegraphics[trim = 0 0 0 0, clip, width = 55mm]{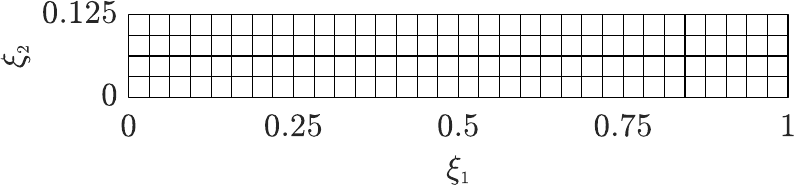}
\caption{Computational Domain}
\end{subfigure}
\caption{Orthogonal curvilinear grids for a sector between two concentric circular cylinders via conformal mapping between the Cartesian coordinates ($x_1, x_2$) for the physical domain and the GOC ($\xi_1, \xi_2$) for the computational domain.}
\label{fig:1_annularsector_conformal_mapping}
\end{figure}

Let's first introduce a complex variable $Z$ in the physical plane (in the Cartesian coordinates) and another complex variable $W$ in the computational plane (in the GOC), which can be represented as
\begin{align}\label{eq:1_annularsector}
  Z&=x_1 +i\;x_2, \quad\quad\quad W =\tilde{\xi}_1 +i\;\tilde{\xi}_2,
\end{align}
where the tilde over the GOC variables are introduced to denote that they are tentative and their final forms will be achieved via appropriate shifting and scaling (see below). For generating orthogonal curvilinear grids around a circular cylinder of radius $a$ (such as in Fig.~\ref{fig:1_cylinder_conformal_mapping}), one can use the following conformal mapping between these two complex variables:
\begin{equation*}
Z = a\exp(\pi W).
\end{equation*}
However, in the two benchmark flow problems mentioned above, it is necessary to consider just an annular sector between two concentric full circles that can result in significant savings in the gridding requirements. In this regard, we introduce the following conformal mapping that transforms an annular circular sector that is bounded by two circles of inner and outer radii of $r = a$ and $r = R_o$, respectively, and the two lines with the polar angles  $\theta = -\pi/N_\theta$ and $\theta = \pi/N_\theta$:
\begin{align}\label{eq:2_annularsector}
Z&= a \exp\left(\frac{\pi}{N_\theta} W \right),
\end{align}
where $N_\theta$ is a parameter that divides a full circle into an annular sector of angle $2\pi/N_\theta$. This maps the Cartesian coordinates representing the annulus region into a rectangle in the computational domain with the GOC bounded by $0 \le \tilde{\xi}_1 \le \tilde{\xi}_{1,o} $ and $-1 \le \tilde{\xi}_2 \le 1$ as shown in Fig.~\ref{fig:1_annularsector_conformal_mapping}, where $\hat{\xi}_{1,o}$ will depend on the associated geometric parameters $a$, $R_o$ and $N_\theta$ in what follows.

Then, substituting Eq.~\eqref{eq:1_annularsector} into Eq.~\eqref{eq:2_annularsector}, we get the mapping relations between the two coordinate systems in their component form, which read as
\begin{equation} \label{eq:3_annularsector}
x_{1} = a \exp\left( \frac{\pi}{N_\theta} \tilde{\xi}_{1}\right) \cos\left( \frac{\pi}{N_\theta} \tilde{\xi}_{2}\right),\qquad\qquad
x_{2} = a \exp\left( \frac{\pi}{N_\theta} \tilde{\xi}_{1}\right) \sin\left( \frac{\pi}{N_\theta} \tilde{\xi}_{2}\right).
\end{equation}
From this last equation, we can obtain an explicit expression for the outer limit of the GOC $\tilde{\xi}_{1,o}$, which corresponds to $r = R_o$ via $ r= R_o = (x_1^2 + x_2^2)^{1/2}$ in the Cartesian coordinates when $\tilde{\xi}_{1} = \tilde{\xi}_{1,o}$. The result can be written as
\begin{equation}\label{eq:xi1outerlimit}
\tilde{\xi}_{1,o} = \frac{N_\theta}{\pi}\ln\left(\frac{R_o}{a}\right).
\end{equation}
Based on these, the annulus sectorial region in the physical plane in the Cartesian coordinates ($x_1, x_2$), which can be represented more compactly in the following ranges for their equivalent polar representation ($r,\theta$) and the ranges for the GOC ($\tilde{\xi}_1, \tilde{\xi}_2$) in the computational plane:
\begin{align*}
\begin{rcases}
a \le r \le R_{\infty}\\
-\dfrac{\pi}{N_\theta} \le \theta \le \dfrac{\pi}{N_\theta}
\end{rcases}
&\quad\text{Physical Plane} \quad\quad\quad\quad\quad
\begin{rcases}
0 \le \tilde{\xi}_{1} \le \tilde{\xi}_{1,o}\\
-1 \le \tilde{\xi}_{2}  \le 1
\end{rcases}
\quad\text{Computational Plane}
\end{align*}
However, the ranges for the GOC ($\tilde{\xi}_1, \tilde{\xi}_2$) are not in a convenient form to directly implement the GOC-LBM in the usual lattice units, which can be accomplished via the following considerations. If we resolve the GOC encompassing the annular sector with $N_{\xi_1}$ and $N_{\xi_2}$ grid points along the $\xi_1$ and $\xi_2$, respectively, their corresponding length scales in the lattice units are $L_{\xi_1}=N_{\xi_1}-1$ and $L_{\xi_2}=N_{\xi_2}-1$. Based on this we require the usual target ranges for the GOC as $0 \le \xi_1 \le L_{\xi_1}$ and $0 \le \xi_2 \le L_{\xi_2}$. In other words, we need some transformations that take the strained coordinates ($\tilde{\xi}_1, \tilde{\xi}_2$) to the usual coordinates ($\xi_1, \xi_2$) in the computational plane, which can be formally expressed as
\begin{align*}
\begin{rcases}
0 \le \tilde{\xi}_{1} \le \tilde{\xi}_{1,o}\\
-1 \le \tilde{\xi}_{2}  \le 1
\end{rcases}
&\Longrightarrow \quad
\begin{rcases}
0 \le \xi_1 \le L_{\xi_1}\\
0 \le \xi_2 \le L_{\xi_2}
\end{rcases}
\end{align*}
Such transformations can be readily achieved via the following shifting and scaling-based mappings:
\begin{align}\label{eq:5_annularsector}
  \tilde{\xi}_{1}&= \frac{\tilde{\xi}_{1,o}}{L_{\xi_1}}\xi_1 \quad\quad\quad \tilde{\xi}_{2}= -1 + \frac{2}{L_{\xi_2}}\xi_2,
\end{align}
where $\tilde{\xi}_{1,o}$ is given in Eq.~(\ref{eq:xi1outerlimit}).
Then, substituting this last equation in Eq.~\eqref{eq:5_annularsector} along with  Eq.~(\ref{eq:xi1outerlimit}) into Eq.~\eqref{eq:3_annularsector}, we finally get the desired coordinate transformations that map a sectorial region between two concentric circles of inner and outer radii $a$ and $R_o$, respectively, in the Cartesian coordinates ($x_1, x_2$) to the GOC ($\xi_1, \xi_2$) which are amenable for implementation with the GOC-LBM:
\begin{subequations}
\begin{align}
x_{1}&= a \exp \bigg[\ln \left(\frac{R_{o}}{a}\right)\frac{\xi_1}{L_{\xi_1}}\bigg] \cos\bigg[\frac{\pi}{N_\theta} \left(2\frac{\xi_2}{L_{\xi_2}}-1\right)\bigg],\\
x_{2}&= a \exp \bigg[\ln \left(\frac{R_{o}}{a}\right)\frac{\xi_1}{L_{\xi_1}}\bigg] \sin\bigg[\frac{\pi}{N_\theta} \left(2\frac{\xi_2}{L_{\xi_2}}-1\right)\bigg],
\end{align}
\end{subequations}
We complete this section by obtaining the analytic formulas for the metric factors and the curvature coefficient matrix elements from their definitions given in Eqs.~(\ref{eq:metricfactor}) and~(\ref{eq:curvaturematrix}), respectively. The results read as
\begin{eqnarray}
&&h\sbs{1}=\frac{a}{L\sbs{\xi\sbs{1}}}\ln \left(\frac{R_o}{a}\right)\exp\left[ \ln \left(\frac{R_o}{a}\right) \frac{\xi\sbs{1}}{L\sbs{\xi\sbs{1}}}  \right],\qquad
h\sbs{2}=\frac{2\pi a}{N_\theta L\sbs{\xi\sbs{2}}}\exp\left[ \ln \left(\frac{R_o}{a}\right) \frac{\xi\sbs{1}}{L\sbs{\xi\sbs{1}}}  \right],\\
&&\theta\sbs{11} = \theta\sbs{21} = \frac{1}{a} \exp \left[-\ln \left(\frac{R_o}{a}\right)\frac{\xi\sbs{1}}{L\sbs{\xi\sbs{1}}}\right], \quad\quad\quad
\theta\sbs{12} = \theta\sbs{22} = 0.
\end{eqnarray}
Thus, $L_{\xi_1}$ (via the number of grid points $N_{\xi_1}$ as $N_{\xi_1} = L_{\xi_1} + 1$) resolves the annular sector in the radial direction such that $L_{\xi_1}= R_{o}-a$, and $L_{\xi_2}$ (via the number of grid points $N_{\xi_2}$ as $L_{\xi_2} = N_{\xi_2} - 1$) resolves the annular sector in the azimuthal direction (in the range $-\pi/N_\theta \le \theta \le \pi/N_\theta$), i.e., the sector then consists of one of the $N_\theta$ parts of dividing the fully circular domain.

\subsection{Simulation Results of Taylor-Couette Flow Between Two Concentric Rotating Cylinders}\label{subsec:Taylor-CouetteFlow}

\begin{figure}[htbp]
\centering
\begin{subfigure}{0.45\textwidth}
\includegraphics[trim = 0 0 0 0, clip, width =50mm]{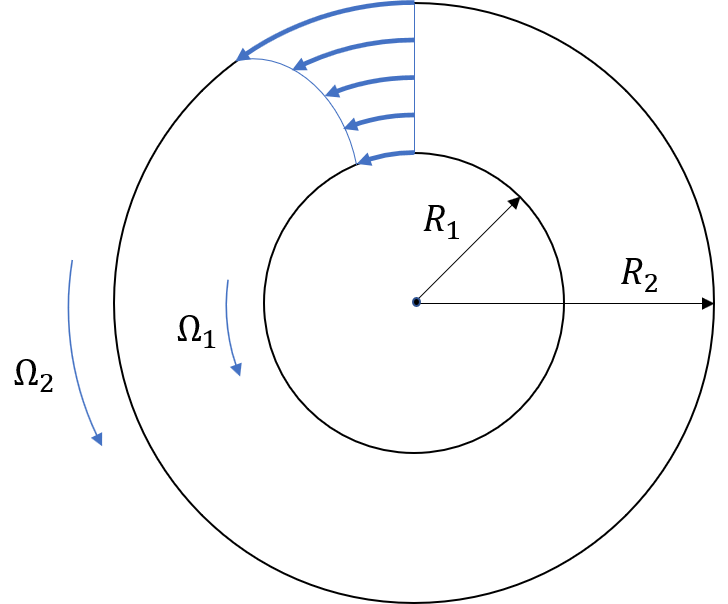}
\end{subfigure}
\caption{Schematic of circular Taylor-Couette flow between two concentric cylinders of radii $R_1$ and $R_2$ rotating at rates $\Omega_1$ and $\Omega_2$, respectively.}
\label{circflow}
\end{figure}

The benchmark flow case study here involves two concentric cylinders with radii $R_1$ and $R_2$ for the inner and outer cylinders, respectively, each rotating with angular velocities $\Omega_1$ and $\Omega_2$, which sets up shear driven flow within the annulus region as shown schematically in Fig.~\ref{circflow}.
This Taylor-Couette flow has the following analytical solution for the azimuthal velocity component $u\sbs{\theta}$:
\begin{equation}
u\sbs{\theta}(r) = \frac{1}{1-(R\sbs{1}/R\sbs{2})\sps{2}}\left\{\left[\Omega\sbs{2}-\Omega\sbs{1}\left(\frac{R\sbs{1}}{R\sbs{2}}\right)\sps{2} \right]r + \frac{R\sbs{1}\sps{2}}{r}\left(\Omega\sbs{1}-\Omega\sbs{2}\right) \right\},
\end{equation}
which will then be used to compare with the results obtained using our GOC-LBM for this curved geometry case using the body-fitted orthogonal curvilinear grids as discussed above.

\begin{figure}[htbp]
\centering
\begin{subfigure}{0.45\textwidth}
\includegraphics[trim = 0 0 0 0, clip, width =70mm]{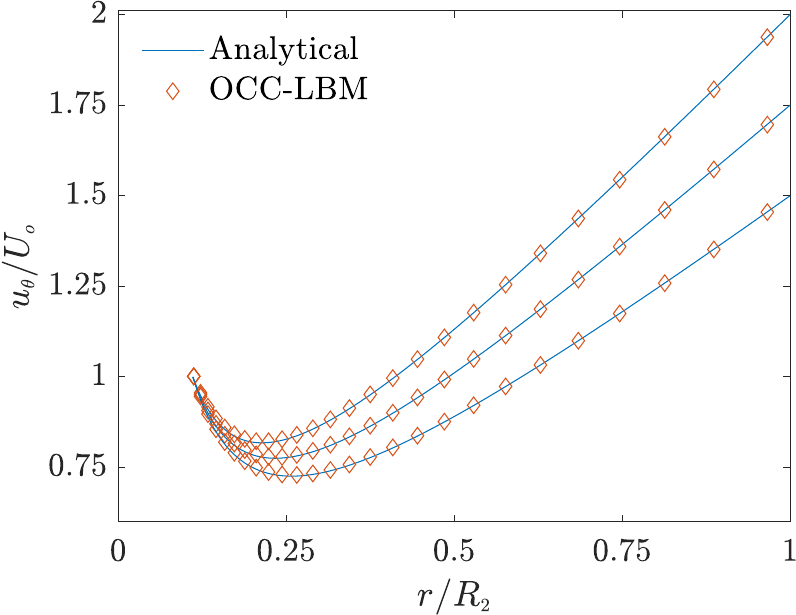}
\caption{}
\label{case1}
\end{subfigure}
\begin{subfigure}{0.45\textwidth}
\includegraphics[trim = 0 0 0 0, clip, width =70mm]{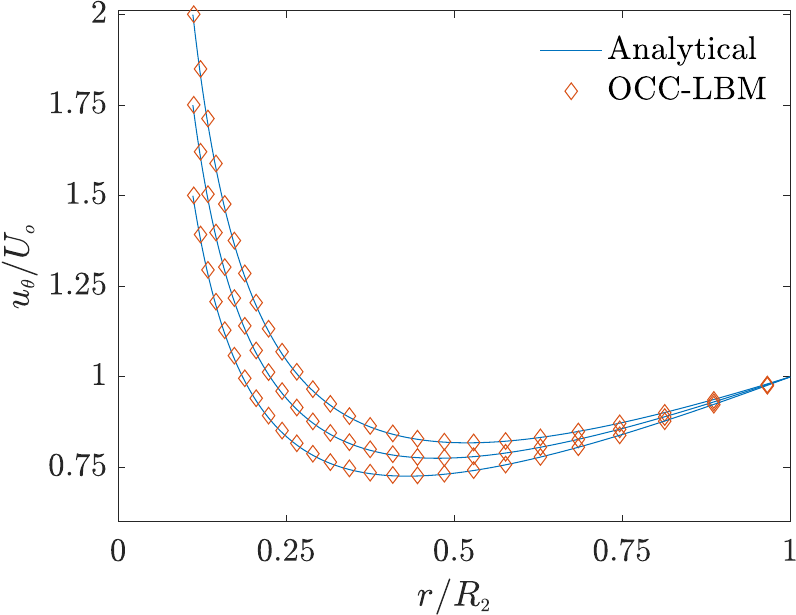}
\caption{}
\label{case2}
\end{subfigure}
\caption{Comparison of the velocity profiles $u_\theta(r)$ for the Taylor-Couette flow computed using the GOC-LBM with the analytical solution. (a) The inner cylinder rotates at a fixed linear velocity $U_1 = \Omega_1 R_1 = U_o$ while the outer cylinder rotates at the following three linear velocities: $U_2 = \Omega_2 R_2 = 1.5U_o, 1.75U_o$, and $2.0U_o$. (b)  The outer cylinder rotates at a fixed linear velocity $U_2 = \Omega_2 R_2 = U_o$ while the inner cylinder rotates at the following three linear velocities: $U_1 = \Omega_1 R_1 = 1.5U_o, 1.75U_o$, and $2.0U_o$.}
\end{figure}

In the numerical implementation, the upper and lower boundaries in the computational domain along the $\xi_2$ direction (see Fig.~\ref{fig:1_annularsector_conformal_mapping}) are taken to be periodic since the flow is invariant along the $\theta$ direction, while the no-slip conditions at the left and right boundaries along the $\xi_1$ directions representing the moving solid walls with linear velocities $\Omega_1 R_1$ and $\Omega_2 R_2$ are imposed via momentum-augmented half-way bounce back scheme (see Appendix~\ref{sec:initial_boundary_conditions}). We used the following choices for the parameters to set up the numerical simulations: the number of grid points along the radial and azimuthal directions are $N_{\xi_1} = 256$ and $N_{\xi_2} = 6$, respectively, the number of sectors obtained by dividing the fully circular domain is $N_\theta = 60$ where only one of such $N_\theta = 60$ sectors encompassing the computational domain, the inner and outer radii are $R_1 = 32$, and $R_2 = R_1 + N_{\xi_1}$, respectively, the velocity scale to impose the boundary conditions is $U_0 = 0.01$, and the fluid viscosity (and the relaxation rates) are specified by setting the Reynolds number $\mbox{Re} = U_0 (R_2 - R_1)/\nu$. Figure~\ref{case1} shows comparisons of the velocity profiles $u_\theta(r)$ for the Taylor-Couette flow computed using the GOC-LBM with the analytical solution, when the inner cylinder rotates at a fixed linear velocity $U_1 = \Omega_1 R_1 = U_o$ while the outer cylinder rotates at the following three linear velocities: $U_2 = \Omega_2 R_2 = 1.5U_o, 1.75U_o$, and $2.0U_o$. On the other hand, Fig.~\ref{case2} presents similar comparisons when the outer cylinder rotates at a fixed linear velocity $U_2 = \Omega_2 R_2 = U_o$ while the inner cylinder rotates at the following three linear velocities: $U_1 = \Omega_1 R_1 = 1.5U_o, 1.75U_o$, and $2.0U_o$. In all the cases considered the numerically predicted velocity profiles using the GOC-LBM are in excellent agreement with the analytical solution, which validates its ability to effectively simulate flow through a curved geometry resolved with body-fitted orthogonal curvilinear grids.

\subsection{Simulation Results of Shear-driven Polar Cavity Flow}\label{sec:polar-cavity}
As a final case study, we consider the shear-driven polar cavity flow by the motion of a side wall, which is a well-established benchmark for evaluating numerical methods applied to incompressible flow in curved coordinate systems. Fuchs and Tillmark~\cite{FuchsTillmark1985} studied this flow problem both numerically and experimentally and provided visualization of the flow patterns as well as quantitative data for the velocity profiles. The former was carried out using a 2D multigrid-based finite-difference (MG-FD) scheme, while the latter via laser-Doppler velocity measurements for Reynolds numbers of about 60 and 350. Since the flow field in a polar cavity is fully 2D, it provides a more rigorous test case than that in Sec.~\ref{subsec:Taylor-CouetteFlow} to assess the accuracy of the GOC-LBM in the presence of curved confining boundaries.

In the present study, simulations are performed using the GOC-LBM at $\mbox{Re}=60$ and $\mbox{Re}=350$, corresponding to the Reynolds-number regimes for which Fuchs and Tillmark reported their numerical and experimental data~\cite{FuchsTillmark1985}. We will make direct comparisons of our results for the radial and azimuthal velocity profiles with this benchmark data while also presenting the 2D flow structures that arise for the cases under consideration. The geometry consists of an annular sector bounded by inner and outer circular walls and two radial sidewalls. The inner cylindrical boundary rotates with a constant tangential velocity, while the outer circular boundary and the two radial boundaries remain stationary. No-slip boundary conditions for the velocity field are imposed for all the stationary walls as well as the moving inner cylindrical wall. The motion of the inner cylinder then transfers momentum to the fluid and generates a circulating flow within the confined annular section. At relatively low Reynolds numbers, the flow is dominated by a primary recirculation region. As the Reynolds number is increased, inertial effects become increasingly important and the flow structure becomes more complex, with secondary counter-rotating vortices becoming increasingly prominent near the corners of the cavity~\cite{FuchsTillmark1985,RosenfeldKwakVinokur1991}.

The simulations using the GOC-LBM are performed in the computational domain representing the annular region consisting of $256\times256$ lattice nodes. The inner radius of the polar cavity is specified as $R_i = 256$, while the outer radius is taken to be twice the inner radius, $R_o = 2R_i = 512$. Then, the characteristic length scale is defined as the radial gap between the inner and outer cylindrical boundaries, i.e., $L_o = R_o-R_i = 256$; the characteristic velocity is prescribed as $U_o = 0.01$. Based on these considerations, the Reynolds number is defined as $\mbox{Re} = U_o L_o/\nu$, from which the shear viscosity $\nu$ is specified in setting up the simulations for $\mbox{Re}=60$ and $350$.

\begin{figure}[H]
    \centering
    \begin{subfigure}[b]{0.48\textwidth}
        \centering
        \includegraphics[
            width=0.75\linewidth,
            trim={0 0 0 0},
            clip
        ]{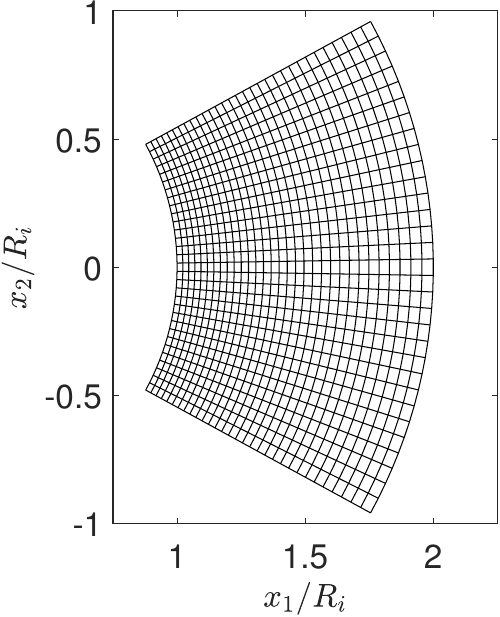}
        \caption{Physical domain}
        \label{fig:cartesian-grid}
    \end{subfigure}
    \hfill
    \begin{subfigure}[b]{0.48\textwidth}
        \centering
        \includegraphics[
            width=0.75\linewidth,
            trim={0 0 0 0},
            clip
        ]{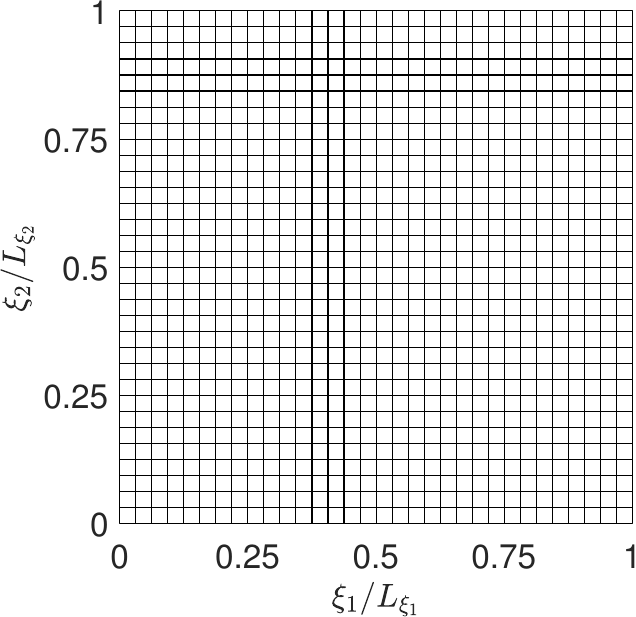}
        \caption{Computational domain}
        \label{fig:occ-grid}
    \end{subfigure}
    \caption{Mapping of a uniformly spaced computational lattice grid
    to an orthogonal curvilinear grid representing an annular sector for the polar cavity flow benchmark.}
    \label{fig:polar-grid-mapping}
\end{figure}

The annular sector that comprises the polar cavity is represented using a body-fitted orthogonal curvilinear grid based on the conformal mapping approach given earlier in Sec.~\ref{sec: curvedbenchmark-conformal-mapping}. The opening angle of the polar cavity is set to be 1 radian from which the parameter $N_\theta$ used in that section is specified as $2\pi$ which then results in the polar angle $\theta$ varying within the $-1/2 \leq \theta \leq 1/2$. The transformation maps $\xi_1=0$ to the inner circular boundary, $r=R_i$, and $\xi_1=L_{\xi_1}$ to the outer circular boundary, $r=R_o$. The boundaries $\xi_2=0$ and $\xi_2=L_{\xi_2}$ correspond to
the two radial sidewalls of the cavity. Thus, a uniformly spaced lattice in the computational space is transformed into the
curvilinear annular-sector geometry in the physical space by such a mapping (see Fig.~\ref{fig:polar-grid-mapping} for illustration).

\begin{figure}[H]
    \centering
    \begin{subfigure}[b]{0.48\textwidth}
        \centering
        \includegraphics[
            width=\textwidth,
	        angle=0,
            trim={60 0 60 0},
            clip
        ]{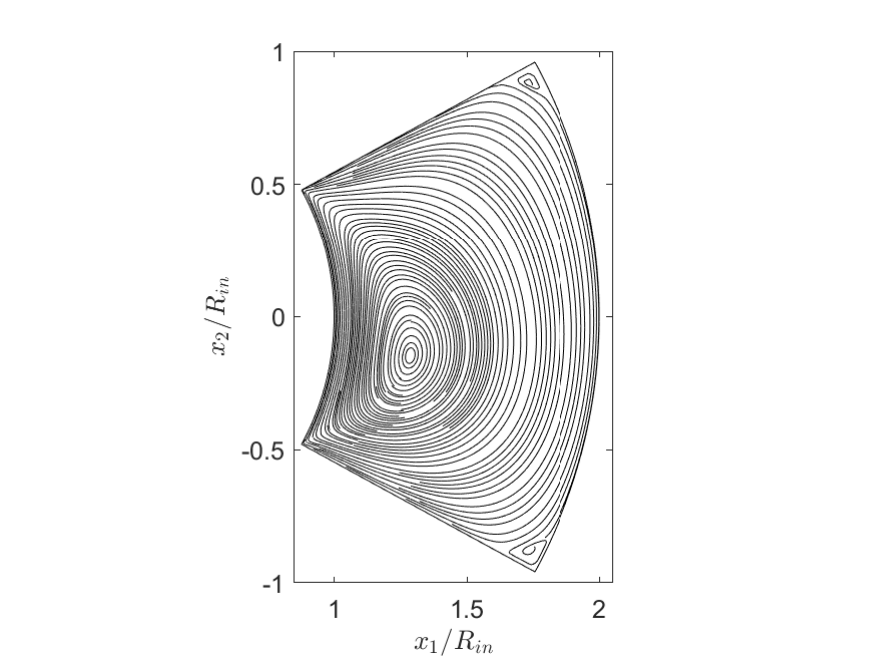}
        \caption{$\mbox{Re}=60$}
        \label{fig:polar-streamlines-Re60}
    \end{subfigure}
    \hfill
    \begin{subfigure}[b]{0.48\textwidth}
        \centering
        \includegraphics[
            width=\textwidth,
	        angle=0,
            trim={60 0 60 0},
            clip
        ]{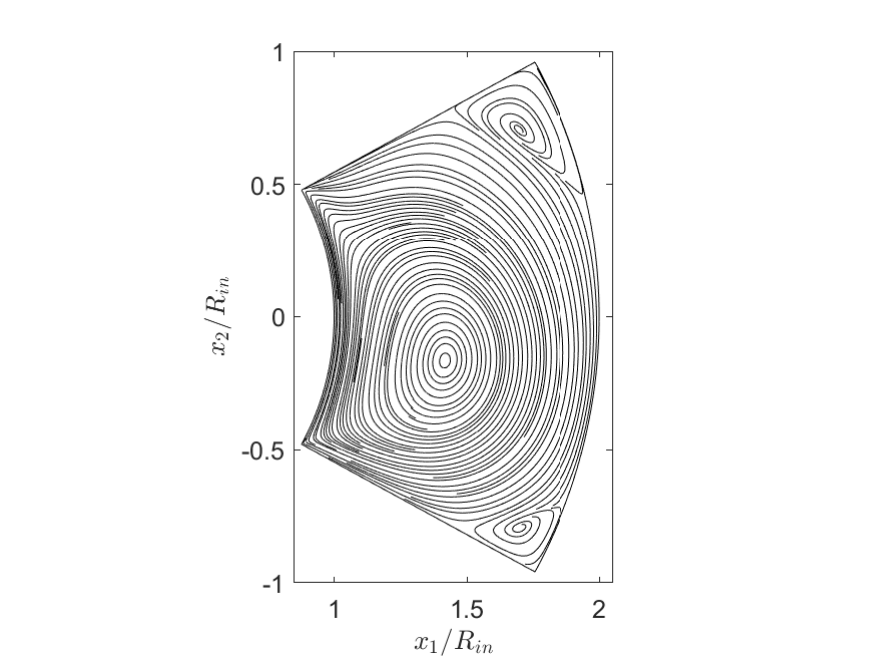}
        \caption{$\mbox{Re}=350$}
        \label{fig:polar-streamlines-Re350}
    \end{subfigure}
    \caption{Streamlines for the shear-driven polar cavity flow computed using the GOC-LBM at (a) $\mathrm{Re}=60$ and (b) $\mathrm{Re}=350$.}
    \label{fig:polar-streamlines}
\end{figure}

The streamline patterns obtained with the GOC-LBM for $\mbox{Re} = 60$ and $350$ shown in Fig.~\ref{fig:polar-streamlines} exhibit flow characteristics that are strikingly consistent with the experimental flow visualizations reported by Fuchs and Tillmark~\cite{FuchsTillmark1985}. For both Reynolds numbers, the computed flow structures reproduce the dominant recirculating motion observed experimentally. At the higher Reynolds number, the secondary corner vortices become substantially more pronounced, which is in agreement with the benchmark results of Fuchs and Tillmark showing the ability of the GOC-LBM to faithfully reproduce the overall 2D shear-driven polar cavity flow patterns.

\begin{figure}[H]
    \centering
    \begin{subfigure}[b]{0.48\textwidth}
        \centering
        \includegraphics[
            width=\textwidth,
            trim={0 0 0 14},
            clip
        ]{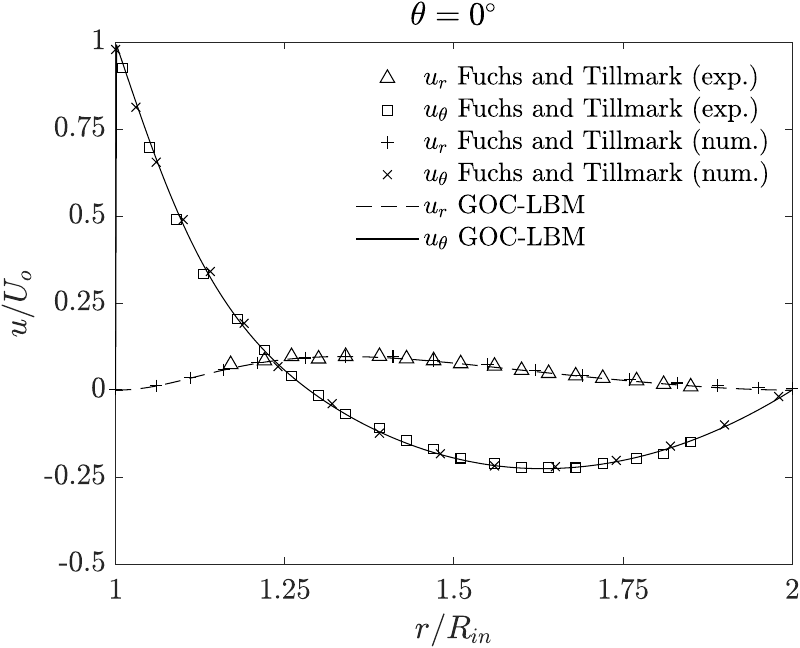}
        \caption{$\mbox{Re}=60$}
        \label{fig:polar-centerline-Re60}
    \end{subfigure}
    \hfill
    \begin{subfigure}[b]{0.48\textwidth}
        \centering
        \includegraphics[
            width=\textwidth,
            trim={0 0 0 14},
            clip
        ]{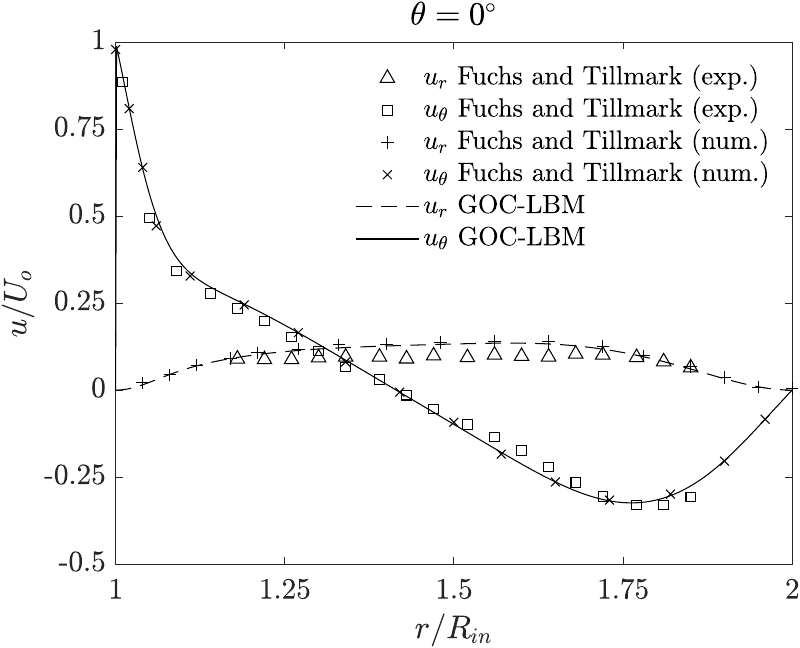}
        \caption{$\mbox{Re}=350$}
        \label{fig:polar-centerline-Re350}
    \end{subfigure}
    \caption{Comparison of the centerline radial and azimuthal velocity components $u_r$ and $u_\theta$, respectively, at the polar angle
    $\theta=0$ for (a) $\mathrm{Re}=60$ and (b) $\mathrm{Re}=350$ computed using the GOC-LBM with the benchmark numerical and experimental data of Fuchs and Tillmark~\cite{FuchsTillmark1985}.}
    \label{fig:polar-centerline}
\end{figure}

In order to provide a quantitative validation of the GOC-LBM for this curvilinear geometry~-~based flow problem, Fig.~\ref{fig:polar-centerline} shows comparisons of the centerline radial and azimuthal velocity components $u_r$ and $u_\theta$, respectively, at the polar angle $\theta=0$ for $\mbox{Re}=60$ and $\mbox{Re}=350$ computed using the GOC-LBM with the benchmark numerical and experimental data of Fuchs and Tillmark~\cite{FuchsTillmark1985}. The velocity components along both the directions computed using the GOC-LBM show very good agreement with the 2D MG-FD-based numerical results reported by Fuchs and Tillmark~\cite{FuchsTillmark1985} for both Reynolds numbers considered. The GOC-LBM profiles closely reproduce both the shape and the magnitude of the velocity components along both the directions across the entire span of the centerline, including the locations of the principal extrema and zero crossings. The agreement with the experimental measurements of Fuchs and Tillmark is also reasonable, although some deviations are observed in comparison with their numerical solution based on the 2D MG-FD scheme. These differences are consistent with the experimental uncertainty and the 3D effects in experiments reported in the original study, particularly at the higher Reynolds number. Overall, the results obtained for the shear-driven polar cavity benchmark demonstrate the validity of the two-dimensional GOC-LBM for simulating complex two-dimensional flow in curved geometries. The method accurately reproduces the principal recirculation structure, secondary vortices, and centerline velocity profiles over the Reynolds numbers considered, highlighting one of the primary advantages of the GOC-LBM formulation: the ability to simulate fluid flows using body-fitted orthogonal curvilinear grids in the physical domain, while retaining the standard collide-and-stream implementation in the computational domain.

\section{Summary and Conclusions} \label{sec: summary}
For accurate and efficient computation of boundary layer/shear layer flows and those involving curved boundaries, the ability to use grid clustering that conforms with the multiscale flow features as well as the use of body-fitted grids are highly important. However, the standard LBM is restricted to uniform grids. In this paper, in order to address this limitation and enable the use of orthogonally clustered/curvilinear grids, we have formulated general orthogonal curvilinear coordinates (GOC)-based LBM for the solution of the 2D Navier-Stokes equations (NSE) using the D2Q9 lattice. Our approach is general and can be used with any collision model and we have indeed illustrated as examples of its construction for the collision models using the SRT and MRT based on raw moments and that based on central moments, with the latter being the most robust in simulating flows at relatively high Reynolds numbers. It involves a careful design of the moment equilibria and the source term moments parameterized by the metric factors with additional equilibrium corrections that also depend on the curvature coefficient matrix elements and the normal velocity and density gradients derived via a top-down Chapman-Enskog analysis that restores an accurate representation of the normal stresses and recovers the NSE in the GOC. Moreover, interestingly, analogous to the standard LBM, the GOC-LBM relates the shear stresses to the non-equilibrium second order moments without further corrections in a time-implicit formulation. The equilibrium corrections used to restore the normal stresses in the GOC are free of the non-Galilean cubic velocity artifacts thereby generalizing the previous corrections developed for uniform grids.

As such, the resulting GOC-LBM maintains the simplicity of the standard LBM by preserving the use of the collide-and-stream-based steps while endowing them with the flexibility associated with the classical CFD methods, viz., the use of clustered/curvilinear grids. As a special case where the metric factors are set everywhere to unity, the GOC-LBM reduces to the standard LBM for uniform grids and is thus modular in construction. Our approach is further extended to the development of a new vector-based GOC-LBM for the solution of the magnetic induction equation for magnetohydrodynamics that are characterized by the presence of relatively thin boundary layer regions near walls when the Lorentz force arising from the interaction of the magnetic field with the fluid motions dominate over the viscous forces (i.e., at large Hartmann numbers). Numerical simulations of a variety of flow benchmark problems, including those involving curvilinear geometries validate the new GOC-LBM, confirm its ability to maintain Galilean invariance and demonstrate significant improvements in efficiently resolving thin boundary layers and in naturally representing flows in curved geometries via body-fitted grids.

Further extensions of the GOC-LBM to efficiently simulate the NSE in 3D as well as other complex flows, such as thermal convective flows, are currently in progress and will be reported in future studies. Moreover, the approach developed in this work serves as a natural framework for further extensions in a number of other directions. These include enabling the use of nonuniform (clustered) multiblock grids in the GOC, developing an approach for incorporating the overset (chimera) grids, i.e., composite overlapping meshes via appropriate specifications of the metric coefficients and the curvature coefficient matrix (see e.g.,~\cite{steger1987use, chesshire1990composite, petersson1999algorithm, meakin1999composite}) and/or the use of numerically generated orthogonal curvilinear grids in the LBM, and developing more robust LBM with non-orthogonal general curvilinear grids which are subjects for future investigations.

\section*{Acknowledgements}
The first and the third authors (EY and WS) thank the Department of Mechanical Engineering at the University of Colorado Denver for various forms of support during their doctoral research work and studies. The second author (KNP) acknowledges the U.S. National Science Foundation for support of the development of a computer cluster infrastructure through the project ``CC* Compute: Accelerating Science and Education by Campus and Grid Computing'' under Award 2019089. All the authors thank Hassan Hwisa for his help with the implementation of the Galilean invariance test in the GOC-LBM and the anonymous reviewers for the various insightful comments that helped improve the quality of this work.

\appendix
\section{Local Computation of the Shear Stress in the GOC using a Non-Equilibrium Moment}\label{sec:Appendix_shear_stress_local}
As shown via a C-E analysis in Eq.~(\ref{eq:72a}), the diagonal component of the second order non-equilibrium moment $n_5^{(1)}$ is related to the shear stress $\tau_{12} = \tau_{21}$. Now, this non-equilibrium moment based on its definition is given by
\begin{equation}\label{eq:noneqmoment5defn}
n_5^{(1)} = k_{11}^{\prime}+\frac{1}{2}\sigma_{11}^{\prime}\Delta t -k_{11}^{eq\prime} = k_{11}^{\prime}+\frac{1}{2}\sigma_{11}^{\prime}\Delta t-\rho U_1U_2,
\end{equation}
where we have included a contribution due to the body force $\sigma_{11}^{\prime}\Delta t/2$ when evaluating the moment $k_{11}^{\prime}$ (see the paragraph below Eq.~(\ref{eq:relaxationratematrix}) for details). In this last equation, the source moment $\sigma_{11}^{\prime}$ is related to the effective body force and the velocity field -- see Eq.~(\ref{eq:GOCmomentsources}). That is,
\begin{equation}\label{eq:sourcemoment11}
\sigma_{11}^{\prime} = F_1 \frac{U_2}{h_2} + F_2\frac{U_1}{h_1}.
\end{equation}
Now, inspecting the effective body force component equations given in Eq.~(\ref{eq:bodyforce1}) and (\ref{eq:bodyforce2}), we see that they themselves depend on the unknown shear stress components $\tau_{12} = \tau_{21}$. Isolating them from these equations, we can conveniently redefine the effective body force equations as
\begin{equation}\label{eq:redefinedeffbodyforce}
F_1 = \widetilde{F}_1 + 2h_2\theta_{12}\tau_{12}, \qquad F_2 = \widetilde{F}_2 + 2h_1\theta_{21}\tau_{12}.
\end{equation}
where $\tilde{F}_1$ and $\tilde{F}_2$ are the effective body force components without the presence of the shear stress, i.e.,
\begin{equation}\label{eq:modifiedbodyforce1}
   \widetilde{F}_1 = h_2 \bigg[F_{ext,1}+ \left( \theta_{21}- \theta_{11}\right)P + \theta_{11}\left(\tau_{11} -  \rho U_1^2\right)+ 2 \theta_{12}\left(0 -  \rho U_1 U_2\right)-  \theta_{21}\left(\tau_{22} -  \rho U_2^2\right)\bigg],
\end{equation}
\begin{equation}\label{eq:modifiedbodyforce2}
   \widetilde{F}_2 = h_1 \bigg[F_{ext,2}+ \left( \theta_{12}-  \theta_{22}\right)P + \theta_{22}\left(\tau_{22} -  \rho U_2^2\right)+ 2 \theta_{21}\left(0 -  \rho U_1 U_2\right)-  \theta_{12}\left(\tau_{11} -  \rho U_1^2\right)\bigg],
\end{equation}
which are obtained by replacing $\tau_{12}$ and $\tau_{21}$ with $0$ in $F_1$ and $F_2$, respectively.

Then, substituting Eq.~(\ref{eq:noneqmoment5defn}), (\ref{eq:sourcemoment11}), and (\ref{eq:redefinedeffbodyforce}) into Eq.~(\ref{eq:72a}), i.e.,
\begin{equation*}
\tau_{12} = -\left(1-\frac{\omega_5}{2}\right)n_5^{(1)},
\end{equation*}
the resulting equation is \emph{implicit} in the shear stress $\tau_{12}$, which can be solved by simply rearranging it to finally obtain
\begin{equation}\label{eq:shearlocallocalcomputation}
\tau_{12} = \frac{-\left(1-\dfrac{\omega_5}{2}\right)\left[k_{11}^{\prime}+\dfrac{1}{2}\left(\widetilde{F}_1\dfrac{U_2}{h_2}+\widetilde{F}_2\dfrac{U_1}{h_1}\right)-\rho U_1U_2\right]}{1+\left(1-\dfrac{\omega_5}{2}\right) (\theta_{12}U_2 + \theta_{21}U_1) },
\end{equation}
where we have set $\Delta t = 1$ to account for the use of the usual lattice units in computations. As such, Eq.~(\ref{eq:shearlocallocalcomputation}) provides a \emph{local} approach for the computation of the shear stress with the GOC-LBE, which generalizes similar result in the case of the standard LBE using uniform grids for orthogonally clustered/curvilinear grids.

\section{Simplifications of the Coefficients of Correction Terms and the Effective Body for the GOC-LBE under Some Limiting Special Cases}\label{sec:Appendix_B}
Here, we summarize some simplified forms of the coefficients of the moment equilibria corrections and the effective body force derived in Sec.~\ref{sec:GOC-LBM} under some limiting special cases.

\begin{enumerate}[label=\textbf{(\roman*)}]
\item \textbf{Non-Curvilinear (Straight) Orthogonal Clustered Grids}
In this case, the metric coefficients and the curvature coefficient matrix elements are defined as
\begin{eqnarray*}
&& h_{1} = h_1(\xi_1), \qquad h_{2} = h_2(\xi_2),\nonumber\\
&& \theta_{11} \ne 0, \qquad \theta_{22} \ne 0, \qquad \theta_{12} = 0, \qquad \theta_{21} = 0
\end{eqnarray*}

For this special case, the coefficients $C_3$ and $C_4$ associated with the moment equilibria corrections (see Eqs.~(\ref{eq:83}) and (\ref{eq:85})) simplify to
\begin{subequations}
\begin{align}
C_3&=- \rho \zeta \Bigg\{-2 U_1 \left(\frac{ U_1^2}{c_{s}^2} +1 \right)\frac{h_2}{h_1} \theta_{11}  -2 U_2 \left(\frac{ U_2^2}{c_{s}^2} +1 \right)\frac{h_1}{h_2} \theta_{22}\Bigg\},\\
C_4&= - \rho \nu \Bigg\{-2 U_1 \left(\frac{ U_1^2}{c_{s}^2} +1 \right)\frac{h_2}{h_1} \theta_{11} +2 U_2 \left(\frac{ U_2^2}{c_{s}^2} +1 \right)\frac{h_1}{h_2} \theta_{22}\Bigg\}.
\end{align}
\end{subequations}
The rest of the coefficients $D_{3,k}$ and $D_{4,k}$ with $k = 1,2,3,4$, remains the same as those given in Eqs.~(\ref{eq:84}) and (\ref{eq:85}). Moreover, the effective body force components $F_1$ and $F_2$ are reduced to
\begin{subequations} \label{eq:Special1_Force_terms}
\begin{align}
&F_1 = h_2 \Big\{F_{ext,1}- \rho c_s^2  \theta_{11} + \left(\tau_{11} -  \rho U_1^2\right) \theta_{11}\Big\},\\
&F_2 = h_1 \Big\{F_{ext,2}- \rho c_s^2 \theta_{22}+\left(\tau_{22} -  \rho U_2^2\right)\theta_{22}\Big\},
\end{align}
\end{subequations}
and the associated normal stress components to
\begin{equation} \label{eq:Special1_Force_terms_tau}
\hspace{-0.3cm}\tau_{11 }= \rho \left( \zeta + \nu \right) \frac{1}{h_1} \partial_ {\xi_1} U_1 + \rho \left( \zeta- \nu  \right) \frac{1}{h_2}\partial_ {\xi_2} U_2, \quad \tau_{22}= \rho \left( \zeta + \nu \right) \frac{1}{h_2} \frac{\partial U_2}{\partial \xi_2} + \rho \left( \zeta- \nu \right) \frac{1}{h_1} \frac{\partial U_1}{\partial \xi_1}.
\end{equation}

\item \textbf{Non-Curvilinear Orthogonal Clustered Grids with Negligible Non-Galilean Invariant Cubic Velocity Corrections for Low Mach Number Flow Simulations}

Considering the previous case and with an additional assumption that the non-Galilean Invariant cubic velocity corrections are negligible which are applicable for the simulations of flows at relatively low Mach numbers, all the coefficients given in Eqs.~(\ref{eq:83}) and (\ref{eq:85}) further simplify as follows:
\begin{subequations}\label{eq:special_2_D3}
\begin{align}
D_{3,1}&= \rho \bigg[ \zeta \frac{h_2}{c_{s}^2} - \left(\zeta- \nu  \right)\frac{1}{h_2}- \left(2\zeta +\nu \right)\frac{h_2}{h_1^2} \bigg],& D_{3,3}=0, \\
D_{3,2}&= \rho \bigg[ \zeta \frac{ h_1}{c_{s}^2} - \left(\zeta- \nu  \right)\frac{1}{h_1}- \left(2\zeta +\nu \right)\frac{h_1}{h_2^2}\bigg],& D_{3,4}=0,\\
C_3&= - \rho \zeta \bigg[-2 U_1 \frac{h_2}{h_1} \theta_{11}  -2 U_2 \frac{h_1}{h_2} \theta_{22} \bigg],
\end{align}
\end{subequations}
and
\begin{subequations}\label{eq:special_2_D4}
\begin{align}
D_{4,1}&= \rho \bigg[ \nu \frac{h_2}{c_{s}^2} + \left(\zeta- \nu  \right)\frac{1}{h_2}- \left(2\nu +\zeta \right)\frac{h_2}{h_1^2} \bigg],& D_{4,3}=0, \\
D_{4,2}&= \rho \bigg[ -\nu \frac{ h_1}{c_{s}^2} - \left(\zeta- \nu  \right)\frac{1}{h_1}+ \left(2\nu +\zeta \right)\frac{h_1}{h_2^2}\bigg],& D_{4,4}=0,\\
C_4&= - \rho \nu \bigg[-2 U_1 \frac{h_2}{h_1} \theta_{11}  + 2 U_2 \frac{h_1}{h_2} \theta_{22} \bigg].
\end{align}
\end{subequations}
The simplified form of the effective body force and the associated normal stress components here are the same are the same as those given in Eqs.~\eqref{eq:Special1_Force_terms} and~\eqref{eq:Special1_Force_terms_tau}.

\item \textbf{Non-Curvilinear Orthogonal Clustered Grids with Negligible Non-Galilean Invariant Cubic Velocity Corrections for Parallel Flow Simulations}

In this limiting case, we make a further assumption of a parallel flow, such as, for example, in Poiseuille-type flows, where the following conditions hold: \[ \partial_{\xi_1}U_1 =0 \quad\quad \text{and}\quad\quad  \partial_{\xi_2}U_2 =0 \].
Then, all the coefficients associated with the gradients simply drop out. That is,
\[D_{3,k}=0 \quad\quad \text{and}\quad\quad  D_{4,k}=0 \] with $k=1,2,3,4$. The coefficients $C_3$ and $C_4$ are the same as those given above in Eqs.~\eqref{eq:special_2_D3} and~\eqref{eq:special_2_D4}, respectively. Moreover, the effective body force components with $\tau_{11} = \tau_{22} = 0$ further simplify to
\begin{align}\label{eq:Special2_Force_terms}
F_1&= h_2 \Big\{F_{ext,1}- \rho c_s^2 \theta_{11} -  \rho U_1^2 \theta_{11}\Big\},\\
F_2&= h_1 \Big\{F_{ext,2}- \rho c_s^2 \theta_{22} -  \rho U_2^2 \theta_{22}\Big\}.
\end{align}
\end{enumerate}

\section{Implementations of Initial and Boundary Conditions in GOC-LBM}\label{sec:initial_boundary_conditions}
\subsection{Initial Conditions in GOC-LBM}\label{sec:init_condition_GOC-LBM}
Since the GOC-LBM is specified in terms of its natural settings based on raw moments, the associated initial conditions can be conveniently introduced in terms of suitably chosen initial moments of the distribution functions with appropriate scalings/corrections based on metric factors. To that end, let's assume that at some initial time $t=t^{i}$, the density and the velocity field be given as $\rho^{i}$ and $\bm{U}^{i}= (U_1^{i}, U_2^{i})$, respectively, everywhere in the computational domain. Equivalently, the initial pressure field can be provided rather than the density field since $P^{i} =  c_s^2\rho^{i}$.

Using these given initial hydrodynamic fields, the initial moment equilibria, which includes the base equilibria (Eq.~(\ref{eq:GOCmomentequilibria})) together with the associated corrections derived using the C-E analysis in the previous section, we set them equal to initial raw moments (Note that, if necessary, non-equilibrium moments could be included as well by readily augmenting them with additional terms obtained in the previous section). Formally, we can express this as $\mathbf{n}^{i} = \mathbf{n}^{eq,eff}(\rho^{i}, \bm{U}^{i})$ or equivalently written in terms of the bare forms of the initial moments, i.e., $\mathbf{m}^{i} = \mathbf{m}^{eq,eff}(\rho^{i}, \bm{U}^{i})$. In its component form, the resulting initial moments can be written as
\begin{equation*}
\mathbf{m}^{i}=\left(k_{00}^{i\prime},k_{10}^{i\prime},k_{01}^{i\prime},k_{20}^{i\prime},k_{02}^{i\prime}, k_{11}^{i\prime}, k_{21}^{i\prime}, k_{12}^{i\prime}, k_{22}^{i\prime}\right)^{\dag}.
\end{equation*}
Then, the initial distribution functions $\mathbf{f}^{i} = \left( f_0^i, f_1^i, f_2^i, f_3^i, f_4^i, f_5^i, f_6^i, f_7^i, f_8^i\right)^\dag$ can be constructed readily from $\mathbf{m}^i$ via the inverse mapping $\mathbf{f}^i = \tensor{P}^{-1}\mathbf{m}^i$, or, in explicit form, they read as
\begin{align}\label{eq:initial_distribution_functions}
&f_{0}^i= k_{00}^{i\prime}-k_{20}^{i\prime}- k_{02}^{i\prime} + k_{22}^{i\prime},\nonumber  \\
&f_{1}^i= \frac{1}{2}\left(k_{10}^{i\prime} + k_{20}^{i\prime} - k_{12}^{i\prime}- k_{22}^{i\prime}\right),
\quad \quad f_{5}^i = \frac{1}{4}\left(k_{11}^{i\prime} + k_{21}^{i\prime} + k_{12}^{i\prime} + k_{22}^{i\prime}\right),\nonumber\\
&f_{2}^i= \frac{1}{2}\left(k_{01}^{i\prime} + k_{02}^{i\prime} - k_{21}^{i\prime}- k_{22}^{i\prime}\right), \quad \quad
f_{6}^i = \frac{1}{4}\left(-k_{11}^{i\prime}+ k_{21}^{i\prime} - k_{12}^{i\prime} + k_{22}^{i\prime}\right), \nonumber\\
&f_{3}^i= \frac{1}{2}\left(-k_{10}^{i\prime} + k_{20}^{i\prime} + k_{12}^{i\prime}- k_{22}^{i\prime}\right), \quad \quad f_{7} = \frac{1}{4}\left(k_{11}^{i\prime} - k_{21}^{i\prime} - k_{12}^{i\prime} + k_{22}^{i\prime}\right),\nonumber\\
&f_{4}^i= \frac{1}{2}\left(-k_{01}^{i\prime} + k_{02}^{i\prime} + k_{21}^{i\prime}- k_{22}^{i\prime}\right),  \quad \quad
f_{8}^i = \frac{1}{4}\left(-k_{11}^{i\prime}- k_{21}^{i\prime} + k_{12}^{i\prime} + k_{22}^{i\prime}\right).
\end{align}

\subsection{Boundary Conditions On Moving Walls: Momentum-Augmented Bounce Back Scheme in GOC}\label{sec:B_conditions}
Next, let's illustrate the implementation of a simple, momentum-augmented half-way bounce back scheme to specify a moving wall boundary condition for a case with $U_1 = U_{wall}$ with $U_2 = 0$ for the GOC-LBM, where $U_{wall}$ is the specified wall velocity. Let $\bm{\xi}_f$ be the nearest fluid node located half-way from a wall node at $\bm{\xi}_w$. Taking $\bm{e}_\alpha$ as the outgoing particle direction and $\bm{e}_{\overline{\alpha}}$ as the incoming direction from the wall, i.e., with $\bm{e}_{\overline{\alpha}}=-\bm{e}_\alpha$, the distribution functions for the incoming directions based on the corresponding outgoing post-collision distribution functions (denoted by a tilde over the symbol) along with the correction to incorporate the moving wall effects can then be written as follows:
\begin{equation}\label{eq:101}
f_{\overline{\alpha}}(\bm{\xi}_f,t+\Delta t)=\widetilde{f}_{\alpha}(\bm{\xi}_f,t)-\left(f_{\alpha}^{eq}(\bm{\xi}_w) -f_{\overline{\alpha}}^{eq}(\bm{\xi}_w)\right),
\end{equation}
Here, as in the implementation of the initial conditions above, the equilibrium distribution functions are obtained via an inverse mapping of the respective raw moments evaluated based on the moving wall conditions. That is, from $\mathbf{m}^{w} = \mathbf{m}^{eq,eff}(\rho_w, U_{wall}, 0)$, we obtain the equilibrium distribution functions as $\mathbf{f}^{eq}(\bm{\xi}_w) = \tensor{P}^{-1} \mathbf{m}^{w}$. Here, the wall density $\rho_w$ is usually approximated by extrapolating it from the density for the nearest fluid node, i.e., $\rho_w = \rho_f$.

Consider the case of the top moving wall as an example, where the three unknowns for the incoming directions $\overline{\alpha}=\{4,7,8\}$ can then be evaluated using Eq.~(\ref{eq:101}) via the inverse transformation to get the equilibrium distribution functions with the wall condition as indicated above. This then yields the following closure relations to implement the unknown distribution functions coming from the wall into the fluid for the GOC-LBM:
\begin{subequations}\label{eq:103}
\begin{eqnarray}
 f_4(\bm{\xi}_f, t+\Delta{t})&=&\widetilde{f}_2(\bm{\xi}_f,t), \\
 f_7(\bm{\xi}_f, t+\Delta{t})&=&\widetilde{f}_5(\bm{\xi}_f,t) -\dfrac{ \rho_w c_s^2  U_{wall}}{2 h_{2wall}}, \\
 f_8(\bm{\xi}_f, t+\Delta{t})&=&\widetilde{f}_6(\bm{\xi}_f,t)+\dfrac{ \rho_w c_s^2  U_{wall}}{2 h_{2wall}},
\end{eqnarray}
\end{subequations}
which are parameterized by the metric factor $h_2$ evaluated on the wall, i.e., $h_{2wall}$. Note that when $h_{2wall} = 1$, this formulation recovers the momentum-augmented, half-way moving boundary scheme for the standard LBM using uniform grids.

\section{GOC-LBM Based on Raw Moments and MRT for NSE: Summary of Algorithm) }\label{sec: Appendix_A}
In this section, we will provide the step-by-step details of implementing the GOC-LBM using the raw moments and MRT as given in a matrix form in Eq.~(\ref{eq:LBErawmomentmatrixform}) in Sec.~\ref{sec:Reformulation GOC-LB}. First, let's outline the following preparatory step, which is performed just once at the beginning prior to the implementation of the GOC-LBM algorithm:
\begin{itemize}
  \item \textbf{Preparatory step (a):} Compute the local metric factors $h_1$ and $h_2$ and the curvature coefficient matrix elements $\theta_{11}$, $\theta_{22}$, $\theta_{12}$ and $\theta_{21}$, which define the orthogonally clustered/curvilinear grids for the given flow geometry (see Secs.~\ref{sec:Transformations} and~\ref{sec: curvedbenchmark-conformal-mapping} for examples).
  \item  \textbf{Preparatory step (b):} Set the initial conditions are then specified using the strategy given in Appendix~\ref{sec:initial_boundary_conditions}.
  \item \textbf{Preparatory step (c):} Set the speed of sound $c_s$ for the GOC-LBE via a parametrization using the metric factors based on the CFL condition according to Eq.~(\ref{eq:GOC-LBE_CFL}).
  \item \textbf{Preparatory step (d):} Selection of the relaxation rate parameters should be based on the following considerations. Choose the relaxation rates $\omega_3$ based on the kinematic bulk viscosity $\zeta$ according to Eq.~(\ref{eq:76}) and the relaxation parameters $\omega_4 = \omega_5$ based on the kinematic shear viscosity $\nu$ according to Eqs.~(\ref{eq:73}) and (\ref{eq:76}). The rest of the relaxation parameters $\omega_j$ for $j = 0, 1, 2, 6, 7$ and $8$ can be chosen freely based on stability considerations and often they are set to be unity.
\end{itemize}
Then, the following sequence of steps are implemented to obtain the hydrodynamic fields in the computational domain such as $\rho$, $U_1$ and $U_2$, which satisfy the NSE in the GOC, via executing Eq.~(\ref{eq:LBErawmomentmatrixform}) over a time step $\Delta t$.

\begin{itemize}
\item \textbf{\underline{Step 1}: Compute pre-collision raw moments}\label{itemone}

The distribution functions $f_\alpha = f_\alpha(\bm{\xi},t)$ for $\alpha = 0, 1,\ldots, 8$ are mapped from the velocity space to the raw moment space by performing $\mathbf{m}=\tensor{P}\mathbf{f}$, where the transformation matrix $\tensor{P}$ is shown in Eq.~\eqref{eq:Q-momentbasis}. The result reads as
\begin{align*}\label{eq:1A}
  &\Kps{00}= f_0 +f_1 +f_2 + f_3 +f_4 +f_5 +f_6 +f_7 +f_8,\\
  &\Kps{10} = f_1  -f_3  +f_5 -f_6 -f_7 +f_8, \quad\quad\quad\quad \Kps{11}= f_5 -f_6 +f_7 -f_8, \\
  &\Kps{01} = f_2  -f_4  +f_5 +f_6 -f_7 -f_8, \quad\quad\quad\quad \Kps{21}= f_5 +f_6 -f_7 -f_8, \\
  &\Kps{20} = f_1  +f_3  +f_5 +f_6 +f_7 +f_8, \quad\quad\quad\quad \Kps{12}= f_5 -f_6 -f_7 +f_8, \\
  &\Kps{02} = f_2  +f_4  +f_5 +f_6 +f_7 +f_8, \quad\quad\quad\quad \Kps{22}= f_5 +f_6 +f_7 +f_8.
\end{align*}

\item \textbf{\underline{Step 2}: Perform Collision Step: Relax Raw moments to their Equilibria and Augment with Source Moments}
\begin{enumerate}
  \item Compute the normal velocity derivatives $\partial_{\xi_1}U_1$ and $\partial_{\xi_2}U_2$ via second-order finite difference schemes.
  \item Compute the normal stress components $\tau_{11}$ and $\tau_{22}$ using $\partial_{\xi_1}U_1$ and $\partial_{\xi_2}U_2$ based on Eqs.~(\ref{eq:viscousstress_OCC_modified3_normal11}) and (\ref{eq:viscousstress_OCC_modified3_normal22}), where the associated curvature contributions $\tau_{11}^c$ and $\tau_{22}^c$ are given in Eq.~(\ref{eq:tauc}).
  \item Compute the shear stress $\tau_{12}$ locally using the off-diagonal non-equilibrium part of the moment $k_{11}^\prime$ via Eq.~(\ref{eq:shearlocallocalcomputation}), where the reduced body force components (i.e., without $\tau_{12}$) $\widetilde{F}_1$ and $\widetilde{F}_2$ are given in Eqs.~(\ref{eq:modifiedbodyforce1}) and (\ref{eq:modifiedbodyforce2}), respectively (see Appendix~\ref{sec:Appendix_shear_stress_local}).
  \item Compute the effective body force components in their full forms, i.e., $F_1$ and $F_2$ using Eqs.~(\ref{eq:bodyforce1}) and (\ref{eq:bodyforce2}), respectively.
  \item Compute the nine independent source moments $\sigma_{00}^\prime$, $\sigma_{10}^\prime$, $\sigma_{01}^\prime$, $\sigma_{20}^\prime + \sigma_{02}^\prime$, $\sigma_{20}^\prime-\sigma_{02}^\prime$, $\sigma_{11}^\prime$, $\sigma_{21}^\prime$, $\sigma_{12}^\prime$ and $\sigma_{22}^\prime$ using Eq.~(\ref{eq:GOCmomentsources}). Label the combined diagonal second order source moments for convenience (which amounts to applying the operator $\tensor{B}$ to the bare form of the source vector $\Phi$ to get $\Psi$ for the combined form of the source moment vector in Eq.~(\ref{eq:LBErawmomentmatrixform})) as
      \begin{equation*}
      \sigma_{2s}^\prime = \sigma_{20}^\prime + \sigma_{02}^\prime, \qquad \sigma_{2d}^\prime = \sigma_{20}^\prime - \sigma_{02}^\prime.
      \end{equation*}
  \item Compute the nine independent base equilibrium moments $k_{00}^{eq\prime}$, $k_{10}^{eq\prime}$, $k_{01}^{eq\prime}$, $k_{20}^{eq\prime} + k_{02}^{eq\prime}$, $k_{20}^{eq\prime} - k_{02}^{eq\prime}$, $k_{11}^{eq\prime}$, $k_{21}^{eq\prime}$, $k_{12}^{eq\prime}$, $k_{22}^{eq\prime}$ using Eq.~(\ref{eq:GOCmomentequilibria}). Then apply the necessary corrections to the
      combined diagonal second order base equilibrium moments to restore the full recovery of the NSE in the GOC and relabel them for convenience (which amounts to applying the operator $\tensor{B}$ to the bare form of the equilibrium vector $\mathbf{m}^{eq}$ to get $\mathbf{n}^{eq}$ for the combined form of the equilibrium moment vector in Eq.~(\ref{eq:LBErawmomentmatrixform})) as
      \begin{equation*}
      k_{2s}^{eq\prime} = \left(k_{20}^{eq\prime} + k_{02}^{eq\prime}\right) + \underline{k_{2s}^{eq(1)\prime}}\Delta t, \qquad k_{2d}^{eq\prime} = \left(k_{20}^{eq\prime} - k_{02}^{eq\prime}\right) + \underline{k_{2d}^{eq(1)\prime}}\Delta t,
      \end{equation*}
      where $\underline{k_{2s}^{eq(1)\prime}}= n_3^{eq(1)}$ and $\underline{k_{2d}^{eq(1)\prime}}= n_4^{eq(1)}$ are the moment equilibria corrections given in Eqs.~(\ref{eq:82}) and (\ref{eq:84}), respectively. That is,
      \begin{eqnarray*}
        &&\underline{k_{2s}^{eq(1)\prime}}\Delta t = n_3^{eq(1)} \Delta t =D_{3,1} \partial_{\xi_1} U_1 + D_{3,2} \partial_{\xi_2} U_2+ D_{3,3} \partial_{\xi_1} \rho +D_{3,4} \partial_{\xi_2} \rho + C_3,\\
        &&\underline{k_{2d}^{eq(1)\prime}}\Delta t = n_4^{eq(1)} \Delta t =D_{4,1} \partial_{\xi_1} U_1 + D_{4,2} \partial_{\xi_2} U_2+ D_{4,3} \partial_{\xi_1} \rho +D_{4,4} \partial_{\xi_2} \rho + C_4.
      \end{eqnarray*}
      Here, the coefficients $D_{3, k}$ for $k = 1, 2, 3$ and $4$ and $C_3$ are given in Eq.~(\ref{eq:83}) and, similarly, the other set of coefficients $D_{4, k}$ for $k = 1, 2, 3$ and $4$ and $C_4$ are provided in Eq.~(\ref{eq:85}).
      \item Label the combined diagonal second order moments for convenience (which amounts to applying the operator $\tensor{B}$ to the bare form of the moment vector $\mathbf{m}$ to get $\mathbf{n}$ for the combined form of the moment vector in Eq.~(\ref{eq:LBErawmomentmatrixform})) as
      \begin{equation*}
      k_{2s}^\prime= k_{20}^\prime + k_{02}^\prime, \qquad k_{2d}^\prime = k_{20}^\prime - k_{02}^\prime.
      \end{equation*}

  \item Relax all the nine independent moments to their corresponding equilibria and augmented with source moments to obtain their post-collision states, which are represented as follows:
      \begin{eqnarray*}
      &\tilde{k}_{00}^\prime = k_{00}^{\prime}+\omega_{0}\left(k_{00}^{eq\prime}-k_{00}^\prime\right)+\left(1-\dfrac{\omega_{0}}{2}\right)\sigma_{00}^{\prime}\Delta t,\nonumber \\
      &\tilde{k}_{10}^\prime = k_{10}^{\prime}+\omega_{1}\left(k_{10}^{eq\prime}-k_{10}^\prime\right)+\left(1-\dfrac{\omega_{1}}{2}\right)\sigma_{10}^{\prime}\Delta t,\nonumber \\
      &\tilde{k}_{01}^\prime = k_{01}^{\prime}+\omega_{2}\left(k_{01}^{eq\prime}-k_{01}^\prime\right)+\left(1-\dfrac{\omega_{2}}{2}\right)\sigma_{01}^{\prime}\Delta t,\nonumber \\
      &\tilde{k}_{2s}^\prime = k_{2s}^{\prime}+\omega_{3}\left(k_{2s}^{eq\prime}-k_{2s}^\prime\right)+\left(1-\dfrac{\omega_{3}}{2}\right)\sigma_{2s}^{\prime}\Delta t,\nonumber \\
      &\tilde{k}_{2d}^\prime = k_{2d}^{\prime}+\omega_{4}\left(k_{2d}^{eq\prime}-k_{2d}^\prime\right)+\left(1-\dfrac{\omega_{4}}{2}\right)\sigma_{2d}^{\prime}\Delta t,\nonumber \\
      &\tilde{k}_{11}^\prime = k_{11}^{\prime}+\omega_{5}\left(k_{11}^{eq\prime}-k_{11}^\prime\right)+\left(1-\dfrac{\omega_{5}}{2}\right)\sigma_{11}^{\prime}\Delta t,\nonumber \\
      &\tilde{k}_{21}^\prime = k_{21}^{\prime}+\omega_{6}\left(k_{21}^{eq\prime}-k_{21}^\prime\right)+\left(1-\dfrac{\omega_{6}}{2}\right)\sigma_{21}^{\prime}\Delta t,\nonumber \\
      &\tilde{k}_{12}^\prime = k_{12}^{\prime}+\omega_{7}\left(k_{12}^{eq\prime}-k_{12}^\prime\right)+\left(1-\dfrac{\omega_{7}}{2}\right)\sigma_{12}^{\prime}\Delta t,\nonumber \\
      &\tilde{k}_{22}^\prime = k_{22}^{\prime}+\omega_{8}\left(k_{22}^{eq\prime}-k_{22}^\prime\right)+\left(1-\dfrac{\omega_{8}}{2}\right)\sigma_{22}^{\prime}\Delta t.\nonumber \\
      \end{eqnarray*}

  \item Split the post-collision combined second-order diagonal moments to their bare forms (which amounts to applying the operator $\tensor{B}^{-1}$ to $\tilde{\mathbf{n}}$ to get $\tilde{\mathbf{m}}$ for the post-collision moment vector in Eq.~(\ref{eq:LBErawmomentmatrixform})) as
      \begin{align*}
        &\tilde{k}_{20}^{\prime}= \frac{1}{2}\left(\tilde{k}_{2s}^{\prime} + \tilde{k}_{2d}^{\prime}\right), \quad\quad\quad \tilde{k}_{02}^{\prime} = \frac{1}{2}\left(\tilde{k}_{2s}^{\prime} - \tilde{k}_{2d}^{\prime}\right),
      \end{align*}

\end{enumerate}

\item \textbf{\underline{Step 3}: Compute Post-Collision Distribution Functions}

The post-collision distribution functions $\tilde{f}_\alpha = \tilde{f}_\alpha(\bm{\xi},t)$ for $\alpha = 0, 1,\ldots, 8$ are mapped from the post-collision raw moments by performing $\tilde{\mathbf{f}}=\tensor{P}^{-1}\tilde{\mathbf{m}}$, where the $\tensor{P}$ matrix is given in Eq.~\eqref{eq:Q-momentbasis}. The result reads as
\begin{align*}\label{eq:3A}
&\tilde{f}_{0}= \tilde{k}_{00}^\prime-\tilde{k}_{20}^\prime- \tilde{k}_{02}^\prime + \tilde{k}_{22}^\prime,\nonumber  \\
&\tilde{f}_{1}= \frac{1}{2}\left(\tilde{k}_{10}^\prime+ \tilde{k}_{20}^\prime - \tilde{k}_{12}^\prime- \tilde{k}_{22}^\prime\right), \quad \quad \quad \tilde{f}_{5} = \frac{1}{4}\left(\tilde{k}_{11}^\prime+ \tilde{k}_{21}^\prime + \tilde{k}_{12}^\prime+ \tilde{k}_{22}^\prime\right),\nonumber\\
&\tilde{f}_{2}= \frac{1}{2}\left(\tilde{k}_{01}^\prime+ \tilde{k}_{02}^\prime - \tilde{k}_{21}^\prime- \tilde{k}_{22}^\prime\right), \quad \quad \quad \tilde{f}_{6} = \frac{1}{4}\left(-\tilde{k}_{11}^\prime+ \tilde{k}_{21}^\prime - \tilde{k}_{12}^\prime+ \tilde{k}_{22}^\prime\right), \nonumber\\
&\tilde{f}_{3}= \frac{1}{2}\left(-\tilde{k}_{10}^\prime+ \tilde{k}_{20}^\prime + \tilde{k}_{12}^\prime- \tilde{k}_{22}^\prime\right), \quad \quad \tilde{f}_{7} = \frac{1}{4}\left(\tilde{k}_{11}^\prime- \tilde{k}_{21}^\prime - \tilde{k}_{12}^\prime+ \tilde{k}_{22}^\prime\right),\nonumber\\
&\tilde{f}_{4}= \frac{1}{2}\left(-\tilde{k}_{01}^\prime+ \tilde{k}_{02}^\prime + \tilde{k}_{21}^\prime- \tilde{k}_{22}^\prime\right),  \quad \quad\tilde{f}_{8} = \frac{1}{4}\left(-\tilde{k}_{11}^\prime- \tilde{k}_{21}^\prime + \tilde{k}_{12}^\prime+ \tilde{k}_{22}^\prime\right).
\end{align*}

\item \textbf{\underline{Step 4}: Stream Distribution Functions}

In this step, the post-collision distribution functions $\tilde{f}_\alpha$ are streamed along the directions of particle characteristics $\bm{e}_\alpha$ to adjacent nodes during the time step $\Delta t$, i.e., perform a perfect shift advection, which reads as
\begin{equation*}\label{eq:streaming_rawmoments}
f_\alpha(\bm{\xi},t+\Delta t)=\widetilde{f}_\alpha(\bm{\xi}-\bm{e}_\alpha\Delta t,t).
\end{equation*}
In addition, as necessary, this step also incorporates the boundary conditions to update incoming the distribution functions from the boundaries (see e.g., Eq.~(\ref{eq:103}) for the momentum-augmented half-way bounce back scheme for moving walls for the GOC-LBM.

\item \textbf{\underline{Step 5}: Update Hydrodynamic Fields}

Finally, the hydrodynamic fields such as the fluid density and the velocity are updated via the zeroth and first moments, respectively, of the distribution functions computed in the previous step. That is,
\begin{align*}
\rho =\frac{1}{h_1 h_2}\sum_{\alpha=0}^{8} f_\alpha, \quad   U_1 = \frac{1}{\rho h_2}\left(\sum_{\alpha=0}^{8} f_{\alpha} e_{\alpha_1} + \frac{1}{2} F_1\Delta t\right), \quad U_2 = \frac{1}{\rho h_1}\left(\sum_{\alpha=0}^{8} f_{\alpha} e_{\alpha_2} + \frac{1}{2} F_2\Delta t\right).
\end{align*}

\end{itemize}
Note that, when necessary, the post-processing of the results in the Cartesian coordinates based physical domain can be performed via first mappings any vectors and tensors computed using the GOC to the Cartesian coordinates using the transformation relations given in Eqs.~(\ref{eq:vectormapping}) and (\ref{eq:tensormapping}), respectively, whereas any scalar remains invariant under coordinate transformations.

\section{GOC-LBM Based on Central Moments and MRT for NSE: Summary of Algorithm}\label{sec:Appendix_C}
When the collision step is constructed using the relaxation of the central moments of the distribution function, where the particle velocities are shifted by the local fluid velocity, rather than in terms of the raw moments, it offers significant additional improvements in numerical stability while naturally preserving the Galilean invariance of the independent set of moments for a given lattice (see e.g.,~\cite{ning2016numerical,CHAVEZMODENA2018397}). Our numerical investigations based on the model developed in this section shows that such an advantage carries over from the use of uniform grids to the orthogonally clustered/curvilinear grids using GOC-LBM.

A key idea that we exploit in designing a GOC-LBM based on central moments is that the second order non-equilibrium central moments and the raw moments are identical to one another, which follows from their definitions and relating one with the other via the binomial expansions and by using the fact that zeroth and first order moments are the collision invariants; as a result, the non-equilibrium moments and the extended moment equilibria corrections that we derived in the case of the raw moments in Sec.~\ref{sec:GOC-LBM} simply carry over to the central moment-based GOC-LBM. Then, the matrix formulation presented in Sec.~\ref{sec:Reformulation GOC-LB} just needs to be modified accordingly by using central moments rather than raw moments when executing the collision step. This requires the construction of the base equilibria in central moments for the D2Q9 lattice via the binomial expansions of the corresponding known raw moments, which are given in Sec.~\ref{sec:MRT-LBM}, and similarly for the source central moments from their raw moments counterparts. These will be carried out, along with presenting a summary of the resulting algorithm, in what follows, but first, let's define the central moments of the distribution functions $k_{mn}$, their equilibria $k_{mn}^{eq}$, and the source terms $\sigma_{mn}$, which read as
\begin{eqnarray}
  &&k_{mn}= \sum_{\alpha=0}^{8} f_{\alpha} \;(e_{\alpha \xi_1} -U_1)^m  (e_{\alpha \xi_2} -U_2)^n, \qquad k_{mn}^{eq}= \sum_{\alpha=0}^{8} f_{\alpha}^{eq} \;(e_{\alpha \xi_1} -U_1)^m  (e_{\alpha \xi_2} -U_2)^n,\nonumber\\
  &&\sigma_{mn}= \sum_{\alpha=0}^{8} S_{\alpha} \;(e_{\alpha \xi_1} -U_1)^m  (e_{\alpha \xi_2} -U_2)^n. \label{eq:9B}
\end{eqnarray}
Note that our notation in this paper uses the prime to distinguish raw moments $k_{mn}^\prime$ defined earlier in Eq.~(\ref{eq:9A}) from the central moments (e.g., $k_{mn}$) shown here in Eq.~(\ref{eq:9B}). Based on these, similar to Eq.~(\ref{baremoments-meq-phi}), we collect together all the 9 independent central moments of the distribution functions, their equilibria, and the source terms, as the following respective 9-dimensional vectors, which will be useful in the description of the algorithmic implementation later:
\begin{subequations} \label{barecentralmoments-meq-phic}
\begin{eqnarray}
&\mathbf{m}^c=\left(k_{00},k_{10},k_{01},k_{20},k_{02}, k_{11}, k_{21}, k_{12}, k_{22}\right)^{\dag},\\
&\mathbf{m}^{eq,c}=\left(k_{00}^{eq},k_{10}^{eq},k_{01}^{eq},k_{20}^{eq},k_{02}^{eq}, k_{11}^{eq}, k_{21}^{eq}, k_{12}^{eq}, k_{22}^{eq}\right)^{\dag},\\
&\mathbf{\Phi}^c=\left(\sigma_{00},\sigma_{10},\sigma_{01},\sigma_{20},\sigma_{02}, \sigma_{11}, \sigma_{21}, \sigma_{12}, \sigma_{22}\right)^{\dag}.
\end{eqnarray}
\end{subequations}
Note that, in particular, during implementations, it is necessary to map the raw moments $\mathbf{m}$ given in Eq.~(\ref{baremoments-meq-phi}) to the central moments $\mathbf{m}^c$ given above in Eq.~(\ref{barecentralmoments-meq-phic}) and vice versa. They can be accomplished via the following transformations:
\begin{eqnarray}\label{eq:m-mc-F-mappings}
\mathbf{m}^c= \tensor{F} \mathbf{m}, \qquad \mathbf{m}=\tensor{F}^{-1} \mathbf{m}^c,
\end{eqnarray}
where $\tensor{F}$ is designated as the frame transformation matrix that is obtained from the binomial transforms of the 9 moments of different orders supported by the D2Q9 lattice and given by
\begin{equation} \label{eq:Fmatrix}
  \tensor{F} =
  \begin{bmatrix}
   1 &  0 &  0 &  0 &  0 &  0 &  0  &  0  &  0\\[4pt]
  -U_1  &  1 &  0 &  0 &  0 &  0 &  0  &  0  &  0\\[4pt]
  -U_2  &  0  & 1  & 0  &  0   & 0 & 0 & 0 &   0 \\[4pt]
   U_1^2  &  -2 U_1  & -2 U_2  & 1   &  0   & 0 & 0 & 0 &   0 \\[4pt]
   U_2^2   &  -2 U_1  & 2 U_2 & 0   & 1   & 0 & 0 & 0 &   0
  \\[4pt]
   U_1 U_2  &  -U_2  & -U_1  & 0  &  0   & 1 & 0 & 0 &  0 \\[4pt]
  -U_1 ^2 U_2    &  2 U_1 U_2  & U_1 ^2   & - \frac{1}{2} U_2  &  - \frac{1}{2} U_2  & -2 U_1 & 1 & 0 &  0 \\[4pt]
  -U_1 U_2 ^2   & U_2 ^2   & 2 U_1 U_2  & - \frac{1}{2} U_1  &  \frac{1}{2} U_1  & -2 U_2 & 0 & 1 &  0 \\[4pt]
  U_1^2 U_2 ^2   & -2 U_1 U_2 ^2   & -2 U_1^2 U_2  &  U_2^2  &  U_1^2  & 4 U_1 U_2 & -2 U_2 & -2 U_1 &  1.
  \end{bmatrix}
\end{equation}
We point out here that if $\tensor{F}=\tensor{F}(U_1, U_2)$, then this lower triangular matrix has a special structural property arising from the nature of the underlying binomial expansions involved that its inverse satisfies the following simple relation~\cite{yahia2021central}: $\tensor{F}^{-1} = \tensor{F}(-U_1, -U_2)$. That is, there is no need to perform an explicit matrix inversion of Eq.~(\ref{eq:Fmatrix}) to obtain $\tensor{F}^{-1}$, but rather be read-off directly from $\tensor{F}$ via the variable substitutions $U_1 \rightarrow - U_1$ and $U_2 \rightarrow - U_2$.

\subsection{Construction of Central Moment Equilibria and Source Terms}
The raw moments of the base equilibria given earlier in Eq.~(\ref{eq:GOCmomentequilibria}) can be used to construct the corresponding base central moment equilibria via applying the definition given in Eq.~(\ref{eq:9B}) and subsequently using appropriate binomial expansions. Let's illustrate it for a few cases to fix the main ideas: $k_{00}^{eq} = k_{00}^{eq\prime}$, $k_{10}^{eq} = k_{10}^{eq\prime} - U_1k_{00}^{eq\prime}$ and $k_{20}^{eq} = k_{20}^{eq\prime}- 2U_1k_{10}^{eq\prime} + U_1^2 k_{00}^{eq\prime}$. We carry out such mappings from the raw moment equilibria to the central moment equilibria for up to the third order moments, which appear in the C-E analysis for the derivation of the NSE in the GOC. However, one is free to choose the remaining fourth order moment $k_{22}^{eq}$ for the D2Q9 lattice as it does not influence the macroscopic fluid dynamical equations. Based on stability considerations, we prescribe it to be same as that for the standard uniform grid-based case, where it is obtained by matching it to the corresponding fourth order moment of the continuous Maxwell distribution, i.e., $k_{22}^{eq} = c_s^4 \rho$.

Then, defining for convenience, the following coefficients
\begin{eqnarray}
&& q_{\scriptscriptstyle 20} = \dfrac{h_2}{h_1}(h_1 - 1)^2, \qquad q_{\scriptscriptstyle 02} = \dfrac{h_1}{h_2}(h_2 - 1)^2, \nonumber\\
&& q_{\scriptscriptstyle 11} = (h_1 - 1)(h_2 - 1), \nonumber\\
&& q_{\scriptscriptstyle 21} = \dfrac{1}{h_1}(h_1 - 1)^2(h_2 - 1), \qquad q_{\scriptscriptstyle 12} = \dfrac{1}{h_2}(h_1 - 1)(h_2 - 1)^2, \label{eq:coeffscmeq},
\end{eqnarray}
and based on the above considerations, we can finally express the 9 independent central moments of the base equilibria as
\begin{eqnarray}
  && \hspace{-0.8cm}k_{00}^{eq} = h_1h_2\rho, \qquad k_{10}^{eq} = (1-h_1)h_2\rho U_1, \qquad k_{01}^{eq} = (1-h_2)h_1\rho U_2,\nonumber\\
  && \hspace{-0.8cm}(k_{20}^{eq} + k_{02}^{eq}) = q_{\scriptscriptstyle 20}\rho U_1^2 + q_{\scriptscriptstyle 02}\rho U_2^2 +\left(\frac{h_2}{h_1} + \frac{h_1}{h_2}\right)\rho c_s^2, \nonumber\\
  && \hspace{-0.8cm}(k_{20}^{eq} - k_{02}^{eq}) = q_{\scriptscriptstyle 20}\rho U_1^2 - q_{\scriptscriptstyle 02}\rho U_2^2 +\left(\frac{h_2}{h_1} - \frac{h_1}{h_2}\right)\rho c_s^2, \qquad k_{11}^{eq} = q_{\scriptscriptstyle 11}\rho U_1U_2\nonumber\\
  && \hspace{-0.8cm}k_{21}^{eq} = q_{\scriptscriptstyle 21} \rho U_1^2 U_2 + (1-h_2)\rho c_s^2\frac{U_2}{h_1}, \qquad k_{\scriptscriptstyle 12}^{eq} = q_{12} \rho U_1 U_2^2 + (1-h_1)\rho c_s^2\frac{U_1}{h_2},\nonumber\\
  && \hspace{-0.8cm}k_{22}^{eq} = c_s^4 \rho. \label{eq:cmequilibriaGOC}
\end{eqnarray}
Then, in order to restore the correct representation of the normal stress components and recover the NSE in the GOC, the second order combined diagonal central moments need to incorporate the necessary corrections, just like in the case of the raw moment-based GOC-LBM. That is, following the key idea outlined in the first paragraph of this appendix, we require
\begin{equation}
      k_{2s}^{eq} = \left(k_{20}^{eq} + k_{02}^{eq}\right) + \underline{k_{2s}^{eq(1)}}\Delta t, \qquad k_{2d}^{eq} = \left(k_{20}^{eq} - k_{02}^{eq}\right) + \underline{k_{2d}^{eq(1)}}\Delta t, \label{eq:cmeqcorrections}
\end{equation}
where $\underline{k_{2s}^{eq(1)}}= n_3^{eq(1)}$ and $\underline{k_{2d}^{eq(1)}}= n_4^{eq(1)}$ are the moment equilibria corrections given in Eqs.~(\ref{eq:82}) and (\ref{eq:84}), respectively, obtained via a C-E analysis. Then defining $k_{2s} = k_{20} + k_{02}$ and $k_{2d} = k_{20} - k_{02}$, we require $k_{2s}$ to relax to the above $k_{2s}^{eq}$ and analogously $k_{2d}$ to relax to the above $k_{2d}$ and along with the relaxations of the other central moments to the equilibria given in Eq.~(\ref{eq:cmequilibriaGOC}), each generally at a different rate, under collision.

Moreover, since the source central moments $\sigma_{mn}$ up to the second order appear in the analysis based on C-E analysis, they can be obtained via the definition given in Eq.~(\ref{eq:9B}) and then applying binomial expansions to relate them to the various source raw moments given in Eq.~(\ref{eq:GOCmomentsources}). The third and higher order source central moments are set to zero for simplicity. Then, the resulting 9 independent source central moments can be summarized as follows:
\begin{eqnarray}
&& \sigma_{00} = 0, \qquad \sigma_{10}= F_1, \qquad \sigma_{01}= F_2, \nonumber\\
&& \sigma_{20} + \sigma_{02} = 2 \left( \frac{1}{h_1}-1 \right)F_1 U_1 + 2 \left( \frac{1}{h_2}-1 \right)F_2 U_2, \nonumber\\
&& \sigma_{20} - \sigma_{02} = 2 \left( \frac{1}{h_1}-1 \right)F_1 U_1 - 2 \left( \frac{1}{h_2}-1 \right)F_2 U_2,\nonumber\\
&& \sigma_{11} =  \left( \frac{1}{h_2}-1 \right)F_1 U_2 + \left( \frac{1}{h_1}-1 \right)F_2 U_1, \nonumber\\
&& \sigma_{21} = 0, \qquad \sigma_{12} = 0, \qquad \sigma_{22} = 0. \label{eq:cmsourcesGOC}
\end{eqnarray}

\subsection{Matrix Formulation of the Central Moment GOC-LBM}
Then, analogous to Eq.~\eqref{eq:GOC-LBM-equivalent}, the central moment-based GOC-LBE can be formulated as
\begin{equation}\label{eq:GOC-equivalent-centralmoment}
 \mathbf{f} (\bm{\xi}+\mathbf{e}\Delta t, t+\Delta t) = \tensor{P}^{-1}\tensor{F}^{-1}
 \Big[\mathbf{m}^c + \tensor{B}^{-1}\tensor{\Lambda}\;\left(\; \tensor{B}\mathbf{m}^{eq,c}-\tensor{B}\mathbf{m}^c \;\right) + \tensor{B}^{-1}\left(\tensor{I} - \frac{\tensor{\Lambda}}{2}\right)  \tensor{B}\mathbf{\Phi}^c \Delta t \Big].
\end{equation}
Next, the above central moment GOC-LBE can be conveniently divided into different substeps to facilitate its implementation, which reads as
\begin{eqnarray}
\mathbf{m} &=& \tensor{P}\mathbf{f},\nonumber\\
\mathbf{m}^c&=&\tensor{F}\mathbf{m},\nonumber\\
\tilde{\mathbf{m}}^c&=&\mathbf{m}^c + \tensor{B}^{-1}\left\{\tensor{\Lambda}\;\left(\; \tensor{B}\mathbf{m}^{c,eq}-\tensor{B}\mathbf{m}^c \;\right) + \left(\tensor{I} - \frac{\tensor{\Lambda}}{2}\right)  \tensor{B}\mathbf{\Phi}^c\Delta t\right\},\nonumber\\
\tilde{\mathbf{m}} &=& \tensor{F}^{-1}\tilde{\mathbf{m}}^c,\nonumber\\
\tilde{\mathbf{f}} (\bm{\xi},t) &=&  \tensor{P}^{-1}\tilde{\mathbf{m}},\nonumber\\
 \mathbf{f} (\bm{\xi}+\mathbf{e}\Delta t, t+\Delta t)&=&\tilde{\mathbf{f}} (\bm{\xi},t).\label{eq:GOC-LBEcentralmoment}
\end{eqnarray}
Clearly, compared to the raw moment version given in Eq.~(\ref{eq:LBErawmomentmatrixform}), the central moment GOC-LBM given in Eq.~(\ref{eq:GOC-LBEcentralmoment}) requires extra pre- and post-collision mappings between raw moments and central moments via applying $\tensor{F}$ and $\tensor{F}^{-1}$, respectively. However, overall, the central moment GOC-LBM was found to be significantly more stable in achieving higher Reynolds numbers and/or finer grid clustering than the raw moment GOC-LBM in numerical investigations with various flow benchmarks, which is consistent with our earlier work with rectangular/cuboid LB formulations~\cite{yahia2021central, yahia2021three, yahia2022preconditioned}.

\subsection{Summary of the Algorithmic Steps of the Central Moment GOC-LBM}
Let's now discuss the step-by-step implementation details of Eq.~(\ref{eq:GOC-LBEcentralmoment}). Note that the preparatory steps to set up the orthogonal clustering/curvilinear grids, and the specification of the initial conditions, the speed of sound, and the relaxation rate parameters are identical to the respective preparatory steps given at the beginning of Appendix~\ref{sec: Appendix_A} and hence are not repeated here.

\begin{itemize}
\item \textbf{\underline{Step 1}: Compute pre-collision raw moments}
The expanded form of applying $\mathbf{m}=\tensor{P}\mathbf{f}$ in Eq.~(\ref{eq:GOC-LBEcentralmoment}) reads as
\begin{align*}
  &\Kps{00}= f_0 +f_1 +f_2 + f_3 +f_4 +f_5 +f_6 +f_7 +f_8,\\
  &\Kps{10} = f_1  -f_3  +f_5 -f_6 -f_7 +f_8, \quad\quad\quad\quad \Kps{11}= f_5 -f_6 +f_7 -f_8, \\
  &\Kps{01} = f_2  -f_4  +f_5 +f_6 -f_7 -f_8, \quad\quad\quad\quad \Kps{21}= f_5 +f_6 -f_7 -f_8, \\
  &\Kps{20} = f_1  +f_3  +f_5 +f_6 +f_7 +f_8, \quad\quad\quad\quad \Kps{12}= f_5 -f_6 -f_7 +f_8, \\
  &\Kps{02} = f_2  +f_4  +f_5 +f_6 +f_7 +f_8, \quad\quad\quad\quad \Kps{22}= f_5 +f_6 +f_7 +f_8.
\end{align*}

\item \textbf{\underline{Step 2}: Compute pre-collision central moments}

The mapping $\mathbf{m}^c=\tensor{F}\mathbf{m}$ in Eq.~(\ref{eq:GOC-LBEcentralmoment}) results in the central moments $k_{mn}$ from its relevant raw moments $k_{pq}^{\prime}$, which are given as follows:
\begin{eqnarray*}\label{eq:64}
k_{00} &=& k_{00}^\prime, \nonumber\\
k_{10} &=& k_{10}^\prime - U_1  k_{00}^\prime , \nonumber\\
k_{01} &=& k_{01}^\prime - U_2  k_{00}^\prime, \nonumber\\
k_{20} &=& k_{20}^\prime - 2 U_1  k_{10}^\prime + U_1^2 k_{00}^\prime , \nonumber\\
k_{02} &=& k_{02}^\prime - 2 U_2 k_{01}^\prime + U_2^2 k_{00}^\prime, \nonumber\\
k_{11} &=& k_{11}^\prime - U_2  k_{10}^\prime - U_1 k_{01}^\prime+ U_1 U_2 k_{00}^\prime , \nonumber\\
k_{21} &=& k_{21}^\prime - 2 U_1 k_{11}^\prime + U_1^2 k_{01}^\prime- U_2 k_{20}^\prime + 2 U_1 U_2 k_{10}^\prime - U_1^2 U_2 k_{00}^\prime , \nonumber\\
k_{12} &=& k_{12}^\prime - 2 U_2 k_{11}^\prime + U_2^2 k_{10}^\prime- U_1 k_{02}^\prime + 2 U_1 U_2 k_{01}^\prime - U_1 U_2^2 k_{00}^\prime , \nonumber\\
k_{22} &=& k_{22}^\prime - 2 U_1 k_{12}^\prime + U_1^2 k_{02}^\prime- 2 U_2 k_{21}^\prime + 4 U_1 U_2 k_{11}^\prime - 2 U_1^2 U_2 k_{01}^\prime + U_2^2 k_{20}^\prime - 2 U_1 U_2^2 k_{10}^\prime \nonumber\\
           &&+ U_1^2 U_2^2 k_{00}^\prime.
\end{eqnarray*}

\item \textbf{\underline{Step 3}: Perform Collision Step: Relax Central moments to their Equilibria and Augment with Source Central Moments}
\begin{enumerate}
  \item Compute the normal velocity derivatives $\partial_{\xi_1}U_1$ and $\partial_{\xi_2}U_2$ via second-order finite difference schemes.
  \item Compute the normal stress components $\tau_{11}$ and $\tau_{22}$ using $\partial_{\xi_1}U_1$ and $\partial_{\xi_2}U_2$ based on Eqs.~(\ref{eq:viscousstress_OCC_modified3_normal11}) and (\ref{eq:viscousstress_OCC_modified3_normal22}), where the associated curvature contributions $\tau_{11}^c$ and $\tau_{22}^c$ are given in Eq.~(\ref{eq:tauc}).
  \item Compute the shear stress $\tau_{12}$ locally using the off-diagonal non-equilibrium part of the moment $k_{11}^\prime$ via Eq.~(\ref{eq:shearlocallocalcomputation}), where the reduced body force components (i.e., without $\tau_{12}$) $\widetilde{F}_1$ and $\widetilde{F}_2$ are given in Eqs.~(\ref{eq:modifiedbodyforce1}) and (\ref{eq:modifiedbodyforce2}), respectively (see Appendix~\ref{sec:Appendix_shear_stress_local}).
  \item Compute the effective body force components in their full forms, i.e., $F_1$ and $F_2$ using Eqs.~(\ref{eq:bodyforce1}) and (\ref{eq:bodyforce2}), respectively.
  \item Compute the nine independent source moments $\sigma_{00}$, $\sigma_{10}$, $\sigma_{01}$, $\sigma_{20} + \sigma_{02}$, $\sigma_{20}-\sigma_{02}$, $\sigma_{11}$, $\sigma_{21}$, $\sigma_{12}$ and $\sigma_{22}$ using Eq.~(\ref{eq:cmsourcesGOC}). Label the combined diagonal second order source moments for convenience (which amounts to applying the operator $\tensor{B}$ to the bare form of the source moments $\mathbf{\Phi}^c$ in Eq.~(\ref{eq:GOC-LBEcentralmoment}) as
      \begin{equation*}
      \sigma_{2s} = \sigma_{20} + \sigma_{02}, \qquad \sigma_{2d} = \sigma_{20} - \sigma_{02}.
      \end{equation*}

  \item Compute the nine independent base equilibrium central moments $k_{00}^{eq}$, $k_{10}^{eq}$, $k_{01}^{eq}$, $k_{20}^{eq} + k_{02}^{eq}$, $k_{20}^{eq} - k_{02}^{eq}$, $k_{11}^{eq}$, $k_{21}^{eq}$, $k_{12}^{eq}$, $k_{22}^{eq}$ using Eq.~(\ref{eq:cmequilibriaGOC}). Then apply the necessary corrections to the
      combined diagonal second order base equilibrium moments to restore the full recovery of the NSE in the GOC and relabel them for convenience (which amounts to applying the operator $\tensor{B}$ to the bare form of the equilibrium vector $\mathbf{m}^{eq,c}$ in Eq.~(\ref{eq:GOC-LBEcentralmoment}) as
      \begin{equation*}
      k_{2s}^{eq} = \left(k_{20}^{eq} + k_{02}^{eq}\right) + \underline{k_{2s}^{eq(1)}}\Delta t, \qquad k_{2d}^{eq} = \left(k_{20}^{eq} - k_{02}^{eq}\right) + \underline{k_{2d}^{eq(1)}}\Delta t,
      \end{equation*}
      where $\underline{k_{2s}^{eq(1)}}= n_3^{eq(1)}$ and $\underline{k_{2d}^{eq(1)}}= n_4^{eq(1)}$ are the moment equilibria corrections given in Eqs.~(\ref{eq:82}) and (\ref{eq:84}), respectively. That is,
      \begin{eqnarray*}
        &&\underline{k_{2s}^{eq(1)}}\Delta t = n_3^{eq(1)} \Delta t =D_{3,1} \partial_{\xi_1} U_1 + D_{3,2} \partial_{\xi_2} U_2+ D_{3,3} \partial_{\xi_1} \rho +D_{3,4} \partial_{\xi_2} \rho + C_3,\\
        &&\underline{k_{2d}^{eq(1)}}\Delta t = n_4^{eq(1)} \Delta t =D_{4,1} \partial_{\xi_1} U_1 + D_{4,2} \partial_{\xi_2} U_2+ D_{4,3} \partial_{\xi_1} \rho +D_{4,4} \partial_{\xi_2} \rho + C_4.
      \end{eqnarray*}
      Here, the coefficients $D_{3, k}$ for $k = 1, 2, 3$ and $4$ and $C_3$ are given in Eq.~(\ref{eq:83}) and, similarly, the other set of coefficients $D_{4, k}$ for $k = 1, 2, 3$ and $4$ and $C_4$ are provided in Eq.~(\ref{eq:85}).
      \item Label the combined diagonal second order central moments for convenience (which amounts to applying the operator $\tensor{B}$ to the bare form of the central moment vector $\mathbf{m}^c$ to get its combined form in Eq.~(\ref{eq:GOC-LBEcentralmoment})) as
      \begin{equation*}
      k_{2s} = k_{20} + k_{02}, \qquad k_{2d} = k_{20} - k_{02}.
      \end{equation*}

  \item Relax all the nine independent central moments to their corresponding equilibria and augmented with source central moments to obtain their post-collision states, which are represented as follows:
      \begin{eqnarray*}
      &\tilde{k}_{00} = k_{00}+\omega_{0}\left(k_{00}^{eq}-k_{00}\right)+\left(1-\dfrac{\omega_{0}}{2}\right)\sigma_{00}\Delta t,\nonumber \\
      &\tilde{k}_{10} = k_{10}+\omega_{1}\left(k_{10}^{eq}-k_{10}\right)+\left(1-\dfrac{\omega_{1}}{2}\right)\sigma_{10}\Delta t,\nonumber \\
      &\tilde{k}_{01} = k_{01}+\omega_{2}\left(k_{01}^{eq}-k_{01}\right)+\left(1-\dfrac{\omega_{2}}{2}\right)\sigma_{01}\Delta t,\nonumber \\
      &\tilde{k}_{2s} = k_{2s}+\omega_{3}\left(k_{2s}^{eq}-k_{2s}\right)+\left(1-\dfrac{\omega_{3}}{2}\right)\sigma_{2s}\Delta t,\nonumber \\
      &\tilde{k}_{2d} = k_{2d}+\omega_{4}\left(k_{2d}^{eq}-k_{2d}\right)+\left(1-\dfrac{\omega_{4}}{2}\right)\sigma_{2d}\Delta t,\nonumber \\
      &\tilde{k}_{11} = k_{11}+\omega_{5}\left(k_{11}^{eq}-k_{11}\right)+\left(1-\dfrac{\omega_{5}}{2}\right)\sigma_{11}\Delta t,\nonumber \\
      &\tilde{k}_{21} = k_{21}+\omega_{6}\left(k_{21}^{eq}-k_{21}\right)+\left(1-\dfrac{\omega_{6}}{2}\right)\sigma_{21}\Delta t,\nonumber \\
      &\tilde{k}_{12} = k_{12}+\omega_{7}\left(k_{12}^{eq}-k_{12}\right)+\left(1-\dfrac{\omega_{7}}{2}\right)\sigma_{12}\Delta t,\nonumber \\
      &\tilde{k}_{22} = k_{22}+\omega_{8}\left(k_{22}^{eq}-k_{22}\right)+\left(1-\dfrac{\omega_{8}}{2}\right)\sigma_{22}\Delta t.\nonumber \\
      \end{eqnarray*}

  \item Split the post-collision combined second-order diagonal central moments to their bare forms (which amounts to applying the operator $\tensor{B}^{-1}$ to get $\tilde{\mathbf{m}}^c$ for the post-collision moment vector in Eq.~(\ref{eq:GOC-LBEcentralmoment})) as
      \begin{align*}
        &\tilde{k}_{20}= \frac{1}{2}\left(\tilde{k}_{2s} + \tilde{k}_{2d}\right), \quad\quad\quad
        \tilde{k}_{02} = \frac{1}{2}\left(\tilde{k}_{2s} - \tilde{k}_{2d}\right),
      \end{align*}

\end{enumerate}

\item \textbf{\underline{Step 4}: Compute Post-Collision Raw Moments}

Performing $\tilde{\mathbf{m}}=\tensor{F}^{-1}\tilde{\mathbf{m}}^c$ in Eq.~(\ref{eq:GOC-LBEcentralmoment})) to get the post-collision raw moments $\tilde{k}_{mn}^\prime$ from the relevant post-collision central moments $\tilde{k}_{pq}$, i.e.,
\begin{align*}
  &\tilde{k}_{00}^\prime = \tilde{k}_{00}, \nonumber\\
  &\tilde{k}_{10}^\prime = \tilde{k}_{10} + U_1\tilde {k}_{00}, \nonumber\\
  &\tilde{k}_{01}^\prime = \tilde{k}_{01} + U_2\tilde{k}_{00}, \nonumber\\
  &\tilde{k}_{20}^\prime = \tilde{k}_{20} + 2 U_1\tilde{k}_{10} + U_1^2 \tilde{k}_{00}, \nonumber\\
  &\tilde{k}_{02}^\prime = \tilde{k}_{02} + 2 U_2 \tilde{k}_{01} + U_2^2 \tilde{k}_{00}, \nonumber\\
  &\tilde{k}_{11}^\prime = \tilde{k}_{11} + U_2  \tilde{k}_{10} + U_1  \tilde{k}_{01} + U_1 U_2\tilde{k}_{00}, \nonumber\\
  &\tilde{k}_{21}^\prime = \tilde{k}_{21} + 2 U_1  \tilde{k}_{11} + U_1^2 \tilde{k}_{01} + U_2  \tilde{k}_{20} + 2 U_1 U_2 \tilde{k}_{10} + U_1^2 U_2 \tilde{k}_{00}, \nonumber\\
  &\tilde{k}_{12}^\prime = \tilde{k}_{12} + 2 U_2  \tilde{k}_{11} + U_2^2 \tilde{k}_{10} + U_1  \tilde{k}_{02} + 2 U_1 U_2 \tilde{k}_{01} + U_1 U_2^2 \tilde{k}_{00}, \nonumber\\
  &\tilde{k}_{22}^\prime = \tilde{k}_{22} + 2 U_1  \tilde{k}_{12} + U_1^2 \tilde{k}_{02} + 2 U_2  \tilde{k}_{21} + 4 U_1 U_2 \tilde{k}_{11} + 2 U_1^2 U_2 \tilde{k}_{01} + U_2^2 \tilde{k}_{20} + 2 U_1 U_2^2 \tilde{k}_{10} + U_1^2 U_2^2 \tilde{k}_{00}. \nonumber
\end{align*}

\item \textbf{\underline{Step 5}: Compute Post-Collision Distribution Functions}

Mapping the post-collision raw moments to the distribution functions via $\tilde{\mathbf{f}}=\tensor{P}^{-1}\tilde{\mathbf{m}}$ in Eq.~(\ref{eq:GOC-LBEcentralmoment})) yields
\begin{align*}\label{eq:3A}
&\tilde{f}_{0}= \tilde{k}_{00}^\prime-\tilde{k}_{20}^\prime- \tilde{k}_{02}^\prime + \tilde{k}_{22}^\prime,\nonumber  \\
&\tilde{f}_{1}= \frac{1}{2}\left(\tilde{k}_{10}^\prime+ \tilde{k}_{20}^\prime - \tilde{k}_{12}^\prime- \tilde{k}_{22}^\prime\right), \quad \quad \quad\tilde{f}_{5} = \frac{1}{4}\left(\tilde{k}_{11}^\prime+ \tilde{k}_{21}^\prime + \tilde{k}_{12}^\prime+ \tilde{k}_{22}^\prime\right),\nonumber\\
&\tilde{f}_{2}= \frac{1}{2}\left(\tilde{k}_{01}^\prime+ \tilde{k}_{02}^\prime - \tilde{k}_{21}^\prime- \tilde{k}_{22}^\prime\right), \quad \quad \quad\tilde{f}_{6} = \frac{1}{4}\left(-\tilde{k}_{11}^\prime+ \tilde{k}_{21}^\prime - \tilde{k}_{12}^\prime+ \tilde{k}_{22}^\prime\right), \nonumber\\
&\tilde{f}_{3}= \frac{1}{2}\left(-\tilde{k}_{10}^\prime+ \tilde{k}_{20}^\prime + \tilde{k}_{12}^\prime- \tilde{k}_{22}^\prime\right), \quad \quad \tilde{f}_{7} = \frac{1}{4}\left(\tilde{k}_{11}^\prime- \tilde{k}_{21}^\prime - \tilde{k}_{12}^\prime+ \tilde{k}_{22}^\prime\right),\nonumber\\
&\tilde{f}_{4}= \frac{1}{2}\left(-\tilde{k}_{01}^\prime+ \tilde{k}_{02}^\prime + \tilde{k}_{21}^\prime- \tilde{k}_{22}^\prime\right),  \quad \quad\tilde{f}_{8} = \frac{1}{4}\left(-\tilde{k}_{11}^\prime- \tilde{k}_{21}^\prime + \tilde{k}_{12}^\prime+ \tilde{k}_{22}^\prime\right).
\end{align*}

\item \textbf{\underline{Step 6}: Stream Distribution Functions}

Performing lockstep advection of the post-collision distribution functions to the neighboring nodes finally updates the distribution functions as
\begin{equation*}
f_\alpha(\bm{\xi},t+\Delta t)=\widetilde{f}_\alpha(\bm{\xi}-\bm{e}_\alpha\Delta t,t).
\end{equation*}
This step is supplemented with the implementation of boundary conditions for the boundary nodes as necessary.

\item \textbf{\underline{Step 7}: Update Hydrodynamic Fields}

Taking the zeroth and first moments, respectively, of the distribution functions, we finally update the fluid density and velocity fields as
\begin{align*}
\rho =\frac{1}{h_1 h_2}\sum_{\alpha=0}^{8} f_\alpha, \quad   U_1 = \frac{1}{\rho h_2}\left(\sum_{\alpha=0}^{8} f_{\alpha} e_{\alpha_1} + \frac{1}{2} F_1\Delta t\right), \quad U_2 = \frac{1}{\rho h_1}\left(\sum_{\alpha=0}^{8} f_{\alpha} e_{\alpha_2} + \frac{1}{2} F_2\Delta t\right).
\end{align*}

\end{itemize}

\section{GOC-LBM Based on Collisions in Velocity Space and SRT for NSE: Summary of Algorithm}\label{sec:Appendix_D}
Let's now discuss the simplest version of the GOC-LBM based on performing collisions in the velocity space and using a single relaxation time (SRT) model. That is, we use a single relaxation parameter $\tau$, which is equivalent to assuming the relaxation rates $\omega_j=\dfrac{1}{\tau}$ for $j=0,1,2,...,8$, and thus the bulk viscosity $\zeta$ reduces to be equal to the shear viscosity $\nu$, i.e., $\zeta=\nu$. As a result, the corrections associated with the combined diagonal components of the second-order moment equilibria $k_{2s}^{eq(1)\prime}$ and $k_{2d}^{eq(1)\prime}$ derived in the case of raw moments based GOC-LBM in Sec.~\ref{sec:GOC-LBM}, when adopted for constructing the GOC-LBM using SRT, are considerably simplified. That is, the simplified second order combined diagonal equilibrium moments read as
\begin{subequations} \label{Eq:1D}
\begin{eqnarray}
  k_{2s}^{eq\prime} = k_{20}^{eq\prime}+k_{02}^{eq\prime}+k_{2s}^{eq(1)}\Delta t &=& \frac{h_2}{h_1}\rho U_1^2 + \frac{h_1}{h_2}\rho U_2^2+\left(\frac{h_2}{h_1}+ \frac{h_1}{h_2}\right)\rho c_s^2\nonumber\\
  &&+D_{3,1} \partial_{\xi_1}U_1 +D_{3,2} \partial_{\xi_2}U_2+ D_{3,3} \partial_{\xi_1}\rho+ D_{3,4} \partial_{\xi_2}\rho + C_3,\\
  k_{2d}^{eq\prime} = k_{20}^{eq\prime}-k_{02}^{eq\prime}+k_{2d}^{eq(1)}\Delta t &=& \frac{h_2}{h_1}\rho U_1^2 - \frac{h_1}{h_2}\rho U_2^2+\left(\frac{h_2}{h_1}- \frac{h_1}{h_2}\right)\rho c_s^2\nonumber\\
  &&+D_{4,1} \partial_{\xi_1}U_1 +D_{4,2} \partial_{\xi_2}U_2+ D_{4,3} \partial_{\xi_1}\rho+ D_{4,4} \partial_{\xi_2}\rho + C_4,
\end{eqnarray}
\end{subequations}
where with $\nu=\zeta$, the resulting coefficients $D_{3,k}$, $D_{4,k}$ for $j=1,2,3$, and $4$, and $C_3$ and $C_4$ reduce as follows:
\begin{eqnarray}
&&D_{3,1}= \rho \left[\nu \frac{h_2}{c_s^2} - 3 \nu \left(1 + \frac{U_1^2}{c_s^2}\right)\frac{h_2}{h_1^2} \right], \quad\quad\quad
D_{3,3} = -\nu \left[ 3  \frac{h_2}{h_1^2} - \frac{h_2}{c_s^2} \right] U_1, \label{Eq:2D}\\
&&D_{3,2}= \rho \left[\nu \frac{h_1}{c_s^2}- 3 \nu \left(1 + \frac{U_2^2}{c_s^2}\right)\frac{h_1}{h_2^2} \right], \quad\quad\quad
D_{3,4}= -\nu \left[ 3  \frac{h_1}{h_2^2} - \frac{h_1}{c_s^2} \right] U_2, \nonumber\\
&&C_{3} = -\left( \frac{h_2}{h_1} \tau_{11}^c + \frac{h_1}{h_2} \tau_{22}^c \right)
- \rho \nu \bigg\{-2 U_1 \left( \frac{U_1^2}{c_s^2} + 1 \right)\frac{h_2}{h_1} \theta_{11}
-2 U_2 \left( \frac{U_2^2}{c_s^2} + 1 \right)\frac{h_1}{h_2} \theta_{22}\nonumber\\
&& \quad\quad\;\; +  U_1\left[\left(3 +\frac{U_1^2}{c_s^2}\right) \frac{h_2}{h_1}+ 2 \frac{h_1}{h_2} - \frac{h_1 h_2}{c_s^2}\right] \theta_{21}       + U_2\left[ \left(3 +\frac{U_2^2}{c_s^2}\right) \frac{h_1}{h_2}+ 2 \frac{h_2}{h_1} - \frac{h_1 h_2}{c_s^2}\right] \theta_{12}
\bigg\}.\nonumber
\end{eqnarray}
and
\begin{eqnarray}
&&D_{4,1}= \rho \left[\nu \frac{h_2}{c_s^2} - 3 \nu \left(1 + \frac{U_1^2}{c_s^2}\right)\frac{h_2}{h_1^2} \right], \quad\quad\quad  D_{4,3} =-\nu \left[ 3  \frac{h_2}{h_1^2} - \frac{h_2}{c_s^2} \right] U_1, \label{Eq:3D}\\
&&D_{4,2}= \rho \left[-\nu \frac{h_1}{c_s^2}+ 3 \nu \left(1 + \frac{U_2^2}{c_s^2}\right)\frac{h_1}{h_2^2} \right], \quad\quad\quad  D_{4,4}= \nu \left[ 3  \frac{h_1}{h_2^2} - \frac{h_1}{c_s^2} \right] U_2, \nonumber\\
&&C_{4} = -\left( \frac{h_2}{h_1} \tau_{11}^c  - \frac{h_1}{h_2} \tau_{22}^c\right)
- \rho \nu \bigg\{ -2 U_1 \left( \frac{U_1^2}{c_s^2}+1 \right)\frac{h_2}{h_1} \theta_{11}
+ 2 U_2 \left( \frac{U_2^2}{c_s^2}+1 \right)\frac{h_1}{h_2} \theta_{22}\nonumber\\
&& \quad\quad\;\; + U_1 \left[\left(3 +\frac{U_1^2}{c_s^2}\right) \frac{h_2}{h_1}-2 \frac{h_1}{h_2} -
   \frac{h_1 h_2}{c_s^2}\right] \theta_{21}
   + U_2 \left[ -\left(3 +\frac{U_2^2}{c_s^2}\right) \frac{h_1}{h_2}+ 2 \frac{h_2}{h_1} + \frac{h_1 h_2}{c_s^2}\right] \theta_{12}
   \bigg\}.\nonumber
\end{eqnarray}
Here, the contributions from the curvature coefficient matrix elements to the normal stress components $\tau_{11}^c$ and $\tau_{22}^c$, and the shear viscosity $\nu$ appearing in the last two equations also reduce to
\begin{equation}
  \tau_{11}^c = 2 \rho \nu U_2 \theta_{12}, \qquad \tau_{22}^c = 2 \rho \nu U_1 \theta_{21} \qquad \nu= c_s^2 \left(\tau -\dfrac{1}{2}\right) \Delta t,
\end{equation}
and the speed of sound $c_s$ is parameterized based on the metric factors and using the CFL condition according to Eq.~(\ref{eq:GOC-LBE_CFL}) just as in the case with all other collision models. Here, for reference, we note that the simplified forms of the normal stress components $\tau_{11}$ and $\tau_{22}$ that we will exploit in the implementation strategy of GOC-LBM using SRT can be written as
\begin{equation} \label{eq:reducednormalstress}
  \tau_{11} = \frac{2 \rho \nu}{h_1} \partial_{\xi_1}U_1+ \tau_{11}^c, \qquad \tau_{22} = \frac{2 \rho \nu}{h_2} \partial_{\xi_2}U_2+ \tau_{22}^c.
\end{equation}
Here, by contrast, we emphasize that the shear stress $\tau_{12} = \tau_{21}$ is obtained locally based on the off-diagonal second order non-equilibrium moment using the procedure given in Appendix~\ref{sec:Appendix_shear_stress_local} as its computation is independent of the collision model used.

Based on these considerations, for the SRT-based GOC-LBM with $\zeta =\nu$, there is no need to separately relax the combined diagonal second moments $k_{2s}^{\prime}$ and $k_{2d}^{\prime}$. Rather, we just need the bare equilibrium mments $k_{20}^{eq\prime}$ and $k_{02}^{eq\prime}$ which can then be converted to their equivalent contributions in the velocity space for an algorithmic implementation. In order to accomplish this, by invoking the definitions of the bare diagonal moments in terms of their combinations
\begin{align*}
k_{20}^{eq\prime}&= \frac{1}{2}(k_{2s}^{eq\prime}+ k_{2d}^{eq\prime}),&
k_{02}^{eq\prime} = \frac{1}{2}(k_{2s}^{eq\prime}- k_{2d}^{eq\prime}),
\end{align*}
and using Eqs.\eqref{Eq:1D}-\eqref{Eq:3D}, we get
\begin{align}\label{Eq:4D}
  &k_{20}^{eq\prime} = \frac{h_2}{h_1}\rho (U_1^2 + c_s^2)+G_1 \partial_{\xi_1}U_1 +G_2 \partial_{\xi_2}U_2+ G_3 \partial_{\xi_1}\rho +G_4 \partial_{\xi_2}\rho + G_5,\\
  &k_{02}^{eq\prime} = \frac{h_1}{h_2}\rho (U_2^2 + c_s^2)+H_1 \partial_{\xi_1}U_1 +H_2 \partial_{\xi_2}U_2+ H_3 \partial_{\xi_1}\rho +H_4 \partial_{\xi_2}\rho + H_5.
\end{align}
Here, we define the coefficients $G_j$ and $H_j$ as
\begin{equation}
G_j=\dfrac{1}{2}(D_{3,j} + D_{4,j}), \qquad H_j=\dfrac{1}{2}(D_{3,j} - D_{4,j}), \qquad j = 1, 2, 3, 4, 5
\end{equation}
Then, by utilizing Eqs.~\eqref{Eq:2D} and \eqref{Eq:3D}, we can obtain the following dramatically simplified explicit expressions for these coefficients:
\begin{eqnarray}
&&\hspace{-1.2cm} G_1= \rho \nu \left[\frac{h_2}{c_s^2} - 3  \left(1 + \frac{U_1^2}{c_s^2}\right)\frac{h_2}{h_1^2} \right], \qquad G_2= 0,
  \qquad G_3 =-\nu \left[ 3  \frac{h_2}{h_1^2} - \frac{h_2}{c_s^2} \right] U_1, \qquad G_4= 0, \nonumber\\
&&\hspace{-1.2cm} G_5 = - \frac{h_2}{h_1} \tau_{11}^c  - \rho \nu \bigg\{-2 U_1 \left( \frac{U_1^2}{c_s^2}+1 \right)\frac{h_2}{h_1} \theta_{11} + U_1 \left[\left(3 +\frac{U_1^2}{c_s^2}\right) \frac{h_2}{h_1} -\frac{h_1h_2}{c_s^2}\right]\theta_{21} + 2 U_2\frac{h_2}{h_1}\theta_{12}\bigg\}, \label{Eq:5D}
\end{eqnarray}
and
\begin{eqnarray}
&&\hspace{-1.2cm} H_1= 0, \qquad H_2= \rho \nu \left[ \frac{h_1}{c_s^2} - 3 \left(1 + \frac{U_2^2}{c_s^2}\right)\frac{h_1}{h_1^2} \right], \qquad H_3 = 0, \qquad H_4= -\nu \left[ 3  \frac{h_1}{h_2^2} - \frac{h_1}{c_s^2} \right] U_2, \nonumber\\
&&\hspace{-1.2cm} H_5 = - \frac{h_1}{h_2} \tau_{22}^c - \rho \nu \bigg\{-2 U_2 \left( \frac{U_2^2}{c_s^2}+1 \right)\frac{h_1}{h_2} \theta_{22}  + U_2\left[\left(3 +\frac{U_2^2}{c_s^2}\right)\frac{h_1}{h_2}- \frac{h_1 h_2}{c_s^2}\right] \theta_{12}+ 2 U_1\frac{h_1}{h_2} \theta_{21} \bigg\}. \label{Eq:6D}
\end{eqnarray}
Similarly, the second order combined diagonal raw moments of the source terms $\sigma_{2s}^{\prime}$ and $\sigma_{2d}^{\prime}$, i.e., $\sigma_{20}^{\prime}= \left(\sigma_{2s}^{\prime}+ \sigma_{2d}^{\prime}\right)/2$ and $\sigma_{02}^{eq\prime}=\left(\sigma_{2s}^{\prime}- \sigma_{2d}^{\prime}\right)/2$, which are obtained in Sec.~\ref{sec:GOC-LBM}, can be resolved into their bare components as
\begin{align}
&\sigma_{20}^{\prime}= 2 F_1 \frac{U_1}{h_1}, \quad\quad \sigma_{02}^{\prime}= 2 F_2 \frac{U_2}{h_2},
\end{align}
which are adequate for implementing the GOC-LBM using SRT.

\subsection{Construction of the GOC-LBE using SRT}
For convenience, based on the above considerations, we now summarize the raw moments of the equilibrium distribution functions and the source terms, which serve as the key elements in constructing the SRT-based GOC-LBM as follows:
\begin{align}\label{Eq:14D}
  &k_{00}^{eq\prime} = \rho h_1 h_2 ,  \quad \quad \quad k_{10}^{eq\prime} = \rho U_1 h_2,\quad \quad \quad k_{01}^{eq\prime} = \rho U_2 h_1, \nonumber\\
  &k_{20}^{eq\prime} = \frac{h_2}{h_1}\rho (U_1^2 + c_s^2)+ G_1 \partial_{\xi_1}U_1 +G_2 \partial_{\xi_2}U_2+ G_3 \partial_{\xi_1}\rho +G_4 \partial_{\xi_2}\rho + G_5,\nonumber\\
  &k_{02}^{eq\prime} = \frac{h_1}{h_2}\rho (U_2^2 + c_s^2)+H_1 \partial_{\xi_1}U_1 +H_2 \partial_{\xi_2}U_2+ H_3 \partial_{\xi_1}\rho +H_4 \partial_{\xi_2}\rho + H_5, \nonumber\\
  &k_{11}^{eq\prime} = \rho U_1 U_2,  \quad \quad \quad\;\;\;\;\;\; k_{21}^{eq\prime} =\rho\frac{U_2 }{h_1} (c_s^2+U_1^2), \qquad k_{12}^{eq\prime} =\rho\frac{ U_1 }{h_2} (c_s^2+U_2^2), \nonumber\\
  &k_{22}^{eq\prime} =\rho c_{s}^4 + \rho c_s^2 (U_1^2 + U_2^2) + \rho U_1^2 U_2^2.
\end{align}
where $G_j$ and $H_j$ are given in Eqs. \eqref{Eq:5D} and \eqref{Eq:6D}, respectively, and
\begin{align}\label{Eq:14D-source}
\sigma_{00}^{\prime}&=0,           &  \sigma_{10}^{\prime} &=F_1,             &  \sigma_{01}^{\prime}&=F_2,\nonumber\\
\sigma_{20}^{\prime}&=2 F_1 \frac{U_1}{h_1},         &  \sigma_{02}^{eq\prime}&=2 F_2 \frac{U_2}{h_2},   &  \sigma_{11}^{\prime}&=2 F_1 \frac{U_2}{h_2} + F_2 \frac{U_1}{h_1},\nonumber\\
\sigma_{21}^{\prime}&=0   &  \sigma_{12}^{\prime}&=0          &  \sigma_{22}^{\prime}&=0.
\end{align}

Then, to construct a SRT GOC-LBE, let's recall the following 9-dimensional vectors: $\mathbf{f}^{eq}=\left(f_0^{eq},f_1^{eq},f_2^{eq},\ldots,f_8^{eq}\right)^{\dag}$ and $\mathbf{S}=\left({S}_{0},{S}_{1},{S}_{2},\ldots,{S}_{8}\right)^{\dag}$ for the equilibria and the source terms, respectively, in the velocity space, and similarly $\mathbf{m}^{eq}=\left(k_{00}^{\eqp},k_{10}^{\eqp},k_{01}^{\eqp},k_{20}^{\eqp},k_{02}^{\eqp}, k_{11}^{\eqp}, k_{21}^{\eqp}, k_{12}^{\eqp}, k_{22}^{\eqp}\right)^{\dag}$~and~ $\mathbf{\Psi}=\left(\sigma_{00}^{\eqp},\sigma_{10}^{\eqp},\sigma_{01}^{\eqp},\sigma_{20}^{\eqp},\sigma_{02}^{\eqp}, \sigma_{11}^{\eqp}, \sigma_{21}^{\eqp}, \sigma_{12}^{\eqp}, \sigma_{22}^{\eqp}\right)^{\dag}$ for the raw moments of the equilibria and the source terms, respectively. Then, the derive the explicit expressions of the equilibrium distribution functions and the source terms in the velocity space for all the 9 discrete directions of the D2Q9 lattice, we apply the following inverse mappings
 \begin{equation}
\mathbf{f}^{eq}=\tensor{Q}^{-1} \mathbf{m}^{eq}, \;\;\;\;\;\;\;\;\quad \mathbf{S}=\tensor{Q}^{-1}\mathbf{\Psi}.
\end{equation}
Evaluating these inverse transformations, we finally obtain the required key expressions for the equilibrium distribution functions and the source terms, which read as
\begin{align}
   f_0^{eq}&=\left({k}_{00}^{\eqp} - {k}_{20}^{\eqp} - {k}_{02}^{\eqp} + {k}_{22}^{\eqp}\right), \nonumber\\
   f_1^{eq}&=\frac{1}{2} \left({k}_{10}^{\eqp} + {k}_{20}^{\eqp} - {k}_{12}^{\eqp} - {k}_{22}^{\eqp} \right), &
   f_5^{eq}&=\frac{1}{4} \left({k}_{11}^{\eqp} + {k}_{21}^{\eqp} + {k}_{12}^{\eqp} + {k}_{22}^{\eqp} \right), \nonumber\\
   f_2^{eq}&=\frac{1}{2} \left({k}_{01}^{\eqp} + {k}_{02}^{\eqp} - {k}_{21}^{\eqp} - {k}_{22}^{\eqp} \right), &
   f_6^{eq}&=\frac{1}{4} \left(-{k}_{11}^{\eqp} + {k}_{21}^{\eqp} - {k}_{12}^{\eqp} + {k}_{22}^{\eqp} \right), \nonumber\\
   f_3^{eq}&=\frac{1}{2} \left(-{k}_{10}^{\eqp} + {k}_{20}^{\eqp} + {k}_{12}^{\eqp} - {k}_{22}^{\eqp} \right), &
   f_7^{eq}&=\frac{1}{4} \left({k}_{11}^{\eqp} - {k}_{21}^{\eqp} - {k}_{12}^{\eqp} + {k}_{22}^{\eqp} \right), \nonumber\\
   f_4^{eq}&=\frac{1}{2} \left(-{k}_{01}^{\eqp} + {k}_{02}^{\eqp} + {k}_{21}^{\eqp} - {k}_{22}^{\eqp} \right), & f_8^{eq}&=\frac{1}{4} \left(-{k}_{11}^{\eqp} - {k}_{21}^{\eqp} + {k}_{12}^{\eqp} + {k}_{22}^{\eqp} \right), \label{eq:feqGOCLBE}
\end{align}
and
\begin{align}
S_0&= \left({\sigma}_{00}^{\prime} - {\sigma}_{20}^{\prime} - {\sigma}_{02}^{\prime} + {\sigma}_{22}^{\prime}\right), \nonumber\\
S_1&=\frac{1}{2} \left({\sigma}_{10}^{\prime} + {\sigma}_{20}^{\prime} - {\sigma}_{12}^{\prime} - {\sigma}_{22}^{\prime} \right),  &
S_5&=\frac{1}{4} \left({\sigma}_{11}^{\prime} + {\sigma}_{21}^{} + {\sigma}_{12}^{\prime} + {\sigma}_{22}^{\prime} \right),\nonumber\\
S_2&=\frac{1}{2} \left({\sigma}_{01}^{\prime} + {\sigma}_{02}^{\prime} - {\sigma}_{21}^{\prime} - {\sigma}_{22}^{\prime} \right),   &  S_6&=\frac{1}{4} \left(-{\sigma}_{11}^{\prime} + {\sigma}_{21}^{\prime} - {\sigma}_{12}^{\prime} + {\sigma}_{22}^{\prime} \right)\nonumber\\
S_3&=\frac{1}{2} \left(-{\sigma}_{10}^{\prime} + {\sigma}_{20}^{\prime} + {\sigma}_{12}^{\prime} - {\sigma}_{22}^{\prime} \right),   &  S_7&=\frac{1}{4} \left({\sigma}_{11}^{\prime} - {\sigma}_{21}^{\prime} - {\sigma}_{12}^{\prime} + {\sigma}_{22}^{\prime} \right), \nonumber\\
S_4&=\frac{1}{2} \left(-{\sigma}_{01}^{\prime} + {\sigma}_{02}^{\prime} + {\sigma}_{21}^{\prime} - {\sigma}_{22}^{\prime} \right),   &  S_8&=\frac{1}{4} \left(-{\sigma}_{11}^{\prime} - {\sigma}_{21}^{\prime} + {\sigma}_{12}^{\prime} + {\sigma}_{22}^{\prime} \right), \label{eq:SGOCLBE}
\end{align}
where the moments of the equilibria and the source terms, i.e., $k_{mn}^{eq\prime}$ and $\sigma_{mn}^\prime$, are given in Eqs.~(\ref{Eq:14D}) and (\ref{Eq:14D-source}), respectively. This immediately can be utilized to write the GOC-LBE in SRT as
\begin{equation}\label{eq:GOC-LBE_SRT}
f_{\alpha}\left(\bm{\xi}+\bm{e}_{\alpha} \Delta t, t+ \Delta t \right) = f_{\alpha}+ \frac{1}{\tau}\left( f_{\alpha}^{eq} - f_{\alpha} \right)+ \left( 1- \frac{1}{2\tau}  \right) S_{\alpha} \Delta t, \quad \alpha = 0, 1, \ldots, 8,
\end{equation}
where $f^{eq}_\alpha$ and $S_\alpha$ are given in Eqs.~(\ref{eq:feqGOCLBE}) and (\ref{eq:SGOCLBE}), which are, in turn, based on Eqs.~(\ref{Eq:14D}) and (\ref{Eq:14D-source}); the coefficients of the correction terms in Eq.~(\ref{Eq:14D}) are given in Eq.~(\ref{Eq:5D}) and (\ref{Eq:6D}). Equation~(\ref{eq:GOC-LBE_SRT}) is precisely of the standard form, but is endowed with using orthogonally clustered/curvilinear grids. While it is simpler than the GOC-LBE based on the raw moments and the central moments developed in the previous two sections since it does not require any pre- or post-collision mappings to other spaces, it is susceptible to numerical instabilities when simulating flow at relatively high Reynolds numbers and/or with high degree of grid clustering/coarsening when the metric factors $h_j$ for $j = 1$ or $2$ are significantly different from $1$. Nevertheless, it can be useful in the simulations of flows with moderate/low Reynolds number flows and with moderate grid clustering/coarsening.

\subsection{Summary of the Algorithmic Steps of the SRT-based GOC-LBM}
The implementation of the SRT-based GOC-LBM given in Eq.~(\ref{eq:GOC-LBE_SRT}) is straightforward, but will be outlined here for completeness. In this regard, the preparatory steps provided at the beginning of Appendix~\ref{sec: Appendix_A} for setting up the clustered/curvilinear grids, and the specification of the initial conditions, and the speed of sound are the same, with the exception that only a single relaxation rate $\omega = 1/\tau$ is used and the bulk and shear viscosities are equal to each other (i.e., $\zeta = \nu$).

\begin{itemize}
\item \textbf{\underline{Step 1}: Perform Collision Step: Relax Distribution Functions to their Equilibria and Augment with Source Terms}

Executing the collision step via relaxing the distributions to their equilibria, which is then augmented with the source terms in the velocity space yields
\begin{align*}
  &\tilde{f}_{\alpha}\left(\bm{\xi}, t\right)=f_{\alpha}+ \frac{1}{\tau}\left( f_{\alpha}^{eq} - f_{\alpha} \right)+ \left( 1- \frac{1}{2\tau}  \right) S_{\alpha} \Delta t, \qquad \alpha = 0, 1,\ldots, 8.
\end{align*}
Here, $f^{eq}_\alpha$ and $S_\alpha$ are provided in Eqs.~(\ref{eq:feqGOCLBE}) along with (\ref{Eq:14D}), and (\ref{eq:SGOCLBE}) along with (\ref{Eq:14D-source}), respectively, while the coefficients of the correction terms appearing in Eq.~(\ref{Eq:14D}) are given in Eq.~(\ref{Eq:5D}) and (\ref{Eq:6D}).

\item \textbf{\underline{Step 2}: Stream Distribution Functions}

Performing perfect-shift displacement of the post-collision distribution functions to the neighboring nodes updates the distribution functions as
\begin{equation*}
f_\alpha(\bm{\xi},t+\Delta t)=\widetilde{f}_\alpha(\bm{\xi}-\bm{e}_\alpha\Delta t,t),
\end{equation*}
which is often used along with specifying the boundary conditions for the boundary nodes as needed.

\item \textbf{\underline{Step 3}: Update Hydrodynamic Fields}

The macroscopic fluid density and velocity components are then obtained by taking the zeroth and first moments, respectively, of the distribution functions. That is,
\begin{align*}
\rho =\frac{1}{h_1 h_2}\sum_{\alpha=0}^{8} f_\alpha, \quad   U_1 = \frac{1}{\rho h_2}\left(\sum_{\alpha=0}^{8} f_{\alpha} e_{\alpha_1} + \frac{1}{2} F_1\Delta t\right), \quad U_2 = \frac{1}{\rho h_1}\left(\sum_{\alpha=0}^{8} f_{\alpha} e_{\alpha_2} + \frac{1}{2} F_2\Delta t\right).
\end{align*}

\end{itemize}

\section{GOC-LBM for the Magnetic Induction Equation for Magnetohydrodynamics}\label{sec:Appendix_E}
In this section, let's illustrate a further development of our approach beyond hydrodynamics, and consider the magnetohydrodynamics (MHD) of electrically conducting fluids. In MHD, the application of a magnetic field results in thin boundary layers around the bounding walls that are also known as the Hartmann layers, and the magnitude of the force arising due to the interaction of such a magnetic field with the flow field, which is also known as the Lorentz force, relative to the viscous force determines the extent of this thin layer region. By increasing this force ratio, also known as the Hartmann number, the boundary layer can be made progressively thinner. Hence, an implementation of our formulation using the GOC to MHD provides an excellent case study to analyze the efficacy of the use of orthogonally clustered grids.

\subsection{Magnetic Induction Equation in the GOC}
In MHD, we solve a coupled system of equations involving the spatio-temporal changes in the density and momentum fields governed by the Navier-Stokes equations (NSE), which includes the Lorentz force and the space and time variations in the magnetic field governed by the magnetic induction equation (MIE). The MIE can be derived from the Maxwell's equations of electrodynamics by combining the Faraday's law, the Ampère-Maxwell's law, and the Ohm's law under the constraint of a solenoidal magnetic field, which excludes any magnetic monopoles. Accordingly, the MIE, which is a vector equation, can be expressed compactly in a symbolic form as
\begin{equation}\label{eq:1_C}
\partial_t \bm{B} + {\bm{\nabla}}\cdot(\bm{U}\bm{B}-\bm{B}\bm{U}) = \bm{\nabla} \cdot \left( \eta \bm{\nabla}\bm{B}\right),
\end{equation}
where, as before, $\bm{U}$ is the fluid velocity field, $\bm{B}$  is the magnetic field, and $\eta$ is the magnetic resistivity, which is related to the electrical conductivity of the fluid $\sigma$ and the magnetic permeability $\mu_m$ via $\eta = 1/(\sigma \mu_m)$. A SRT-LBE for the MIE that is applicable for uniform grids was proposed in an earlier work~\cite{dellar2002lattice}. In this section, we will significantly extend such a formulation with a flexibility to handle orthogonally clustered/curvilinear grids.

In this regard, in order to first express the MIE in Eq.~(\ref{eq:1_C}) in the GOC, we denote the dyadic products appearing in the flux term in its LHS, $\bm{U}\bm{B} - \bm{B}\bm{U}$ as an antisymmetric tensor $\tensor{A}$, i.e.,
\begin{equation}\label{eq:2_C}
  \tensor{A}= \bm{U}\bm{B}-\bm{B}\bm{U},\;\; \mbox{or}\quad A_{ij} = U_j B_i - B_j U_i.
\end{equation}
Then, using the identity of the divergence of a tensor in the GOC given in Eq.\eqref{eq:vectoridentity}, by setting the tensor $\tensor{T} = \tensor{A}$, we get
\begin{equation}\label{eq:3_C}
  \bm{\nabla}\cdot \tensor{A} = \Big[ \frac{1}{h}\frac{\partial}{\partial \xi_j}\left(  \frac{h}{h_j} A_{ij}\right) + \left( A_{ij} \theta_{(i)j} - A_{jj} \theta_{(j)i}   \right) \Big] \hat{\xi_i},
\end{equation}
where $h= h_1 h_2$, and $\theta_{ij}$ is a component of the curvature coefficient matrix $\tensor{\theta}$ shown in Eq.~(\ref{eq:curvaturematrix}). Moving on to the RHS of the MIE, designating the magnetic field gradient tensor associated with the diffusion term as
\begin{equation}\label{eq:magneticfieldgradtensor}
\tensor{M} = \bm{\nabla}\bm{B},\;\; \mbox{or}\quad M_{ij} = \partial_{\xi_j} B_i,
\end{equation}
and then using $\bm{\nabla}\bm{B}=\hat{\xi}_j\cdot\bm{\nabla}\bm{B}\hat{\xi}_i$ and setting $(\bm{A})_j = \hat{\xi}_j$ and $(\bm{B})_i = B_i$ in the identity of a product of a vector with a tensor $\bm{A}\cdot\bm{\nabla}\bm{B}$ given in Eq.~(\ref{eq:vectoridentity}) and following the Appendix C of Ref.~\cite{fitzpatrick2017theoretical} (its Eq.~(C.26)) by invoking $(\hat{\xi}_i)_j = \hat{\xi}_i\cdot\hat{\xi}_j = \delta_{ij}$, we get
\begin{equation}\label{eq:magneticfieldgradtensoridentity}
\bm{\nabla}\bm{B} = \frac{1}{h_j}\frac{\partial B_i}{\partial \xi_j} + \delta_{ij}B_k\theta_{(i)k} - B_j\theta_{(j)i}.
\end{equation}
From this last equation (Eq.~(\ref{eq:magneticfieldgradtensoridentity})), for convenience, we introduce the following two tensors
\begin{equation}\label{eq:magneticfieldgradtensortwoparts}
R_{ij} = \frac{1}{h_j}\frac{\partial B_i}{\partial \xi_j}, \qquad Q_{ij} = \delta_{ij}B_k\theta_{(i)k} - B_j\theta_{(j)i}.
\end{equation}
Then, the magnetic field gradient tensor $\tensor{M} = \bm{\nabla}\bm{B}$ introduced in Eq.~(\ref{eq:magneticfieldgradtensor}) can be explicitly written in the following component form
\begin{equation}\label{eq:magneticfieldgradtensorcomponentform}
\tensor{M} = \tensor{R} + \tensor{Q}, \qquad M_{ij} = R_{ij} + Q_{ij}.
\end{equation}
In view of these developments, the diffusion term in Eq.~(\ref{eq:1_C}) can be expressed in the index notation as
\begin{equation}\label{eq:diffusiontermMIE}
\bm{\nabla} \cdot \left( \eta \bm{\nabla}\bm{B}\right) = \bm{\nabla} \cdot \left( \eta \tensor{M}\right) = \Big[ \frac{1}{h}\frac{\partial}{\partial \xi_j}\left(\eta\frac{h}{h_j} M_{ij}\right) + \eta\left( M_{ij} \theta_{(i)j} - M_{jj} \theta_{(j)i}   \right) \Big] \hat{\xi_i},
\end{equation}
where $h=h_1h_2$ and $M_{ij}$ is given in Eqs.~(\ref{eq:magneticfieldgradtensorcomponentform}) and (\ref{eq:magneticfieldgradtensortwoparts}).

Then, for a 2D magnetic field $\bm{B}= B_1\hat{\xi}_1 + B_2 \hat{\xi}_2$, the MIE in Eq.~(\ref{eq:1_C}) using Eqs.~(\ref{eq:3_C}) and (\ref{eq:diffusiontermMIE}) and invoking $A_{ii} = 0$ can be transformed into the GOC, which reads as
\begin{equation}\label{eq:MIEGOCform}
\frac{\partial B_i}{\partial t} + \frac{1}{h}\frac{\partial}{\partial {\xi_j}} \left(\frac{h}{h_j} A_{ij}\right) + A_{ij}\theta_{(i)j} =  \frac{1}{h}\frac{\partial}{\partial {\xi_j}} \left(\eta\frac{h}{h_j} M_{ij}\right)+\eta\left( M_{ij} \theta_{(i)j} - M_{jj} \theta_{(j)i}\right).
\end{equation}
In order to put this last equation in a conservative form, we first multiply it by $h = h_1h_2$ on both sides, and express the attendant source term as
\begin{equation}\label{eq:6ab_C}
Y_i= h_1 h_2 \left[ A_{ij} \theta_{(i)j}- \eta\left( M_{ij} \theta_{(i)j} - M_{jj} \theta_{(j)i}\right)\right],
\end{equation}
and then further rearraning the resulting equation, we finally obtain the MIE in the GOC in the following form that serves as the basis for the development a vector-based GOC-LBE in what follows:
\begin{equation}\label{eq:5_C}
\frac{\partial}{\partial t}(h_1 h_2 B_i) + \frac{\partial}{\partial {\xi_j}} \left(\frac{h}{h_j} A_{ij}\right) = \frac{\partial}{\partial {\xi_j}} \left(\eta\frac{h}{h_j} M_{ij}\right) - Y_i.
\end{equation}
Equation~\eqref{eq:5_C} is of advection-diffusion-with a source term type, albeit in a vector form, with $(h_1h_2/h_j) A_{ij}$ serving as the unusual components of the advection since $A_{ij}$ is an antisymmetric tensor (see Eq.~(\ref{eq:2_C})), $\eta(h_1h_2/h_j) M_{ij}$ being the anisotropic diffusive flux terms, and where its source term $-Y_i$ is given in Eq.~(\ref{eq:6ab_C}) that reflects the curvature effects of the GOC on the interactions between the magnetic and velocity fields. Then, our goal is to construct a GOC-LBM to solve the MIE as given in Eq.~(\ref{eq:5_C}), where the magnetic field gradients $\partial_{\xi_j}B_i$ appearing in $M_{ij}$ via $R_{ij}$ (see Eqs.~(\ref{eq:magneticfieldgradtensortwoparts}) and~(\ref{eq:magneticfieldgradtensorcomponentform})) will be evaluated locally based on non-equilibrium moments. We will illustrate this by developing it for the simplest collision model based on the SRT. Its extensions to more complex moments-based formulations can be achieved based on approaches used for the NSE given in Secs.~\ref{sec:MRT-LBM} and~\ref{sec:GOC-LBM} and various appendices, which are left for a future work. Moreover, since the MIE is structurally simpler than the NSE, we perform a C-E analysis of the GOC-LBE for the MIE directly based on the distribution functions rather than using their moments as done earlier in Secs.~\ref{sec:MRT-LBM} and~\ref{sec:GOC-LBM}.

\subsection{GOC-LBE for the Magnetic Induction Equation}
While our formulation can be used for any appropriate lattice set, including the D2Q9 lattice, based on simplicity and efficiency considerations, as in Ref.~\cite{dellar2002lattice}, we use the two-dimensional, five velocity (D2Q5) lattice for the SRT-based GOC-LBE for the MIE, and its particle velocity components read as
\begin{subequations}
\begin{eqnarray}
\ket{\mathbf{e}_{\xi_1}} &=& (0,1, 0,-1, 0)^\dag, \label{eq:D2Q5a}\\
\ket{\mathbf{e}_{\xi_2}} &=& (0,0, 1, 0,-1)^\dag,\label{eq:D2Q5b}
\end{eqnarray}
\end{subequations}
on which we will use the vector distribution functions $g_{\alpha i}$, where $\alpha = 0, 1, \ldots, 4$ and $i = \xi_1$ and $\xi_2$ are the particle directions and the principal directions of the GOC, respectively. The effect of collisions is modeled as relaxation of $g_{\alpha i}$ to their respective equilibrium distribution functions $g_{\alpha i}^{eq}$ over a timescale $\tau_m$ and the effect of the source term $Y_i$ in the MIE will be incorporated via a corresponding source term contribution in the velocity space $S_{\alpha i}^m$. For the time discretization of the latter, we will use the standard trapezoidal rule along the particle characteristics. Hence, the resulting SRT-based GOC-LBE can be written as
\begin{equation}\label{eq:7_C}
  g_{\alpha i}\left(\bm{\xi}+\bm{e_{\alpha}} \Delta t, t+ \Delta t\right)-  g_{\alpha i}\left(\bm{\xi}, t\right)= -\frac{1}{\tau_m}\left( g_{\alpha i} - g_{\alpha i}^{eq}  \right)+ \frac{1}{2}\left[S_{\alpha i}^m\left(\bm{\xi}+\bm{e_{\alpha}} \Delta t, t+ \Delta t\right) +  S_{\alpha i}^m\left(\bm{\xi}, t\right) \right]\Delta t.
\end{equation}
As in the case of the GOC-LBE for the NSE, but using the velocity space or the distribution functions rather than moments, we will represent the effective equilibria $g_{\alpha i}^{eq}$ in terms of a base equilibria $g_{\alpha i}^{eq(0)}$ which is augmented with certain corrections $g_{\alpha i}^{eq,corr}$ that will be determined via a C-E analysis in order to consistently recover the MIE (see Eq.~(\ref{eq:5_C})), especially its diffusive flux terms, in the following. That is,
\begin{equation}\label{eq:effectiveEDF-MIE}
g_{\alpha i}^{eq}= g_{\alpha i}^{eq(0)}+ g_{\alpha i}^{eq,corr}\Delta t.
\end{equation}
Here, the base equilibrium distribution functions $g_{\alpha i}^{eq(0)}$ can be readily constructed from an inspection of the MIE shown in Eq.~(\ref{eq:5_C}): The zeroth velocity moment of $g_{\alpha i}^{eq(0)}$ should recover the effective magnetic field inside its time derivative, i.e., $h_1h_2B_i$ and its first velocity moments with components $e_{\alpha\xi_1}$ and $e_{\alpha\xi_2}$ should represent the unusual advection terms $(h/h_1) A_{i1}$ and $(h/h_2) A_{i2}$, respectively, $h=h_1h_2$. Then, based on these three constraints, and introducing a speed of sound $c_{sm}$ for the GOC-LBE for the MIE, whose choice will be discussed in what follows, we can write the base equilibrium distribution functions as
\begin{equation}\label{eq:EDF-MIE-base}
g_{\alpha i}^{eq}= W_{\alpha} \left[ h_1 h_2 B_i + \dfrac{e_{\alpha j}}{c_{sm}^2} \left( \frac{h_1h_2}{h_j} A_{ij} \right) \right],
\end{equation}
where $W_\alpha$ is a weighting factor that should satisfy the following constraints:
\begin{align}\label{eq:weightingfactor-EDF-MIE}
  & \sum_{\alpha=0}^{4} W_{\alpha} = 1,  \quad \quad \sum_{\alpha=0}^{4} W_{\alpha} e_{\alpha \xi_i} =0, \quad \quad \quad  \sum_{\alpha=0}^{4} W_{\alpha} e_{\alpha \xi_i}e_{\alpha \xi_j} = c_{sm}^2 \delta_{ij}.
\end{align}
Respecting the symmetry of the lattice, i.e., $W_1=W_2=W_3=W_4$, these constraints imply $W_0 + 4W_1 = 1$ and $W_1 = c_{sm}^2/2$, which upon solving them yields the following explicit expressions for the weighting factors:
\begin{equation}\label{eq:10_a_C}
W_\alpha=
 \begin{cases}
     1- 2 c_{sm}^2   & \quad \alpha = 0\\
     \dfrac{c_{sm}^2}{2} & \quad \alpha = 1,2,3,4
  \end{cases}
\end{equation}
Note that for the special case when the speed of sound $c_{sm}=1/\sqrt{3}$, the weighting factors used in the LBE for the MIE using the uniform grid~\cite{dellar2002lattice} are recovered. For use with the orthogonally clustered/curvilinear grids, the speed of sound $c_{sm}$ should satisfy the CFL condition just like the speed of sound $c_s$ used in the NSE given in Eq.~(\ref{eq:GOC-LBE_CFL}) (see the discussion around Eq.~(\ref{eq:GOC-LBE_CFL}) for further details). Hence, we can write
\begin{equation}\label{eq:GOC-LBE_CFL-MIE}
c_{sm} = qc_{s*},  \qquad\mbox{where}\qquad q = \mbox{min}(h_{1min}, h_{2min}), \quad c_{s*} = 1/\sqrt{3}.
\end{equation}
Moreover, we can relate the source term $Y_i$ in Eq.~(\ref{eq:5_C}) by requiring it to match with the zeroth velocity moment of the source term in the velocity space $S_{\alpha i}^m$ that appears in the GOC-LBE for MIE in Eq.~(\ref{eq:7_C}). Based on this requirement and by noting the constraints on the weighting factors in Eq.~(\ref{eq:weightingfactor-EDF-MIE}), we then get
\begin{equation}\label{eq:8a_C}
   S_{\alpha i}^m = - W_{\alpha} Y_i.
\end{equation}
Finally, we note that the zeroth velocity moment of the distribution functions $g_{\alpha \xi_i}$ is a collision invariant in Eq.~(\ref{eq:7_C}) that recovers the magnetic field via
\begin{equation}\label{eq:9_C}
  h_1 h_2 B_i = \sum_{\alpha=0}^{4} g_{\alpha i} = \sum_{\alpha=0}^{4} g_{\alpha i}^{eq}.
\end{equation}
Once the magnetic field $\bm{B}= B_1 \hat{\xi}_1 + B_2 \hat{\xi}_2$ is computed, the local current density $\bm{J}= (1/\mu_m) \bm{\nabla} \times \bm{B}$ results in the Lorentz force $\bm{F}_{Lorentz} = \bm{J} \times \bm{B}$, which act on the fluid retarding its motion and hence appears as a body force thereby coupling the MIE with the NSE. We will discuss a local computational strategy for obtaining the current density and hence the Lorentz force later in this section.

In order to develop a computational procedure based on Eq.~(\ref{eq:7_C}), which is implicit due to the presence of the source term $S_{\alpha i}^m$ at time $t+\Delta t$, we first make it effectively explicit by applying the standard variable transformation given by
\begin{equation}\label{eq:11_C}
  \bar{g}_{\alpha i}= g_{\alpha i} - \frac{1}{2}  S_{\alpha i}^m \Delta t,
\end{equation}
which then modifies Eq.~\eqref{eq:7_C} to
\begin{equation}\label{eq:12_C}
  \bar{g}_{\alpha i}\left(\bm{\xi}+\bm{e_{\alpha}} \Delta t, t + \Delta t\right)-  \bar{g}_{\alpha i}\left(\bm{\xi}, t\right)= -\frac{1}{\tau_m}\left( \bar{g}_{\alpha i} - g_{\alpha i}^{eq}  \right)+ \left( 1- \frac{1}{2 \tau_m}\right)S_{\alpha i}^m \Delta t.
\end{equation}
Accordingly, using $ \sum_{\alpha=0}^{4} S_{\alpha i}^m = - Y_i$, then the magnetic field can be computed from the transformed variable via Eqs.~\eqref{eq:9_C} and~\eqref{eq:11_C} as
\begin{equation}\label{eq:13_C}
  h_1 h_2 B_i = \sum_{\alpha=0}^{4} \bar{g}_{\alpha i}+ \cfrac{1}{2} \left( - Y_i \right) \delta_t,
\end{equation}

\subsection{Chapman-Enskog Analysis}
We will now perform a C-E analysis with the goal of recovering the MIE in the GOC given in Eq.~\eqref{eq:5_C} while determining the associated moment equilibria corrections $g_{\alpha i}^{eq,corr}$ (see Eq.~(\ref{eq:effectiveEDF-MIE})). Denoting the Lagrangian time derivative as $D_t= \partial_t +\bm{e}_{\alpha} \cdot \bm{\nabla}$ and using it to compactly write the Taylor series in the LHS of Eq.~\eqref{eq:12_C} so that the latter becomes
\begin{equation*}
  \sum_{j=1}^{\infty}\frac{\epsilon^j}{j!} D_t^j \bar{g}_{\alpha i}= -\frac{1}{\tau_m}\left( \bar{g}_{\alpha i} - g_{\alpha i}^{eq}  \right)+ \left( 1- \frac{1}{2 \tau_m}\right)S_{\alpha i}^m \Delta t,
\end{equation*}
where we take $\epsilon =\Delta t$. Now, expanding the LHS of this last equation up to $O(\epsilon^2)$ and neglecting the higher order terms, we get
\begin{equation*}
 \epsilon D_t \bar{g}_{\alpha i} + \frac{\epsilon^2}{2} D_t^2 \bar{g}_{\alpha i}= -\frac{1}{\tau_m}\left( \bar{g}_{\alpha i} - g_{\alpha i}^{eq}  \right)+ \left( 1- \frac{1}{2 \tau_m}\right)S_{\alpha i}^m \epsilon,
\end{equation*}
Then, rewriting $\bar{g}_{\alpha i}$ in terms of the original variable $g_{\alpha i}$ via \eqref{eq:11_C}, we finally arrive at the following that serves as the starting point of applying the usual C-E expansions in what follows:
\begin{equation}\label{eq:15_C}
 \epsilon D_t g_{\alpha i} - \frac{1}{2} D_t S_{\alpha i}^m \epsilon^2 + \frac{\epsilon^2}{2} D_t^2 g_{\alpha i}  = -\frac{1}{\tau_m}\left( g_{\alpha i} - g_{\alpha i}^{eq}  \right)+ S_{\alpha i}^m \epsilon.
\end{equation}

Now, applying the C-E expansions
\begin{equation} \label{eq:C-EMHD}
g_{\alpha i} = g_{\alpha i}^{eq(0)}+\epsilon g_{\alpha i}^{(1)} + \epsilon^2 g_{\alpha i}^{(2)}+ \ldots, \qquad \partial_t=\partial_{t_0} +\epsilon \partial_{t_1} + \epsilon^2 \partial_{t_2}+ \ldots,
\end{equation}
and using the expression for the effective equilibrium distribution functions given in Eq.~(\ref{eq:effectiveEDF-MIE}), i.e., $g_{\alpha i}^{eq}= g_{\alpha i}^{eq(0)}+ g_{\alpha i}^{eq,corr}\Delta t$ and substituting them into Eq.~\eqref{eq:15_C}, and then grouping terms of the same order in $O(\epsilon^k)$ for $k = 1, 2$ together, we get the fast time (i.e., $O(\epsilon)$) scale and slow time (i.e., $O(\epsilon^2)$) scale variations of the GOC-LBE for the MIE, which read as
\begin{subequations}
\begin{eqnarray}\label{eq:17_18_C}
&&O (\epsilon^1 ):\quad D_{t_0} g_{\alpha i}^{eq(0)} = -\cfrac{1}{\tau_m}\left( g_{\alpha i}^{(1)} - g_{\alpha i}^{eq,corr}  \right)+  S_{\alpha i}^m, \label{eq:17_c} \\
&&O (\epsilon^2 ):\quad \partial_{t_1} g_{\alpha i}^{eq(0)}+ D_{t_0}g_{\alpha i}^{(1)}
-\cfrac{1}{2} D_{t_0} S_{\alpha i}^m + \cfrac{1}{2} D_{t_0}^2 g_{\alpha_i}^{eq(0)}= -\cfrac{1}{\tau_m} g_{\alpha_i}^{(2)}. \label{eq:18_c}
\end{eqnarray}
\end{subequations}
In the above equations, terms of order $O(\epsilon^3)$ are neglected and we also employed the compact notation $D_{t_n}= \partial_{t_n} +\bm{e}_{\alpha} \cdot \bm{\nabla}$. For further simplification, we substitute the expression for $D_{t_0} g_{\alpha i}^{eq(0)}$ from Eq.~\eqref{eq:17_c} into Eq.~\eqref{eq:18_c} and rearrange the resulting equation to get
\begin{equation*}\label{eq:18_c_modefied}
 O (\epsilon^2 ): \qquad \partial_{t_1} g_{\alpha i}^{eq(0)}+ \left( 1-  \cfrac{1}{2 \tau_m} \right)D_{t_0}g_{\alpha i}^{(1)} +  \cfrac{1}{2\tau_m} D_{t_0} g_{\alpha_i}^{eq,corr}= -\cfrac{1}{\tau_m} g_{\alpha_i}^{(2)}.
\end{equation*}
Thus, in summary, the temporal and spatial variations of the GOC-LBE for the MIE at $O(\epsilon^1)$ and $O(\epsilon^2)$ are, respectively, given by
\begin{subequations}\label{eq:21_C}
\begin{eqnarray}
 \centering
&&O (\epsilon^1 ):\qquad D_{t_0} g_{\alpha i}^{eq(0)} = -\cfrac{1}{\tau_m}\left( g_{\alpha i}^{(1)} - g_{\alpha i}^{eq,corr}  \right)+  S_{\alpha i}^m, \label{eq:21a_c} \\
&&O (\epsilon^2 ):\qquad  \partial_{t_1} g_{\alpha i}^{eq(0)}+ \left( 1-  \cfrac{1}{2 \tau_m} \right)D_{t_0}g_{\alpha i}^{(1)} +  \cfrac{1}{2\tau_m} D_{t_0} g_{\alpha_i}^{eq,corr}= -\cfrac{1}{\tau_m} g_{\alpha_i}^{(2)}. . \label{eq:21b_c}
\end{eqnarray}
\end{subequations}
Notice the presence of the equilibria corrections $g_{\alpha i}^{eq,corr}$ appearing in these systems of equations, which will need to be determined in such a way that their overall moment representation of this system matches with the MIE in the GOC given in Eq.~\eqref{eq:5_C}.

Then, to proceed further, we introduce the following solvability constraints on the C-E expansions for the various moments of $g_{\alpha i}^{eq(0)}$, $g_{\alpha i}^{eq,corr}$, $g_{\alpha i}^{(n)}$, and $S_{\alpha i}^{m}$, which read as
\begin{align}\label{eq:22_C}
& \sum_{\alpha} g_{\alpha i}^{eq(0)} = h_1 h_2 B_i, \quad \quad \sum_{\alpha} g_{\alpha i}^{eq,corr} = 0, \quad \quad  \sum_{\alpha} g_{\alpha i}^{(n)} = 0   \quad \text{for}\; n\ge 1, \quad \quad  \sum_{\alpha} S_{\alpha i}^m = - Y_i.
\end{align}
For convenience, we write the first velocity moment $e_{\alpha \xi_p}$ of the vector distribution function $g_{\alpha q}^{eq(0)}$  as the following rank-2 tensor $\Lambda_{pq}^{eq(0)}$ (and similarly for the corrections and the non-equilibrium distribution functions):
\begin{align}\label{eq:23_C}
\Lambda_{pq}^{eq(0)}=\sum_{\alpha} e_{\alpha \xi_p}g_{\alpha q}^{eq(0)},\qquad  \Lambda_{pq}^{eq,corr}=\sum_{\alpha} e_{\alpha \xi_p}g_{\alpha q}^{eq,corr}, \qquad \Lambda_{pq}^{(n)}=\sum_{\alpha} e_{\alpha \xi_p}g_{\alpha q}^{(n)}.
\end{align}
Moreover, we have for the first velocity moment of the source terms
\begin{equation}
\sum_{\alpha} e_{\alpha \xi_p} S_{\alpha q}^{m}=0.\label{eq:firstmomentsource-MIE}
\end{equation}
In view of these notations in Eq.~(\ref{eq:23_C}) and the moment constraints in Eq.~(\ref{eq:22_C}),we can rewrite the base equilibrium distribution functions $g_{\alpha i}^{eq(0)}$ and its corrections $g_{\alpha i}^{eq,corr}$ in terms of their respective moments as,
\begin{eqnarray}\label{eq:24_C}
&&g_{\alpha i}^{eq(0)}=W_\alpha\left[h_1 h_2 B_i+\cfrac{e_{\alpha \xi_j}}{c_{sm}^2}\Lambda_{ji}^{eq(0)}\right],\label{eq:24a_C} \\
&&g_{\alpha i}^{eq,corr}=W_\alpha\left[\cfrac{e_{\alpha \xi_j}}{c_{sm}^2}\Lambda_{ji}^{eq,corr}\right].\label{eq:24b_C}
\end{eqnarray}
Thus, the equilibrium corrections are determined if the respective rank-2 moment tensor $\Lambda_{ji}^{eq,corr}$ is obtained, and the tensor $\Lambda_{ji}^{eq(0)}$ in the base equilibrium distribution functions is chosen in such a way as to recover the MIE in the GOC.

In order to achieve this, we start by taking the zeroth velocity moment of the system of equations in Eq.~\eqref{eq:21_C}, which when using the solvability constraints and the moment notations given in Eqs.~\eqref{eq:22_C}~-~\eqref{eq:24_C}, yield
\begin{subequations}
\begin{eqnarray}\label{Eq:26_27_C}
&&O (\epsilon^1 ):\qquad \partial_{t_0}\left( h_1 h_2 B_i \right) + \partial_j \Lambda_{ji}^{eq(0)}= - Y_i, \label{Eq:26_C}\\
&&O (\epsilon^2 ):\qquad \partial_{t_1}\left( h_1 h_2 B_i \right) + \left( 1- \dfrac{1}{2 \tau_m}  \right)\partial_j \Lambda_{ji}^{(1)} + \dfrac{1}{2 \tau_m}\partial_j \Lambda_{ji}^{eq,corr}= 0.\label{Eq:27_C}
\end{eqnarray}
\end{subequations}
Then, taking $\partial_t= \partial_{t_0}+ \epsilon \partial_{t_1}$, we combine the conserved moment equation at the fast time scale $t_0$ given in Eq.~\eqref{Eq:26_C} with the corresponding equation at the slow time scale $t_1$ in Eq.~\eqref{Eq:27_C} after multiplying the latter by $\epsilon = \Delta t$. As a result, we get the emergent macroscopic equation of the variation of the effective magnetic field from the GOC-LBE for the MIE, which reads as
\begin{equation}\label{eq:29_C}
\partial_{t}\left( h_1 h_2 B_i \right) + \partial_j \Lambda_{ji}^{eq(0)} = \partial_j \bigg\{- \left( 1- \cfrac{1}{2 \tau_m}  \right)  \Lambda_{ji}^{(1)}\Delta t- \cfrac{1}{2 \tau_m}\Lambda_{ji}^{eq,corr}\Delta t \bigg\}- Y_i.
\end{equation}
Then, matching the moment tensors in the LHS of this last equation (Eq.~(\ref{eq:29_C})) with the unusual advection flux terms appearing in the LHS of the target MIE given in Eq.~\eqref{eq:5_C}, we obtain the following expressions for the base equilibrium moment tensor $\Lambda_{ji}^{eq(0)}$:
\begin{equation}\label{eq:30_C}
\Lambda_{ji}^{eq(0)}= \frac{h_1h_2}{h_{j}} A_{ij}.
\end{equation}
Thus, $\Lambda_{1i}^{eq(0)} = h_2 A_{i1}$ and $\Lambda_{2i}^{eq(0)} = h_1 A_{i2}$. Taken together, Eqs.~(\ref{eq:24a_C}) and (\ref{eq:30_C}) are then identical to the base equilibrium distributions given earlier in Eq.~(\ref{eq:EDF-MIE-base}).

Next, in order to fully recover the GOC-based MIE in Eq.~\eqref{eq:5_C}, the non-equilibrium moments $\Lambda_{ji}^{(1)}$ together with the equilibria correction moments $\Lambda_{ji}^{eq,corr}$ appearing in the emergent equation (Eq.~\eqref{eq:29_C}) are required to match with the diffusive flux terms in Eq.~\eqref{eq:5_C}. That is, the resulting constraint relation is then given by
\begin{equation}\label{eq:31_C}
-\left( 1- \dfrac{1}{2 \tau_m}  \right)\Lambda_{ji}^{(1)}  \Delta t - \dfrac{1}{2\tau_m} \Lambda_{ji}^{eq,corr}  \Delta t = \eta \dfrac{h_1h_2}{h_j}M_{ij}.
\end{equation}

Now, in order to determine the equilibrium moment corrections $\Lambda_{ji}^{eq,corr}$ from Eq.~\eqref{eq:31_C}, we need to determine the expressions for the non-equilibrium moments, i.e., $\Lambda_{ji}^{(1)}$. In this regard, we take the first velocity moment of Eq.~\eqref{eq:21a_c}, i.e., apply $\sum_{\alpha} e_{\alpha \xi_j}$ to both sides of this equation as follows:
\begin{equation*}
\partial_{t_0}\Big[ \sum_{\alpha} e_{\alpha \xi_j} g_{\alpha i}^{eq(0)} \Big] +\partial_{k}\Big[ \sum_{\alpha} e_{\alpha \xi_k}e_{\alpha \xi_j} g_{\alpha i}^{eq(0)} \Big] = -\cfrac{1}{\tau_m} \sum_{\alpha} e_{\alpha \xi_j}  g_{\alpha i}^{(1)} + \cfrac{1}{\tau_m} \sum_{\alpha} e_{\alpha \xi_j} g_{\alpha i}^{eq,corr}  + \sum_{\alpha} e_{\alpha \xi_j} S_{\alpha i}^m,
\end{equation*}
Rearranging this last equation using the notations introduced for the rank-2 moment tensors in Eq.~\eqref{eq:23_C} yields,
\begin{equation}\label{eq:32_C}
\partial_{t_0}\Lambda_{ji}^{eq(0)} + \partial_{k} \Gamma_{kji}^{eq(0)} = -\cfrac{1}{\tau_m} \Lambda_{ji}^{(1)}+ \dfrac{1}{\tau_m} \Lambda_{ji}^{eq,corr},
\end{equation}
where $\Gamma_{kji}^{eq(0)}= \sum_{\alpha} e_{\alpha k}e_{\alpha j} g_{\alpha i}^{eq(0)}$. Following Ref.~\cite{dellar2002lattice}, $\partial_{t_0}\Lambda_{ji}^{eq(0)}\sim O(\mbox{Ma}^3)$, which can be deemed small and hence is neglected. Here, evaluating the second velocity moment of the base equilibrium distribution functions given in Eq.~(\ref{eq:EDF-MIE-base}), we get
\begin{equation}\label{eq:33_C}
\Gamma_{kji}^{eq(0)}= c_{sm}^2 \delta_{kj} \left( B_i h_1 h_2  \right),
\end{equation}
Then, substituting this last equation (Eq.~\eqref{eq:33_C}) into Eq.~\eqref{eq:32_C} yields
\begin{equation}\label{eq:34_C}
c_{sm}^2 \partial_{j} \left( B_i h_1 h_2  \right)= -\cfrac{1}{\tau_m} \Lambda_{ji}^{(1)}+ \cfrac{1}{\tau_m} \Lambda_{ji}^{eq,corr},
\end{equation}
which we rearrange and to finally obtain the non-equilibrium moments $\Lambda_{ji}^{(1)}$. The results read as
\begin{equation}\label{eq:35_C}
\Lambda_{ji}^{(1)}= -\tau_m c_{sm}^2 \partial_{\xi_j} \left( h_1 h_2 B_i \right) + \Lambda_{ji}^{eq,corr}.
\end{equation}

Next, for further simplification to determine the moment equilibria corrections, for our GOC-LBE, we relate the magnetic resistivity of the fluid $\eta$ to the relaxation time $\tau_m$ as well as the speed of sound $c_{sm}$, like in the case of the standard LBE for uniform grids using the Cartesian coordinates~\cite{dellar2002lattice}, as
\begin{equation}\label{eq:magneticresistivity}
\eta= c_{sm}^2 \left( \tau_m -\dfrac{1}{2}  \right)\Delta t.
\end{equation}
Then, substituting the non-equilibrium moments obtained in Eq.~(\ref{eq:35_C}) into Eq.~\eqref{eq:31_C}, rearranging and using $M_{ij} = R_{ij}+Q_{ij}$ and substituting for $R_{ij}$, where $R_{ij}$ and $Q_{ij}$ are given in Eq.~(\ref{eq:magneticfieldgradtensortwoparts}), we finally get
\begin{equation}\label{eq:intermediateLambdacorrection}
\eta \partial_{\xi_j}(h_1h_2B_i)-\Lambda_{ji}^{eq, corr}\Delta t = \eta \frac{h_1h_2}{h_j}M_{ij} = \eta\frac{h_1h_2}{h_j^2}\partial_{\xi_j}B_i + \eta\frac{h_1h_2}{h_j}Q_{ij},
\end{equation}
Using this last equation (Eq.~(\ref{eq:intermediateLambdacorrection})) and simplifying, we can then express the required moment equilibria corrections to restore the correct diffusive fluxes and recover the MIE in the GOC, and the results read as
\begin{equation}\label{eq:Lambdacorr}
\Lambda_{ji}^{eq,corr}\Delta t = \eta h_1h_2\left[\left(1-\frac{1}{h_j^2}\right)\partial_{\xi_j}B_i + h_j(\theta_{1j}+\theta_{2j})B_i-\frac{1}{h_j}Q_{ij}\right]
\end{equation}
Moreover, these moment equilibria corrections $\Lambda_{ji}^{eq,corr}$ can be readily transformed into the equilibrium corrections in the velocity space via Eq.~\eqref{eq:24b_C} (appearing in the effective equilibrium distribution functions given in Eq.~(\ref{eq:effectiveEDF-MIE})), which can be explicitly written as follows:
\begin{equation}\label{eq:42_C}
g_{\alpha i}^{eq,corr}=W_\alpha \eta h_1h_2\Bigg\{\frac{e_{\alpha j}}{c_{sm}^2}\left[\left(1-\frac{1}{h_j^2}\right)\partial_{\xi_j}B_i + h_j(\theta_{1j}+\theta_{2j})B_i-\frac{1}{h_j}Q_{ij}\right]\Bigg\},
\end{equation}
where $Q_{ij} = \delta_{ij}B_k\theta_{(i)k} - B_j\theta_{(j)i}$ (see Eq.~(\ref{eq:magneticfieldgradtensortwoparts})).

\subsection{Local Computation of Spatial Derivatives of Magnetic field using Non-equilibrium Moments in the GOC-LBM}
An advantage of our vector-based GOC-LBE for the MIE derived above is that the derivatives $\partial_{\xi_j} B_i$ appearing in Eq.~(\ref{eq:42_C}) for the corrections terms to the equilibria (as well as the current density discussed in the next subsection (see Appendix~\ref{appendix:local_computation_current_density})) can be computed locally based on the non-equilibrium moments of the vector distribution functions. Starting from Eq.~(\ref{eq:35_C}) and substituting for $\partial_{\xi_j}(h_1h_2B_i)$ using
\begin{equation*}
\partial_{\xi_j}(h_1h_2B_i) = h_1h_2\left[\partial_{\xi_j}B_i + h_j(\theta_{1j}+\theta_{2j})B_i\right],
\end{equation*}
which follows from the product rule and the use of the definition of the curvature coefficient matrix element, and then substituting for $\Lambda_{ji}^{eq,corr}\Delta t$ from Eq.~(\ref{eq:Lambdacorr}) and using the expression for the magnetic resistivity $\eta$ in terms of the relaxation time $\tau_m$ (see Eq.~(\ref{eq:magneticresistivity})), and rearranging and simplifying, we finally get magnetic field gradient tensor $\partial_{\xi_j}B_i$ locally in terms of the non-equilibrium moment $\Lambda_{ji}^{(1)}$. The result reads as
\begin{equation}\label{eq:gradB}
\partial_{\xi_j} B_i= \frac{-\left[  \dfrac{\Lambda_{ji}^{(1)}}{h_1h_2}+ \dfrac{c_{sm}^2}{2} h_j \left(\theta_{1j}+ \theta_{2j}\right)B_i + c_{sm}^2\left(\tau_m-\frac{1}{2}\right)\dfrac{1}{h_j}Q_{ij} \right]}{ c_{sm}^2 \left[\dfrac{1}{2}+ \left(\tau_m-\dfrac{1}{2}\right) \dfrac{1}{h_j^2} \right] },
\end{equation}
where $Q_{ij} = \delta_{ij}B_k\theta_{(i)k} - B_j\theta_{(j)i}$ and $ i,j \in \{1,2\} $. Here, the non-equilibrium moment $\Lambda_{ji}^{(1)}$ follows from the definition given in Eq.~\eqref{eq:23_C} with the use of the variable transformation (see Eq.~(\ref{eq:11_C})) along with the constraint on the source moment in Eq.~(\ref{eq:firstmomentsource-MIE}) and is given as
\begin{equation}\label{eq:firstvelmomenttensor}
\Lambda_{ji}^{(1)}=\sum_{\alpha} e_{\alpha \xi_1} g_{\alpha i}^{(1)}=\sum_{\alpha}e_{\alpha \xi_j} \left( g_{\alpha i} -  g_{\alpha i}^{eq(0)} \right) = \sum_{\alpha}e_{\alpha \xi_j} \left( \bar{g}_{\alpha i} -  g_{\alpha i}^{eq(0)} \right).
\end{equation}

\subsection{Local Computation of Current Density and the Lorentz Force in the GOC-LBM}\label{appendix:local_computation_current_density}
The current density $\bm{J}$ is given by $\bm{J}= \dfrac{1}{\mu_m}\bm{\nabla} \times \bm{B}$ and can be rewritten in terms of the GOC in 2D based on the curl of a vector field either from Eq.~(\ref{eq:Identity}) or using the following by setting $h_3 = 1$ and $v_3 = 0$ in $\bm{V} = v_1\hat{\xi}_1+v_2\hat{\xi}_2+v_3\hat{\xi}_3$ and setting $\bm{V}=\bm{B}$:
\begin{equation}\label{eq:53_C}
\bm{\nabla} \times \bm{V}=\dfrac{1}{h_1 h_2 h_3}
  \begin{vmatrix}
h_1 \hat{\xi}_1 & h_2 \hat{\xi}_2 & h_3 \hat{\xi}_3 \\
\partial_{\xi_1} & \partial_{\xi_2} & \partial_{\xi_3}\\
h_1 v_1 & h_2 v_2 & h_3 v_3,   \notag
\end{vmatrix}
\end{equation}
which then yields
\begin{equation*}
\bm{J}=\dfrac{1}{\mu_m}\dfrac{1}{h_1 h_2} \left[ \partial_{\xi_1} \left(  h_2 B_2 \right) - \partial_{\xi_2} \left(  h_1 B_1 \right) \right]\hat{\xi}_3.
\end{equation*}
This last equation can be further simplified by using the identities given by
\begin{align*}
& \partial_{\xi_1}\left(  h_2 B_2 \right)= h_2 \partial_{\xi_1} B_2 + B_2 \partial_{\xi_1} h_2 =h_2 \partial_{\xi_1} B_2 + h_1 h_2 B_2 \theta_{21}  \\
& \partial_{\xi_2}\left(  h_1 B_1 \right)= h_1 \partial_{\xi_2} B_1 + B_1 \partial_{\xi_1} h_1 =h_1 \partial_{\xi_2} B_1 + h_1 h_2 B_1 \theta_{12}.
\end{align*}
so that finally the current density $\bm{J} = J_1\hat{\xi}_1+J_2\hat{\xi}_2+J_3\hat{\xi}_3$ can be written as follows:
\begin{equation}\label{eq:currentdensity-GOC}
J_1 = 0, \qquad J_2 = 0, \qquad J_3 = \dfrac{1}{\mu_m }\dfrac{1}{h_1 h_2} \bigg[ h_2 \partial_{\xi_1} B_2 - h_1 \partial_{\xi_2}  B_1 + h_1 h_2 \left(B_2 \theta_{21}- B_1 \theta_{12}\right) \bigg],
\end{equation}
where the spatial derivatives of the magnetic field can be evaluated locally using Eqs.~(\ref{eq:gradB}) and (\ref{eq:firstvelmomenttensor}).

Then, the Lorentz force $\bm{F}_{Lorentz}=F_{Lorentz, 1}\hat{\xi}_1 + F_{Lorentz, 2}\hat{\xi}_2 + F_{Lorentz, 3}\hat{\xi}_3$, which acts on the fluid resisting its motions and hence coupling the GOC-LBEs for the MIE and the NSE, can be calculated via $\bm{F}_{Lorentz}=\bm{J}\times\bm{B}$, or writing in its components form in the GOC and evaluating using
\begin{equation*}
\bm{F}_{Lorentz} = \bm{J} \times \bm{B}=
\begin{vmatrix}
\hat{\xi}_1 & \hat{\xi}_2 & \hat{\xi}_3 \\
0 & 0 & J_3 \\
B_1 & B_2 & 0 \notag
\end{vmatrix}
\end{equation*}
yields
\begin{equation}\label{eq:Lorentzforce-GOC}
F_{Lorentz, 1} = - J_3 B_2 , \qquad F_{Lorentz, 2} = J_3 B_1, \qquad F_{Lorentz, 3} = 0.
\end{equation}
Here, the current density component $J_3$ is computed from Eq.~(\ref{eq:currentdensity-GOC}) given above.

\subsection{Summary of the Algorithmic Steps of the GOC-LBM for MIE}
The preparatory steps for the GOC-LBM for the MIE are analogous to those for the NSE given earlier in Appendix~\ref{sec: Appendix_A}, but one should take into account of the changes as necessary, such as that the former uses vector distribution functions, involves magnetic resistivity rather than viscosity which is related to a single relaxation time. Here, we summarize the main steps involved in implementing the GOC-LBE for the MIE given in Eq.~(\ref{eq:12_C}) over a single time step $\Delta t$.
\begin{itemize}
  \item \textbf{\underline{Step 1}: Compute the spatial derivatives of the magnetic field}

  That is, calculate $\partial_{\xi_1} B_i$ and $\partial_{\xi_2} B_i$ locally based on non-equilibrium moments using Eqs.~(\ref{eq:gradB}) and~(\ref{eq:firstvelmomenttensor}).
  \item \textbf{\underline{Step 2}: Compute equilibrium distribution functions}

  That is, calculate the base equilibrium distribution functions $g_{\alpha i}^{eq(0)}$ and its corrections $g_{\alpha i}^{eq, corr}$ from Eqs.~(\ref{eq:EDF-MIE-base}) (together with Eq.~(\ref{eq:2_C})) and (\ref{eq:42_C}), respectively, to get the effective equilibria $g_{\alpha i}^{eq}$ from Eq.~(\ref{eq:effectiveEDF-MIE}).

  \item \textbf{\underline{Step 3:} Perform Collision Step: Relax Distribution Functions to their Equilibria and Augment with Source Terms}
  \begin{equation*}
  \tilde{\bar{g}}_{\alpha i} = \bar{g}_{\alpha i}\left(\bm{\xi}, t\right) - \frac{1}{\tau_m}\left( \bar{g}_{\alpha i} - g_{\alpha i}^{eq}  \right)+ \left( 1- \frac{1}{2 \tau_m}\right)S_{\alpha i}^m \Delta t,
  \end{equation*}
  where the source terms $S_{\alpha i}^m$ are obtained from Eq.~(\ref{eq:8a_C}) along with Eqs.~(\ref{eq:6ab_C}) and~(\ref{eq:2_C}).

  \item  \textbf{\underline{Step 4:} Stream Distribution Functions}

  Perform the advection of vector distribution functions to the neighboring nodes along their characteristic directions as
  \begin{equation*}
  \bar{g}_{\alpha i}\left(\bm{\xi}, t+\Delta t\right) = \tilde{\bar{g}}_{\alpha i}\left(\bm{\xi} - \bm{e_{\alpha}} \Delta t, t\right).
  \end{equation*}
  As necessary, include any boundary conditions on the distribution functions -- see the section that follows for a brief discussion on an implementation strategy.

  \item  \textbf{\underline{Step 5:} Update Magnetic Fields and Current Densities}

  That is, compute $\bm{B}$ as the zeroth moment of the distribution functions from Eq.~(\ref{eq:13_C}) and $\bm{J}$ as the first moment of the distribution functions from Eq.~(\ref{eq:currentdensity-GOC}), via Eqs.~(\ref{eq:gradB}) and~(\ref{eq:firstvelmomenttensor}).

  \item \textbf{\underline{Step 6:} Compute Lorentz force and Couple it with the GOC-LBM for NSE}

  That is, compute the components of the Lorentz force $\bm{F}_{Lorentz}$ from Eq.~(\ref{eq:Lorentzforce-GOC}) and include it as part of a body force $\bm{F}_{ext}$ used in the GOC-LBM for the NSE thereby coupling the flow field and magnetic field GOC-LB solvers.
\end{itemize}

\subsection{Boundary Conditions for the Magnetic Field}
To illustrate the main ideas in implementing the boundary conditions (BCs) for the distribution functions for the GOC-LBM for the MIE, let's consider a prototypical example of the Hartmann flow bounded by solid walls that is subject to an imposed magnetic field $B_0$ in the direction normal to the flow. Hence, for this situation, the magnetic field BCs on the walls may be expressed as
\begin{equation*}
B_{1, wall} = 0, \qquad B_{2, wall} = B_0.
\end{equation*}

Since the MIE is analogous to a form of an advection-diffusion-type equation (ADE) with a source term, albeit in a vector form, we exploit the half-way anti-bounce back scheme to impose the Dirichlet BCs (see e.g.,~\cite{kruger2017lattice}). For the flow field aligned direction $\xi_1$, the distribution functions along the incoming directions $\bar{\alpha}$ can be expressed in terms of the post-collision distribution functions (denoted by using the tilde symbol) along the outgoing directions $\alpha$, where $\bm{e}_{\bar{\alpha} \xi_i} = - \bm{e}_{\alpha \xi_i}$, as
\begin{equation}
  g_{\bar{\alpha} 1} = - \tilde{g}_{\alpha 1}.
\end{equation}
For the imposed magnetic field aligned direction $\xi_2$ with a non-zero imposed field, we subject the incoming and outgoing distribution functions to the following constraint:
\begin{equation*}
  g_{\bar{\alpha} 2} + \tilde{g}_{\alpha 2} =  g_{\bar{\alpha} 2, wall}^{eq(0)} + g_{\alpha 2, wall}^{eq(0)},
\end{equation*}
where the subscript `wall' refers to evaluating the equilibrium distribution functions based on the conditions or properties specified at the wall. Evaluating the RHS of this last equation and rearranging, we get the following anti-bounce back scheme with an augmented correction term for the imposed magnetic field along with the dependence on the metric factors at the wall $h_{1, wall}$ and $h_{2, wall}$ that is defined by the choice made for the orthogonal grid clustering:
\begin{equation}
  g_{\bar{\alpha} 2} = - \tilde{g}_{\alpha 2} + 2 W_\alpha h_{wall}  B_0,
\end{equation}
where $h_{wall} = h_{1, wall} h_{2, wall}$ and $W_\alpha$ is the weighting factor given in Eq.~(\ref{eq:10_a_C}).

\end{document}